\documentclass{tufte-book}

\usepackage{amsmath,amssymb,amsth,mathrsfs,hyperref}
\usepackage{wasysym}

\theoremstyle{plain}
\newtheorem{theorem}{Theorem} %[section]
\newtheorem{lemma}{Lemma}

\newtheorem{corollary}{Corollary}
\theoremstyle{definition}
\newtheorem{definition}{Definition}

\let\cleardoublepage\clearpage

\hypersetup{colorlinks}% uncomment this line if you prefer colored hyperlinks (e.g., for onscreen viewing)

\title{Statistical Inference}
\author[K. M. Zuev]{Konstantin M. Zuev}
\publisher{ACM Lecture Notes}

\IfFileExists{bergamo.sty}{\usepackage[osf]{bergamo}}{}% Bembo
\IfFileExists{chantill.sty}{\usepackage{chantill}}{}% Gill Sans

\usepackage{lipsum}

\usepackage{booktabs}

\usepackage{graphicx}
\setkeys{Gin}{width=\linewidth,totalheight=\textheight,keepaspectratio}
\graphicspath{{graphics/}}

\usepackage{fancyvrb}
\fvset{fontsize=\normalsize}

\usepackage{xspace}

\newcommand{\monthyear}{%
  \ifcase\month\or January\or February\or March\or April\or May\or June\or
  July\or August\or September\or October\or November\or
  December\fi\space\number\year
}

\newcommand{\openepigraph}[2]{%
  \begin{fullwidth}
  \sffamily\large
  \begin{doublespace}
  \noindent\allcaps{#1}\\% epigraph
  \noindent\allcaps{#2}% author
  \end{doublespace}
  \end{fullwidth}
}

\usepackage{units}

\newcommand{\hlred}[1]{\textcolor{Maroon}{#1}}% prints in red
\newcommand{\hangleft}[1]{\makebox[0pt][r]{#1}}
\newcommand{\hairsp}{\hspace{1pt}}% hair space
\newcommand{\ie}{\textit{i.\hairsp{}e.}\xspace}
\newcommand{\eg}{\textit{e.\hairsp{}g.}\xspace}
\providecommand{\XeLaTeX}{X\lower.5ex\hbox{\kern-0.15em\reflectbox{E}}\kern-0.1em\LaTeX}

\newcommand{\tuftebs}{\symbol{'134}}% a backslash in tt type in OT1/T1
\newcommand{\doccmddef}[2][]{%
  \hlred{\texttt{\tuftebs#2}}\label{cmd:#2}%
  \ifthenelse{\isempty{#1}}%
    {% add the command to the index
      \index{#2 command@\protect\hangleft{\texttt{\tuftebs}}\texttt{#2}}% command name
    }%
    {% add the command and package to the index
      \index{#2 command@\protect\hangleft{\texttt{\tuftebs}}\texttt{#2} (\texttt{#1} package)}% command name
      \index{#1 package@\texttt{#1} package}\index{packages!#1@\texttt{#1}}% package name
    }%
}% command name -- adds backslash automatically
\newcommand{\doccmd}[2][]{%
  \texttt{\tuftebs#2}%
  \ifthenelse{\isempty{#1}}%
    {% add the command to the index
      \index{#2 command@\protect\hangleft{\texttt{\tuftebs}}\texttt{#2}}% command name
    }%
    {% add the command and package to the index
      \index{#2 command@\protect\hangleft{\texttt{\tuftebs}}\texttt{#2} (\texttt{#1} package)}% command name
      \index{#1 package@\texttt{#1} package}\index{packages!#1@\texttt{#1}}% package name
    }%
}% command name -- adds backslash automatically
\newcommand*{\approxdist}{\mathrel{\vcenter{\offinterlineskip
			\vskip-.25ex\hbox{\hskip.55ex$\cdot$}\vskip-.25ex\hbox{$\sim$}}}}

\usepackage{makeidx}
\makeindex

\begin{document}

%\bigskip
%\newthought{The pages}
%\paragraph{Paragraph}
%\marginnote{}
% \doccmd{sidenote}[\docopt{number}][\docopt{offset}]\{\docarg{Sidenote text.}\}

% Front matter
%\frontmatter

% r.1 blank page
%\blankpage

% v.2 epigraphs
\iffalse
\newpage\thispagestyle{empty}
\openepigraph{%
The public is more familiar with bad design than good design.
It is, in effect, conditioned to prefer bad design, 
because that is what it lives with. 
The new becomes threatening, the old reassuring.
}{Paul Rand%, {\itshape Design, Form, and Chaos}
}
\vfill
\openepigraph{%
A designer knows that he has achieved perfection 
not when there is nothing left to add, 
but when there is nothing left to take away.
}{Antoine de Saint-Exup\'{e}ry}
\vfill
\openepigraph{%
\ldots the designer of a new system must not only be the implementor and the first 
large-scale user; the designer should also write the first user manual\ldots 
If I had not participated fully in all these activities, 
literally hundreds of improvements would never have been made, 
because I would never have thought of them or perceived 
why they were important.
}{Donald E. Knuth}
\fi

% r.3 full title page
\maketitle

% v.4 copyright page
\iffalse
\newpage
\begin{fullwidth}
~\vfill
\thispagestyle{empty}
\setlength{\parindent}{0pt}
\setlength{\parskip}{\baselineskip}
Copyright \copyright\ \the\year\ \thanklessauthor

\par\smallcaps{Published by \thanklesspublisher}

\par\smallcaps{tufte-latex.googlecode.com}

\par Licensed under the Apache License, Version 2.0 (the ``License''); you may not
use this file except in compliance with the License. You may obtain a copy
of the License at \url{http://www.apache.org/licenses/LICENSE-2.0}. Unless
required by applicable law or agreed to in writing, software distributed
under the License is distributed on an \smallcaps{``AS IS'' BASIS, WITHOUT
WARRANTIES OR CONDITIONS OF ANY KIND}, either express or implied. See the
License for the specific language governing permissions and limitations
under the License.\index{license}

\par\textit{First printing, \monthyear}
\end{fullwidth}
\fi

% r.5 contents
\tableofcontents

\listoffigures

%\listoftables

% r.7 dedication
\iffalse
\cleardoublepage
~\vfill
\begin{doublespace}
\noindent\fontsize{18}{22}\selectfont\itshape
\nohyphenation
Dedicated to those who appreciate \LaTeX{} 
and the work of \mbox{Edward R.~Tufte} 
and \mbox{Donald E.~Knuth}.
\end{doublespace}
\vfill
\vfill
\fi

% r.9 introduction
\cleardoublepage
\chapter{Preface}  

%\vspace{-3mm}
\begin{marginfigure}
%	\vspace{1mm}	
	\centerline{\includegraphics[width=0.9\linewidth]{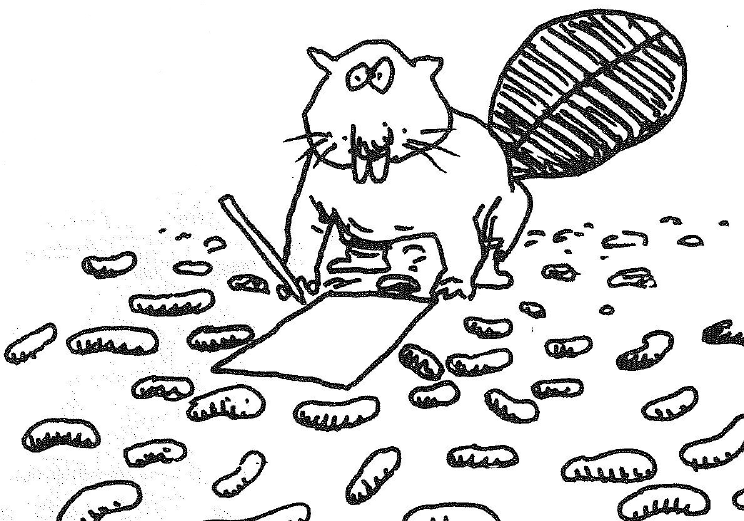}} \caption{Illustration by Larry Gonick, \textit{The Cartoon Guide to Statistics}.}\label{fig:beave1}
\end{marginfigure}
\section{What is Statistics?} 
\newthought{Opinions vary}. In fact, there is a continuous spectrum of attitudes toward statistics ranging from pure theoreticians, proving asymptotic efficiency and searching for most powerful tests, to wild practitioners, blindly reporting $p$-values\footnote{\href{http://fivethirtyeight.com/features/statisticians-found-one-thing-they-can-agree-on-its-time-to-stop-misusing-p-values/}{What is a $p$-value?}} and claiming statistical significance for scientifically insignificant results. Even among most prominent statisticians there is no consensus: some discuss the relative importance of the core goals of statistical inference\footnote{G.~Shmueli (2010) ``\href{https://projecteuclid.org/euclid.ss/1294167961}{To explain or to predict?}'' \textit{Statistical Science}, 25(3): 289-310.}, others comment of the differences between  ``mathematical'' and ``algorithmic'' cultures of statistical modeling\footnote{L.~Breiman (2001) ``\href{http://projecteuclid.org/euclid.ss/1009213726}{Statistical modeling: the two cultures},'' \textit{Statistical Science}, 16(3): 199-231.}, yet others argue that mathematicians should not even teach statistics\footnote{D.S.~Moore (1988) ``\href{http://www.jstor.org/stable/2686686}{Should mathematicians teach statistics?}'' \textit{The College Mathematics Journal}, 19(1): 3-7.}. The absence of a unified view on the subject led to different approaches and philosophies: there is frequentist and Bayesian statistics, parametric and nonparametric, mathematical, computational, applied, etc. To complicate the matter, machine learning, a modern subfield of computer science, is bringing more and more new tools and ideas for data analysis.    

Here we view  statistics as a branch of \textit{mathematical engineering},\footnote{To the best of our knowledge, this formulation is due to Cosma Shalizi.}  that studies ways of extracting reliable information from limited data for learning, prediction, and decision making in the presence of uncertainty.   Statistics is not mathematics \textit{per se} because it is intimately related to real data. Mathematics is abstract, elegant, and can often be useful in applications; statistics is concrete, messy, and always useful.\footnote{The difference between statistics and  mathematics is akin to the difference between a real man and the Vitruvian.} As a  corollary, although present, the proofs are not of paramount importance in these notes. Their main role is to provide intuition and rationale behind the corresponding methods. On the other hand, statistics is not simply a toolbox that contains answers for all data related questions. Almost always, as in solving engineering problems, statistical analysis of new data requires adjustment of existing tools or even developing completely new methods\footnote{For example, recent years witnessed an explosion of network data for which most of the classical statistical methods and models are simply inappropriate.}.  

\section{What are these Notes?}
These ACM\footnote{\text{Applied \& Computational Mathematics}} Lecture Notes are based on the statistical courses I taught at the University of Southern California in 2012 and 2013, and at the California Institute of Technology in 2016. 

\section{What are the Goals?}

The main goals of these notes are: 
%\vspace{-1mm}
\begin{enumerate}
	\item Provide a logical introduction to statistical inference,
	\item Develop statistical thinking and intuitive feel for the subject,
	\item Introduce the most fundamental ideas, concepts, and methods of statistics, explain how and why they work, and when they don't. 
\end{enumerate}
After working on these notes you should be able to read\footnote{And understand.} most\footnote{Admittedly not all.} contemporary papers that use statistical inference and perform basic statistical analysis yourself.

\section{What are the Prerequisites?}
%\vspace{-1mm}
This is an introductory text on statistical inference. As such, no prior knowledge of statistics is assumed. However, to achieve the aforementioned goals, you will need a firm understanding of probability\footnote{Here is the list of concepts you should know: random variable, cumulative distribution function, probability	mass function, probability density function; specific distributions, such as uniform, Bernoulli, binomial, normal, $\chi^2$, $t$; expectation; variance, standard deviation; joint and conditional distributions; conditional expectations and variances; independence; Markov's inequality, Chebyshev's inequality; law of large numbers, central limit theorem. %If most of these sounds unfamiliar, you are not ready to take this course. Please take a probability class and return next time.
	}, which is --- in the context of statistics --- a language for describing variability in the data and uncertainty associated with the phenomenon of interest. %The key part of the course is homework, where you will get experience in using the learned methods and models in applications via simulations in Matlab. So, some familiarity with Matlab and programming is desired, but it is not very critical.  

\section{Why Prerequisites are Important?}
%\vspace{-1mm} 
Because without knowing probability, the best you could hope for is to memorize several existing concepts and methods without understanding why they work. This would increase the risk of an unfortunate event of turning into a ``wild practitioner'' mentioned above.

\section{How to read these Lecture Notes?}
%\vspace{-1mm}
I would suggest to read each lecture note twice. First time: glancing through, ignoring footnotes, examining figures, and trying to get the main idea and understand  a big picture of what is going on. Second time: with a pencil and paper, working through all details, constructing examples, counterexamples, finding errors and typos\footnote{Please look for them. There are many, I promise. Please, inform me of those you find by sending an email to kostia@caltech.edu.}, and blaming me for explaining easy things in a complicated way.

\section{What is Missing?}
A lot by any standards. Bayesian inference, causal inference, decision theory, simulation methods are not covered at all. I hope to expand these notes in the feature. This is simply the first draft.

\section{Acknowledgment}
I wish to express my sincere thanks to Professor Mathieu Desbrun of Caltech for granting me a teaching-free fall term in 2015. This allowed me to bite the bullet and write these notes.

\section{References} 
These notes, which I tried to make as self-contained as possible, are heavily based on the following texts:
\bigskip
 \begin{fullwidth}
\begin{tabular}{l  l}
	\text{[CB]} & G. Casella \& R.L. Berger (2002) \textit{\href{https://books.google.com/books/about/Statistical_Inference.html?id=0x_vAAAAMAAJ}{Statistical Inference}}. \\ 
	\text{[CH]} & L. Chihara \& T. Hesterberg (2011) \textit{\href{https://books.google.com/books/about/Mathematical_Statistics_with_Resampling.html?id=9KRHFDKDV84C}{Mathematical Statistics with Resampling and R}}.  
	 \\  
	 \text{[D]} & A.C. Davison (2003) \textit{\href{https://books.google.com/books/about/Statistical_Models.html?id=LSp4ngEACAAJ}{Statistical Models}}.\\
	 \text{[FPP]} & D. Freedman, R. Pisani, \& R. Purves (2007) \textit{\href{https://books.google.com/books/about/Statistics.html?id=mviJQgAACAAJ}{Statistics}}.\\
	 \text{[La]} & M. Lavine (2013) \textit{\href{http://people.math.umass.edu/~lavine/Book/book.html}{Introduction to Statistical Thought}}.\\
	 \text{[Lo]} & S.L. Lohr (2009) \textit{\href{https://books.google.com/books/about/Sampling_Design_and_Analysis.html?id=aSXKXbyNlMQC}{Sampling: Design and Analysis}}.\\
	 \text{[MPV]} & D.C. Montgomery, E.A. Peck, \& G.G. Vining (2012) \textit{\href{https://books.google.com/books?id=0yR4KUL4VDkC}{Introduction to Linear Regression Analysis}}.\\
	 \text{[NS]} & D. Nolan \& T. Speed (2000) \textit{\href{https://www.stat.berkeley.edu/~statlabs/}{Stat Labs: Mathematical Statistics Through Applications}}.\\
	 \text{[Wa]} & L.A. Wasserman (2005) \textit{\href{http://www.stat.cmu.edu/~larry/all-of-statistics/}{All of Statisitcs: A Concise Course in Statistical Inference}}.\\
	 \text{[We]} & S. Weisberg (2005) \textit{\href{https://books.google.com/books?id=xd0tNdFOOjcC}{Applied Linear Regression}}.
\end{tabular}
\end{fullwidth}
\bigskip\bigskip
\begin{marginfigure}
	\vspace{12mm}
	\includegraphics[width=\linewidth]{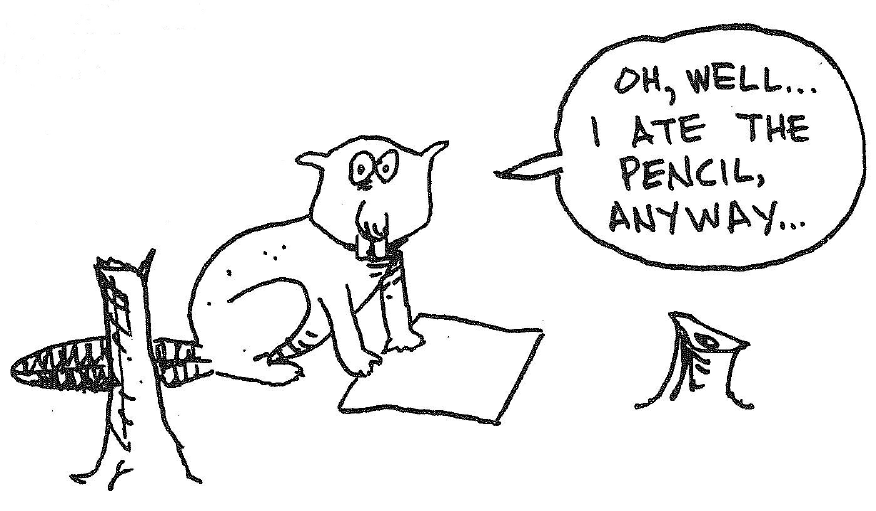} \caption{Illustration by Larry Gonick, \textit{The Cartoon Guide to Statistics}.}\label{fig:beaver2}
\end{marginfigure}
\bigskip
\bigskip
\begin{flushright}
	Konstantin M. Zuev\\
	Pasadena, California,\\
	March 15, 2016
\end{flushright}

\chapter{Summarizing Data}\label{ch:Summarizing Data}
\newthought{Data} is\footnote{For ``data is'' vs "data are" see \href{http://grammarist.com/usage/data/}{grammarist.com}} at the heart of statistics.  The most basic element of data is a single observation, $x$, a number. Usually real data comes in the form of a (very long)  list of numbers. Even if the original data is more complex --- a text, curve, or image --- we will assume that we can always convert it to a set of $n$ numerical observations $x_1,\ldots,x_n$, called a \textit{sample}.

To get a better feel for the data in hand, it is often useful (especially if the sample size $n$ is large) to summarize it numerically or graphically. This can bring some insights about the data. In this lecture, we discuss several kinds of summary statistics\footnote{In statistics, any (possibly vector valued) quantity $s=s(x_1,\ldots,x_n)$ that can be calculated from data is called a \textit{statistic}.}. 

\section{The Histogram}
If you Google images for ``statistics,'' you will see something like this: 
\begin{figure}
	\includegraphics[width=\linewidth]{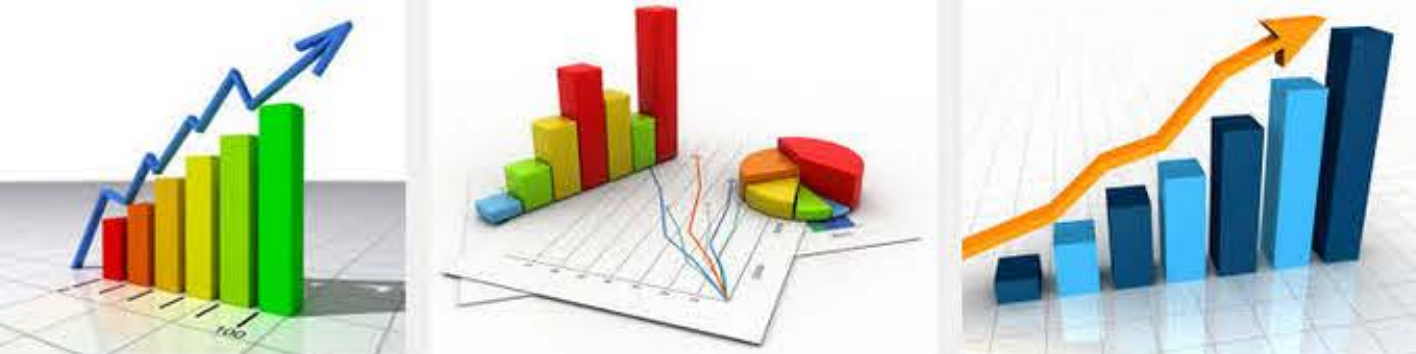}\caption{Googled histograms.}\label{sec:histogram}
\end{figure}

These graphs, called histograms,  are perhaps the best-known statistical plots. To construct a histogram from data $x_1,\ldots,x_n$:
\begin{enumerate}
	\item Divide the horizontal axis into disjoint \textit{bins}, the intervals $I_1,\ldots,I_K$.
	\item Denote the number of observation in $I_k$ by $n_k$, so that $\sum_{k=1}^Kn_k=n$.
	\item For each bin, plot a column over it so that the area of the column is the proportion $\frac{n_k}{n}$ of
	the data in the bin\footnote{This makes the total area of the histogram equal to 1. Such histograms are called normalized. Sometimes not normalized histograms are used, where the area of a column over $I_k$ is simply the number of observations $n_k$. In this case, the total area of the histogram is $n$.}. The height $h_k$ of the column over $I_k$ is therefore $h_k=\frac{n_k/n}{|I_k|}$.
\end{enumerate}

\textit{Question:} How to chose bins? 

There is no unique recipe for choosing bins: a good choice depends on the data. Let us try to understand what ``good'' means. The main purpose of the histogram is to represent the \textit{shape} of the sample: symmetry (bell-shaped? uniform?), skewness (right-skewed? left-skewed?), modality (unimodal? multimodal?). Let us assume for simplicity that all bins have equal width\footnote{Sometimes it might be better to vary the bin width, with narrower bins in the center of the data, and wider ones at the tails.} $w$: 
\begin{equation}
\begin{split}
I_1&=[x_{(1)},x_{(1)}+w),\\
I_2&=[x_{(1)}+w,x_{(1)}+2w),\\
&\ldots\\
I_K&=[x_{(n)}-w,x_{(n)}],
\end{split}
\end{equation}
where $x_{(1)}$ and $x_{(n)}$ are respectively the minimum and maximum of the sample, $x_{(1)}=\min\{x_1,\ldots,x_n\}$ and $x_{(n)}=\max\{x_1,\ldots,x_n\}$. In this case, the total number of bins is
\begin{equation}
K=\frac{x_{(n)}-x_{(1)}}{w}.
\end{equation}

The number of bins $K$ in a histogram can drastically affect its appearance. If $K$ is too large ($w$ is too small), then the histogram looks too rough. On the other hand, if $K$ is too small, then the histogram is too smooth. This effect is illustrated below with the normally distributed sample.
\begin{figure}
	\includegraphics[width=\linewidth]{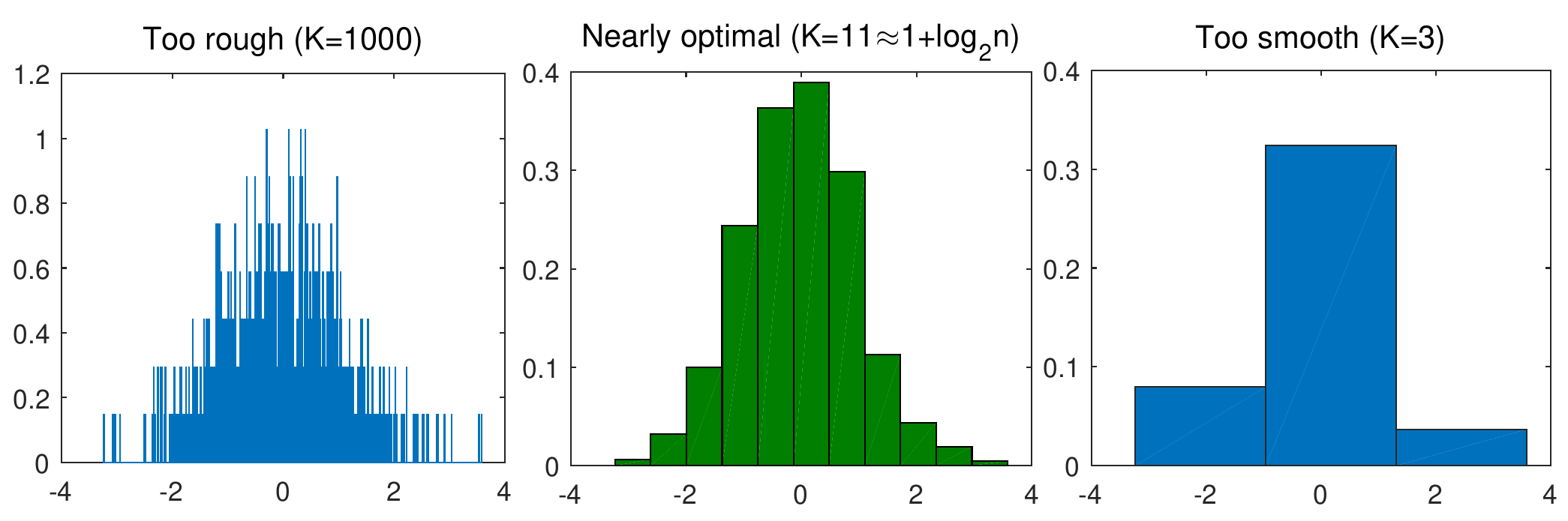} \caption{Histograms of the normally distributed data $x_1,\ldots,x_n$, $n=1000$.}\label{fig:histograms}
\end{figure}

\vspace{-5mm}
Thus, either too few or too many bins can obscure structure in the data. There are several simple heuristic rules for the approximate number if bins. For example, if the sample $x_1,\ldots,x_n$ appears to be approximately normally distributed\footnote{That is we expect the histogram is bell-shaped, \ie looks similar to this: }, then we can use Sturges' formula:
\begin{equation}\label{eq:sturges}
K\approx 1+\log_2 n.
\end{equation} 
\begin{marginfigure}
	\includegraphics[width=\linewidth]{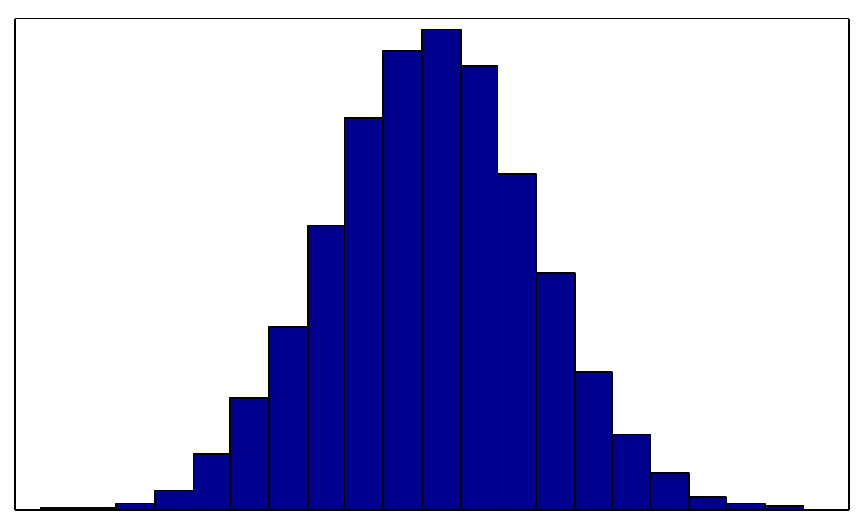} \caption{Normal bell-shaped histogram.}\label{fig:norm_hist}
\end{marginfigure}
In general, exploring the data using histograms with different numbers of bins and different \textit{cut points} between bins is useful in understanding the shape of the data, and heuristics like (\ref{eq:sturges}) can be used as a starting point of exploration. But this exploration should not be confused with manipulation of the data for presentation purposes!

\section{Numerical Summaries}  
Numerical summaries provide quantitative information about the data. Two basic features of a sample are its location and spread.
\subsection{Measures of location}
A measure of location is a statistic that represents the center of the sample\footnote{If $x_1,\ldots,x_n$ are different measurements of the same quantity (say, measurements of temperature obtained from different thermometers), a measure of location is often used as an estimate of the quantity in the hope that it is more accurate than any single measurement.}. One such measure is the \textit{sample mean}, which is simply the average of the sample:
\begin{equation}
\bar{x}=\frac{1}{n}\sum_{i=1}^nx_i.
\end{equation}
The main drawback of the mean is that it is sensitive to \textit{outliers}. An outlier is an observation $x^*$ that is distant from other observations in the sample $x_1,\ldots,x_n$\footnote{Outliers are often easy to spot in histograms.}. An outlier may be due to variability in the data\footnote{For example, Bill Gates will be an outlier in the study of people's wealth.} or it may indicate measurement error. For example, by changing only the value of $x_1$ we can make the mean $\bar{x}$ arbitrary small or large, and, in this case, it will be a poor measure of the sample center. 

An alternative measure of location, which is \textit{robust}\footnote{Insensitive.} to outliers, is the \textit{median}. The median $\tilde{x}$ is the point that divides the sample in half.  To calculate the median, we need to order the data. The \textit{order statistics} of $x_1,\ldots,x_n$ are their values
put in increasing order, which we denote 
\begin{equation}
x_{(1)}\leq x_{(2)}\leq\ldots\leq x_{(n)}.
\end{equation}
The median is then defined as follows\footnote{Convince yourself that $\tilde{x}$ defined this way indeed splits the data in half.}:
\begin{equation}
\tilde{x}=\begin{cases}
x_{(\frac{n+1}{2})}, & \mbox{if } n \mbox{ is odd},\\
\frac{1}{2}\left(x_{(\frac{n}{2})}+x_{(\frac{n}{2}+1)}\right), & \mbox{if } n \mbox{ is even}.
\end{cases}
\end{equation}

The main drawback of the median is the opposite of the drawback of the mean: it is too insensitive to the change in the sample values. Suppose for simplicity, that the sample size is odd, $n=2k-1$, then the median is the $k^{th}$ order statistic, $\tilde{x}=x_{(k)}$. Making the values of the right half of the sample $x_{(k+1)},\ldots,x_{(n)}$ arbitrary large does not affect the median. Similar effect holds for the left half od the sample. 

\textit{Question:} Can we find a compromise between $\bar{x}$ and $\tilde{x}$? 

A compromise between the mean and the median is a \textit{trimmed mean}. 
The mean is the average of all observations. We can think of the median as the average of the middle one or two observations as if the rest observations were discarded. The $\alpha$-trimmed mean $\bar{x}_\alpha$ is defined as follows: discard $100\alpha\%$ of the observations on each side of the ordered sample $x_{(1)}\leq x_{(2)}\leq\ldots\leq x_{(n)}$ and take the mean of the remaining middle $100(1-2\alpha)\%$ of the observations. Mathematically, 
\begin{equation}
\overline{x}_{\alpha}=\frac{x_{([n\alpha]+1)}+\ldots+x_{(n-[n\alpha])}}{n-2[n\alpha]},
\end{equation} 
where $[s]$ denotes the greatest integer less than or equal to $s$. Then $0$-trimmed mean is the standard sample mean, and the median can be thought of as the $0.5$-trimmed mean. If the trimmed mean is a slowly varying function of $\alpha$, then the sample has a well defined center.

\subsection{Measures of spread}
A measure of location is often accompanied by a measure of spread that gives an idea as to how far an individual value $x_i$ may vary from the center of the data (``scatteredness'' of the sample). The simplest measure of spread is the \textit{range}, which is the difference between the largest and smallest values,
\begin{equation}
r=x_{(n)}-x_{(1)}.
\end{equation}
The range ignores most of the data and is very sensitive to outliers. 

One of the most popular measures of spread in statistics is the \textit{sample standard deviation}\footnote{Sometimes it is defined as $s_x=\sqrt{\frac{1}{n-1}\sum_{i=1}^n(x_i-\bar{x})^2}$ for reasons we  discuss later. But if $n$ is large the difference between the two versions is negligible.}:
\begin{equation}
s_x=\sqrt{\frac{1}{n}\sum_{i=1}^n(x_i-\bar{x})^2}.
\end{equation} 
Although it is also sensitive to outliers but it is more robust than the range. 

The measure of spread that typically accompanies the median is the \textit{interquartile range} (IQR), which is the difference between the upper and lower quartiles of the sample,
\begin{equation}
IQR=Q_3-Q_1,
\end{equation}
where $Q_1$ is is first (lower) quartile that splits lowest $25\%$ of the sample and $Q_3$ is the third (upper)  quartile that splits highest $25\%$ of the sample\footnote{What is the second quartile $Q_2$?}.
\subsection{Five-Number Summary}
The five-number summary provides simultaneously a measure of location and spread. The five numbers are: 
the minimum $x_{(1)}$,
the first quartile $Q_1$,
the median $\tilde{x}=Q_2$,
the third quartile $Q_3$, and
the maximum $x_{(n)}$.

\subsection{Boxplots}
A boxplot is a graph that visualizes the five-number summary, gives a good idea of the shape of the data, and shows potential outliers. To create a boxplot from data $x_1,\ldots,x_n$:
\begin{enumerate}
	\item Draw a box with the bottom end placed at the first quartile $Q_1$ and the top end placed at the third quartile $Q_3$. Thus about a half of the data lie in the box, and its hight is IQR.
	\item Draw a horizontal line through the box at the median $\tilde{x}=Q_2$.
	\begin{marginfigure}
	%	\vspace{-60mm}
		\includegraphics[width=\linewidth]{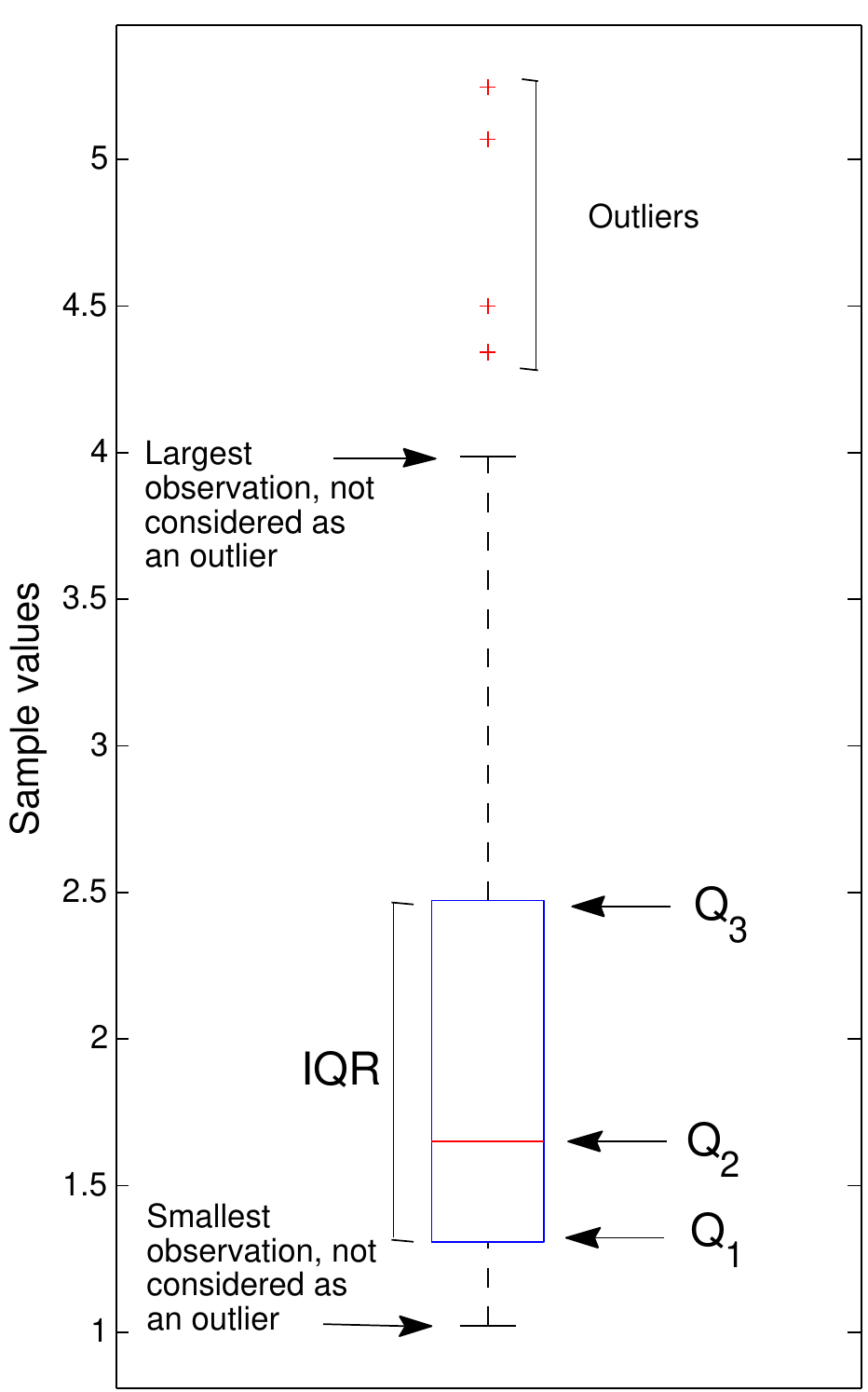} \caption{Boxplot. We can clearly see that the sample is skewed to the right.}\label{fig:boxplot}
	\end{marginfigure}
	\item Place a cap at the largest observation that is less than or equal to $Q_3+1.5IQR$. Similarly, place a cap at the smallest observation that is greater than or equal to $Q_1-1.5IQR$.
	\item Extend whiskers (dashed lines in Fig.~\ref{fig:boxplot}) from the edges of the box to the caps. Observations that lie between the caps are not considered as outliers.
	\item The observations that fall outside the caps are considered as outliers\footnote{Outliers are often defined as points which fall more than $k>0$ times the interquartile range above $Q_3$ or below $Q_1$ with $k=1.5$ as a usual choice.}. Plot them individually with $\cdot$, $+$, $*$, or your favorite symbol.  
\end{enumerate}
A box plot is a semi-graphical and  semi-numerical summary of data. It contains more information than a five-number summary, but it is less informative than a histogram: from a boxplot it is not possible to ascertain whether there are gaps in the data or multiple modes.

Boxplots are especially useful when comparing related samples. For examples, household incomes in different states, lengths of the flight delays of different airlines, heights of males and females, etc.  

\section{Empirical CDF}
The basic problem of statistical inference is: given the data $x_1,\ldots,x_n$, what can we say about the process that generated the data? Probabilistically, we model the sample $x_1,\ldots,x_n$ as \textit{realizations} of a \textit{random variable} $X$ with (unknown and to be inferred) \textit{cumulative distribution function} (CDF) $F_X$, which is the theoretical model for the data. 

\begin{marginfigure}
	\includegraphics[width=\linewidth]{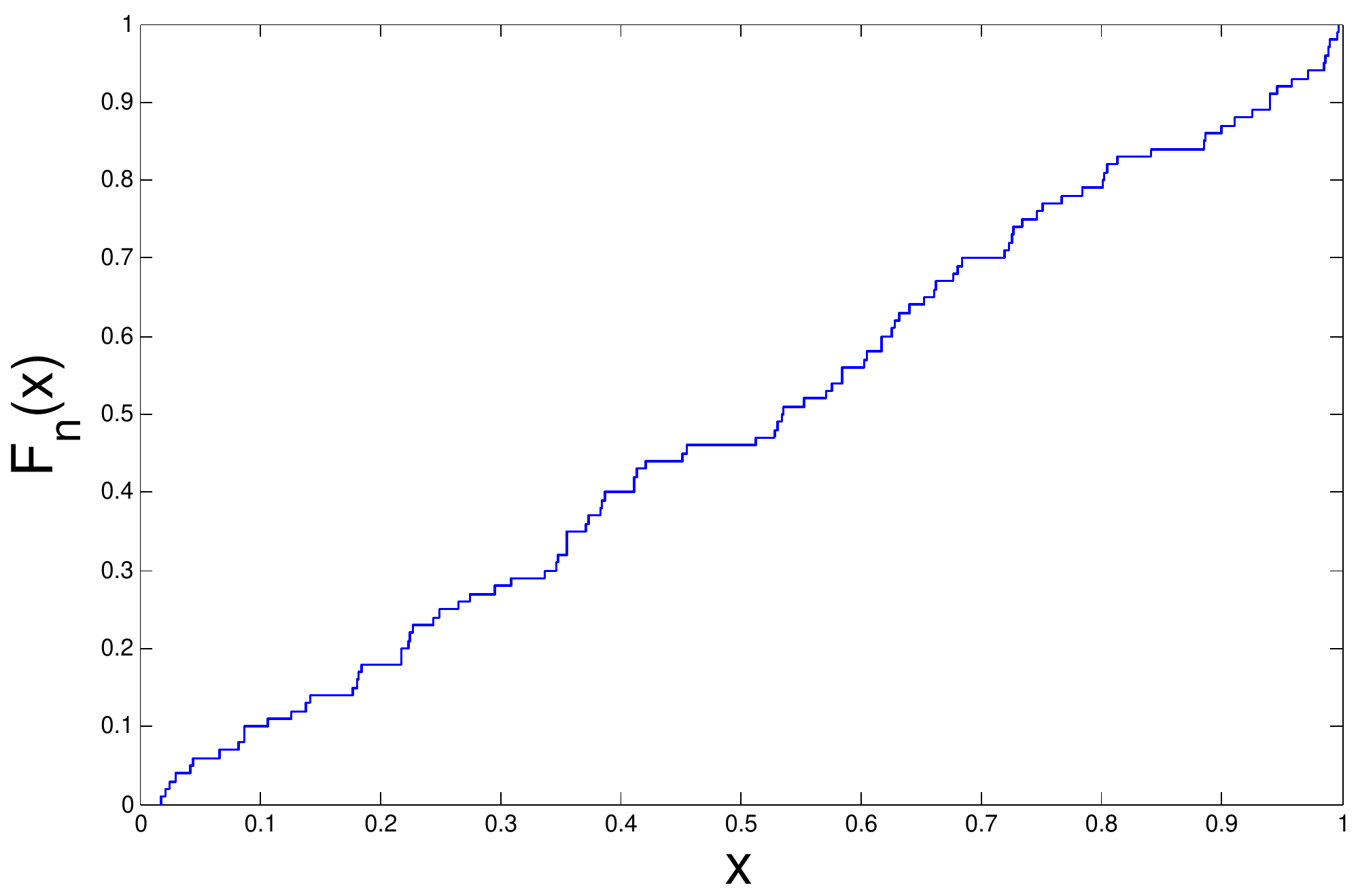} \caption{Empirical CDF for sample of size $n=100$ drawn from the uniform distribution on $[0,1]$.}\label{fig:uniform_ecdf}
\end{marginfigure}

The empirical CDF (eCDF) is the ``data analogue'' of the CDF of a random variable. Recall that $F_X(x)=\mathbb{P}(X\leq x)$. The eCDF of $x_1,\ldots,x_n$ is defined as follows:
\begin{equation}
\begin{split}
F_n(x)&=\frac{\mbox{number of observations less than or equal to }x}{n}\\
&=\frac{1}{n}\sum_{i=1}^n H(x-x_i),
\end{split}
\end{equation}
where $H(x)$ is the Heaviside function\footnote{This is one of those ``standard'' functions that you are likely to meet in any math/science/engineering course.} that puts mass one at zero:
\begin{equation}
H(x)=\begin{cases}
0, & x<0,\\
1, & x\geq1.
\end{cases}
\end{equation}
The eCDF is thus a step function that jumps by $\frac{1}{n}$ at each of the $x_i$\footnote{If the value of $x_i$ appears $k$ times in the sample, then the eCDF jumps by $\frac{k}{n}$ at this value.}.
Figure~\ref{fig:uniform_ecdf} shows how it looks for the sample drawn from the uniform distribution on $[0,1]$. Notice how closely the eCDF resembles the true uniform CDF, $F_n(x)\approx F_X(x)$\footnote{This is not a coincidence, and we will make this statement more precise in the subsequent lectures.}. 

The eCDF is a graphical display that conveniently summarizes the sample. It is more detailed than the histogram, but perhaps a bit more difficult to read and conveys less information about the shape of the data. The eCDF plays an important role in estimating statistical functionals, and we will come back to it in the future. 

\section{Q-Q plots}
Remarkably, many histograms follow the normal curve\footnote{A quick reminder on the normal distribution is given in Appendix.}. Visually, this means that a histogram has a single peak (mode), its mean and median are approximately equal, and its symmetric about the center (see Fig.~\ref{fig:norm_hist}). Examples include histograms of heights of people, errors in measurements, and marks on a test. 

Suppose we wish to check if the data $x_1,\ldots,x_n$ come the normal distribution, that is if the standard normal CDF $\Phi(z)$ is a good theoretical model for the data. We could start from plotting the histogram and see if it is bell-shaped. But the problem with this approaches is that usually histograms do not allow to see clearly what happens in the \textit{tails} of the data distribution, \ie around $x_{(1)}$ and $x_{(n)}$: do the tails decay faster (``short'' tails) or slower (``long'' tails) than the normal tails?\footnote{Why do the tails matter? Think of using the inferred $F_X$ for prediction.} Therefore, we need a more accurate procedure. A quantile-quantile (Q-Q) plot is a graphical method that allows to assess the normality of the sample, and, more generally, to compare the sample $x_1,\ldots,x_n$ with any theoretical model~$F_X$. 

\begin{marginfigure}
	\includegraphics[width=\linewidth]{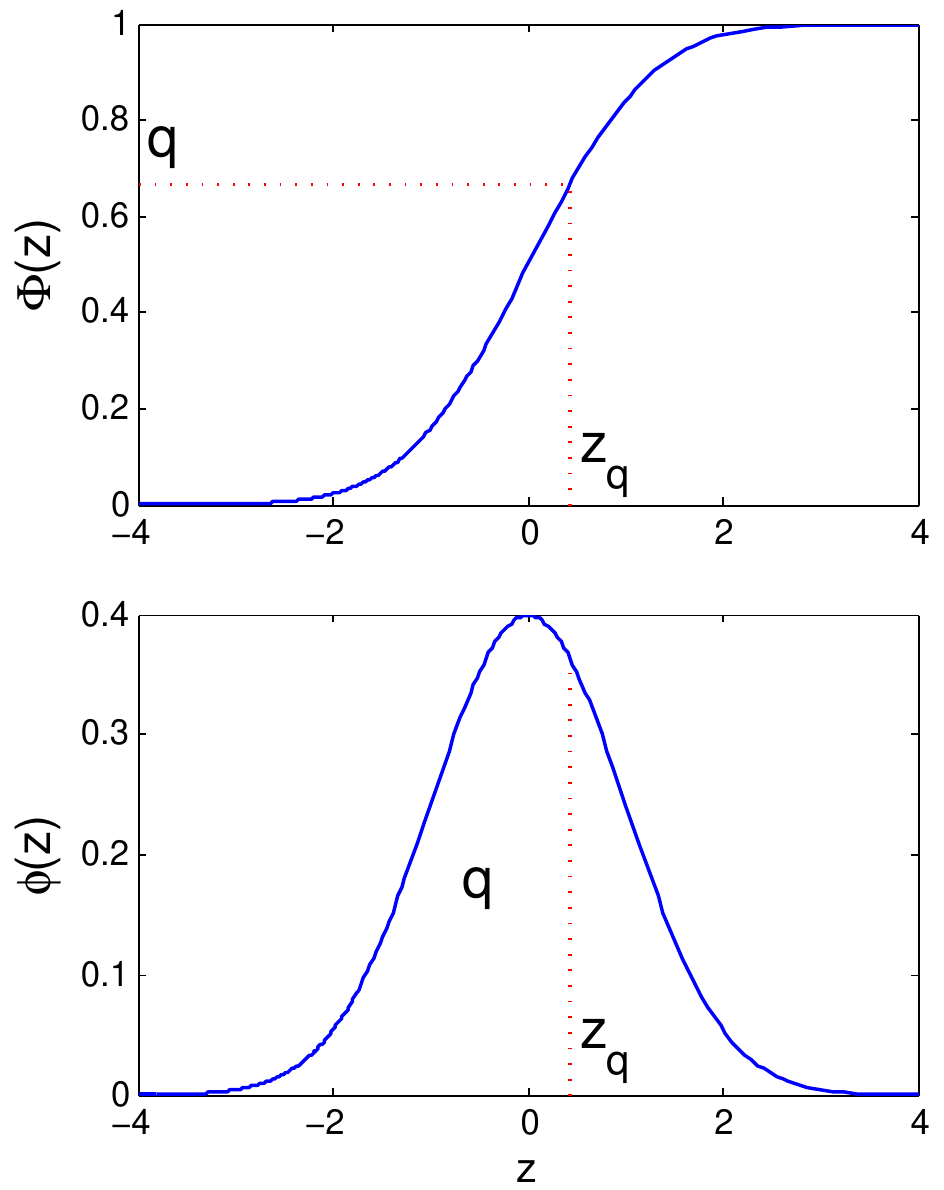} \caption{The standard normal quantile $z_q$ in term of the CDF $\Phi(z)$ (top) and PDF $\phi(z)$ (bottom).}\label{fig:norm_q}
\end{marginfigure}

The $q^{th}$ \textit{quantile}\footnote{Sometime the term ``percentile'' is used in the literature.} of the standard normal distribution is a number $z_q$ such that 
\begin{equation}
\Phi(z_q)=q, \hspace{5mm} \mbox{where  } 0<q<1.
\end{equation} 
In other words, $z_q$ is a number such that the probability mass supported by the interval $(-\infty,z_q)$ is exactly $q$. Figure~\ref{fig:norm_q} clarifies this definition. For example, the median, lower, and upper quartile are, respectively, the 0.5, 0.25, and 0.75 quantiles.  

If the sample $x_1,\ldots,x_n$ is approximately normally distributed, then we expect that $F_n(x)\approx\Phi(x)$, and, therefore, the the corresponding quantiles should also match. Notice that 
\begin{equation}
F_n\left(x_{(1)}\right)=\frac{1}{n}, \ldots, F_n(x_{(k)})=\frac{k}{n},\ldots, F_n(x_{(n)})=\frac{n}{n}=1.
\end{equation}
Therefore, the $k^{th}$ order statistics $x_{(k)}$ should be a good approximation for the $(\frac{k}{n})^{th}$ standard normal quantile $z_{\frac{k}{n}}$. There is a little technical problem: if $k=n$, then $z_1=+\infty$. There are may ways to get around this. One consists of taking $z_{\frac{k}{n+1}}$ instead of $z_{\frac{k}{n}}$, to make sure that $q<1$\footnote{Some software packages use $z_{\frac{k-0.375}{n+0.25}}$.}.

\begin{figure}
	\includegraphics[width=\linewidth]{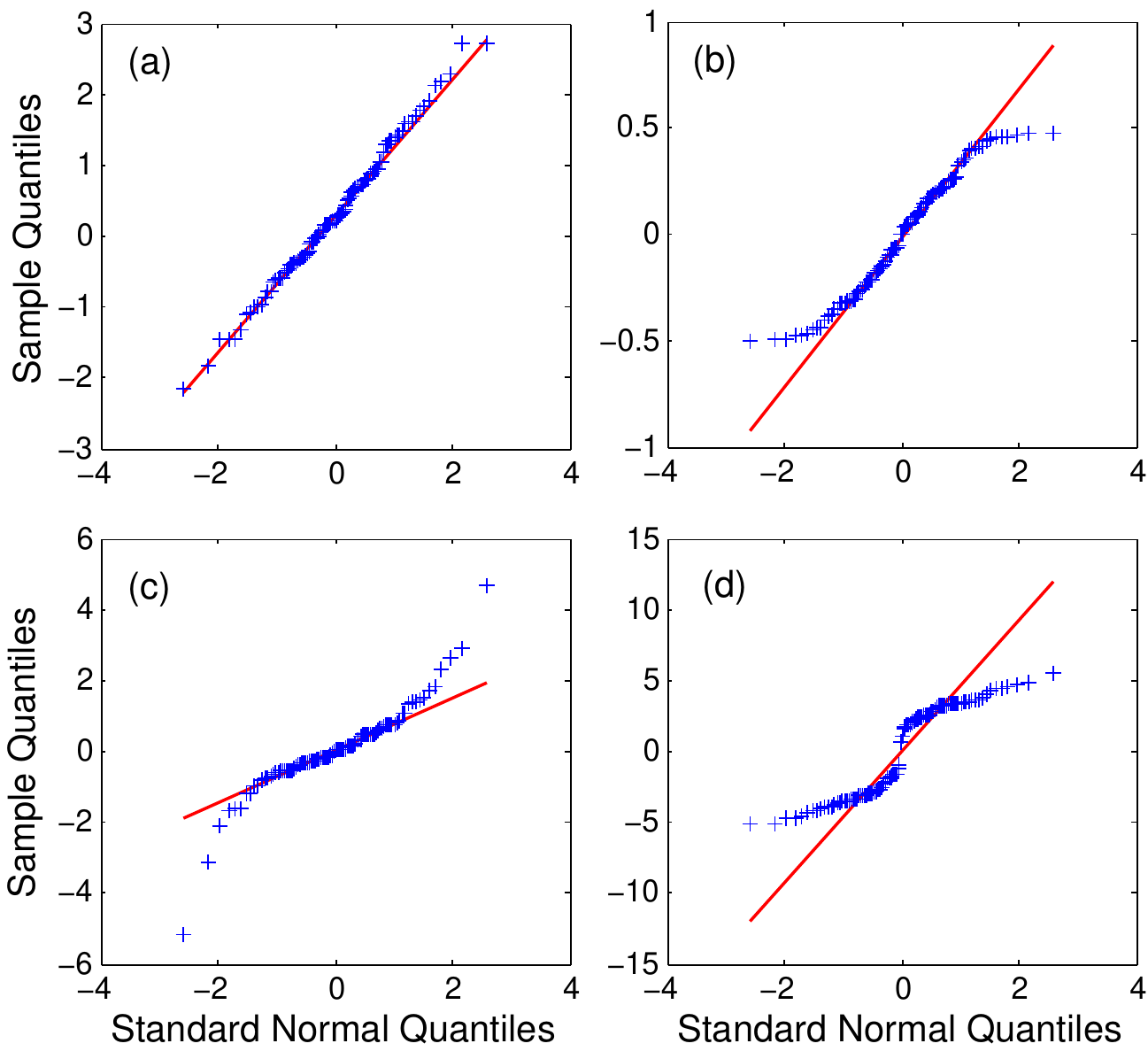} \caption{Normal-quantile plots for the data $x_1,\ldots,x_n$ sampled from (a) the standard normal distribution, (b) uniform distribution on $[-1/2,1/2]$ (short tails), (c) the Laplace distribution $f(x)\propto e^{-|x|}$ (long tails), and (d) a bimodal distribution (a mixture of two well-separated Gaussians). Sample size in all examples is $n=100$.}\label{fig:qqplots}
\end{figure}

The \textit{normal-quantile plot} graphs the pairs
\begin{equation}
\left(z_{\frac{k}{n+1}},x_{(k)}\right),\hspace{5mm} \mbox{for } k=1,\ldots,n.
\end{equation} 
If the plotted points fall roughly on the line $y=x$, then it indicates that the data have an approximate standard normal distribution. As an illustration, Fig~\ref{fig:qqplots}(a) shows the normal-quantile plot for the data $x_1,\ldots,x_n$ sampled from the standard normal distribution. 
\vspace{-5mm}

\textit{Question:} What if the points fall on the line $y=ax+b$?

Departures from normality are indicated by systematic departures from a straight line. Examples of different departures are illustrated in Fig.~\ref{fig:qqplots}(b), (c), and (d).

Q-Q plots can be made for any probability distribution, not necessarily normal, which is considered as a theoretical model for the process that generated the data. For example, we can construct a uniform-quantile plot, exponential-quantile plot, etc. To compare two different samples $x_1,\ldots,x_n$ and $y_1,\ldots,y_n$, we can also create a Q-Q plot by pairing their respective sample quantiles $(x_{(k)},y_{(k)})$\footnote{What would you do if the samples have different sizes $x_1,\ldots,x_n$ and $y_1,\ldots,y_m$, where $m\neq n$?}. Again, a departure from a straight line indicates a difference in the shapes of the two samples. 

\section{Further Reading}

\begin{enumerate}
	\item~[FPP, Part II] gives a very intuitive description of histograms, the mean, and the standard deviation. It is a lengthy but easy and enjoyable read.
	\begin{marginfigure}
		\includegraphics[width=.7\linewidth]{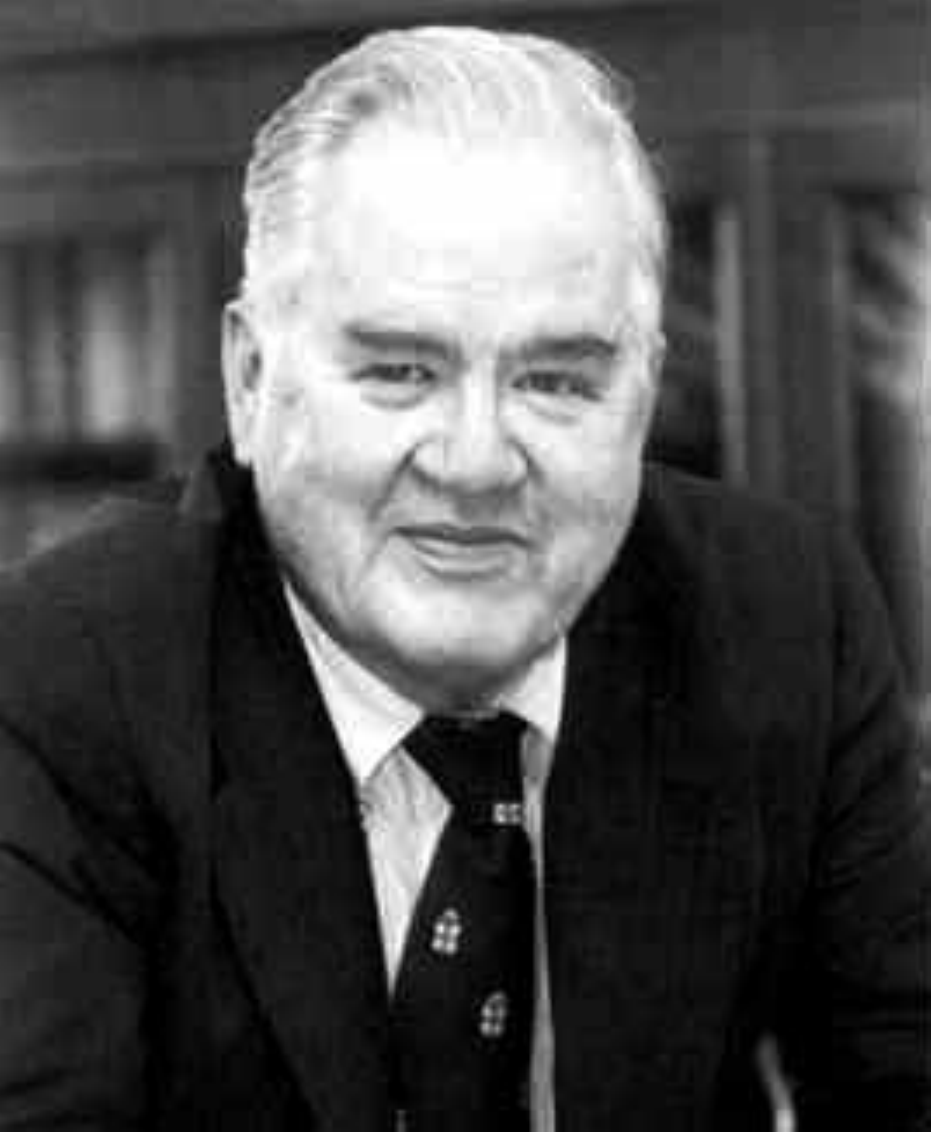} \caption{John Turky. Photo source: \href{https://en.wikipedia.org/wiki/John_Tukey}{wikipedia.org}.}\label{fig:Tukey}
	\end{marginfigure}
	\item~Summarizing data is a part of \textit{Exploratory Data Analysis} (EDA), an approach for data analysis introduced and promoted by John~Tukey. His seminal work Tukey (1977) ``\href{http://www.amazon.com/Exploratory-Data-Analysis-John-Tukey/dp/0201076160}{Exploratory Data Analysis}'' remains one of the best texts on EDA.	
\end{enumerate}

\section{What is Next?}
We discussed how to summarize data, but how to get the data in first place? Perhaps the most popular way is to conduct a survey. In the next three lectures we will discuss arguably the most classical subjects of statistical inference: survey sampling.

\section{Appendix: Normal Distribution}
The \textit{standard normal curve}, known as the bell curve or the Gaussian curve\footnote{V.I.~Arnold's principle states that if a notion bears a personal name, then this name is not the name of the discoverer. The Arnold Principle is applicable to itself as well as to the Gaussian distribution: the standard normal curve was discovered around 1720 by Abraham de Moivre.}, is defined as 
\begin{equation}
\phi(z)=\frac{1}{\sqrt{2\pi}}\exp\left(-\frac{z^2}{2}\right).
\end{equation}
The normal curve is unimodal, symmetric around zero, and follows the so-called ``68-95-99.7 rule'':  approximately 68\% of the area under the curve is within 1 unit of its center\footnote{Mathematically, $\int_{-1}^1\varphi(z)dz\approx0.68.$},  95\% is within 2 units, and 99.7\% is within 3 units. 
\begin{figure}\label{fig.rule}
	\includegraphics[width=\linewidth]{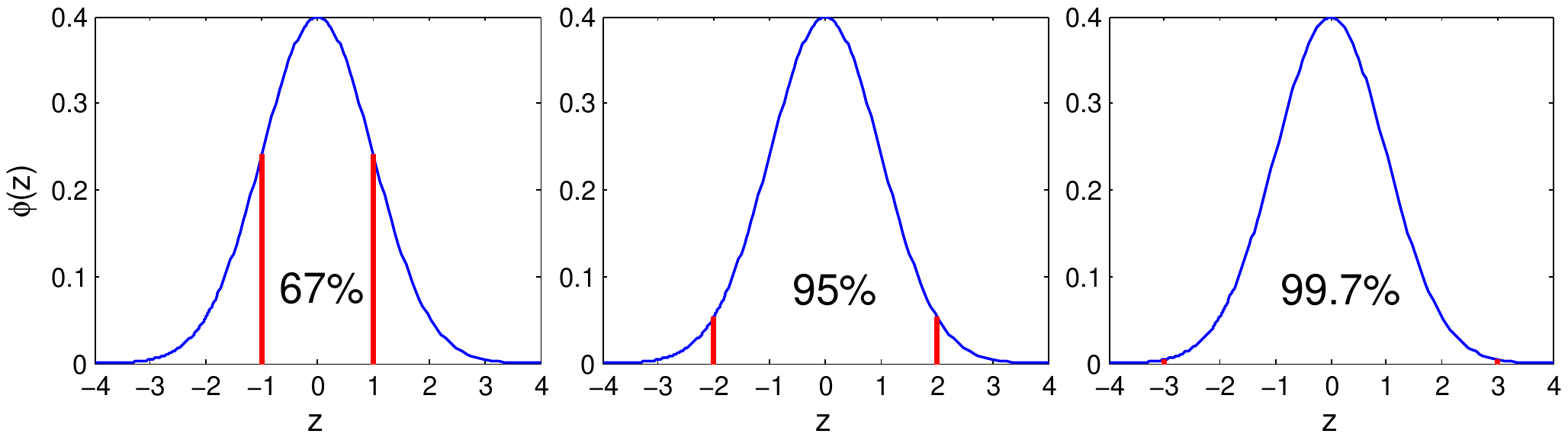}\caption{The 68-95-99.7 rule for the standard normal curve.}
\end{figure}

The coefficient $\frac{1}{\sqrt{2\pi}}$ does not have any sacral meaning, it is simply a normalizing constant that guarantees that the full area under the curve is exactly one, 
\begin{equation}
\int_{-\infty}^{+\infty}\phi(z)dz=1.
\end{equation}
This allows to interpret $\phi(z)$ as the probability density function (PDF) of a random variable\footnote{Recall that, any non-negative function $p(x)$ that integrates to one can be viewed as a PDF. The associated random variable $X$ is fully defined by $p(x)$: $$\mathbb{P}(X\in A)=\int_Ap(x)dx.$$}. This random variable, often denoted by $Z$,  is called  \textit{standard normal}. Thanks to the \textit{Central Limit Theorem} (CLT), $\phi(z)$ is the single most important distribution in probability and statistics\footnote{Intuitively (and very roughly), CLT states that the properly shifted and scaled sum $\sum_{i=1}^NX_i$ of more or less any (!) random variables $X_1,\ldots,X_n$ is approximately standard normal. Many phenomena in Nature can be accurately modeled by sums of of random variables. This partially explains why many histograms (which are can be viewed as approximations for the underlying PDFs) follow the normal curve.}. 

Traditionally, $\Phi(z)$ denotes the cumulative distribution function (CDF), whose value at $z$ is the area under the standard normal curve to the left of $z$,
\begin{equation}
\Phi(z)=\int_{-\infty}^z\phi(z)dz.
\end{equation}
See the top panel of Fig.~\ref{fig:norm_q}. 

The standard normal random variable $Z$ has zero mean and unit variance:
\begin{equation}
\begin{split}
\mu&=\mathbb{E}[Z]=\int_{-\infty}^{+\infty}z\phi(z)dz=0,\\
\sigma^2&=\mathbb{V}[Z]=\mathbb{E}[(Z-\mu)^2]=\int_{-\infty}^{+\infty}z^2\phi(z)dz=1.
\end{split}
\end{equation}
The random variable $X$ is called \textit{normal} with mean $\mu$ and variance $\sigma^2$, denoted $X\sim\mathcal{N}(\mu,\sigma^2)$, if its PDF is\footnote{The 68-95-99.7 rule holds for any normal variable, we need just to replace intervals $[-1,1]$, $[-2,2]$ and $[-3,3]$ with $[-\sigma,\sigma]$, $[-2\sigma,2\sigma]$, and $[-3\sigma,3\sigma]$.} 
\begin{equation}
p(x)=\frac{1}{\sqrt{2\pi}\sigma}\exp\left(-\frac{(x-\mu)^2}{2\sigma^2}\right).
\end{equation}   
Here are some useful facts:
\begin{enumerate}
	\item If $X\sim\mathcal{N}(\mu,\sigma^2)$, then $Z=\frac{X-\mu}{\sigma}\sim\mathcal{N}(0,1)$.
	\item If $X\sim\mathcal{N}(\mu,\sigma^2)$, then $\mathbb{P}(a<X<b)=\Phi\left(\frac{b-\mu}{\sigma}\right)-\Phi\left(\frac{a-\mu}{\sigma}\right)$.
	\item If $X_1\sim\mathcal{N}(\mu_1,\sigma_1^2)$, $X_2\sim\mathcal{N}(\mu_2,\sigma_2^2)$, and $X_1$ and $X_2$ are \textit{independent}, then $X=X_1+X_2\sim\mathcal{N}(\mu_1+\mu_2,\sigma_1^2+\sigma_2^2)$.
\end{enumerate}

\chapter{ Simple Random Sampling}\label{ch:SRS}

	\newthought{Sample surveys} are used to obtain information
	about a large population. The purpose of survey sampling is to reduce the cost and the amount of work that it would take to survey the entire population. Familiar examples of survey sampling include taking a spoonful of soup to determine its taste (a cook does not need to eat the entire pot) and making a blood test to measure the red blood cell count (a medical technician does not need to drain you of blood). In this lecture we learn how to estimate the population average and how to assess the accuracy of the estimation using \textit{simple random sampling}, the most basic rule for selecting a subset of a population. 
		\begin{marginfigure}
			\vspace{-30mm}
			\centerline{\includegraphics[width=.5\linewidth]{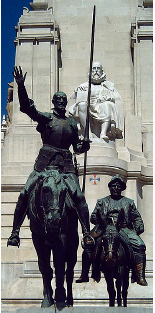}} \caption{\textit{By a small sample we may
					judge of the whole piece}, Miguel de Cervantes
				``Don Quixote.'' Photo source: \href{https://en.wikipedia.org/wiki/Don_Quixote}{wikipedia.org}.}\label{fig:Don}
		\end{marginfigure}
		
	\section{A Bit of History}
		\begin{marginfigure}
			\centerline{\includegraphics[width=0.5\linewidth]{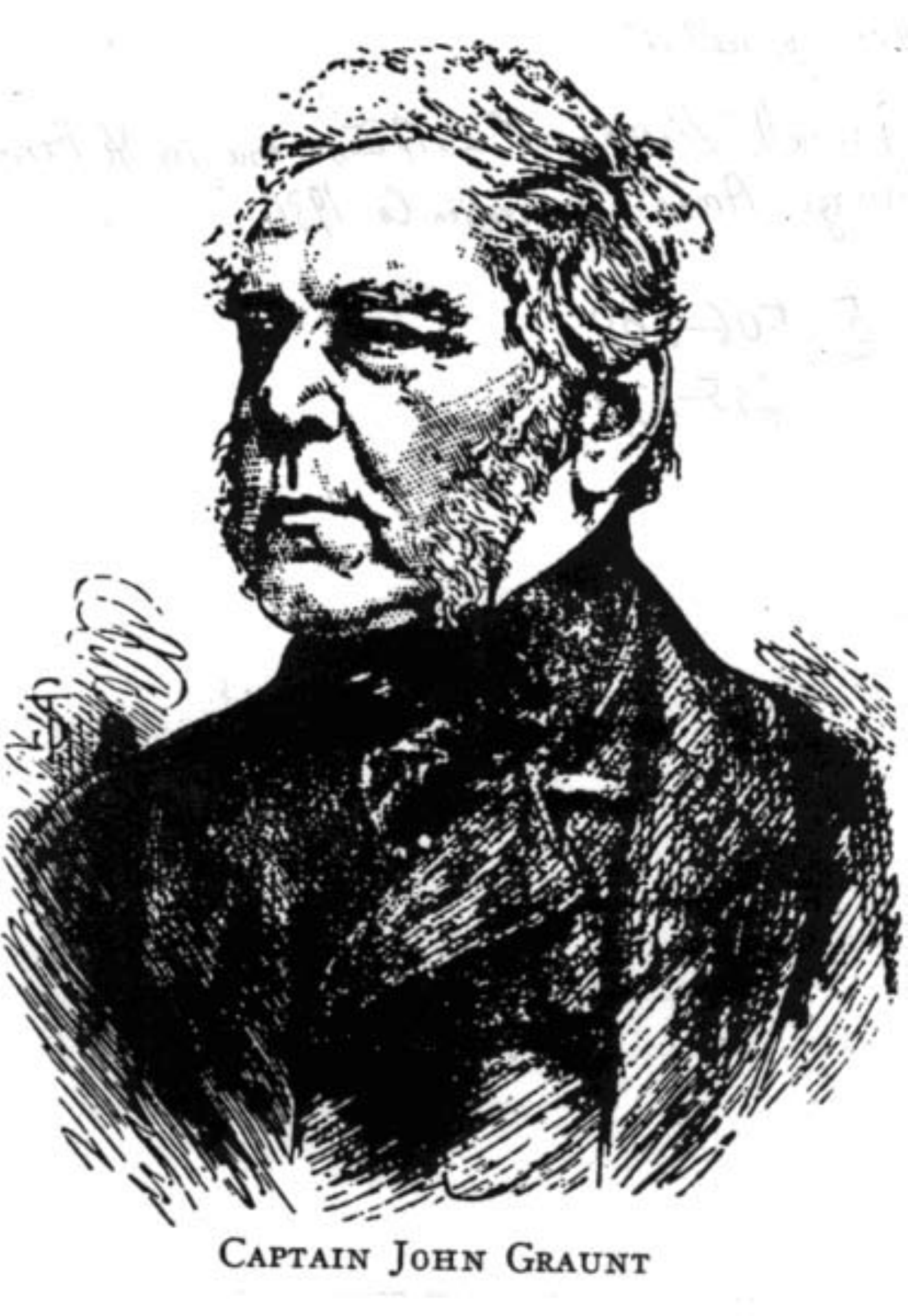}} \caption{Captain John Graunt. Photo source: \href{http://www.goodreads.com/author/show/1004775.John_Graunt}{goodreads.com}}\label{fig:Graunt}
		\end{marginfigure}
		\begin{marginfigure}
			\centerline{\includegraphics[width=0.5\linewidth]{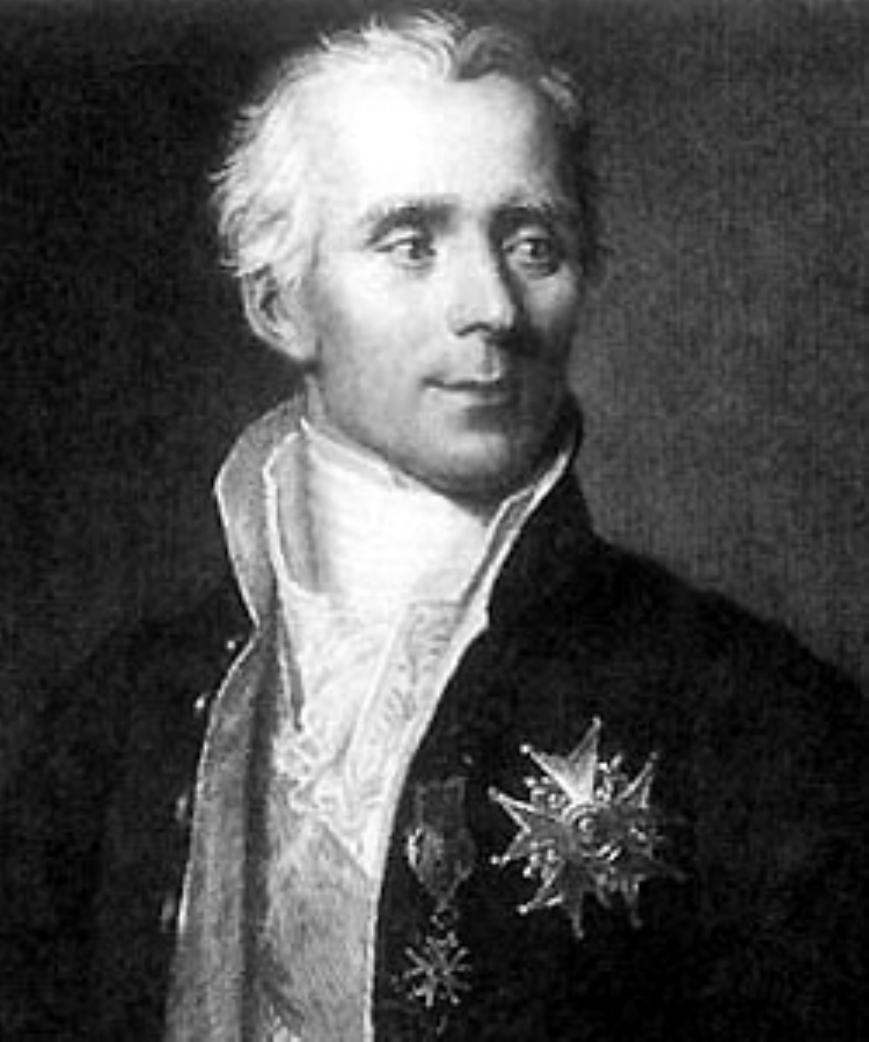}} \caption{Pierre-Simon Laplace. Photo source: \href{https://en.wikipedia.org/wiki/Pierre-Simon_Laplace}{wikipedia.org}. }\label{fig::Laplace}
		\end{marginfigure}
		The first known attempt to make  statements about a population using only information about part of it was made by the English merchant John Graunt. In his famous tract (Graunt, 1662) he describes a method to estimate the population of London based on partial information. John Graunt has frequently been merited as the founder of demography.
		
		The second time a survey-like method was applied was more than a
		century later. Pierre-Simon Laplace realized that it was
		important to have some indication of the accuracy of the estimate of
		the French population (Laplace, 1812).
		
\section{Terminology}
Let us begin by introducing some key terminology. 
\begin{itemize}
	\item {\it Target population:} The group that we want to know more about. Often called ``population'' for brevity\footnote{Defining the target population may be nontrivail. For example, in a political poll, should the target population be all adults eligible to vote, all registered voters, or all persons who voted in the last election?}. 
	\item {\it Population unit:} A member of the target population. In studying human populations,	observation units are often individuals.
	\item {\it Population size:} The total number of units in the population\footnote{For very large populations, the exact size is often not known.}. Usually denoted by $N$. 
	\item {\it Unit characteristic:} A specific piece of information about each member of the population\footnote{For example, age, weight, income, etc.}.  For unit $i$, we denote the numerical value of the characteristic by $x_i$, $i=1,\ldots,N$.
	\item {\it Population parameter:} A summary of the characteristic for all units in the population. One could be interested in various parameters, but here are the four examples that are used most often:
	\begin{enumerate}
		\item Population mean (our focus in this lecture):  
		\begin{equation}\label{eq:mu}
		\mu=\frac{1}{N}\sum_{i=1}^Nx_i. 
		\end{equation}
		\item Population total:
		\begin{equation}
		\tau=\sum_{i=1}^Nx_i=N\mu.
		\end{equation}
		\item Population variance (our focus in the next lecture):
		\begin{equation}
		\sigma^2=\frac{1}{N}\sum_{i=1}^N(x_i-\mu)^2.
		\end{equation}
		\item Population standard deviation
		\begin{equation}
		\sigma=\sqrt{\frac{1}{N}\sum_{i=1}^N(x_i-\mu)^2}.
		\end{equation}
	\end{enumerate}
\end{itemize}

In an ``ideal survey,'' we take the entire target population, measure the value of the characteristic of interest for all units, and compute the corresponding parameter. This ideal (as almost all ideals) is rarely met in practice: either population is too large, or measuring $x_i$ is too expensive, or both. In practice, we select a \textit{subset} of the target population and \textit{estimate} the population parameter using this subset. 

\begin{itemize}
	\item {\it Sample:} A subset of the target population.
	\item {\it Sample unit:} A member of the population selected for the sample. 
	\item {\it Sample size:} The total number of units in the sample. Usually denoted by $n$. Sample size is often much less than the population size,  $n\ll N$. 
\end{itemize}

Let $\mathcal{P}=\{1,\ldots,N\}$ be the target population and $\mathcal{S}=\{s_1,\ldots,s_n\}$ be a sample from $\mathcal{P}$\footnote{$s_i\in\{1,\ldots,N\}$ and $s_i\neq s_j$.}. When it is not ambiguous, we will identify $\mathcal{P}$ and $\mathcal{S}$ with the corresponding values of the characteristic of interest, that is 
\begin{equation}
\mathcal{P}=\{x_1,\ldots, x_N\}\hspace{3mm}\mbox{and}\hspace{3mm}
\mathcal{S}=\{x_{s_1},\ldots, x_{s_n}\}.
\end{equation}   
To avoid cluttered notation, we denote $x_{s_i}$ simply by $X_i$, and thus, 
\begin{equation}
\mathcal{S}=\{X_1,\ldots,X_n\}\subset\{x_1,\ldots,x_N\}=\mathcal{P}.
\end{equation}
\begin{itemize}
	\item {\it Sample statistic:} A numerical summary of the characteristic of the sampled units\footnote{Essentially any function of $X_1,\ldots,X_n$.}. The statistic estimates the population parameter. For example, a reasonable sample statistic for the population mean $\mu$ in (\ref{eq:mu}) is the \textit{sample mean}: 
	\begin{equation}\label{eq:sample_mean}
	\overline{X}_n=\frac{1}{n}\sum_{i=1}^nX_i.
	\end{equation}
	\item {\it Selection Rule:} The method for choosing a sample  from the target population. 
\end{itemize}

Many selection rules used in practice are probabilistic, meaning that $X_1,\ldots,X_n$ are selected  at \textit{random} according to some probability method.  Probabilistic selection rules are important because they allow to quantify the difference between the population parameters and their estimates obtained from the randomly chosen samples. There is a number of different probability methods for selecting a sample. Here we consider the simplest: \textit{simple random sampling}\footnote{More advanced methods include stratified random sampling, cluster sampling, and systematic sampling.}. 

\section{Simple Random Sampling}
In simple random sampling (SRS), \textit{every} subset of $n$ units in the population has the \text{same} chance of being the  sample\footnote{This chance is $1/\binom{N}{n}$.}. Intuitively, we first  mix up the population and then grab $n$ units. Algorithmically, to draw a simple random sample from $\mathcal{P}$, we 
\begin{enumerate}
	\item Select $s_1$ from $\{1,\ldots,N\}$ uniformly at random.
	\item Select $s_2$ from $\{1,\ldots,N\}\setminus\{s_1\}$ uniformly at random.
	\item Select $s_3$ from $\{1,\ldots,N\}\setminus\{s_1, s_2\}$ uniformly at random. 
	\item Proceed like this till $n$ units $s_1,\ldots,s_n$ are sampled.
\end{enumerate}
In short, we draw $n$ units one at a time \textit{without replacement}\footnote{SRS with replacement is discussed in S.L.~Lohr \textit{\href{https://books.google.com/books/about/Sampling_Design_and_Analysis.html?id=aSXKXbyNlMQC}{Sampling: Design and Analysis}}.}. 

\textit{Questions:} What is the probability that unit $\#1$ is the first to be selected for the sample\footnote{\ie what is $\mathbb{P}(s_1=1)$, or, equivalently, what is $\mathbb{P}(X_1=x_1)$?}? What is the probability that unit $\#1$ is the second to be selected for the sample? What is the probability that unit $\#1$ is selected for the sample? How about unit $\#k$?

So, let $X_1,\ldots,X_n$ be the SRS sample drawn from the population $\mathcal{P}$, and let us consider the sample mean $\overline{X}_n$ in (\ref{eq:sample_mean}) as an estimate of the population mean $\mu$ in (\ref{eq:mu}). 

\textit{Our goal:} to investigate how accurately $\overline{X}_n$ approximates $\mu$. 

Before we proceed, let me reiterate a very important point: $x_i$, and therefore $\mu$, are \textit{deterministic}; $X_i$, and therefore $\overline{X}_n$, are \textit{random}. 

Since $\overline{X}_n=\frac{1}{n}\sum X_i$, it is natural to start our investigation from the properties of a single sample element $X_i$. Its distribution is fully described by the following Lemma.
\begin{lemma}
	Let $\xi_1,\ldots,\xi_m$ be the distinct values assumed by the population units\footnote{For example, if $x_1=1, x_2=1, x_3=2, x_4=3,$ and $x_5=3$, then there are $m=3$ distinct values:  $\xi_1=1, \xi_2=2, \xi_3=3.$}.  Denote the number of population units that have the value $\xi_i$ by $n_i$. Then	$X_i$ is a discrete random variable with probability mass function
	\begin{equation}
	\mathbb{P}(X_i=\xi_j)=\frac{n_j}{N},\hspace{3mm} j=1,\ldots,m,
	\end{equation}
	and its expectation and variance are
	\begin{equation}
	\mathbb{E}[X_i]=\mu \hspace{5mm}\mbox{and} \hspace{5mm} \mathbb{V}[X_i]=\sigma^2.
	\end{equation}
\end{lemma}
As an immediate corollary, we obtain the following result:
\begin{theorem}
	With simple random sampling,
	\begin{equation}
	\mathbb{E}[\overline{X}_n]=\mu.
	\end{equation}
\end{theorem}
Intuitively, this result tells us that ``on average'' $\overline{X}_n=\mu$\footnote{This is good news and justifies the characteristic ``reasonable estimate'' of $\mu$ that we gave to $\overline{X}_n$ above.}. The property of an estimator being equal to the estimated quantity on average is so important that it deserves a special name and a definition.

\begin{definition} Let $\theta$ be a population parameter and $\hat{\theta}=\hat{\theta}(X_1,\ldots,X_n)$ be a sample statistic that estimates $\theta$. We say that $\hat{\theta}$ is \textit{unbiased} if 
	\begin{equation}
	\mathbb{E}[\hat{\theta}]={\theta}.
	\end{equation}
\end{definition}	
Thus, $\overline{X}_n$ is an unbiased estimate of $\mu$. The next step is to investigate \textit{how variable}  $\overline{X}_n$ is. As a measure of the dispersion of $\overline{X}_n$ about
$\mu$, we will use the standard deviation of
$\overline{X}_n$\footnote{Standard deviations of estimators are often called \textit{standard errors} (se). Hence the notation in Eq. (\ref{eq:SE}).}
\begin{equation}\label{eq:SE}
\mathrm{se}[\overline{X}_n]=\sqrt{\mathbb{V}[\overline{X}_n]}.
\end{equation}
Let us find the variance\footnote{If sampling were done \textit{with replacement} then $X_i$ would be \textit{independent}, and we would
	have: $\mathbb{V}[\overline{X}_n]=\frac{1}{n^2}\mathbb{V}\left[\sum_{i=1}^nX_i\right]=\frac{1}{n^2}\sum_{i=1}^n\mathbb{V}[X_i]=	\frac{1}{n^2}\sum_{i=1}^n\sigma^2=\frac{\sigma^2}{n}$. In SRS, however, sampling is done without replacement and this introduces dependence between $X_i$.}:
\begin{equation}
\mathbb{V}[\overline{X}_n]=\mathbb{V}\left[\frac{1}{n}\sum_{i=1}^nX_i\right] =\frac{1}{n^2}\mathbb{V}\left[\sum_{i=1}^nX_i\right]=\frac{1}{n^2}\sum_{i=1}^n\sum_{j=1}^n\mathrm{Cov}(X_i,X_j).
\end{equation}
To continue, we need to compute the correlation. 
\begin{lemma}\label{lemma:2}
	If $i\neq j$, then the covariance between $X_i$ and $X_j$ is
	\begin{equation}
	\mathrm{Cov}(X_i,X_j)=-\frac{\sigma^2}{N-1}.
	\end{equation}
\end{lemma}
And, therefore, we have:
\begin{theorem}
	The variance of $\overline{X}_n$ is given by
	\begin{equation}\label{eq:var}
	\mathbb{V}[\overline{X}_n]=\frac{\sigma^2}{n}\left(1-\frac{n-1}{N-1}\right).
	\end{equation}
\end{theorem}
A few important observations are in order:
\begin{enumerate}
	\item The factor $\left(1-\frac{n-1}{N-1}\right)$ is called \textit{finite population correction}. It is approximately $\left(1-\frac{n}{N}\right)$. The ratio $\frac{n}{N}$ is called the \textit{sampling fraction}.
	\item Finite population correction is always less than one. Therefore, $\mathbb{V}[\overline{X}_n]<\frac{\sigma^2}{n}$. This means that SRS is more efficient than sampling with replacement. 
	\item If the sampling fraction is small, that is if $n\ll N$, then 
	\begin{equation}
	\mathbb{V}[\overline{X}_n]\approx \frac{\sigma^2}{n} \hspace{5mm}\mbox{and}\hspace{5mm}
	\mathrm{se}[\overline{X}_n]\approx \frac{\sigma}{\sqrt{n}}.
	\end{equation}	
	\item To double the accuracy of approximation $\overline{X}_n\approx\mu$\footnote{\ie to reduce $\mathrm{se}[\overline{X}_n]$ by half.}, the sample size $n$ must be quadrupled. 
	\item If $\sigma$ is small\footnote{\ie the population values are not very dispersed.}, then a small sample will be fairly
	accurate. But if $\sigma$ is large, then a larger
	sample will be required to obtain the same accuracy.
\end{enumerate}

\section{Further Reading}
\begin{enumerate}
	\item The history of survey sampling, in particular, how sampling became an accepted scientific method, is described in a nice discussion paper by J. Bethlehem (2009) ``\href{http://www.cbs.nl/NR/rdonlyres/BD480FBC-24CF-42FA-9A0D-BBECD4F53090/0/200915x10pub.pdf}{The rise of survey sampling}.''
\end{enumerate}

\section{What is Next?}
The result (\ref{eq:var}) and the above observations are nice, but we have a serious problem: \textit{we don't know $\sigma$!} In the next lecture, we will learn how to estimate the population variance using SRS.

\chapter{Population Variance and the Bootstrap Method}
\label{ch:PopVarandBoot}

\newthought{Estimating} population variance $\sigma$ is important because of at least two reasons: 

1) it is important population parameter by itself and 

2) it appears in the formula for the standard error of the sample mean $\overline{X}_n$\footnote{Recall that $\overline{X}_n$ is an unbiased estimate of the population mean $\mu$, $\mathbb{E}[\overline{X}_n]=\mu$.}: 
\begin{equation}\label{eq:se}
\mathrm{se}[\overline{X}_n]=\frac{\sigma}{\sqrt{n}}\sqrt{\left(1-\frac{n-1}{N-1}\right)}.
\end{equation}

If we want to compute $\mathrm{se}[\overline{X}_n]$ or to determine the required sample size $n$ to achieve a prescribed value of of error, we must know $\sigma$. In this lecture we learn two things: 

1) how to estimate $\sigma$ and 

2) how to estimate $\mathrm{se}[\overline{X}_n]$ ... without estimating $\sigma$!

\section{Estimation of the Population Variance}
Recall that the population variance is 
\begin{equation}
\sigma^2=\frac{1}{N}\sum_{i=1}^N(x_i-\mu)^2.
\end{equation}
It seems natural to use the following estimate:
\begin{equation}
\hat{\sigma}_n^2=\frac{1}{n}\sum_{i=1}^n(X_i-\overline{X}_n)^2.
\end{equation}
However, this estimate is \textit{biased}.
\begin{theorem}
	The expected value of $\hat{\sigma}_n^2$  is given by
	\begin{equation}
	\mathbb{E}[\hat{\sigma}_n^2]=\sigma^2\frac{Nn-N}{Nn-n}.
	\end{equation}\label{th:1}
\end{theorem}
Since $\frac{Nn-N}{Nn-n}<1$, we have that $\mathbb{E}[\hat{\sigma}_n^2]<\sigma^2$, and thus,  $\hat{\sigma}_n^2$ tends to \textit{underestimate} $\sigma^2$. Theorem~\ref{th:1} helps to construct an unbiased estimate for the population variance:
\begin{corollary}
	An unbiased estimate for the population variance $\sigma^2$ is 
	\begin{equation}\label{eq:s^2}
	s^2=\hat{\sigma}_n^2\frac{Nn-n}{Nn-N}=\left(1-\frac{1}{N}\right)\frac{1}{n-1}\sum_{i=1}^n(X_i-\overline{X}_n)^2.
	\end{equation}	
\end{corollary}
Note that if both population size $N$ and the sample size $n$ are large, then $s^2\approx\hat{\sigma}_n^2$.
Combining (\ref{eq:se}) with (\ref{eq:s^2}) gives the estimate of the standard error:
\begin{equation}\label{eq:analutical}
\mathrm{se}[\overline{X}_n]\approx\widehat{\mathrm{se}}[\overline{X}_n]=\frac{s}{\sqrt{n}}\sqrt{\left(1-\frac{n-1}{N-1}\right)}.
\end{equation} 
Thus, in simple random sampling, we can estimate ($\overline{X}_n$) not only the unknown population parameter ($\mu$), but also obtain the
likely size of the error of the estimate ($\widehat{\mathrm{se}}[\overline{X}_n]$). In other words,
we can obtain the estimate of a parameter as well as the estimate  of the error of that estimate.
\section{The Bootstrap Method for Estimating $\mathrm{se}[\overline{X}_n]$}
Let us take a step back and look at Eq. (\ref{eq:se}).

\textit{Question:} Is there a way to estimate $\mathrm{se}[\overline{X}_n]$ without estimating $\sigma$?\footnote{--- Why should we care? We already know how to estimate $\sigma$! 
	
	--- Because there are many cases when we can construct an unbiased estimate $\hat{\theta}$ of a population parameter $\theta$, but we don't know the analytical formula (like (\ref{eq:se})) for its standard error $\mathrm{se}[\hat{\theta}]$. For example, $s^2$ is an unbiased estimate of $\sigma^2$, but what is the standard error $\mathrm{se}[s^2]$? In such cases, we need an alternative way of estimating $\mathrm{se}$.}

Let us quickly refresh our minds. The sample mean $\overline{X}_n$ is a discrete random variable which is obtained by averaging sample units $\mathcal{S}=\{X_1,\ldots,X_n\}$ which are obtained from the target population $\mathcal{P}=\{x_1,\ldots,x_N\}$ by simple random sampling. 

Now let us forget for the moment about SRS and consider the following problem. Suppose $Y$ is a discrete random variable with the probability mass function $\mathbb{P}$. And suppose we can generate independent realizations of $Y$, that is we can independently sample from $\mathbb{P}$:
\begin{equation}
Y_1,\ldots,Y_B\sim\mathbb{P}.
\end{equation}
How can we estimate the variance of $Y$? Well, we can do this using the \textit{law of large numbers}\footnote{The law of large numbers is one of the main achievements in probability. Intuitively, it says that if $B$ is large, then the sample average $\overline{Y}_B=\frac{1}{B}\sum_{i=1}^BY_i$ is a good approximation for $\mathbb{E}[Y]$. More formally, the weak (strong) LLN states that $\overline{Y}_B$ converges to $\mathbb{E}[Y]$ in probability (almost surely), as $B\rightarrow\infty$.} (LLN). Namely, 
\begin{equation}\label{eq:LLN}
\mathbb{V}[Y]=\mathbb{E}[(Y-\mathbb{E}[Y])^2]\approx \frac{1}{B}\sum_{i=1}^B\left(Y_i-\frac{1}{B}\sum_{j=1}^BY_j\right)^2.
\end{equation}

Now let us apply this to $Y=\overline{X}_n$. To do this, we would have to generate $B$ simple random samples from $\mathcal{P}$:
\begin{equation}\label{eq:B}
\begin{split}
\mathcal{S}^{(1)}&=\{X_1^{(1)},\ldots,X_n^{(1)}\}\subset\mathcal{P},\\
&\ldots\\
\mathcal{S}^{(B)}&=\{X_1^{(B)},\ldots,X_n^{(B)}\}\subset\mathcal{P},
\end{split}
\end{equation}
compute the corresponding sample means:
\begin{equation}\label{eq:means}
\overline{X}_n^{(1)}=\frac{1}{n}\sum_{i=1}^nX_i^{(1)},\hspace{5mm}\ldots\hspace{5mm}\overline{X}_n^{(B)}=\frac{1}{n}\sum_{i=1}^nX_i^{(B)},
\end{equation}
and, finally, estimate $\mathrm{se}[\overline{X}_n]$ by analogy with (\ref{eq:LLN}):
\begin{equation}\label{eq:bootstrap}
\mathrm{se}[\overline{X}_n]\approx\widehat{\mathrm{se}}[\overline{X}_n]=\sqrt{\frac{1}{B}\sum_{i=1}^B\left(\overline{X}_n^{(i)}-\frac{1}{B}\sum_{j=1}^B\overline{X}_n^{(j)}\right)^2}.
\end{equation}

Looks good expect for one thing: the \textit{total} sample size in (\ref{eq:B}) is $nB$, which is much larger than our original sample size $n$, $nB\gg n$. Therefore, this straightforward method for estimating $\mathrm{se}[\overline{X}_n]$ is not really acceptable since we assume that sampling $n$ population units is the maximum we can afford\footnote{After all, if we could afford sampling $nB$ units, we would use $\overline{X}_{nB}$ as an estimate of $\mu$ instead of   $\overline{X}_n$!}. Here is where the \textit{bootstrap principle} comes into play. 

The bootstrap is a very general simulation-based method, introduced by Bradley Efron, for measuring uncertainty of an estimate. It requires no analytical calculations and often used in applications. In Lecture~\ref{ch:Bootstrap}, we will discuss the bootstrap in detail in different contexts. Here is our first encounter with the bootstrap: in the context of survey sampling.
\begin{marginfigure}
	\centerline{\includegraphics[width=\linewidth]{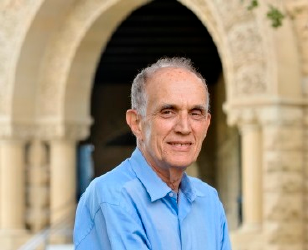}} \caption{Bradley Efron, the father of the bootstrap. Photo source: \href{http://statweb.stanford.edu/~ckirby/brad/}{statweb.stanford.edu}.}\label{fig:Efron}
\end{marginfigure}

The intuition behind the bootstrap is the following. In SRS, our main  underlying assumption is that our sample $\mathcal{S}$ represents the target population $\mathcal{P}$ well.  Based on $\mathcal{S}$, we can then create a new population of size $N$ by simply creating $N/n$ copies of each $X_i$\footnote{For simplicity, we assume here that $N/n$ is an integer. But if it is not, we can always round off $N/n$ to the nearest integer. }. We call it the \textit{bootstrap population}:
\begin{equation}
\mathcal{P}_{boot}=\{\underbrace{X_1,\ldots,X_1}_{N/n},\ldots,\underbrace{X_n,\ldots,X_n}_{N/n}\}.
\end{equation}

\textit{Bootstrap principle:} Use $\mathcal{P}_{boot}$ instead of $\mathcal{P}$ in (\ref{eq:B}).

In other words, instead of sampling the target population, bootstrap\footnote{This method derives its name from the expression ``to pull youself up by your own bootstraps,'' P.~Diaconis and B.~Efron, ``\href{http://www.vanderbilt.edu/psychological_sciences/graduate/programs/quantitative-methods/quantitative-content/diaconis_efron_1983.pdf}{Computer-intensive methods in statistics},'' \textit{Scientific American}, 248(5):116-129, 1983.} says that we can ``reuse'' our original sample $\mathcal{S}=\{X_1,\ldots,X_n\}$. That is, for every $b=1,\ldots,B,$  $\mathcal{S}^{(b)}=\{X_1^{(b)},\ldots,X_n^{(b)}\}$ is a simple random sample from $\mathcal{P}_{boot}$. We call $\mathcal{S}^{(b)}$ a \textit{bootstrap sample}. The rest is exactly as before. The \textit{bootstrap estimate} of the standard error $\mathrm{se}[\overline{X}_n]$ is given by (\ref{eq:bootstrap}).

\section{Example: Gaussian Population}
\begin{figure}
	\centerline{\includegraphics[width=\linewidth]{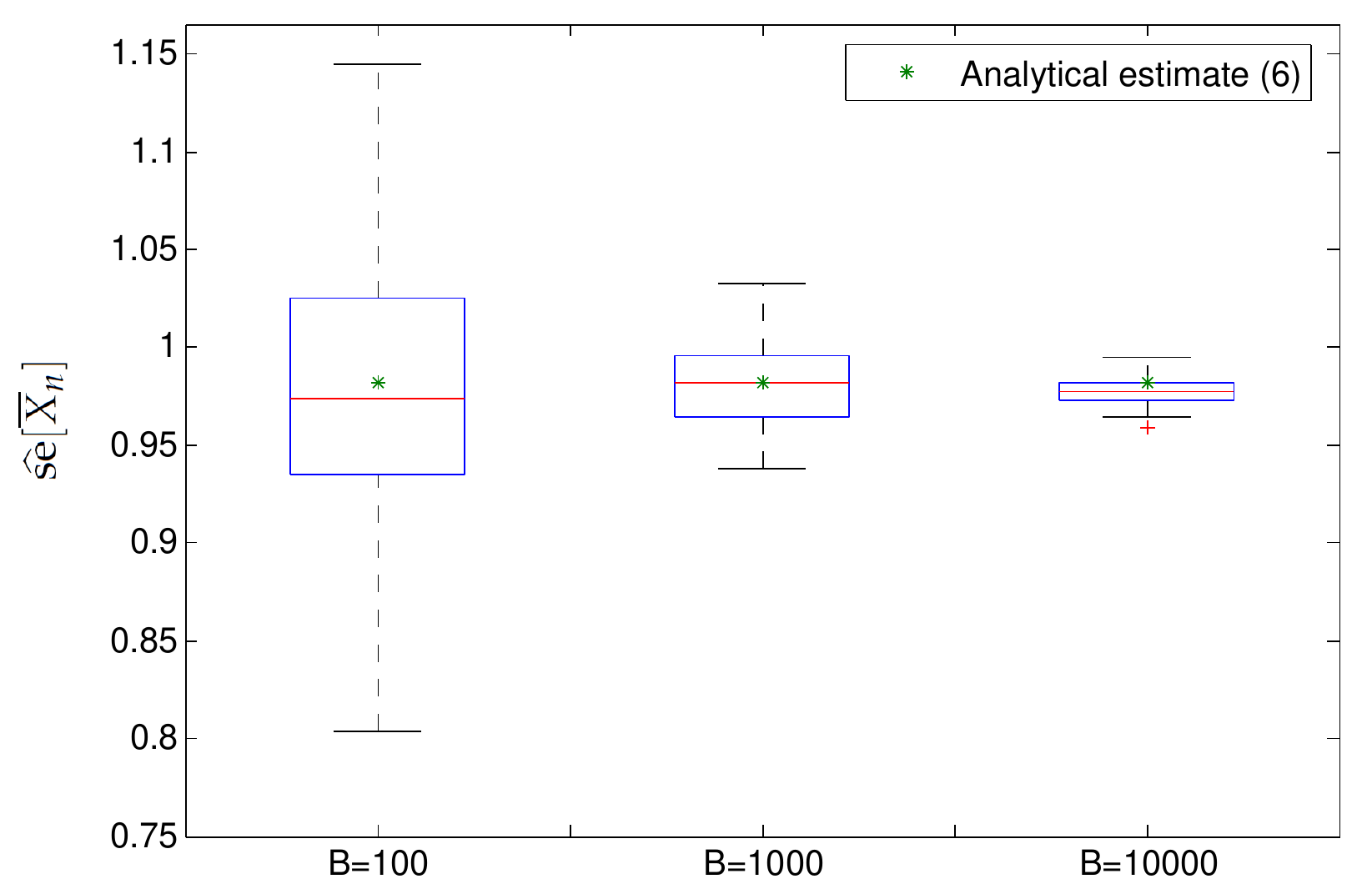}}\caption{Boxplots of bootstrap estimates. Each boxplot is constructed based on $100$ bootstrap estimates. That is, we repeated (\ref{eq:B}) with $\mathcal{P}=\mathcal{P}_{boot}$, (\ref{eq:means}), and (\ref{eq:bootstrap}) 100 times for each value of $B$.}\label{fig1}
\end{figure}

For illustrative purposes, let us consider a ``Gaussian'' population $\mathcal{P}$, where $x_1,\ldots,x_N$ are independently drawn from the normal distribution $\mathcal{N}(\mu_0,\sigma_0^2)$ with $\mu_0=0$, $\sigma_0=10$, and the population size $N=10^4$\footnote{In other words, we generate $N$ realizations of $Z\sim\mathcal{N}(\mu,\sigma^2)$, ``freeze'' them, and denote the obtained values by $x_i$.}. The resulting population mean is $\mu=0.11$ and standard deviation  is $\sigma=10.13$. As expected, they are close to $0$ and $10$, respectively. Let $\mathcal{S}$ be a simple random sample from $\mathcal{P}$ of size $n=10^2$. The obtained value of the sample mean is $\overline{X}_n=0.4$. The \textit{exact} value of the standard error of $\overline{X}_n$ is given by (\ref{eq:se})\footnote{In this example, we can compute the exact value, since we know the population variance $\sigma^2$.}:
\begin{equation}\label{eq:exact}
\mathrm{se}[\overline{X}_n]=1.01.
\end{equation}
Figure~\ref{fig1} shows the boxplots of the bootstrap estimates (\ref{eq:bootstrap}) with $B=10^2, 10^3,$ and $B=10^4$ as well as the analytical estimate (\ref{eq:analutical}) marked by a green star. The larger $B$, the smaller the dispersion of the bootstrap estimates. Both analytical and bootstrap estimates $\widehat{\mathrm{se}}[\overline{X}_n]$ agree with the exact value (\ref{eq:exact}). 

\section{Further Reading}

\begin{enumerate}
	\item Intended for general readership, P.~Diaconis and B.~Efron (1983), ``\href{http://www.vanderbilt.edu/psychological_sciences/graduate/programs/quantitative-methods/quantitative-content/diaconis_efron_1983.pdf}{Computer-Intensive Methods in Statistics},'' \textit{Scientific American}, 248(5):116-129 discusses different applications of bootstrap. 
\end{enumerate}

\section{What is Next?}
The sample mean $\overline{X}_n$ is a \textit{point estimate} (single number) of the population mean $\mu$. In the next lecture, we will learn how to construct \textit{confidence intervals} for $\mu$, which are random intervals that contain $\mu$ with a prescribed probability.

\chapter{Normal Approximation and Confidence Intervals}

\newthought{In the last two lectures} we studied properties of the sample mean $\overline{X}_n$ under SRS. We learned that it is an unbiased estimate of the population mean, 
\begin{equation}
\mathbb{E}[\overline{X}_n]=\mu,
\end{equation} 
derived the formula for its variance, 
\begin{equation}\label{eq:varr}
\mathbb{V}[\overline{X}_n]=\frac{\sigma^2}{n}\left(1-\frac{n-1}{N-1}\right),
\end{equation}
and learned how to estimate it analytically and using the bootstrap. Ideally, however, we would like to know the entire distribution of
$\overline{X}_n$\footnote{A random variable can't be fully described by only first two moments. }, called \textit{sampling distribution}, since it would tell us everything about the accuracy of the estimation $\overline{X}_n\approx\mu$. In this lecture, we discuss the sampling distribution of $\overline{X}_n$ and show how it can be used for constructing \textit{interval estimates} for $\mu$. 

\section{Normal Approximation for $\overline{X}_n$}
First, let us recall one of the most remarkable results in probability: the \textit{Central Limit Theorem} (CLT).  Simply put, the CLT says that if $Y_1,\ldots,Y_n$ are independent and identically distributed  (iid) with mean $\mu$ and variance $\sigma^2$, then $\overline{Y}_n=\frac{1}{n}\sum_{i=1}^nY_i$ has a distribution which is approximately normal with mean $\mu$ and variance $\frac{\sigma^2}{n}$\footnote{The fact that $\mathbb{E}[\overline{Y}_n]=\mu$ and $\mathbb{V}[\overline{Y}_n]=~\frac{\sigma^2}{n}$ is trivial. The remarkable part of the CLT is that the distribution of $\overline{Y}_n$ is normal \textit{regardless} of the distribution of $Y_i$.}:
\begin{equation}
\overline{Y}_n~\dot{\sim}~ \mathcal{N}\left(\mu,\frac{\sigma^2}{n}\right).
\end{equation}
Symbol $\dot{\sim}$ means ``approximately distributed.'' More formally, 
\begin{equation}
\mathbb{P}\left(\frac{\overline{Y}_n-\mu}{\sigma/\sqrt{n}}\leq z\right)\rightarrow
\Phi(z), \hspace{5mm} \mbox{ as } n\rightarrow\infty,
\end{equation}
where $\Phi(z)$ is the CDF of the standard normal $\mathcal{N}(0,1)$. 

\textit{Question:} Can we use the CLT to claim that the sampling distribution of $\overline{X}_n$ under SRS is approximately normal? 

\textit{Answer:} Strictly speaking, no. Since in SRS, $X_i$ are \textit{not independent}\footnote{Recall Lemma~\ref{lemma:2} in Lecture~\ref{ch:SRS}.} (although identically distributed).  Moreover, it makes
no sense to have $n$ tend to infinity while $N$ is fixed.

Nevertheless... it can be shown that if both $n$ and $N$ are large, then  $\overline{X}_n$ is approximately normally distributed: 
\begin{equation}\label{eq:norm_approx}
\overline{X}_n~\dot{\sim}~\mathcal{N}(\mu,\mathbb{V}[\overline{X}_n])\hspace{5mm}\mbox{or}\hspace{5mm} \frac{\overline{X}_n-\mu}{\mathrm{se}[\overline{X}_n]}~\dot{\sim}~\mathcal{N}(0,1).
\end{equation}
The intuition behind this approximation is the following: if both $n,N\gg1$, then $X_i$ are nearly independent, and, therefore, the CLT approximately holds. 

The CLT result in (\ref{eq:norm_approx}) is very powerful: it says that for \textit{any} population, under SRS (for $n\gg1$ and $n\ll N$), the sample mean has an approximate normal distribution. 

\subsection{Example: Birth Weights}
\begin{figure}[h]
	\centerline{\includegraphics[width=.9\linewidth]{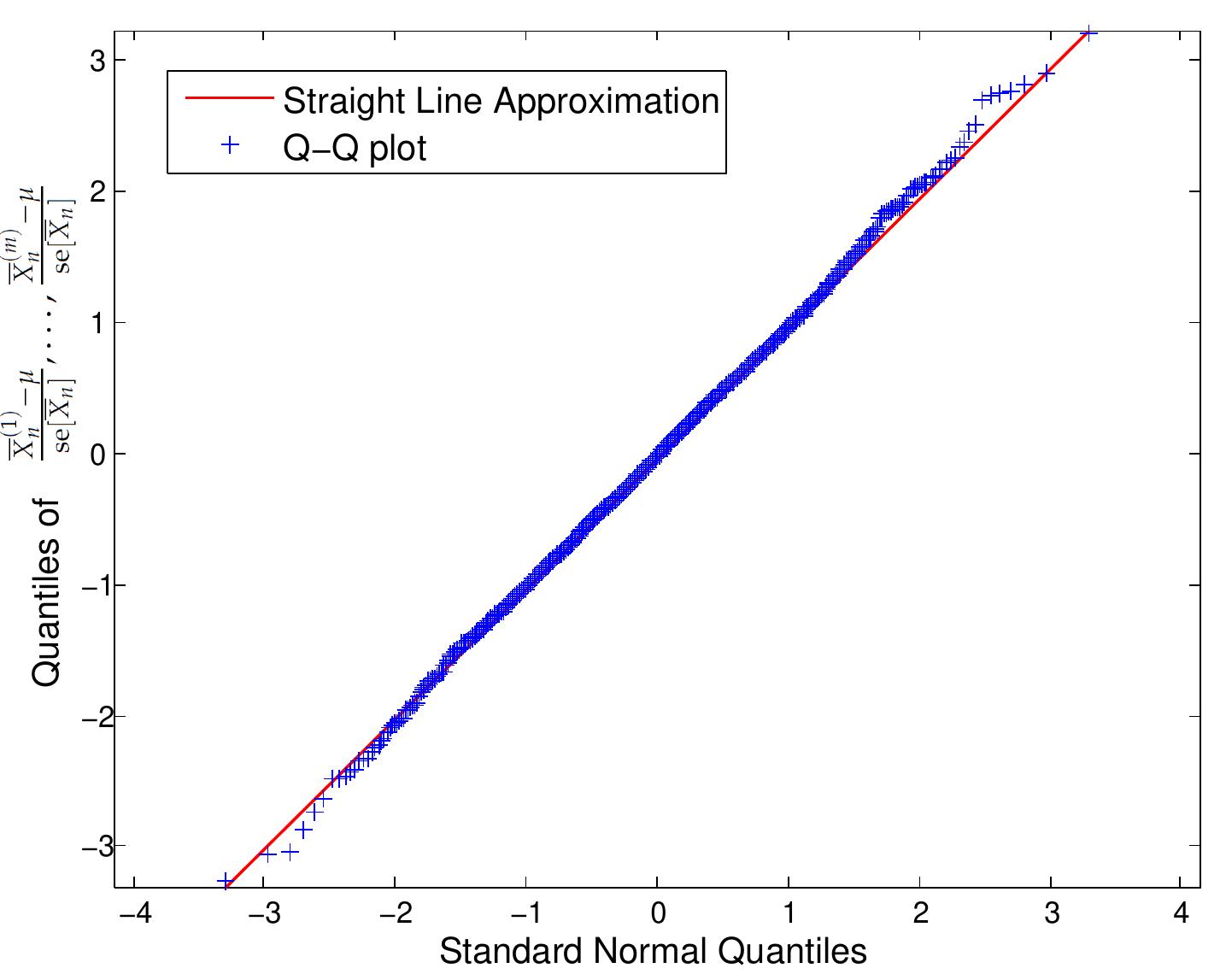}}\caption{Normal-quantile plot for the standardized sample means $\frac{\overline{X}_n^{(1)}-\mu}{\mathrm{se}[\overline{X}_n]},\ldots,\frac{\overline{X}_n^{(m)}-\mu}{\mathrm{se}[\overline{X}_n]}$. The sampling distribution closely follows the standard normal curve.  }\label{fig17}
\end{figure}

Let us consider the example from Lecture 2b, where the target population $\mathcal{P}$ is the set of all birth weights\footnote{The data is available at \href{http://www.its.caltech.edu/~zuev/teaching/2016Winter/birth.txt}{birth.txt}}. The population parameters are: $N=1236$, $\mu=3.39$, and $\sigma=0.52$. Let  $n=100$, and let $\mathcal{S}^{(1)},\ldots,\mathcal{S}^{(m)}$ be the SRS samples from $\mathcal{P}$, $m=10^3.$ Figure~\ref{fig17} shows the normal-quantile plot for the corresponding \textit{standardized}\footnote{If $X$ is a random variable with mean $\mu$ and variance $\sigma^2$, then $\frac{X-\mu}{\sigma}$ is called the standardized variable; it has zero mean and unit variance. This transformation is often used in statistics.} sample means $\frac{\overline{X}_n^{(1)}-\mu}{\mathrm{se}[\overline{X}_n]},\ldots,\frac{\overline{X}_n^{(m)}-\mu}{\mathrm{se}[\overline{X}_n]}$. The normal approximation (\ref{eq:norm_approx}) works well.

In this example, we know both population parameters $\mu$ and $\sigma$, and, therefore, the exact standard error $\mathrm{se}[\overline{X}_n]$ is also known from (\ref{eq:varr}). Suppose now that, in fact, we don't know the entire population, and we have only one simple random sample $\mathcal{S}$ from $\mathcal{P}$.

\textit{Question:} How to check in this case  that the sampling distribution of the sample mean $\overline{X}_n$ follows the normal curve?

\section{Estimating the Probability $\mathbb{P}(|\overline{X}_n-\mu|\leq\epsilon)$}
The normal approximation of the sampling distribution (\ref{eq:norm_approx}) can be used for various purposes. For example, it allows to estimate the probability that the error
made in estimating $\mu$ by $\overline{X}_n$ is less than
$\varepsilon>0$. Namely,
\begin{equation}\label{eq:prob}
\mathbb{P}(|\overline{X}_n-\mu|\leq\varepsilon)\approx
2\Phi\left(\frac{\varepsilon}{\mathrm{se}[\overline{X}_n]}\right)-1,
\end{equation}
where the standard error can be estimated, for example, by the bootstrap. 

\section{Confidence Intervals}
What is the probability that $\overline{X}_n$ exactly equals to $\mu$? Setting $\epsilon=0$ in (\ref{eq:prob}), gives an intuitively expected result: $\mathbb{P}(\overline{X}_n=\mu)\approx0$. But given a simple random sample $X_1,\ldots,X_n$, can we define a \textit{random region}\footnote{As opposed to random number $\overline{X}_n$.} that contains the population mean $\mu$ with high probability? It turns out that yes, the notion of a confidence interval formalizes this idea.

Let $0<\alpha<1$. A $100(1-\alpha)\%$ \textit{confidence interval} for a population parameter $\theta$ is a \textit{random} interval $\mathcal{I}$ calculated from the sample, which contains $\theta$ with probability $1-\alpha$,
\begin{equation}\label{eq:CI0}
\mathbb{P}(\theta\in\mathcal{I})=1-\alpha.
\end{equation}
The value $100(1-\alpha)\%$ is called the \textit{confidence level}\footnote{Usually $90\%$ ($\alpha=0.1$) or $95\%$ ($\alpha=0.05$) levels are used.}.

Let us construct a confidence interval for $\mu$ using the normal approximation (\ref{eq:norm_approx}). Since $\frac{\overline{X}_n-\mu}{\mathrm{se}[\overline{X}_n]}$ is approximately standard normal, 
\begin{equation}\label{eq:1}
\mathbb{P}\left(-z_{1-\frac{\alpha}{2}}\leq\frac{\overline{X}_n-\mu}{\mathrm{se}[\overline{X}_n]}\leq z_{1-\frac{\alpha}{2}}\right)\approx 1-\alpha,
\end{equation}
where $z_q$ is the $q^{\mathrm{th}}$ standard normal quantile, $\Phi(z_q)=q$.  We can rewrite (\ref{eq:1}) as follows:
\begin{equation}\label{eq:CI}
\mathbb{P}\left(\overline{X}_n - z_{1-\frac{\alpha}{2}}\mathrm{se}[\overline{X}_n]\leq \mu\leq \overline{X}_n + z_{1-\frac{\alpha}{2}}\mathrm{se}[\overline{X}_n]\right)\approx 1-\alpha.
\end{equation}
This means that $\mathcal{I}=\overline{X}_n \pm z_{1-\frac{\alpha}{2}}\mathrm{se}[\overline{X}_n] $ is an approximate $100(1-\alpha)\%$ confidence interval for $\mu$. Confidence intervals often have this form:
\begin{equation}
\mathcal{I}=\mbox{statistic } \pm \mbox{ something},
\end{equation}
and the ``something'' is called the \textit{margin of error}\footnote{For the constructed interval for $\mu$, the margin of error is $z_{1-\frac{\alpha}{2}}\mathrm{se}[\overline{X}_n]$.}.

Confidence intervals are often misinterpreted\footnote{Even by professional scientists.}. Suppose that we got a sample $\mathcal{S}=\{X_1,\ldots,X_n\}$ from the target population $\mathcal{P}$, set the confidence level to, say $95\%$, plugged in all the numbers in (\ref{eq:CI}) and obtained that the confidence interval for $\mu$ is, for example, $[0,1]$. Does it mean that $\mu$ belongs to $[0,1]$ with probability $0.95$? No, of course not: $\mu$ is a deterministic (not random) parameter, it either belongs to $[0,1]$ or it does not\footnote{In other words, once a sample is drawn and an interval is calculated, this interval either covers $\mu$ or it does not, it is no longer a matter of probability.}. 

The correct interpretation of confidence intervals is the following. First, it is important to realize that Eq. (\ref{eq:CI0}) is a probability statement about the confidence interval, not the population parameter\footnote{Perhaps, it would be better to rewrite it as $\mathbb{P}(\mathcal{I} \ni \theta)=1-\alpha.$}. It says that if we take many samples $\mathcal{S}^{(1)},\ldots,$ and compute confidence intervals $\mathcal{I}^{(1)},\ldots,$ for each sample, then we expect about $100(1-\alpha)\%$ of these  intervals to contain $\theta$. The confidence level  $100(1-\alpha)\%$ describes the uncertainty associated with a \textit{sampling method}, simple random sampling in our case.

\subsection{Example: Birth Weights}
\begin{figure}
	\centerline{\includegraphics[width=\linewidth]{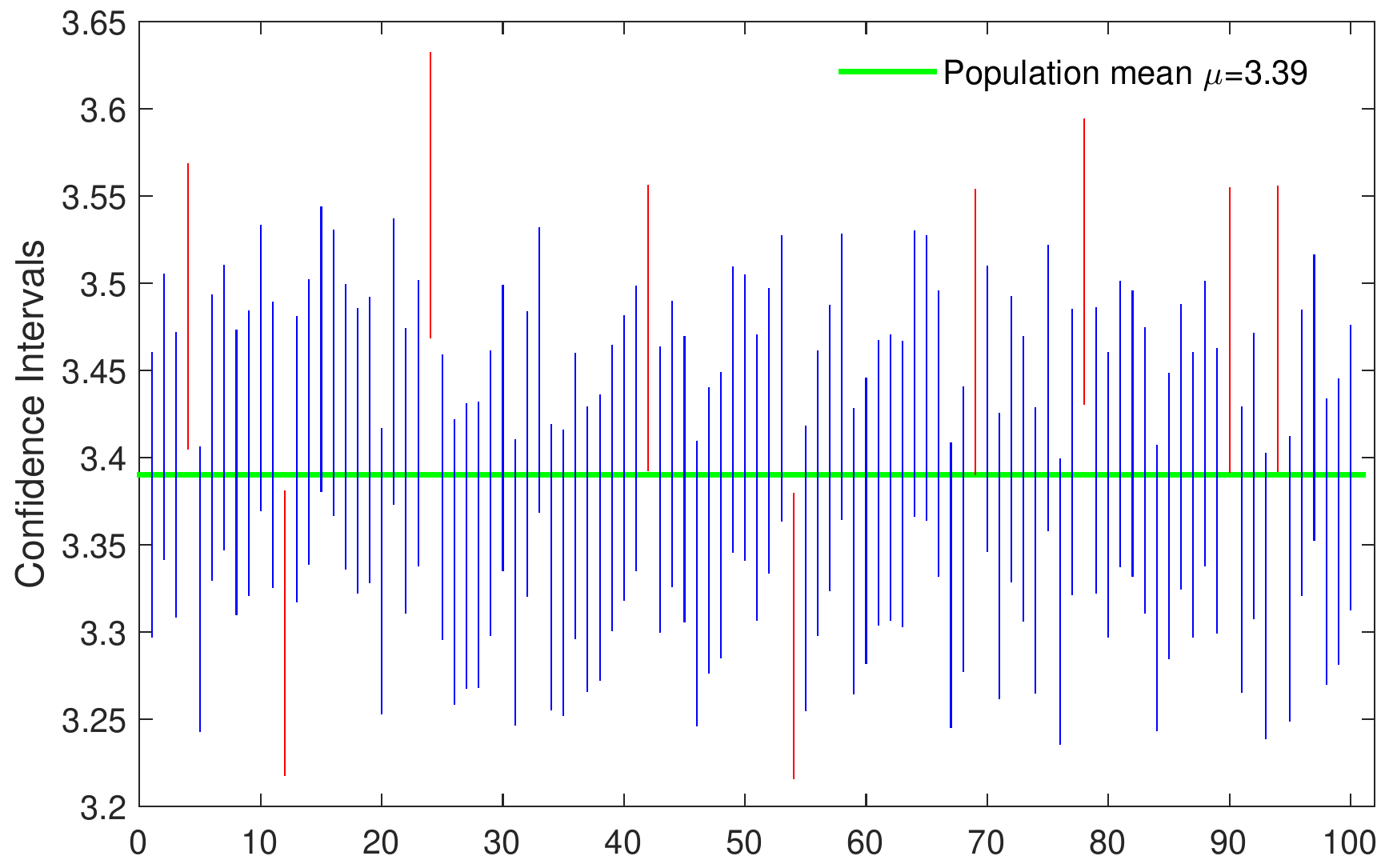}}\caption{$90\%$ confidence intervals for the population mean $\mu=3.39$. Intervals that don't contain $\mu$ are shown in red.}\label{fig2}
\end{figure}

Let us again consider the example with birth weights. Figure~\ref{fig2} shows $90\%$ confidence intervals for $\mu$ computed from $m=100$ simple random samples. Just as different samples lead to different sample means, they also lead to different confidence intervals. We expect that about $90$ out of $100$ intervals would contain $\mu$. In our experiment, $91$ intervals do. 

\section{Survey Sampling: Postscriptum}
We stop our discussion of survey sampling here. The considered SRS is the simplest sampling scheme and provides the basis for more advanced sampling designs, such as \textit{stratified random sampling}, \textit{cluster sample}, \textit{systematic sampling}, etc. For example, in stratified random sampling (StrRS), the population is partitioned into subpopulations, or \textit{strata}, which are then independently sampled using SRS. In many applications, stratification is natural. For example, when studying human populations, geographical areas form natural strata. SrtRs is often used when, in addition to information about the whole population, we are  interested in obtaining information about each natural
subpopulation. Moreover, estimates obtained from StrRS can be considerably more accurate than
estimates from SRS if a)  population units within each stratum are relatively homogeneous and b) there is considerable variation between strata. If the total sample size we could afford is $n$ and there are $L$ strata, then we face an \textit{optimal resource allocation} problem: how to chose the sample sizes $n_k$ for each stratum, so that $\sum_{k=1}^L n_k=n$ and the variance of the corresponding estimator is minimized? This leads to the so-called Neyman allocation scheme, but this is a different story.

\section{Further Reading}

\begin{enumerate}
	
	\item A detailed discussion of survey sampling\footnote{Which contains all the sampling scheme mentioned in the Postscriptum.} is given in the fundamental (yet accessible to students with diverse statistical backgrounds) monograph by S.L. Lohr \textit{\href{https://books.google.com/books/about/Sampling_Design_and_Analysis.html?id=aSXKXbyNlMQC}{Sampling: Design and Analysis}}.  
\end{enumerate}

\section{What is Next?}
Summarizing Data and Survey Sampling constitute the core of classical elementary statistics. In the next lecture, we will draw a big picture of modern statistical inference. 

\chapter{Modeling and Inference: A Big Picture}\label{ch:BigPicture}
\newthought{Suppose} we are interested in studying a certain phenomenon which can be schematically represented as follows:
\vspace{-4mm}
\begin{figure}
	\centerline{\includegraphics[width=0.5\linewidth]{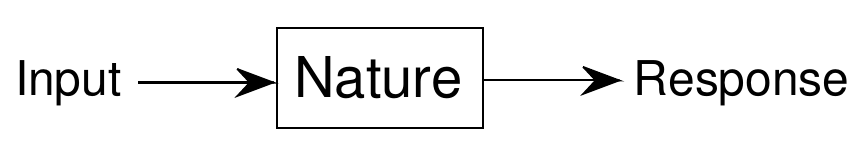}}
\end{figure}

\vspace{-4mm}\hspace{-5mm}
Furthermore, suppose we collected some data $\{(\mbox{inputs, responses})\}$ by observation or experiment.
The most basic question in statistics is: what can we learn, or \textit{infer}, about the phenomenon from data? Generally, there two goals in analyzing the data:
\begin{enumerate}
	\item \textit{Understanding.} To extract some information on how Nature associates the responses to the inputs.
	\item \textit{Prediction.} To be able to predict the response to the future input.
\end{enumerate} 

The main idea of statistical inference is to replace the Nature ``black box'' (\ie the unknown mechanism that Nature uses to associate the responses to the inputs) by a \textit{statistical model}:
\vspace{-3mm}
\begin{figure}
	\centerline{\includegraphics[width=0.6\linewidth]{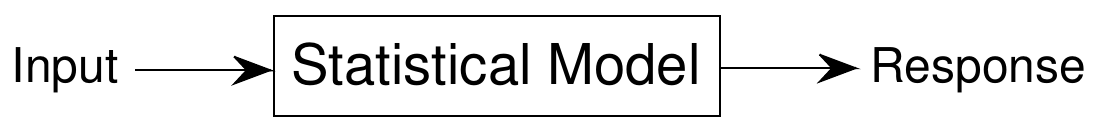}}
\end{figure}

\vspace{-3mm}\hspace{-5mm} The key feature of a statistical model is that the observed \textit{variability} in the data is represented by \textit{probability distributions}, which form the building-blocks of the model. In other words, the data is treated as the outcome of a random experiment, as the realization of random variables. In this lecture, we discuss various statistical models and consider several fundamental concepts of statistical inference.

\section{Statistical Models}

For simplicity, let us first assume that the data consists only from ``responses'' $X_1,\ldots,X_n$.  A statistical model $\mathcal{F}$ is then simply a set of probability distributions $F$ (or probability density functions $f$) for $X_i$. The basic statistical inference problem can then be formulated as follows: given data $X_1,\ldots,X_n$, we \textit{assume}\footnote{It is very important to keep in mind that (\ref{eq:assumption}) is an assumption, which, in fact, can be wrong.} that it is an iid sample from $F$,
\begin{equation}\label{eq:assumption}
X_1,\ldots,X_n\sim F, \hspace{5mm}F\in\mathcal{F},
\end{equation}
and what to infer $F$ or some properties of $F$ (such as its mean). 

There are two big classes of statistical models: \textit{parametric} and \textit{nonparametric}\footnote{Which lead to two subfields of statistics: parametric and nonparametric statistics.}. A statistical model $\mathcal{F}$ is called parametric if it can be parameterized by a finite number of parameters. For example, if we assume that the data comes from a normal distribution, then the model is a two-dimensional parametric model:
\begin{equation}
\mathcal{F}=\left\{f(x;\mu,\sigma^2)=\frac{1}{\sqrt{2\pi}\sigma}e^{\left(-\frac{(x-\mu)^2}{2\sigma^2}\right)}, \mu,\sigma^2\in\mathbb{R}\right\}.
\end{equation}
In general, a parametric model takes the from
\begin{equation}
\mathcal{F}=\{F(x;\theta), \hspace{2mm} \theta\in\Theta\},
\end{equation}
where $\theta$ is an unknown parameter (or vector of parameters) that takes values in the \textit{parameter space} $\Theta\subset\mathbb{R}^d$. In parametric inference, we thus want to learn about $\theta$ from the data.

Quite naturally, a statistical model $\mathcal{F}$ is called nonparametric if it is not parametric, that is if it cannot be parametrized by a finite number of parameters. For example, if we assume that the data comes for a distribution with zero mean, then the model is nonparametric\footnote{If you feel uncomfortable with (\ref{eq:example}), let us take $\mathcal{F}=\left\{f:  \int xf(x) dx=0\right\}$ instead.}:
\begin{equation}\label{eq:example}
\mathcal{F}=\left\{F: \int x d F(x)=0\right\}.
\end{equation}
Taking this example to extreme and throwing away the zero mean assumption, we obtain the most general statistical model,
\begin{equation}\label{eq:ALL}
\mathcal{F}=\{\mbox{all CDFs}\},
\end{equation}  which is of course nonparametric\footnote{At first glance, this model may look silly, but it is not. In fact, this model is often used in practice when nothing is really known about the mechanism that generated the data. Essentially, the model in~(\ref{eq:ALL}) says that all we assume is that $X_1,\ldots,X_n$ is an iid sample from \textit{some} distribution. In the forthcoming lectures we will see that we can learn a lot from the data  even under this seemingly weak assumption.}. 

Historically, parametric models were developed first since most nonparametric methods were not feasible in practice, due to limited computing power. Nowadays, this has changed due to rapid developments in computing science. \vspace{2mm}

\textit{Advantages of parametric models:}
\begin{enumerate}
	\item Parametric models are generally easier to work with.
	\item \textit{If} parametric model is correct\footnote{This means that there exists the value of $\theta_0$, often called the ``true value,'' such that the corresponding distribution $F(x;\theta_0)\in\{F(x;\theta), \hspace{1mm} \theta\in\Theta\}$ indeed adequately describes the data.}, then parametric methods are more efficient than their nonparametric counterparts. 
	\item Sometimes parametric models are easier to interpret.
\end{enumerate}
\textit{Advantages of nonparametric models:}
\begin{enumerate}
	\item Sometimes it is hard to find a suitable parametric model.
	\item Nonparametric methods are often less sensitive to outliers.
	\item Parametric methods have a high risk of \textit{mis-specification}\footnote{Mis-specification is the choice of the model $\mathcal{F}$ that in fact does not contain a distribution that adequately describes the modeled data.}.
\end{enumerate}

The art of statistical modeling is based on a proper incorporation of the scientific knowledge about the underlying phenomenon into the model and on finding a balance between the model complexity on one hand and the ability to analysis the model analytically or numerically on the other hand. The choice of the model also depends  on the problem and the answer required, so that different models may be appropriate for a single set of data. 

\subsection{Example: Darwin and Corn}
Charles Darwin wanted to compare the heights of self-fertilized and cross-fertilized corn plants. To this end, he planted $n=15$ pairs of self- and cross-fertilized plants in different pots, trying to make all other characteristics of the plants in each pair the same (descended from the same parents, planted at the same time, etc).
\begin{marginfigure}\vspace{-32mm}
	\centerline{\includegraphics[width=0.5\linewidth]{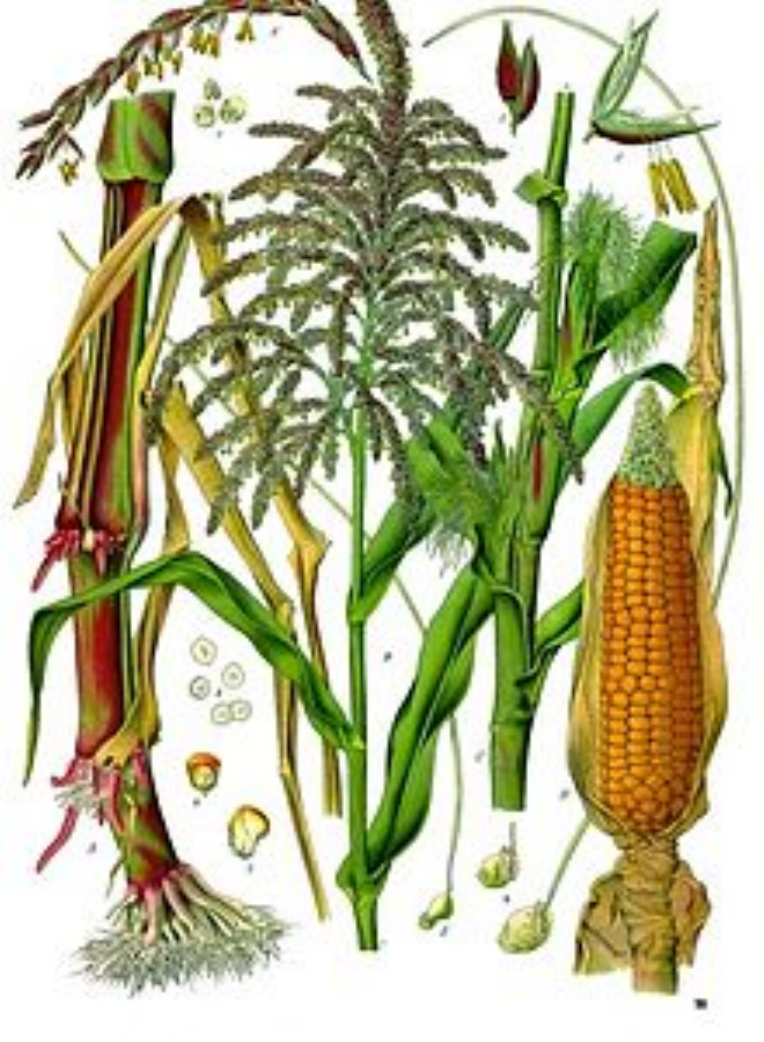}\includegraphics[width=0.5\linewidth]{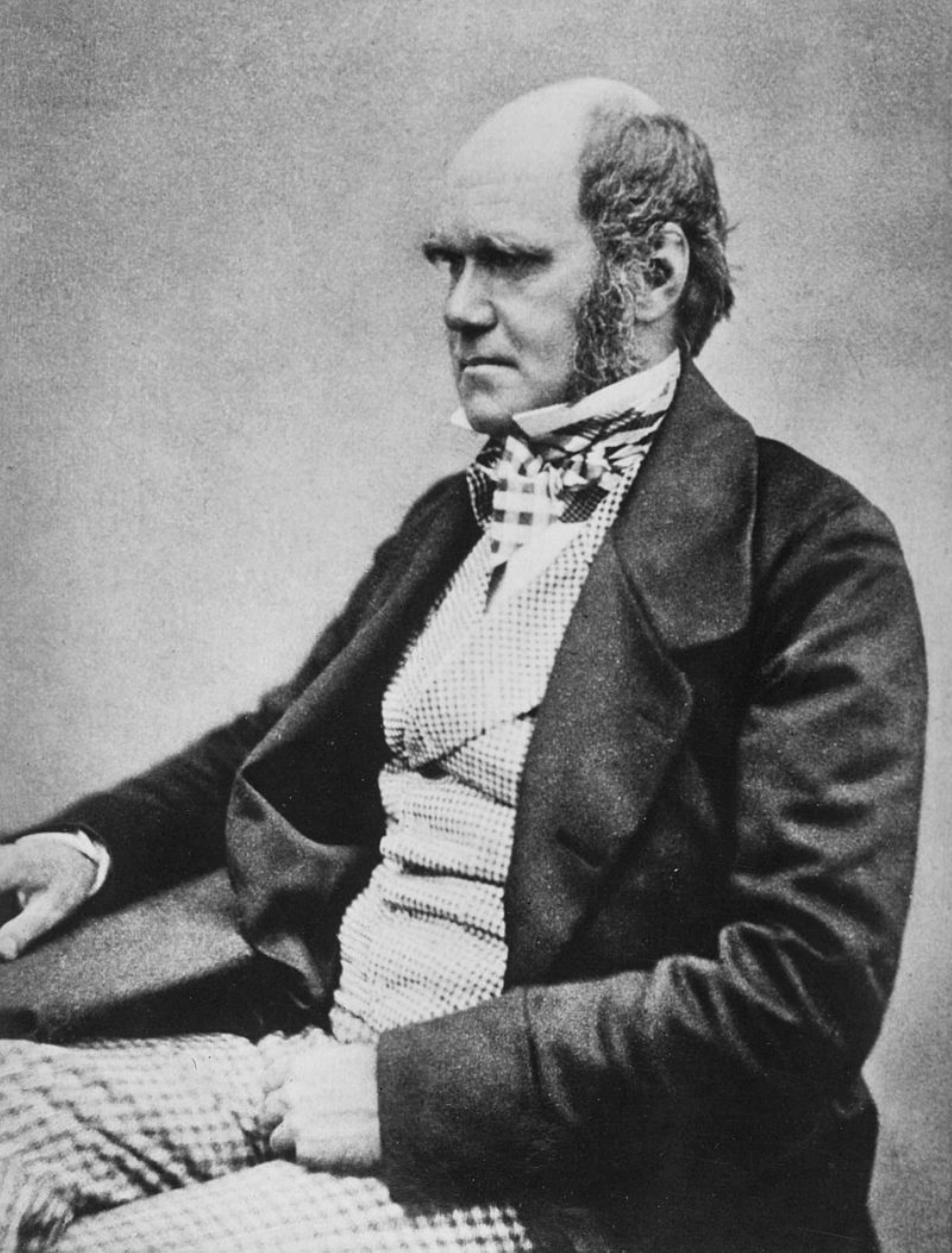}}\caption{Charles Darwin studying corn. Photo source: \href{https://en.wikipedia.org/wiki/Charles_Darwin}{wikipedia.org}.}.\label{fig21}
\end{marginfigure} 
\vspace{-2mm}
\begin{figure}
	\centerline{\includegraphics[width=0.75\linewidth]{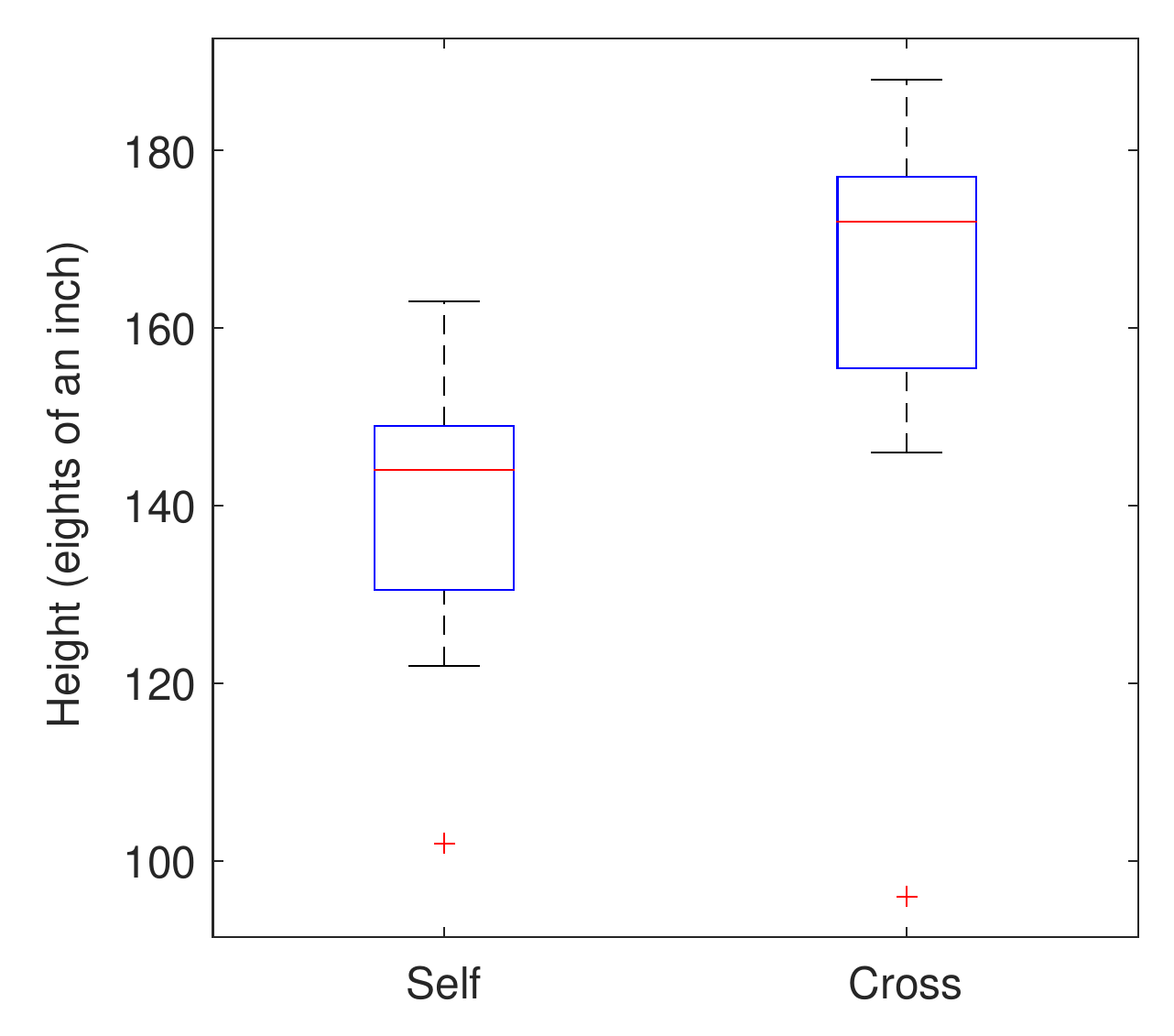}}\caption{The heights of corn plants for two different types of fertilization.}\label{fig22}
\end{figure}
\vspace{-4mm}
Figure~\ref{fig22} summarizes the results in terms of boxplots. Cross-fertilized plants seem generally higher then self-fertilized ones. At the same time, there is a variation of heights within each group, and one could model this variability in terms of probability distributions\footnote{It might be possible, to construct a mechanistic model for plant	growth that could explain all the variation in such data. This would take into account genetic variation, soil and moisture conditions, ventilation, lighting, etc,	through a vast system of equations requiring numerical solution. For most purposes, however, a deterministic model of this sort is  unnecessary, and it is simpler to express variability in terms of probability distributions.}. But if  the spread of heights within each group is modeled by random variability, the same cause will also generate variation between groups. So Darwin asked his cousin, Francis Galton, whether the difference in heights between the types of plants was too large to have occurred by chance, and was in fact due to the effect of fertilization. If so, he wanted to estimate the average height increase.

%\vspace{-1.5cm}
Galton proposed an analysis based  on the following model. The height of a self-fertilized plant is modeled as 
\begin{marginfigure} %\vspace{-5mm}	
	\centerline{\includegraphics[width=0.5\linewidth]{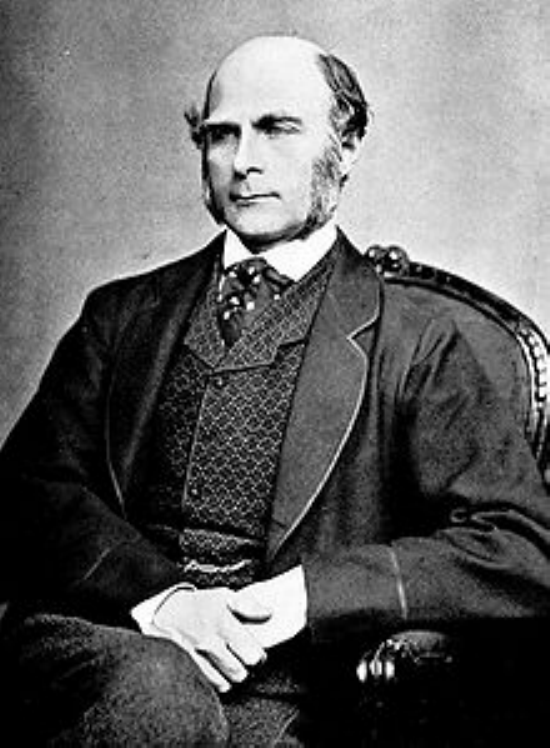}}\caption{Sir Francis Galton, Darwin's cousin. Among many other things, he developed the concept of correlation. Photo source: \href{https://en.wikipedia.org/wiki/Francis_Galton}{wikipedia.org}.}.\label{fig23}
\end{marginfigure}
\vspace{-2mm}
\begin{equation}\label{eq:Xs}
X_s=\mu+\sigma\epsilon,
\end{equation}
where $\mu$ and $\sigma$ are fixed unknown parameters, and $\epsilon$ is a random variable with zero mean and unit variance. Thus, $\mathbb{E}[X_s]=\mu$ and $\mathbb{V}[X_s]=\sigma^2$.  The height of a cross-fertilized plant is modeled as 
\begin{equation}\label{eq:Xc}
X_c=\mu+\eta+\sigma\epsilon,
\end{equation}
where $\eta$ is another unknown parameter. Therefore, $\mathbb{E}[X_c]=\mu+\eta$ and $\mathbb{V}[X_c]=\sigma^2$. In the model (\ref{eq:Xs}) \& (\ref{eq:Xc}), variation within the groups is accounted for by the randomness of $\epsilon$, whereas variation between groups is modeled deterministically by $\eta$, the difference between the means of $X_c$ and $X_f$. Under this model\footnote{By the way, is this model parametric or nonparametric?}, the questions asked by Darwin are:
\begin{itemize}
	\item[a)] Is $\eta\neq0$?
	\item[b)] Can we estimate $\eta$ and state the uncertainty of our estimate?
\end{itemize}

\section{Fundamental Concepts in Inference} 

Many inferential problems can be identified as being one of the three types: estimation, confidence sets, or hypothesis testing\footnote{For example, Darwin's problems a) and b) are, respectively, hypothesis testing and estimation.}. In this lecture we will consider all of these problems. Here we give a brief introduction to the ideas and illustrate them with the iconic coin flipping example. 

\subsection{Point Estimation}
Point estimation refers to providing a single ``best guess'' for some quantity of interest, which could be a population parameter\footnote{For instance, the sample mean $\overline{X}_n$ is a point estimate for the population mean $\mu$.}, a parameter in a parametric model, a CDF $F$, a probability density function $f$, a regression function\footnote{See below.} $r$, to name a few. By convention, we denote a point estimate of $\theta$ by $\hat{\theta}$.

Let $X_1,\ldots, X_n$ be data which is modeled as an iid sample from a distribution $F(x;\theta)\in\mathcal{F}$, where $\mathcal{F}$ is a parametric model.  A point estimate $\hat{\theta}_n$ of a parameter $\theta$ is some function of the data:
\begin{equation}
\hat{\theta}_n=s(X_1,\ldots,X_n).
\end{equation}
Thus, $\theta$ is a fixed \textit{deterministic} unknown quantity, and $\hat{\theta}_n$ is a \textit{random} variable. The distribution of $\hat{\theta}_n$ is called the \textit{sampling distribution}. The standard deviation of $\hat{\theta}_n$ is called the \textit{standard error}\footnote{Notice that these definitions mirror the corresponding definitions for $\overline{X}_n$ that we discussed in the context of survery sampling.},
\begin{equation}
\mathrm{se}[\hat{\theta}_n]=\sqrt{\mathbb{V}[\hat{\theta}_n]}.
\end{equation}
To access how good a point estimate is on average, we introduce \textit{bias}:
\begin{equation}
\mathrm{bias}[\hat{\theta}_n]=\mathbb{E}[\hat{\theta}_n]-\theta.
\end{equation}
We say that $\hat{\theta}_n$ is \textit{unbiased} if $\mathrm{bias}[\hat{\theta}_n]=0$. Unbiasedness  is a good property of an estimator, but its importance should not be overstated: an estimator could be unbiased, but at the same time it could have a very large standard error. Such an estimator is poor since its realizations are likely to be far from  $\theta$, although, on average, the estimator equals to $\theta$. The overall quality of a point estimate is often assessed by the \textit{mean squared error}, or MSE,
\begin{equation}
\mathrm{MSE}[\hat{\theta}_n]=\mathbb{E}[(\hat{\theta}_n-\theta)^2].
\end{equation}
It is straightforward to check that MSE can be written it terms of bias and standard error as follows:
\begin{equation}
\mathrm{MSE}[\hat{\theta}_n]=\mathrm{bias}[\hat{\theta}_n]^2+\mathrm{se}[\hat{\theta}_n]^2.
\end{equation}
This is called the \textit{bias-variance decomposition} for the MSE.

\paragraph{Example:} Let us take a coin and flip it $n$ times. Let $X_i=1$ if we get ``head'' on the $i^{\mathrm{th}}$ toss, and $X_i=0$ if we get ``tail''. Thus, we have the data $X_1,\ldots,X_n$. Since we don't know whether the coin is fair, it is reasonable to model the data by the Bernoulli distribution, which is the probability distribution of a random variable which takes the value $1$ with probability $p$ and the value $0$ with probability of $1-p$, where $p\in[0,1]$ is a model parameter\footnote{If the coin is fair, $p=1/2$.}. So, assume that
\begin{equation}
X_1,\ldots,X_n\sim\mathrm{Bernoulli}(p).
\end{equation}
The goal is to estimate $p$ from the data. It seems reasonable to estimate $p$ by
\begin{equation}
\hat{p}_n=\overline{X}_n=\frac{1}{n}\sum_{i=1}^nX_i.
\end{equation}
This estimate is unbiased, its standard error is $\mathrm{se}[\hat{p}_n]=\sqrt{p(1-p)/n}$, and the means squared error is $\mathrm{MSE}[\hat{p}_n]=p(1-p)/n$. \hfill $\square$

\subsection{Confidence Sets}

We have already encountered confidence intervals in the context of  survey sampling. Here, they are defined similarly. Suppose that $X_1,\ldots,X_n\sim F(x;\theta)$.  A $100(1-\alpha)\%$ \textit{confidence interval} for parameter $\theta$ is a \textit{random} interval $\mathcal{I}$ calculated from the data, which contains $\theta$ with probability $1-\alpha$,
\begin{equation}\label{eq:CI0}
\mathbb{P}(\theta\in\mathcal{I})=1-\alpha.
\end{equation}

If $\theta$ is a vector, then we an interval is replaced by a \text{confidence set}, which can be a cube, a sphere, an ellipsoid, or any other random set that traps $\theta$ with probability $1-\alpha$. 

\paragraph{Example:}  Let us construct a confidence interval for $p$ in the coin example. We can do this using Hoeffding's inequality: if $X_1,\ldots,X_n\sim\mathrm{Bernoulli}(p)$, then, for any $\epsilon>0$,
\begin{equation}\label{eq:H}
\mathbb{P}(|\overline{X}_n-p|>\epsilon)\leq 2e^{-2n\epsilon^2}.
\end{equation}
\begin{marginfigure}
	\centerline{\includegraphics[width=0.6\linewidth]{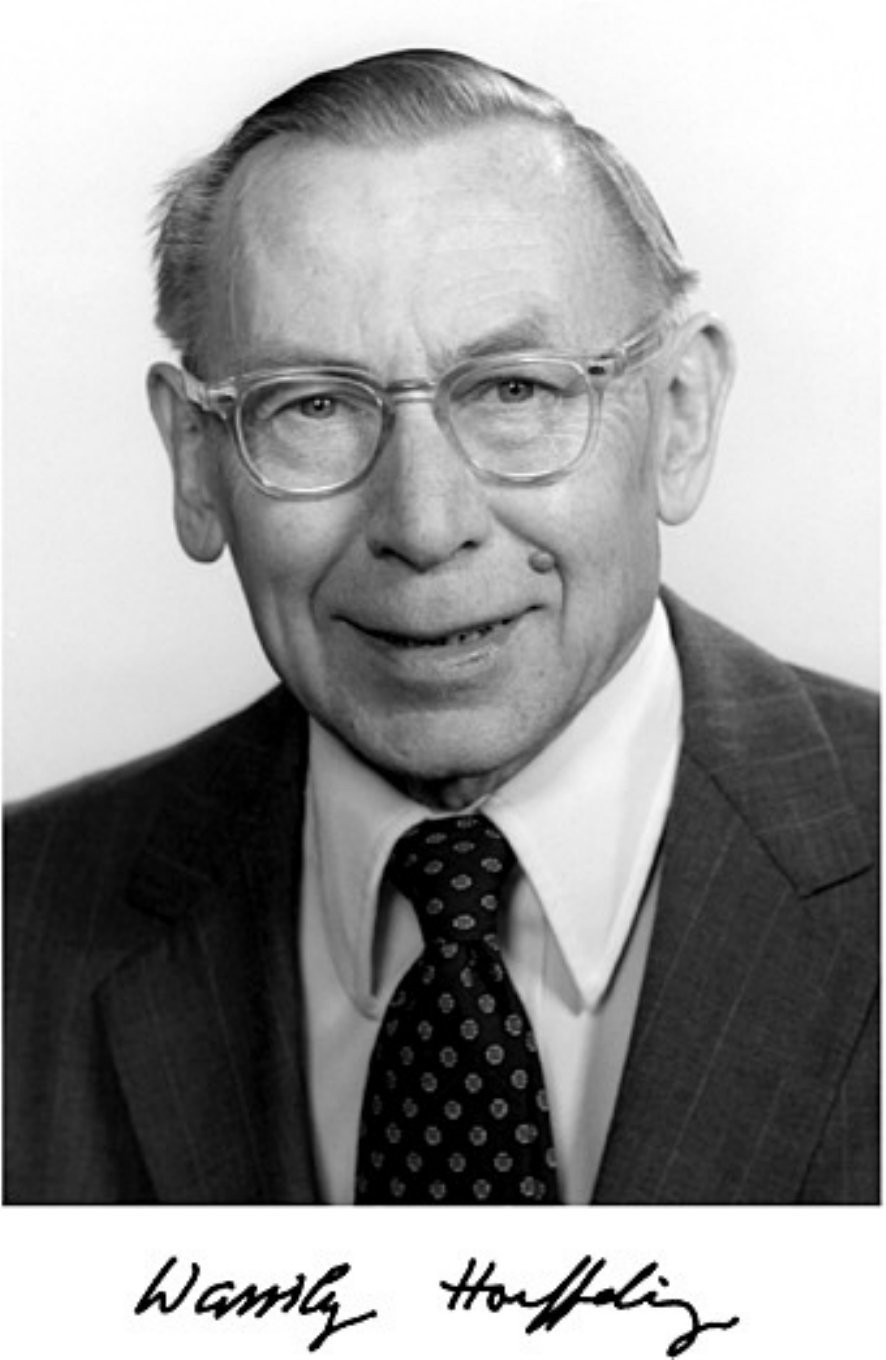}}\caption{Wassily Hoeffding, one of the founders of nonparametric statistics. Photo source: \href{http://www.nap.edu/read/11429/chapter/12}{nap.edu}}.\label{fig24}
\end{marginfigure}
If we set $\epsilon_{n,\alpha}=\sqrt{\frac{1}{2n}\log\frac{2}{\alpha}}$, then (\ref{eq:H}) is equivalent to
\begin{equation}
\mathbb{P}(\overline{X}_n-\epsilon_{n,\alpha}<p<\overline{X}_n+\epsilon_{n,\alpha})>1-\alpha,
\end{equation}
which means that $\overline{X}_n\pm \epsilon_{n,\alpha}$ is at least a $(1-\alpha)100\%$ confidence interval for $p$. \hfill $\square$\bigskip

Clearly, this method for constructing a confidence interval can be used only for data that can be modeled by the Bernoulli distribution. In general, many point estimates turn out to have, approximately, a normal distribution\footnote{Recall the normal approximation for the sample mean $\overline{X}_n$ in SRS.},
\begin{equation}
\frac{\hat{\theta}_n-\theta}{\mathrm{se}[\hat{\theta}_n]}~\dot{\sim}~\mathcal{N}(0,1).
\end{equation} 
This approximation can be used for constructing approximate confidence intervals. %\footnote{Construct a normal-based confidence interval for $p$ in the coin example, and compare it with the Hoeffding interval for different $\alpha$ and $n$.}. 

\subsection{Hypothesis Testing}
While we discussed estimation and confidence intervals in the context of survey sampling, hypothesis testing is something new for us. 

In hypothesis testing, we start with some default theory, called a \textit{null hypothesis}, and we ask if the data provides sufficient evidence to reject the theory. If yes, we reject it; if not, we accept it. 

\paragraph{Example:} Suppose we want to test if the coin is fair. Let $H_0$ denote the null hypothesis that the coin is fair, and let $H_1$ denote the \textit{alternative hypothesis} that the coin is not fair. Under the Bernoulli model, we can write the hypothesis as follows:
\begin{equation}
H_0: p=1/2 \hspace{5mm}\mbox{and}\hspace{5mm} H_1: p\neq 1/2.
\end{equation}
It seems reasonable to reject $H_0$ if  $|\overline{X}_n-1/2|$ is too large. When we discuss hypothesis testing in detail, we will be more
precise about how large the statistic $|\overline{X}_n-1/2|$ should be to reject $H_0$. \hfill $\square$

\section{Prediction, Classification, and Regression}

Suppose now that, in accordance with the schemes in the abstract, our data consists of pairs of observations: $(X_1, Y_1),\ldots, (X_n,Y_n)$, where $X_i$ is an ``input'' and $Y_i$ is the corresponding ``outcome''.  For example, $X_i$ is a the father's height, $Y_i$ in the son's height, and $i$ is the family number. 

The task of predicting the son's height $Y$  based on this father's height $X$ for a new family is called \textit{prediction}. In this context, $X$ is called a \textit{covariate}\footnote{It is also called a \textit{predictor} or \textit{regressor}, or \textit{feature}, or \textit{independent variable}.} and $Y$ is called a \textit{response variable}\footnote{It is also called an \textit{outcome variable} or \textit{dependent variable}.}. If $Y$ is discrete\footnote{For example, $X$ is the lab test results and $Y$ is the presence ($Y=1$) or absence ($Y=0$) of a certain disease.}, $Y\in\{1,\ldots,K\}$, then prediction is called \textit{classification} since it involves
assigning the observation $X$ to a certain class $Y$. 

\textit{Regression} is a method for studying the relationship between a response variable $Y$ and a covariate $X$. It is based on the so-called \textit{regression function}
\begin{equation}
r(x)=\mathbb{E}[Y|X=x],
\end{equation}
which is the expected value of the response given the value of the covariate. In regression, our goal is to estimate the regression function which then can be  used for prediction or classification.  If we \textit{assume} that $r(x)$ is a linear function, 
\begin{equation}
r(x)\in\mathcal{F}=\{r(x)=\beta_0+\beta_1x, \hspace{2mm} \beta_0,\beta_1\in\mathbb{R}\},
\end{equation}
then we have a \textit{liner regression model}. We will discuss regression in the last lectures.

\section{Further Reading}

\begin{enumerate}
	\item A thought provoking and stimulating paper by Leo Breiman ``\href{http://projecteuclid.org/euclid.ss/1009213726}{Statistical modeling: the two cultures},'' \textit{Statistical Science}, 16(3): 199-231., compares stochastic data modeling (which we discussed in this lecture and which is a mainstream in statistical research and practice) with algorithmic modeling which was developed outside statistics (in particular, in computer science) and does not assume any stochastic model for the data. See also the comments on the paper by D.R.~Cox, B.~Efron, B.~Hoadley, and E.~Parzen as well as the rejoinder by Breiman. 
\end{enumerate}

\section{What is Next?}
In the next lecture, we will start discussing the elements of nonparametric inference.

\chapter{Estimating the CDF and Statistical Functionals}
\label{ch:CDF}

\newthought{The basic idea} of nonparametric inference is to use data $X_1,\ldots,X_n$ to infer an unknown quantity of interest $\theta$ while making as few assumptions as possible. Mathematically, ``few assumptions'' means that the statistical model $\mathcal{F}$ used to model the data,
\begin{equation}
X_1,\ldots,X_n~\sim~F, \hspace{3mm} F\in\mathcal{F},
\end{equation}
is large, infinite-dimensional\footnote{A better name for nonparametric inference might be infinite-dimensional inference.}. Here we take $\mathcal{F}=\{\mbox{all CDFs}\}$.

In this lecture we will discuss one of the central problems in nonparametric inference: estimation of a parameter $\theta$ of $F$\footnote{Other problems include density estimation: given $X_1,\ldots,X_n~\sim~F$, estimate $f(x)=F'(x)$; and nonparametric regression: given $(X_1,Y_1),\ldots,(X_n,Y_n)$, estimate the regression function $r(x)=\mathbb{E}[Y|X=x]$. See L.A.~Wasserman, \textit{\href{http://www.stat.cmu.edu/~larry/all-of-nonpar/}{All of Nonparametric Statistics}}.}. Hold on. If $\dim\mathcal{F}=\infty$, the model $\mathcal{F}$ can't be parametrized by a finite number of parameters. So what do we mean by a ``parameter'' of $F$? Let us discuss this.

\section{Functionals and Parameters}
A \textit{statistical functional} is any function of the CDF,
\begin{equation}
t: \mathcal{F}\rightarrow\mathbb{R}, \hspace{5mm} \mathcal{F}\ni F\mapsto t(F)\in\mathbb{R}.
\end{equation}
A \textit{parameter} of a distribution $F$ is the value of a  functional $t$ on $F$, 
\begin{equation}
\theta=t(F).
\end{equation} Examples of $t$ and $\theta$ include:
\begin{enumerate}
	\item $t(F)=\int xdF(x)=\mu_F$, mean\footnote{Notation: if $F$ is discrete with probability mass function $p$, then $\int g(x)dF(x)=\sum g(x_i)p(x_i)$; if $F$ is continuous with PDF $f$, then $\int g(x)dF(x)=\int g(x)f(x)dx$.},
	\item $t(F)=\int (x-\mu_F)^2dF(x)=\sigma_F^2$, variance,
	\item $t(F)=\frac{\left(\int (x-\mu_F)^2dF(x)\right)^{1/2}}{\int xdF(x)}=\frac{\sigma_F}{\mu_F}=\delta_F$, coefficient of variation,
	\item $t(F)=F^{-1}(1/2)=m_F$, median,
	\item $t(F)=\frac{\int (x-\mu_F)^3dF(x)}{\left(\int (x-\mu_F)^3dF(x)\right)^{3/2}}=\kappa_F$, skewness.
\end{enumerate}

So, the problem is the following: given the data $X_1,\ldots,X_n\sim~F$, estimate a parameter of interest $\theta=t(F)$. In this context, the  functional $t$ is known, but $F$, and therefore $\theta$, are unknown. The basic idea is, first, to estimate the CDF, $\hat{F}\approx F$, and then estimate the parameter $\theta$ by $\hat{\theta}=t(\hat{F})$.  

\section{Estimating the CDF}
We will estimate $F$ with the \textit{empirical distribution function} (eCDF)\footnote{We have already encountered the eCDF in lecture 1, in the context of summarizing data. We saw that for the uniform distribution, eCDF$\approx$CDF, and noticed that this is not a coincidence. Here we will explain why this approximation holds for any distribution and why it is good for large $n$.}. Recall that 
the eCDF $\hat{F}_n$ of $X_1,\ldots,X_n$ is the CDF that puts mass $1/n$ at each data point $X_i$. More formally,   
\begin{equation}\label{eq:EDC}
\hat{F}_n(x)=\frac{1}{n}\sum_{i=1}^n H(x-X_i),
\end{equation}
where $H(x)$ is the Heaviside step function. The basic properties of the eCDF are described by the following theorem. 
\begin{theorem}
	For any fixed value of $x$, 
	\begin{enumerate}
		\item  $\hat{F}_n(x)$ is an unbiased estimate of $F(x)$:
		\begin{equation}
		\mathbb{E}[\hat{F}_n(x)]=F(x).
		\end{equation}
		\item The standard error of $\hat{F}_n(x)$ is given by
		\begin{equation}
		\mathrm{se}[\hat{F}_n(x)]=\sqrt{\frac{F(x)(1-F(x))}{n}}.
		\end{equation}
		\item The mean squared error of $\hat{F}_n(x)$ goes to zero as $n$ increases:
		\begin{equation}
		\mathrm{MSE}[\hat{F}_n(x)]\rightarrow0, \hspace{3mm}\mbox{as } n\rightarrow\infty.
		\end{equation}
	\end{enumerate}
\end{theorem}
An estimate $\hat{\theta}_n$ of a quantity of interest  $\theta$ is said to be \textit{consistent}, if it converges to $\theta$ in probability\footnote{See the Appendix at the end of this Lecture for a quick recap on different types of convergence.}:
\begin{equation}\label{eq:convergence}
\hat{\theta}_n\stackrel{\mathbb{P}}{\longrightarrow}\theta, \hspace{3mm}\mbox{as } n\rightarrow\infty.
\end{equation}
It turns out that if $\mathrm{MSE}[\hat{\theta}_n]\rightarrow0$, then an unbiased estimate $\hat{\theta}_n$ is a consistent estimate of $\theta$\footnote{This immediately follows from Chebyshev's inequality.}. Thus, we have:
\begin{theorem}
	For any $x$, $\hat{F}_n(x)$ is a consistent estimate of $F(x)$.
\end{theorem}

Intuitively, this means that, for any $x$, if $n$ is large enough, then $\hat{F}_n(x)$ is very close to $F(x)$ with large probability . This justifies our decision to estimate $F(x)$ with $\hat{F}_n(x)$. 

In fact, there are stronger results about the properties of $\hat{F}_n(x)$  which make it even a more attractive estimate for $F$. First, as it directly follows from the \textit{strong} law of large numbers, $\hat{F}_n(x)$ converges to $F(x)$ almost surely\footnote{Which is stronger than convergence in probability. Again, see the Appendix.},
\begin{equation}
\hat{F}_n(x)\stackrel{\mathrm{a.s.}}{\longrightarrow}F(x), \hspace{3mm}\mbox{as } n\rightarrow\infty.
\end{equation}
The Glivenko-Cantelli theorem strengths this pointwise result by proving the \textit{uniform} convergence: 
\begin{equation}\label{eq:GC}
\sup\limits_{x\in\mathbb{R}}\left|\hat{F}_n(x)-F(x)\right|\stackrel{\mathrm{a.s.}}{\longrightarrow}0, \hspace{3mm} \mbox{as } n\rightarrow\infty.
\end{equation}
Finally, the  Dvoretzky-Kiefer-Wolfowitz (DKW) inequality\footnote{Which strengthens the GK theorem.} says that the convergence in (\ref{eq:GC}) is fast: for any $\epsilon>0$,
\begin{equation}\label{eq:DKW}
\mathbb{P}\left(\sup\limits_{x\in\mathbb{R}}\left|\hat{F}_n(x)-F(x)\right|>\epsilon\right) \leq 2e^{-2n\epsilon^2}.
\end{equation}

The eCDF $\hat{F}_n(x)$ is a point estimate of $F$. The DKW inequality\footnote{Notice the similarity with the Hoeffding inequality.} allows to construct a confidence set for $F$. To construct a confidence set for $F$, we need to find two functions $F_l$ and $F_u$ (construct them from the data) such that 
\begin{equation}
\mathbb{P}(F_l(x)\leq F(x)\leq F_u(x) \mbox{ for all }x)=1-\alpha.
\end{equation}
The DKW inequality implies that we can take 
\begin{equation}
\begin{split}
F_l(x)&=\max\{\hat{F}_n(x)-\epsilon_{n,\alpha}, 0\},\\
F_u(x)&=\min\{\hat{F}_n(x)+\epsilon_{n,\alpha}, 1\},
\end{split}
\end{equation}
where $\epsilon_{n,\alpha}=\sqrt{\frac{1}{2n}\log\frac{2}{\alpha}}$. The set $\{y:  F_l(x)\leq y\leq F_u(x), x\in\mathbb{R}\}$ is called a nonparametric $(1-\alpha)$ \textit{confidence band}.

\section{Plug-In Principle}
The plug-in principle refers to replacing the unknown CDF $F$ with its empirical model $\hat{F}_n$. This principle can be readily used for constructing the \textit{plug-in estimate} of the parameter of interest $\theta=t(F)$:
\begin{equation}\label{eq:plug-in}
\hat{\theta}_n=t(\hat{F}_n).
\end{equation}
For example, if the functional $t$ has the following form:\footnote{Such functionals are called \textit{linear}, because $t(\alpha F+\beta G)=\alpha t(F)+\beta t(G)$.}
\begin{equation}\label{eq:linear functional}
t(F)=\int a(x)dF(x),
\end{equation} 
then the plug-in estimate of $\theta=t(F)$ is simply
\begin{equation}
\hat{\theta}_n=\int a(x)d\hat{F}_n(x)=\frac{1}{n}\sum_{i=1}^na(X_i).
\end{equation}

\paragraph{Example:} The plug-in estimate for the mean is $\hat{\mu}_n=\overline{X}_n$. The plug-in estimator for the variance is $\hat{\sigma}_n^2=\frac{1}{n}\sum_{i=1}^n(X_i-\overline{X}_n)^2$. Note, that $\hat{\sigma}_n^2$ is biased. The unbiased estimate is $s_n^2=\frac{1}{n-1}\sum_{i=1}^n(X_i-\overline{X}_n)^2$. But in practice, there is little difference between $\hat{\sigma}_n^2$ and $s_n^2$. \hfill $\square$\bigskip

Note that using the plug-in principle may not be the best idea\footnote{Nevertheless, it can still be used as a benchmark.} in situations where there is some additional information about $F$ other than provided by the sample $X_1,\ldots,X_n$. In such cases, a better estimate of $F$ may be available. For example, if $F$ is a member of a parametric family $\mathcal{F}=\{F(x;\beta)\}$, replacing $F(x)$ with $F(x;\hat{\beta})$, where $\hat{\beta}$ is an estimate of parameter $\beta$, may be better than replacing it with $\hat{F}_n$. In other words, the estimate $\hat{\theta}=t(F(x;\hat{\beta}))$ may be more accurate than the plug-in estimate $\hat{\theta}_n=t(\hat{F}_n)$.

\section{Further Reading}

\begin{enumerate}
	\item A quick and nice introduction to nonparametric statistics is given in  L.A.~Wasserman (2006), \textit{\href{http://www.stat.cmu.edu/~larry/all-of-nonpar/}{All of Nonparametric Statistics}}.
\end{enumerate}

\section{What is Next?}
We learned how to estimate a parameter of interest non-parametrically using the plug-in principle, $\hat{\theta}_n\approx\theta$, but we saw that a plug-in estimate may be biased (\eg for the variance). In the next lecture, we will learn how to reduce the bias using the jackknife method.

\section{Appendix: Convergence of a Sequence of Random Variables} 
One of the most important questions in probability theory concerns the behavior of sequences of random variables\footnote{This part of probability is called \textit{large sample	theory} or \textit{limit theory} or \textit{asymptotic theory}. It is very important for
	statistical inference.}. The basic question is this: what can we say about the limiting behavior of a sequence of random variables $X_1,\ldots,X_n,\ldots$? In the statistical context, this question can be reformulated as what happens as we gather more and more data?

In Calculus, we say that a sequence of real numbers $x_1, x_2,\ldots$ converges to a limit $x$ if, for every $\epsilon>0$, we can find $N$ such that $|x_n-x|<\epsilon$ for all $n>N$.  In Probability, convergence is more subtle. 

\paragraph{Example:} Suppose that $x_n=1/n$. Then trivially, $\lim_{n\rightarrow\infty}x_n=0$. Consider a
probabilistic version of this example: suppose  that
$X_1,X_2,\ldots$ are independent and $X_n\sim \mathcal{N}(0,1/n)$.
Intuitively, $X_n$ is very concentrated around $0$ for large $n$,
and we are tempted to say that $X_n$ ``converges'' to zero. However,
$\mathbb{P}(X_n=0)=0$ for all $n$! \hfill $\square$\bigskip

There are several types of convergence of a sequence of random variables. One is convergence \textit{in probability}\footnote{This is the type of convergence used in the \textit{weak} law of large numbers: if $X_1,\ldots,X_n\sim F$, then $\overline{X}_n\stackrel{\mathbb{P}}{\longrightarrow}\mu_F$.}. We say that a sequence $\{X_n\}$ converges to a random variable $X$ in probability, written $X_n\stackrel{\mathbb{P}}{\longrightarrow}X$, if for every $\epsilon>0$,
\begin{equation}\label{eq:P}
\lim_{n\rightarrow\infty}\mathbb{P}(|X_n-X|<\epsilon)=1,
\end{equation}
or, in more detail, 
\begin{equation}
\lim_{n\rightarrow\infty}\mathbb{P}(\{\omega\in\Omega: |X_n(\omega)-X(\omega)|<\epsilon\})=1,
\end{equation}
where $\Omega$ is the sample space.  Note that for every $\epsilon$, the sequence $\{\mathbb{P}(|X_n-X|>\epsilon)\}$ is a sequence of numbers, and when  we say  it has zero limit, we understand the limit in the calculus sense. Intuitively, convergence in probability means that, when $n$ is large,  realizations of $X_n$ are very close to the realizations of $X$ with high probability. 

Another important type of convergence is convergence \textit{almost surely}\footnote{This is the type of convergence used in the \textit{strong} law of large numbers: if $X_1,\ldots,X_n\sim F$, then $\overline{X}_n\stackrel{\mathrm{a.s.}}{\longrightarrow}\mu_F$.}. We say that a sequence $\{X_n\}$ converges to a random variable $X$ almost surely, written $X_n\stackrel{\mathrm{a.s.}}{\longrightarrow}X$, if 
\begin{equation}\label{eq:as}
\mathbb{P}(\lim_{n\rightarrow\infty}X_n=X)=1,
\end{equation}
or, in more detail, 
\begin{equation}
\mathbb{P}(\{\omega\in\Omega: \lim_{n\rightarrow\infty}X_n(\omega)=X(\omega))=1.
\end{equation}
Almost sure convergence is stronger, meaning that it implies convergence in probability\footnote{By Fatou's lemma.}
\begin{equation}
X_n\stackrel{\mathrm{a.s.}}{\longrightarrow}X \hspace{3mm}\Rightarrow\hspace{3mm}X_n\stackrel{\mathbb{P}}{\longrightarrow}X.
\end{equation}

There are other types of convergence, \eg convergence in distribution, convergence distribution, convergence in mean, in quadratic mean, but we don't need them here.

\chapter{The Jackknife Method}

	\newthought{The jackknife method} was originally proposed by Maurice Quenouille\footnote{Despite the remarkable influence of the jackknife on the statistical community, I could not find a photo of its inventor on the Internet! Please let me know if you do.} (1949) for estimating the bias of an estimator. A bit later, John Tukey (1956) extended the use the method by showing how to use it for reducing the bias and estimating the variance, and coined the name ``jackknife.'' As a pocket knife, this technique can be used as a ``quick and dirty'' tool that can solve a variety of problems.  
	\begin{marginfigure}
		%\vspace{-4mm}
		\includegraphics[width=\linewidth]{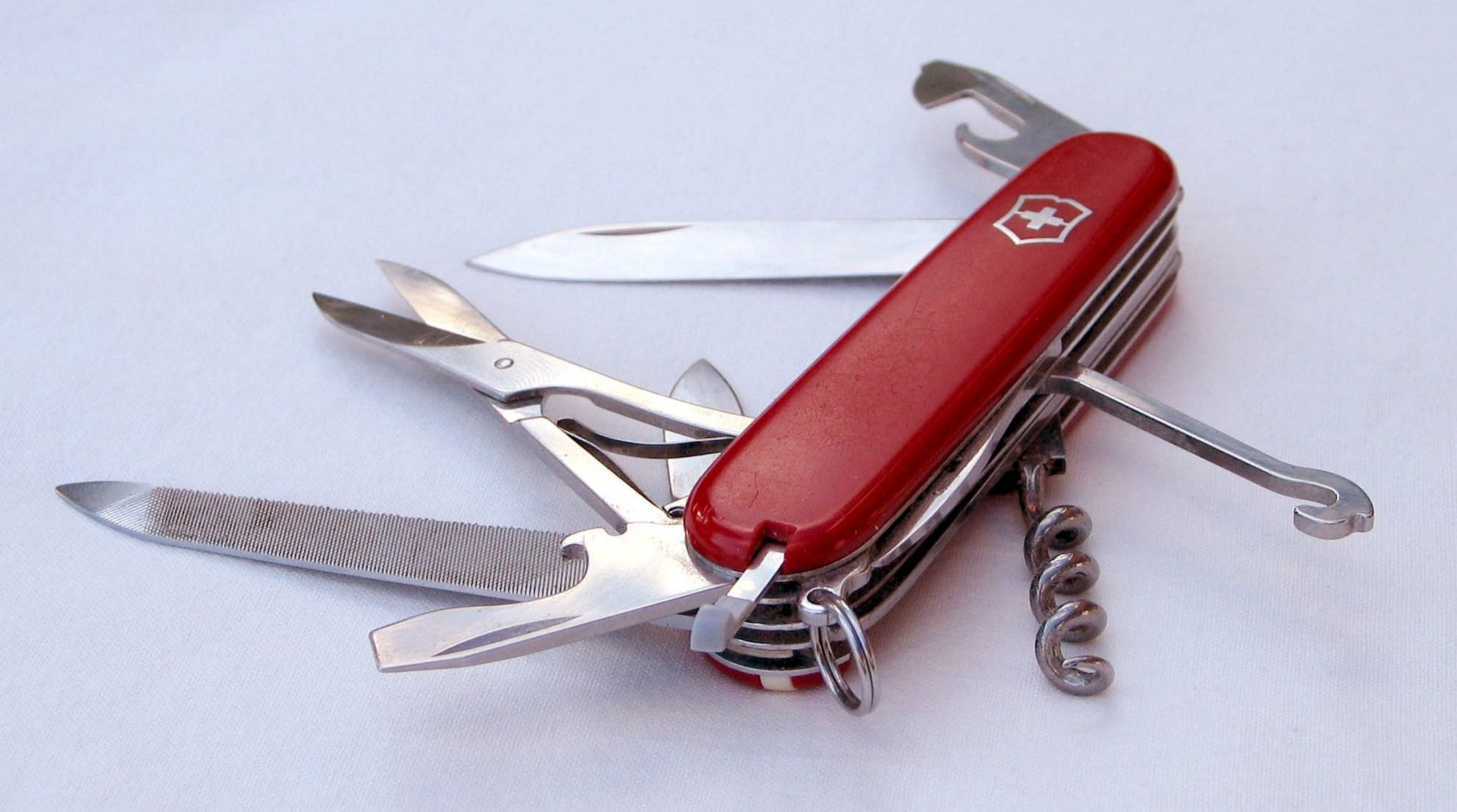}
		\caption{A Victorinox Swiss Army knife. Photo source: \href{https://en.wikipedia.org/wiki/Pocketknife}{wikipedia.org}.}
		\label{fig:jackknife}
	\end{marginfigure}
	
\section{Estimating the Bias}
So, let $X_1,\ldots,X_n$ be the data which is modeled as a sample from a distribution $F$, and let 
\begin{equation}
\hat{\theta}_n=s(X_1,\ldots,X_n)
\end{equation} be an estimate of a parameter of interest $\theta=t(F)$. For example, $\hat{\theta}_n$ could be the plug-in estimate $\hat{\theta}_n=t(\hat{F}_n)$\footnote{But not necessarily. It could be essentially any statistic $s$ that estimate a quantity of interest $\theta$.}. In practice, many estimates are biased. Our ``working'' example is the plug-in estimate of the variance. 

\paragraph{Example:} If $\theta=\sigma^2_F$, then the plug-in estimate is 
\begin{equation}
\hat{\sigma}^2_n=\frac{1}{n}\sum_{i=1}^n(X_i-\overline{X}_n)^2,
\end{equation}
and its bias\footnote{For brevity, let's denote the bias by $\mathbb{B}$.} is 
\begin{equation}\label{eq:bias sigma}
\mathbb{B}[\hat{\sigma}^2_n]=-\frac{\sigma_F^2}{n}. 
\end{equation}
\hfill $\square$ \bigskip

In general, the bias of an estimate $\hat{\theta}_n$ is 
\begin{equation}\label{eq:bias}
\mathbb{B}[\hat{\theta}_n]=\mathbb{E}[\hat{\theta}_n]-\theta.
\end{equation}

\textit{Question:} How to estimate the bias?\footnote{In stead of immediately giving you a ready-to-use formula (like in most textbooks), let me try to provide the intuition behind the jackknife.}

Well, we can estimate $\theta$ simply by $\hat{\theta}_n$. And if we had $m$ iid samples from $F$:
\begin{equation}\label{eq:ideal}
\begin{split}
&X_1^{(1)},\ldots,X_n^{(1)}\sim F,\\
&\ldots\\
&X_1^{(m)},\ldots,X_n^{(m)}\sim F,
\end{split}
\end{equation}
we could, using the law of large numbers, estimate $\mathbb{E}[\hat{\theta}_n]$ by the sample mean:
\begin{equation}\label{eq:ideal2}
\mathbb{E}[\hat{\theta}_n]\approx \frac{1}{m}\sum_{i=1}^m \hat{\theta}_n^{(i)},
\end{equation}
where $\hat{\theta}_n^{(i)}=s(X_1^{(i)},\ldots,X_n^{(i)})$. In particular, if $\hat{\theta}_n$ is the plug-in estimator, then $\hat{\theta}_n^{(i)}=t(\hat{F}_n^{(i)})$, where $\hat{F}_n^{(i)}$ is the eCDF constructed from $X_1^{(i)},\ldots,X_n^{(i)}$. The problem is that we don't have samples (\ref{eq:ideal}). We have only one data set $X_1,\ldots,X_n$\footnote{Recall that the same problem we faced in lecture~\ref{ch:PopVarandBoot}, where we discussed the bootstrap method. The bootstrap approach was to create multiple copies of $X_i$s to mimic the target population. The jackknife also reuses the data, but using a slightly different strategy.}.

The key idea of the jackknife is to emulate (\ref{eq:ideal}) by cooking up $n$ samples of size $n-1$ from the original data by leaving one data point $X_i$ out at a time\footnote{That is why the jackknife is also called a ``leave one out'' procedure.}:
\begin{equation}\label{eq:jack_samples}
\begin{split}
&X_2,\ldots,X_n\sim F,\\
&\ldots\\
&X_1,\ldots,X_{i-1}, X_{i+1},\ldots,X_n\sim F,\\
&\ldots\\
&X_1,\ldots,X_{n-1}\sim F.
\end{split}
\end{equation}
These samples are called the \textit{jackknife samples}. Based on these samples, we compute the \textit{jackknife replications} of $\hat{\theta}_n$: 
\begin{equation}
\hat{\theta}_n^{(-i)}=s(X_1,\ldots,X_{i-1}, X_{i+1},\ldots,X_n).
\end{equation}
Now, similar to (\ref{eq:ideal2}), we can estimate $\mathbb{E}[\hat{\theta}_n]$ by the sample mean of the jackknife replications 
\begin{equation}
\mathbb{E}[\hat{\theta}_n]\approx\bar{\theta}^{J}_n=\frac{1}{n}\sum_{i=1}^n \hat{\theta}_n^{(-i)}.
\end{equation} 
The bias of $\hat{\theta}_n$ in (\ref{eq:bias}) can then be estimated by 
\begin{equation}\label{eq:wrong}
\mathbb{B}[\hat{\theta}_n]\approx\bar{\theta}^{J}_n-\hat{\theta}_n.
\end{equation} 

While intuitively this may feel correct, we have at least two concrete  problems\footnote{In fact, (\ref{eq:wrong}), as it is, is wrong. To make it a good approximation, we need to slightly modify the right-hand side.}:\\
a) the jackknife replications are based on the samples (\ref{eq:jack_samples}) of size $n-1$, not $n$, and therefore $\bar{\theta}^{J}_n$ is more like an estimate of $\mathbb{E}[\hat{\theta}_{n-1}]$,\\
b) more importantly, the jackknife replications $\hat{\theta}_n^{(-i)}$ are not independent, in fact, they are very dependent since any two jackknife samples differ only in two data points. 

It turns out however, that this method will work if we make an assumption about the bias of our estimate $\hat{\theta}_n$: suppose that 
\begin{equation}\label{eq:assumptionJ}
\mathbb{B}[\hat{\theta}_n]=\frac{a}{n}+\frac{b}{n^2}+O\left(\frac{1}{n^3}\right)\hspace{5mm}\mbox{as } n\rightarrow\infty,
\end{equation}
where $a$ and $b$ are some constants. In fact, many estimates have this property, so this assumption is not very strong. For example, (\ref{eq:assumptionJ}) holds for the plug-in variance estimate (\ref{eq:bias sigma}) with $a=-\sigma_F^2$ and $b=0$.

It is straightforward to show\footnote{Here we go: [\href{http://www.its.caltech.edu/~zuev/teaching/2016Winter/bonus.pdf}{pdf}]} that under (\ref{eq:assumptionJ}),
\begin{equation}
\mathbb{E}[\bar{\theta}^{J}_n-\hat{\theta}_n]=\frac{a}{n(n-1)}+\frac{(2n-1)b}{n^2(n-1)^2}+O\left(\frac{1}{n^3}\right)\hspace{5mm}\mbox{as } n\rightarrow\infty.
\end{equation}
This means that we need just to properly rescale $\bar{\theta}^{J}_n-\hat{\theta}_n$ to get an estimate of the bias of $\hat{\theta}_n$. The \textit{jackknife estimate} of the bias is 
\begin{equation}\label{eq:jbias}
\mathbb{B}[\hat{\theta}_n]\approx\hat{\mathbb{B}}_{\mathrm{J}}[\hat{\theta}_n]=(n-1)(\bar{\theta}^{J}_n-\hat{\theta}_n).
\end{equation}
It estimates the bias up to order $O(n^{-2})$:
\begin{equation}\label{eq:EB}
\mathbb{E}[\hat{\mathbb{B}}_{\mathrm{J}}[\hat{\theta}_n]]=\frac{a}{n}+O\left(\frac{1}{n^2}\right)=\mathbb{B}[\hat{\theta}_n]+O\left(\frac{1}{n^2}\right).
\end{equation}

\section{Reducing the Bias}

It is now clear how to reduce the bias of the estimate $\hat{\theta}_n$\footnote{Under the assumption that its bias satisfies (\ref{eq:assumption}).}: we just need to subtract from $\hat{\theta}_n$ its estimated bias\footnote{Careful with notation: $\hat{\theta}_n^J$ and $\bar{\theta}^{J}_n$ are different animals!}:
\begin{equation}\label{eq:J}
\hat{\theta}_n^J=\hat{\theta}_n-\hat{\mathbb{B}}_{\mathrm{J}}[\hat{\theta}_n]=n\hat{\theta}_n-(n-1)\bar{\theta}^{J}_n.
\end{equation}
Using (\ref{eq:EB}), we have:
\begin{equation}
\mathbb{B}[\hat{\theta}_n^J]=\mathbb{E}[\hat{\theta}_n]-\mathbb{E}[\hat{\mathbb{B}}_{\mathrm{J}}[\hat{\theta}_n]]-\theta=O\left(\frac{1}{n^2}\right).
\end{equation}
The bias of $\hat{\theta}_n^J$ is therefore an \textit{order magnitude} smaller than that of $\hat{\theta}_n$.The jackknifed estimate (\ref{eq:J})  is also called the \textit{bias-corrected} estimate. 
An important remark: if the original estimate $\hat{\theta}_n$ is unbiased, then so is the jackknifed estimate:
\begin{equation}
\mathbb{E}[\hat{\theta}_n]=\theta\hspace{2mm}\Rightarrow\hspace{2mm}
\mathbb{E}[\hat{\theta}^J_n]=n\theta-(n-1)\theta=\theta.
\end{equation}

\paragraph{Example:} It can be shown that the bias-corrected estimate of the plug-in estimate of the variance $\hat{\sigma}_n^2$ is simply the usual unbiased estimate:
\begin{equation}
\hat{\sigma}^{2,J}_n=s^2_n=\frac{1}{n-1}\sum_{i=1}^n(X_i-\overline{X}_n)^2.
\end{equation}
\hfill $\square$ \bigskip

Let us look at the definition of the bias-corrected estimate (\ref{eq:J}). It is a linear combination of the original estimate and the mean of its jackknife replications. There is another way to think about the jackknife.

\section{Pseudo-Values}

A straightforward manipulation with (\ref{eq:J}) leads to 
\begin{equation}\label{eq:Jps}
\hat{\theta}_n^J=\frac{1}{n}\sum_{i=1}^n\tilde{\theta}_n^{(i)},
\end{equation}
where 
\begin{equation}
\tilde{\theta}_n^{(i)}=n\hat{\theta}_n-(n-1)\hat{\theta}_n^{(-i)}
\end{equation}
are called \textit{pseudo-values}.
The idea behind pseudo-values is that they allow us to write the bias-corrected estimate as a mean of $n$ ``independent'' data values\footnote{Expect that in general, pseudo-values are not independent.}. Let us consider a couple of examples.

\paragraph{Example:} If $\theta=\mu_F$ and the plug-in estimate is the sample mean $\hat{\theta}_n=\overline{X}_n$, then the pseudo-values are simply $X_i$:
\begin{equation}
\tilde{\theta}_n^{(i)}=\sum_{i=1}^nX_i-\sum_{j\neq i}^nX_j=X_i.
\end{equation}\vspace{-7mm}

\hfill $\square$
\paragraph{Example:} In a more general case of a linear functional $t(F)=\int a(x)dF(x)$, the plug-in estimate is   $\hat{\theta}_n=s(X_1,\ldots,X_n)=\frac{1}{n}\sum_{i=1}^na(X_i)$. The pseudo-values are then 
\begin{equation}
\tilde{\theta}_n^{(i)}=\sum_{i=1}^na(X_i)-\sum_{j\neq i}^na(X_j)=a(X_i).
\end{equation}
This means, in particular, that for linear functionals, the jackknifed estimated coincides with the plug-in estimate, $\hat{\theta}_n^J=\hat{\theta}_n$\footnote{This is expected, since the plug-in estimate $\hat{\theta}_n=\frac{1}{n}\sum_{i=1}^na(X_i)$ is an unbiased estimate of $\theta$.}.
\hfill $\square$\bigskip

In both examples, the pseudo-values are indeed independent. Based on this, Tukey suggested that in general case, we can treat the pseudo-values $\tilde{\theta}_n^{(i)}$ as liner approximations to iid  observations:
\begin{equation}
\mbox{if} \hspace{2mm}\hat{\theta}_n=s(X_1,\ldots,X_n)\approx\frac{1}{n}\sum_{i=1}^na(X_i) \hspace{2mm}\Rightarrow\hspace{2mm} \tilde{\theta}_n^{(i)}\approx a(X_i).
\end{equation}

\section{Estimating the Variance}

Following Tukey's idea of treating the pseudo-values as iid random variables allows to estimate the variance of the bias-corrected estimate $\hat{\theta}_n^{J}$. Indeed, if $\tilde{\theta}_n^{(i)}$ are iid, then from (\ref{eq:Jps}), we have:
\begin{equation}
\mathbb{V}[\hat{\theta}_n^{J}]=\frac{\mathbb{V}[\tilde{\theta}_n^{(i)}]}{n}\approx\frac{\tilde{s}_n^2}{n}=:v_J,
\end{equation}
where $\tilde{s}_n^2$ is the sample variance of the pseudo-values,
\begin{equation}
\tilde{s}_n^2=\frac{1}{n-1}\sum_{i=1}^n(\tilde{\theta}_n^{(i)}-\hat{\theta}_n^J)^2.
\end{equation}
It turns out that under suitable conditions on  statistic $s$, $v_J$ consistently estimates the variance of the original estimate $\hat{\theta}_n=s(X_1,\ldots,X_n)$,
\begin{equation}
v_J\stackrel{\mathbb{P}}{\longrightarrow}\mathbb{V}[\hat{\theta}_n].
\end{equation}
However, there are cases where $v_{J}$ is not a good estimate for the variance of an estimate. This happens when $\hat{\theta}_n$ is not a smooth function of the data $X_1,\ldots,X_n$. For example, if $\hat{\theta}_n$ is the plug-in estimate for the median,  $v_{J}$ is a poor estimate for its variance. 

\section{Further Reading}

\begin{enumerate}
	\item A brief description of the jackknife together with a summary of its underlying theory, advantages, disadvantages, and its general properties is given in Bissell \& Fergusun (1975) ``\href{http://www.jstor.org/stable/2987663}{The Jackknife --- Toy, Tool or Two-edged Weapon?}'' \textit{The Statistician}, 24(2): 79-100.
\end{enumerate}

\section{What is Next?}
We learned how to estimate a parameter of interest non-parametrically using the plug-in principle, $\hat{\theta}_n\approx\theta$, how to reduce its bias and even estimate its variance using the jackknife. On the other hand, we saw that the jackknife works only under appropriate assumptions\footnote{We often hold in practice, but rarely verifiable.}. In the next lecture, we will discuss the \textit{bootstrap method}, which was inspired by the jackknife, and which is a superior technique and can be used pretty much anywhere jackknifing can be used. In some sense the jackknife can be viewed as a linear approximation of the bootstrap. We will learn how to estimate the standard error of an estimate $\hat{\theta}_n$ and how to construct confidence intervals for the parameter of interest $\theta$ using the bootstrap.

\chapter{The Bootstrap Method}\label{ch:Bootstrap}

\newthought{As before}, let $X_1,\ldots,X_n$ be data which we model nonparametrically as a sample from a distribution $F\in\mathcal{F}$, where the statistical model  $\mathcal{F}=\{\mbox{all CDFs}\}$.  Let $\hat{\theta}_n=s(X_1,\ldots,X_n)$ be an estimate of a parameter of interest $\theta=t(F)$ calculated from the data (\eg using the plug-in principle), where $t$ is a given functional. In this lecture, our focus is on the following \\
\textit{Question:} How accurate is $\hat{\theta}_n$? What is its standard error? How to construct a confidence interval for $\theta$?

These questions can be answered by using the bootstrap method\footnote{We already encountered the bootstrap in Lecture~\ref{ch:PopVarandBoot} in the context of survey sampling. Here we will discuss this general method in detail.}. The bootstrap, introduced by Bradley Efron\footnote{Who was inspired by the success of the jackknife, B. Efron (1979) ``\href{http://projecteuclid.org/euclid.aos/1176344552}{Bootstrap methods: another look at the jackknife},'' \textit{The Annals of Statistics}, 7:	1-26.}, is a simulation-based method for measuring uncertainty of an estimate, in particular, for estimating standard errors  and constructing confidence intervals. Its beauty lies in its simplicity and universality: the bootstrap is fully automatic, requires no theoretical calculations, and always available.

\section{Bootstrapping the Standard Error}
If  $\hat{\theta}_n=s(X_1,\ldots,X_n)$ is unbiased\footnote{Recall that $\hat{\theta}$ is unbiased if $\mathbb{E}[\hat{\theta}_n]=\theta$. Plug-in estimates $\hat{\theta}=t(\hat{F}_n)$ are not necessarily unbiased (what if the corresponding functional is linear?), but they tend to have small biases compared to the magnitude of their standard errors.}, then the most common way to assess its statistical accuracy  is to compute the standard error of $\hat{\theta}_n$:
\begin{equation}\label{eq:se2}
\begin{split}
\mathrm{se}_F[\hat{\theta}_n]&=\left(\mathbb{V}_F[\hat{\theta}_n]\right)^{1/2},\\
\mathbb{V}_F[\hat{\theta}_n]&=\int\cdots\int\left(s(x_1,\ldots,x_n)-\mathbb{E}_F[\hat{\theta}_n]\right)^2dF(x_1)\ldots dF(x_n),\\
\mathbb{E}_F[\hat{\theta}_n]&=\idotsint s(x_1,\ldots,x_n)dF(x_1)\ldots dF(x_n).
\end{split}
\end{equation}
We intentionally use the subscript $F$ to emphasize that the standard error, variance, and mean of $\hat{\theta}_n$ do depend on $F$, which is unknown.  

\textit{Question:} How to estimate $\mathrm{se}_F[\hat{\theta}_n]$?

\textit{Bootstrap:} The \textit{ideal} bootstrap estimate of $\mathrm{se}_F[\hat{\theta}_n]$ is a plug-in estimate that uses $\hat{F}_n$ in place of $F$:
\begin{equation}\label{eq:b-se}
\widehat{\mathrm{se}}_B^{\mathrm{ideal}}[\hat{\theta}_n]:=\mathrm{se}_{\hat{F}_n}[\hat{\theta}_n].
\end{equation}

\paragraph{Example:}  Let the parameter of interest $\theta=t(F)$ be the mean $\mu_F$, and the estimate $\hat{\theta}_n$ be the plug-in estimate, $\hat{\theta}_n=t(\hat{F}_n)=\overline{X}_n$. The standard error is then
\begin{equation}\label{eq:se for mean}
\mathrm{se}_F[\hat{\theta}_n]=\left(\mathbb{V}_F\left[\frac{1}{n}\sum_{i=1}^nX_i\right]\right)^{1/2}=\left(\frac{1}{n^2}\sum_{i=1}^n\mathbb{V}_F\left[X_i\right]\right)^{1/2}=\frac{\sigma_F}{\sqrt{n}}.
\end{equation}
The ideal bootstrap estimate is therefore
\begin{equation}\label{eq:b-se-mean}
\widehat{\mathrm{se}}_F[\hat{\theta}_n]=\frac{\sigma_{\hat{F}_n}}{\sqrt{n}}=\frac{\sqrt{\frac{1}{n}\sum_{i=1}^n(X_i-\overline{X}_n)^2}}{\sqrt{n}}=\frac{1}{n}\left(\sum_{i=1}^n(X_i-\overline{X}_n)^2\right)^{1/2}\hfill 
\end{equation}
\hfill $\square$ %\bigskip

This example is very special because essentially only for $\hat{\theta}_n=\overline{X}_n$ explicit calculations in (\ref{eq:se for mean}) are possible.\footnote{Convince yourself by considering other four functionals discussed in Lecture~\ref{ch:CDF}.} Usually the ideal bootstrap estimate (\ref{eq:b-se}) cannot be computed exactly like in (\ref{eq:b-se-mean}) and some approximations are needed.

\subsection{Monte Carlo Simulation}

We can readily compute the approximate numeric value of the bootstrap estimate $\mathrm{se}_{\hat{F}_n}[\hat{\theta}_n]$ by Monte Carlo simulation (\ie using the law of large numbers):
\begin{enumerate}
	\item For $b=1,\ldots,B$, generate a \textit{bootstrap sample} $X_1^{(b)},\ldots,X_n^{(b)}\sim\hat{F}_n$\footnote{How to sample form $\hat{F}_n$? If $X_1,\ldots,X_n$ are all distinct, how many distinct bootstrap samples?} and compute the \textit{bootstrap replication} of $\hat{\theta}_n$, $\hat{\theta}_n^{(b)}=s(X_1^{(b)},\ldots,X_n^{(b)})$.
	\item Estimate $\mathrm{se}_{\hat{F}_n}[\hat{\theta}_n]$ by the sample standard deviation of the $B$ replications:
	\begin{equation}\label{eq:bootstrap_se}
	\widehat{\mathrm{se}}_{B}[\hat{\theta}_n]=\left(\frac{1}{B}\sum_{b=1}^B\left(\hat{\theta}_n^{(b)}-\frac{1}{B}\sum_{b=1}^B\hat{\theta}_n^{(b)}\right)^2\right)^{1/2}.
	\end{equation}
\end{enumerate}
By the law of large numbers, when $B$ is large,  $\widehat{\mathrm{se}}_{B}[\hat{\theta}_n]\approx\mathrm{se}_{\hat{F}_n}[\hat{\theta}_n]$\footnote{The ideal bootstrap estimate $\mathrm{se}_{\hat{F}_n}[\hat{\theta}_n]$ and its Monte Carlo approximation $\widehat{\mathrm{se}}_{B}$ are called  \textit{nonparametric bootstrap} estimates.}.  

\subsection{Schematic Illustration of Nonparametric Bootstrap}
%\vspace{-3mm}
\begin{figure} %[h]
	\includegraphics[width=\linewidth]{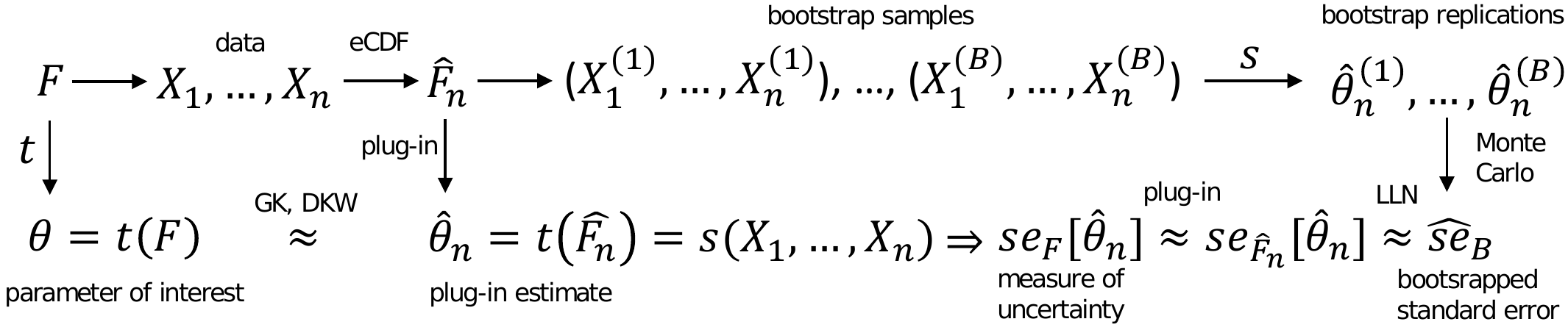}%
\end{figure}
\begin{marginfigure}%
	\vspace{-12mm}
	\includegraphics[width=\linewidth]{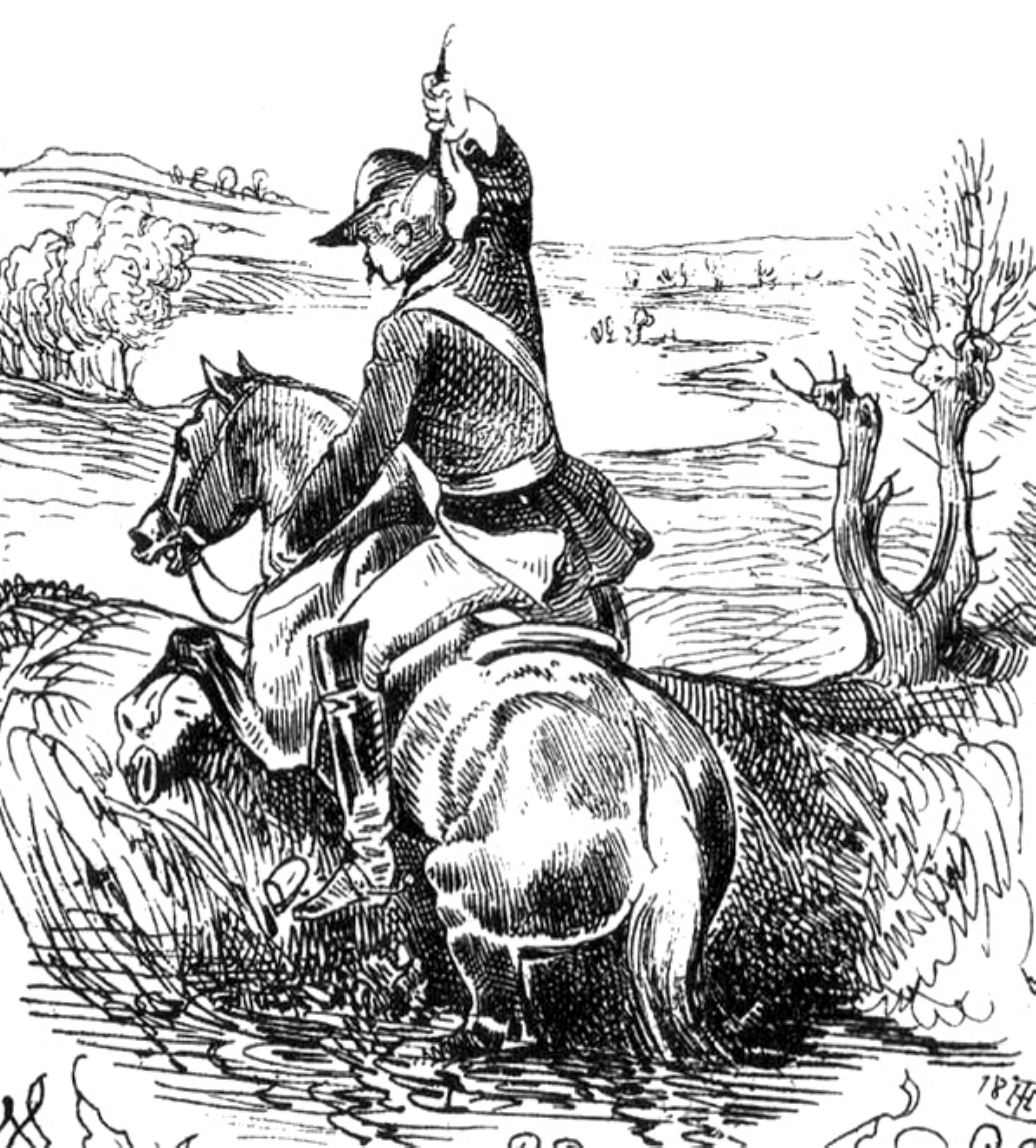}
	\caption{Bootstrap at work. Image source: [\href{https://www.lhup.edu/~dsimanek/museum/themes/BaronMunch.jpg}{jpg}].}
	\label{fig:baron}
\end{marginfigure}
Efron called this method ``bootstrap'' since using data to estimate the uncertainty of an estimate computed from these same  data is akin to the Baron Munchausen's method for getting himself out of a bog by lifting himself by his bootstraps\footnote{In the original version of this tale, Baron lifted himself and his horse by pulling his own hair (Fig.~\ref{fig:baron}).}. It is worth mentioning that in his original paper, Efron was considering even more colorful names such as ``Swiss Army Knife'' and ``Shotgun.''

\section{Errors}

There are two sources of approximation error in  bootstrap:
\begin{itemize}
	\item \textit{Statistical error:} the empirical distribution $\hat{F}_n$ is not exactly the same as the true data-generating process $F$. But they get closer and closer as we have more data (as $n$ increases). 
	\item \textit{Simulation error:}  occurs from using finitely many bootstrap replications $\hat{\theta}_n^{(1)},\ldots,\hat{\theta}_n^{(B)}$. It can be made arbitrary small simply by brute force: take $B$ very large. 
\end{itemize}
Usually we have more control over the simulation error (it is up to us what $B$ to use) than the statistical error (typically data $X_1,\ldots,X_n$ are given and we cannot\footnote{Or it is very expensive.} collect more). In complex models, however, statistic $s$ may be a complicated function of data, and its computation may be time-consuming. It is essential then to reduce the number of $s$-evaluations and the following question becomes relevant:

\textit{Question:} How many replications $B$ should we use?

This is thoroughly discussed by Efron and Tibshirani\footnote{B. Efron \& R.J. Tibshirani (1993), \textit{\href{https://books.google.com/books?id=gLlpIUxRntoC}{An Introduction to the Bootstrap}}.}. In Chapter 19, the formula for the coefficient of variation of $\widehat{\mathrm{se}}_{B}$ is derived which leads to the following rule of thumb: it is very rare when more than $B=200$ replications are needed for estimating a standard error.\footnote{For bootstrap confidence intervals much bigger values are required.} 

A take-home message: the statistical error is larger than the simulation error, provided that the Monte Carlo sampling is properly designed. 

\section{Bootstrap Confidence Intervals}
Recall that a \textit{confidence interval} for a parameter $\theta=t(F)$ is \textit{random} interval $\mathcal{I}$ calculated from the sample $X_1,\ldots,X_n$ that contains $\theta$ with high probability (confidence level\footnote{Confidence level is also called \textit{coverage probability}.}) $1-\alpha$, 
\begin{equation}
\mathbb{P}(\theta\in\mathcal{I})=1-\alpha.
\end{equation}
A point and interval estimates  of  $\theta$  provide the  guess for $\theta$ and how far in error that guess might reasonably be.

Let $\hat{\theta}_n=t(\hat{F}_n)$ be the plug-in estimate and $\widehat{\mathrm{se}}_B$ be the bootstrap estimate of its standard error. There are several ways to construct bootstrap confidence intervals. The simplest and most straightforward is the \textit{normal interval}.\footnote{Sometimes it is called \textit{standard} confidence interval.} 

\subsection{Normal Interval}
Suppose that the parameter of interest is the mean $\theta=\mu_F$. The plug-in estimate of $\theta$ is then $\hat{\theta}_n=\overline{X}_n$. Thanks to the central limit theorem, if the sample size $n$ is large enough, the distribution of  the sample mean $\overline{X}_n$ is approximately normal with mean $\mu_F$ and variance $\frac{\sigma^2_F}{n}=\mathrm{se}_F^2[\overline{X}_n]\approx\widehat{\mathrm{se}}_B^2$. That is $\overline{X}_n\approxdist\mathcal{N}(\mu_F,\widehat{\mathrm{se}}_B^2)$. 

It turns out that, for many reasonable distributions $F$ and functionals $t$, the distribution of $\hat{\theta}_n=t(\hat{F}_n)$ is also approximately normal\footnote{This large-sample result is often true for general statistics $\hat{\theta}=s(X_1,\ldots,X_n)$, not necessarily for plug-in estimates.}, $\hat{\theta}_n\approxdist\mathcal{N}(\theta,\widehat{\mathrm{se}}_B^2)$, or equivalently
\begin{equation}\label{eq:CLT}
\frac{\hat{\theta}_n-\theta}{\widehat{\mathrm{se}}_B}\approxdist\mathcal{N}(0,1).
\end{equation} 
Let $z_\alpha$ denote the $\alpha^{\mathrm{th}}$ quantile of the standard normal distribution\footnote{Recall, that $z_\alpha=\Phi^{-1}(\alpha)$, where $\Phi$ is the standard normal CDF. Obvious, yet useful property: $z_{1-\alpha}=-z_\alpha$.}. Then (\ref{eq:CLT}) results into
\begin{equation}
\mathbb{P}\left(z_{\alpha/2}\leq\frac{\hat{\theta}_n-\theta}{\widehat{\mathrm{se}}_B}\leq z_{1-\alpha/2}\right)\approx \Phi(z_{1-\alpha/2})-\Phi(z_{\alpha/2})=1-\alpha.
\end{equation}
Therefore, 
\begin{equation}\label{eq:normal interval}
\mathcal{I}=\hat{\theta}_n\pm z_{\alpha/2}\widehat{\mathrm{se}}_B.
\end{equation}
is an \textit{approximate} confidence interval for $\theta$ at level $1-\alpha$. This interval is accurate only under assumption (\ref{eq:CLT}) that the distribution of $\hat{\theta}_n$ is close to normal.

\textit{Question:} Can we construct accurate intervals without making normal theory assumptions like (\ref{eq:CLT})? 
The answer is ``yes.'' 

\subsection{Pivotal Interval}

Define the approximate \textit{pivot}\footnote{A pivot is a random variable $\zeta(X,\theta)$ that depends on the sample $X\sim F$ and the unknown parameter $\theta=t(F)$, but whose \textit{distribution} does not depend on $\theta$. The classical example is the so-called $z$-\textit{score}: if $X\sim\mathcal{N}(\mu,\sigma^2)$, then $Z=\frac{X-\mu}{\sigma}\sim\mathcal{N}(0,1)$. It can be shown that $\tilde{\theta}_n=\hat{\theta}_n-\theta$ is an approximate pivot under weak conditions on $t(F)$: the distribution of $\tilde{\theta}_n$ (not necessarily Gaussian) is approximately the same for each value of $\theta$.}
\begin{equation}
\tilde{\theta}_n=\hat{\theta}_n-\theta,
\end{equation}  
and let $G$ be its CDF. We want to find an interval $\mathcal{I}=(a,b)$, such that $\mathbb{P}(a\leq \theta\leq b)=1-\alpha$. Let us rewrite this probability in terms of  $G$:
\begin{equation}
\begin{split}
\mathbb{P}(a\leq \theta\leq b)=&\mathbb{P}(a-\hat{\theta}_n\leq \theta-\hat{\theta}_n\leq b-\hat{\theta}_n)\\
=&\mathbb{P}(\hat{\theta}_n-b\leq \tilde{\theta}_n\leq \hat{\theta}_n-a)\\
=&G(\hat{\theta}_n-a)-G(\hat{\theta}_n-b).
\end{split}
\end{equation}
Therefore, we can achieve the confidence level $1-\alpha$, by setting 
\begin{equation}
a=\hat{\theta}_n-G^{-1}\left(1-\frac{\alpha}{2}\right) \hspace{3mm} \mbox{and} \hspace{3mm} b=\hat{\theta}_n-G^{-1}\left(\frac{\alpha}{2}\right).
\end{equation}
The problem is that $G$, and thus its quantiles $\xi_{1-\alpha/2}=G^{-1}\left(1-\frac{\alpha}{2}\right)$ and $\xi_{\alpha/2}=G^{-1}\left(\frac{\alpha}{2}\right)$, are unknown. 

But we can estimate $G$ using the bootstrap!
\begin{equation}
G(\xi)\approx\hat{G}_B(\xi)=\frac{1}{B}\sum_{b=1}^B H\left(\xi-\tilde{\theta}^{(b)}_n\right),
\end{equation}
where
\begin{equation}
\tilde{\theta}^{(b)}_n=\hat{\theta}_n^{(b)}-\hat{\theta}_n,
\end{equation}  
and $\hat{\theta}^{(b)}_n$ is the bootstrap replication of $\hat{\theta}_n$. Let $\tilde{\xi}_\alpha$ denote the $\alpha^{\mathrm{th}}$ sample quantile of $\tilde{\theta}^{(1)}_n,\ldots,\tilde{\theta}^{(B)}_n$,
\begin{equation}
\tilde{\xi}_\alpha=\inf\{\xi: \hat{G}_B(\xi)\geq\alpha\}.
\end{equation}
Note that $\tilde{\xi}_\alpha=\hat{\xi}_\alpha-\hat{\theta}_n$, where $\hat{\xi}_\alpha$ is the  $\alpha^{\mathrm{th}}$ sample quantile of $\hat{\theta}_n^{(1)},\ldots,\hat{\theta}_n^{(B)}$.
Therefore, the end points $a$ and $b$ of the confidence interval can be approximated  as follows:
\begin{equation}
\begin{split}
a\approx\hat{a}_B=&\hat{\theta}_n-\tilde{\xi}_{1-\alpha/2}=2\hat{\theta}_n-\hat{\xi}_{1-\alpha/2}, \\ b\approx\hat{b}_B=&\hat{\theta}_n-\tilde{\xi}_{\alpha/2}=2\hat{\theta}_n-\hat{\xi}_{\alpha/2}.
\end{split}
\end{equation}
Under weak conditions on $F$ and $t$, $\mathbb{P}(\hat{a}_B\leq\theta\leq\hat{b}_B)\rightarrow1-\alpha$ as $n,B\rightarrow\infty$. The interval $\mathcal{I}=(\hat{a}_B,\hat{b}_B)$ is thus an approximate $1-\alpha$ confidence interval. 

There are other ways to construct the bootstrap confidence intervals, e.g. the \textit{percentile interval}\footnote{This interval is intuitive, but does not have theoretical support. For ``semi-theoretical'' justification, see, for example, Chapter~13 in B.~Efron \& R.J.~Tibshirani (1993), \textit{\href{https://books.google.com/books?id=gLlpIUxRntoC}{An Introduction to the Bootstrap}}.},  \textit{studentized interval}\footnote{Also called ``bootstrap-t.''}, \textit{$BC_a$ interval}\footnote{$BC_a$ stands for ``bias-corrected and accelerated.''}, with many variations. Typically there is a trade-off between accuracy of an interval and the amount of work needed for its construction. Here we described only two basic intervals with the main goal to illustrate the idea of bootstrap. For more advanced methods, see the textbooks listed in section ``Further Reading'' below. 

\section{Example: Enrollment in the U.S. Universities}
\begin{marginfigure} %\vspace{-15.5cm}
	\includegraphics[width=\linewidth]{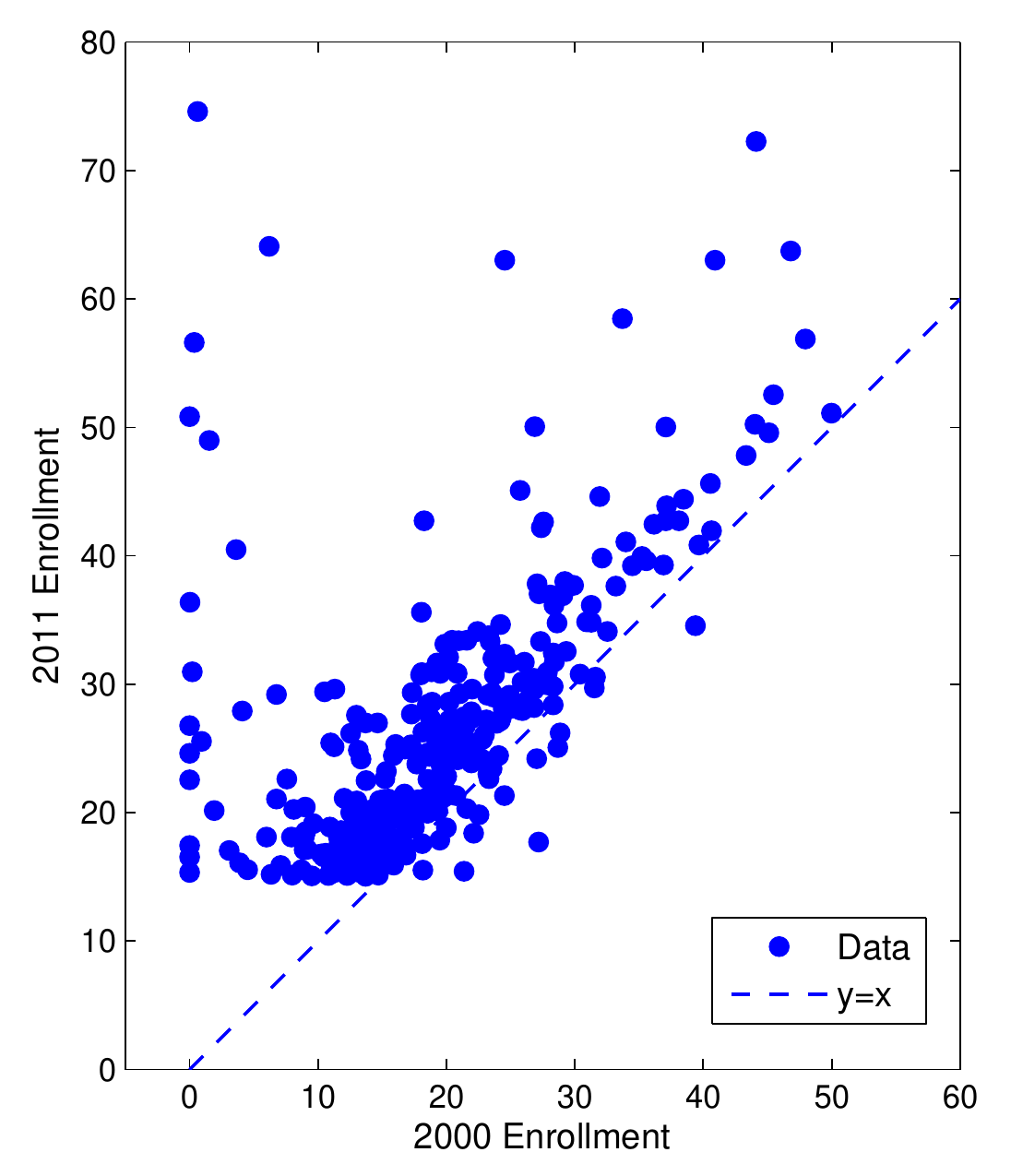}
	\caption{Enrollments (in thousands) of $N=354$ large degree-granting institutions. Data: U.S. Department of Education. Available at \href{http://www.its.caltech.edu/~zuev/teaching/2016Winter/enrollment.xlsx}{enrollment.xlsx}. An ``outlier'' --- University of Phoenix $(14.8,307.9)$ --- is not shown in Fig.~\ref{fig:9fig1}.}
	\label{fig:9fig1}
\end{marginfigure}

Figure~\ref{fig:9fig1} shows $N=354$ data points, each corresponding to a large university in the U.S. The $x$ and $y$ coordinates of each point are the enrollment sizes of the corresponding university in 2000 and 2011. Only universities with 2011 enrollment $\geq15,000$ are considered. 
Let the parameter of interest $\theta$ be the ratio of the means $\bar{y}/\bar{x}$, which is a good proxy\footnote{We ignore small universities.} for the total increase in university enrollment in the country from 2000 to 2011. The distribution $F$ in this case is the bivariate CDF that puts probability $1/N$ at each of the data points $(x_i,y_i)$, and 
\vspace{-3mm}
\begin{equation}
\theta=t(F)=\frac{\int ydF(x,y)}{\int xdF(x,y)}=\frac{\bar{y}}{\bar{x}}.
\end{equation}
\vspace{-1mm}
Given the data in Fig.~\ref{fig:9fig1}, $\theta$ can be computed exactly\footnote{We will use this \textit{true} value is a benchmark.}: 
\vspace{-1mm}
\begin{equation}\label{eq:truetheta}
\theta=1.41.
\end{equation} 

\begin{figure}[h]
	\includegraphics[width=\linewidth]{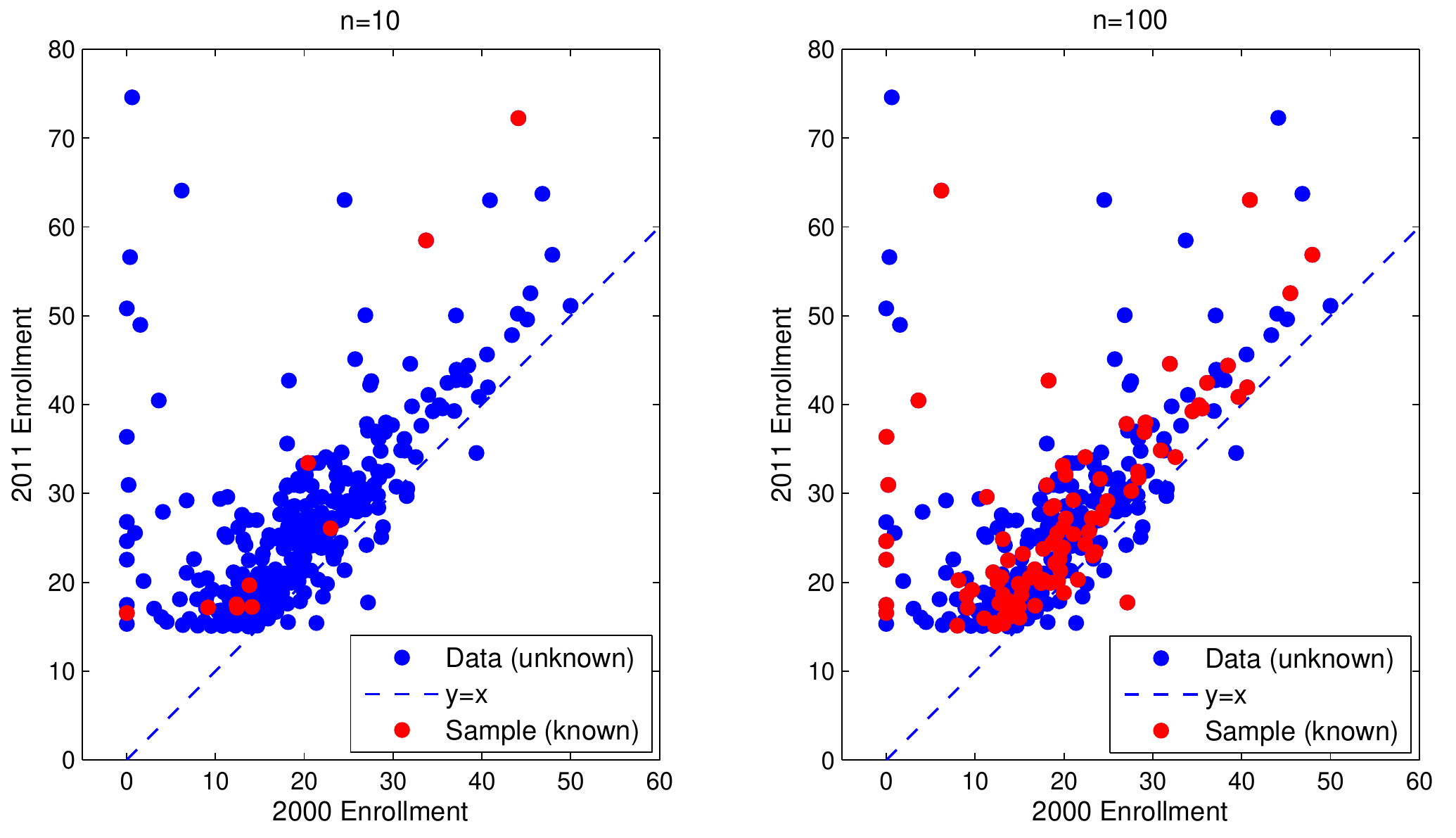}%
	\caption{Random samples of size $n=10$ (left) and $n=100$ (right). Out goal is to estimate $\theta$, compute the standard error of the estimate, and construct  confidence intervals based on these samples.}
	\label{fig:9fig2}%
\end{figure}

Suppose now that we don't have access to the full data, and we can only pick $n<N$ universities at random,  and record their enrollment sizes $(X_1,Y_1),\ldots,(X_n,Y_n)$. Figure~\ref{fig:9fig2} shows the random samples of sizes $n=10$ and $n=100$. \vspace{-0.2cm}
%\vspace{1mm}

Since there is no obvious parametric model for the joint distribution $F$, it is natural to stick to nonparametric estimation. The bivariate eCDF $\hat{F}_n$ puts probability $1/n$ at each sampled pair $(X_i,Y_i)$. The plug-in estimate of $\theta$ is therefore
\begin{equation}\label{eq:theta_n}
\hat{\theta}_n=t(\hat{F}_n)=\frac{\overline{Y}_n}{\overline{X}_n}.
\end{equation}
The estimates computed from the samples depicted in Fig.~\ref{fig:9fig2} are:
\begin{equation}\label{eq:plug}
\hat{\theta}_{10}=1.61 \hspace{3mm} \mbox{and} \hspace{3mm} \hat{\theta}_{100}=1.39.
\end{equation}

Let us compute the bootstrap estimates $\widehat{\mathrm{se}}_B[\hat{\theta}_n]$ of the standard errors $\mathrm{se}_F[\hat{\theta}_n]$. To make the simulation error completely negligible, we use $B=10^4$ bootstrap samples $(X_1^{(b)},Y_1^{(b)}),\ldots,(X_n^{(b)},Y_n^{(b)})\sim\hat{F}_n$.\footnote{We can easily afford this large $B$ since computing the statistic (\ref{eq:theta_n}) is very fast.} The corresponding bootstrap replications are $\hat{\theta}_n^{(b)}=\overline{Y}_n^{(b)}/\overline{X}_n^{(b)}$. The bootstrap estimates obtain from (\ref{eq:bootstrap_se}) are:\footnote{Recall now that we actually know $F$ in this example. How would you compute the true standard errors $\mathrm{se}_F[\hat{\theta}_n]$?}
\begin{equation}\label{eq:boot}
\widehat{\mathrm{se}}_B(\hat{\theta}_{10})=0.14 \hspace{3mm} \mbox{and} \hspace{3mm} \widehat{\mathrm{se}}_B(\hat{\theta}_{100})=0.06.
\end{equation}

\begin{figure}[h]
	\centerline{\includegraphics[width=.9\linewidth]{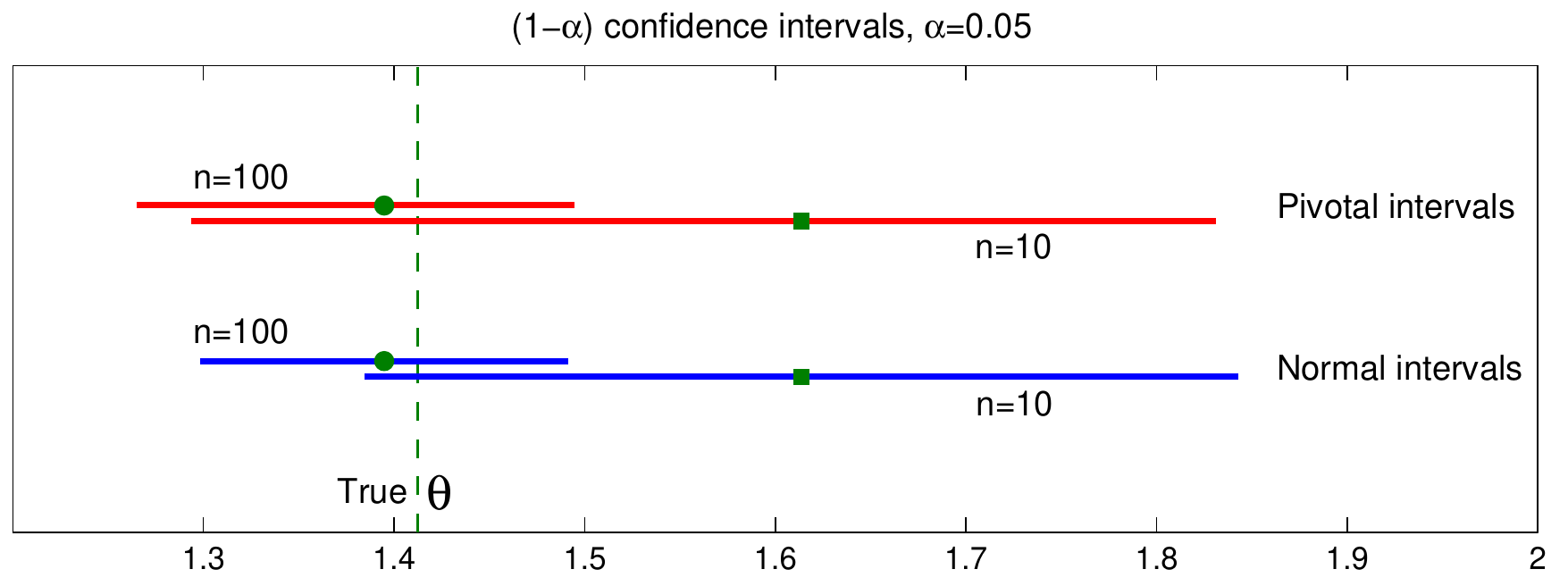}}
	\caption{The normal (blue) and pivotal (red) confidence intervals at level $0.95$. The true value of $\theta$ (\ref{eq:truetheta}) is shown by the dashed line. The estimates $\hat{\theta}_n$ are marked by green circles ($n=100$) and squares ($n=10$).}
	\label{fig:9fig3}%
\end{figure}

Figure~\ref{fig:9fig3} shows the normal and pivotal confidence intervals for $\theta$ constructed from the samples in Fig.~\ref{fig:9fig2}. As expected, intervals constructed from the small sample ($n=10$) are longer. Note also that while the normal intervals are (by definition) symmetric about $\hat{\theta}_n$, the pivotal intervals are not.

It is important to highlight that plug-in estimates (\ref{eq:plug}), bootstrap estimates of their standard errors (\ref{eq:boot}), and confidence intervals in Fig.~\ref{fig:9fig3} are computed based on the two \textit{specific} samples
 $(X_1,Y_1),\ldots,(X_n,Y_n)$  shown in Fig.~\ref{fig:9fig2}. The results of course will change if we get another samples. It is interesting to see the variability of the estimates. Let us repeat all computations for $200$ independent samples from the total population of $N$ universities: $100$ samples of size $n=10$ and $100$ samples of size $n=100$. Figures~\ref{fig:9fig4} and \ref{fig:9fig5} show the simulation results.
{\begin{figure}
		\vspace{-5mm}
	\centerline{\includegraphics[width=\linewidth]{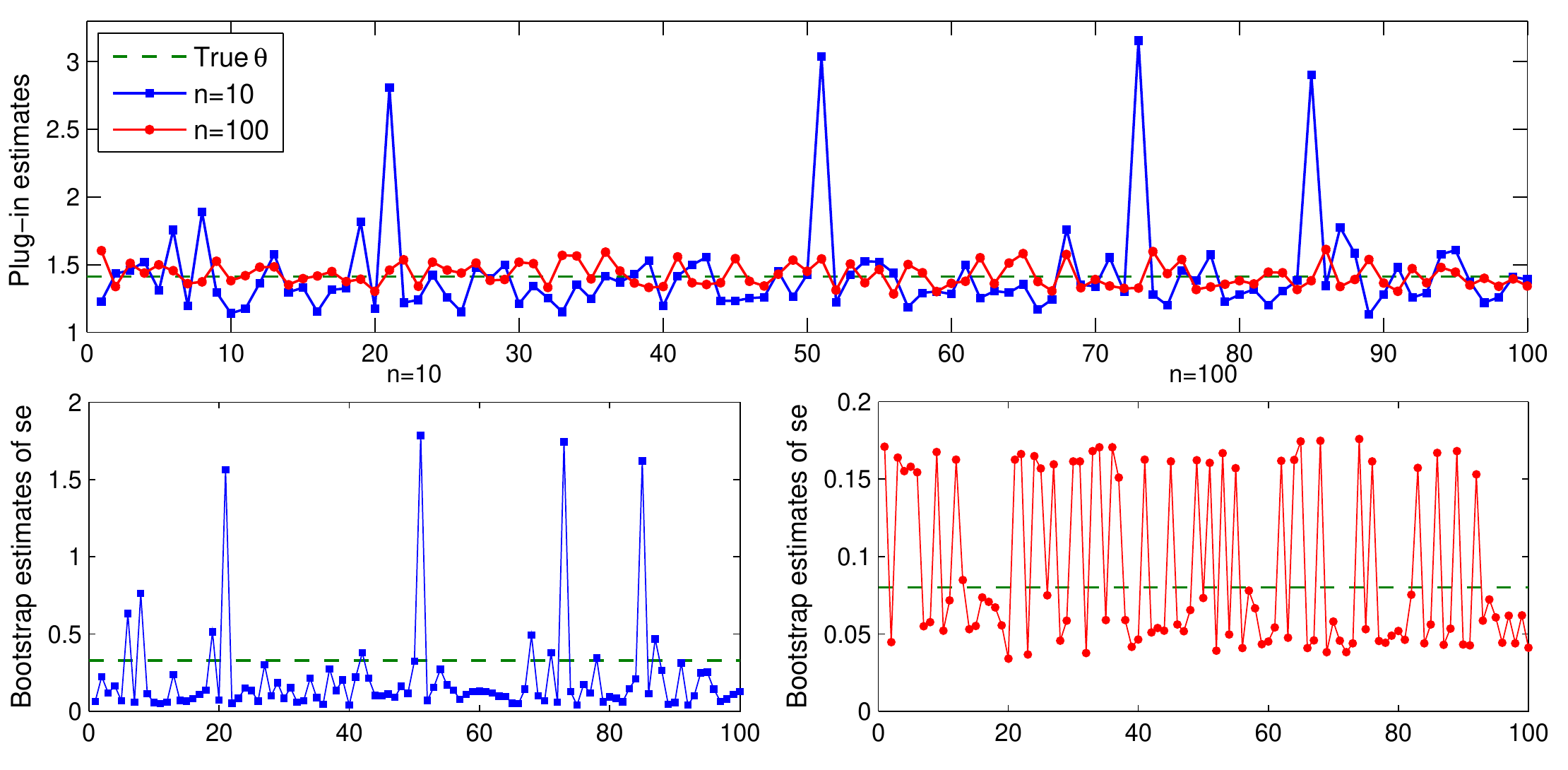}}
	\caption{Top panel illustrates the variability in the values of the plug-in estimates $\hat{\theta}_n$. As expected, on average, the estimate for $n=100$ (red curve) is more accurate. The bottom panel shows the variability of the bootstrap estimates $\widehat{\mathrm{se}}_B[\hat{\theta}_n]$ for $n=10$ (left) and $n=100$ (right). Green dashed lines represent the true values of ${\mathrm{se}}_F[\hat{\theta}]$ computed using $F$ (i.e. using full data).}
	\label{fig:9fig4}%
\end{figure}

\begin{figure}
	\vspace{-15mm}
	\centerline{\includegraphics[width=\linewidth]{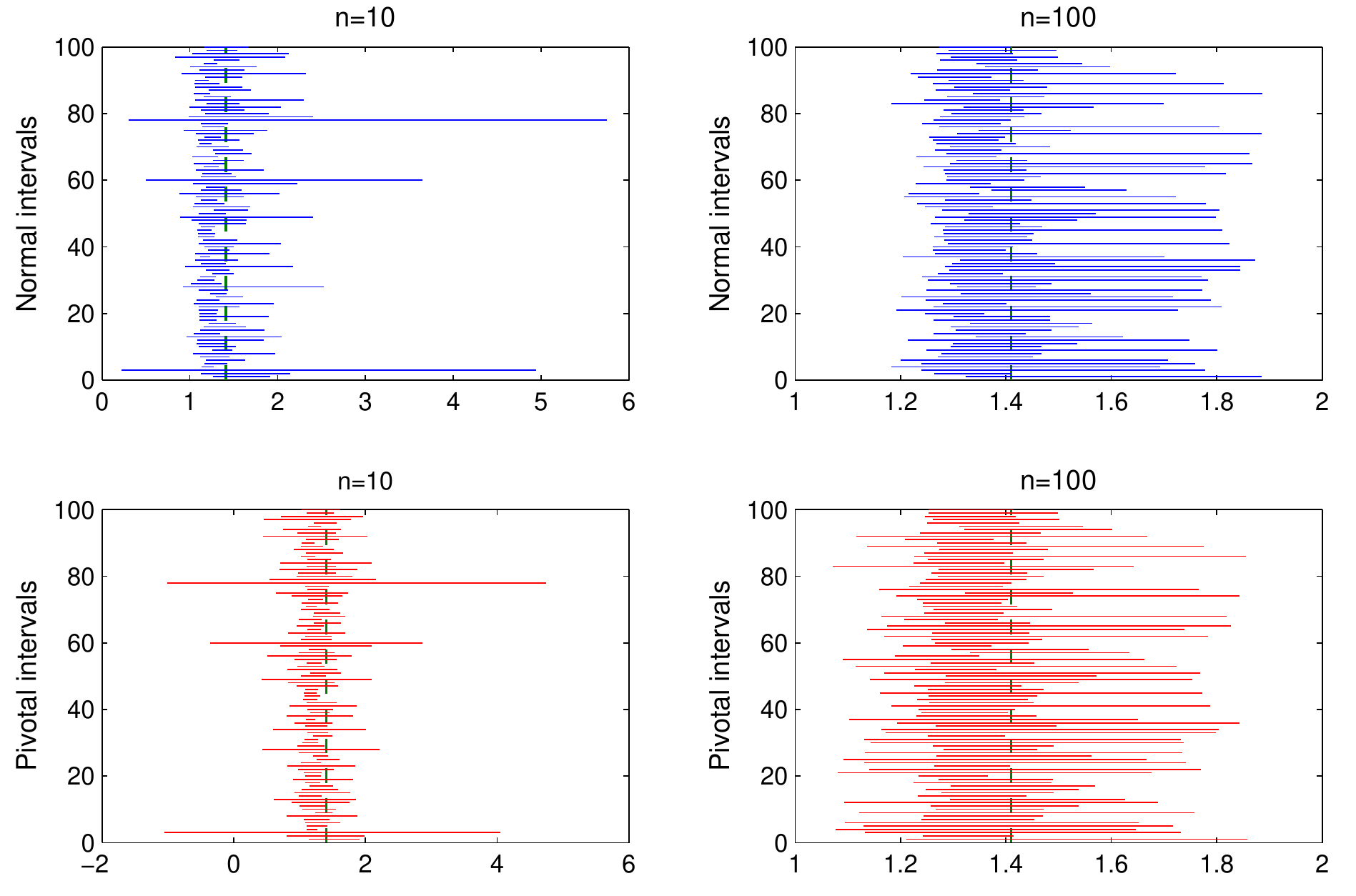}}
	\caption{Here we show approximate 0.95 confidence intervals for $\theta$. Four intervals in Fig.~\ref{fig:9fig3} are ones of these. For $n=10$, only $71$ out of $100$ normal intervals (top left) and $69$ pivotal intervals (bottom left) contain the true value $\theta=1.41$. This means that the approximation is poor, since we expect about $95$ out of $100$ intervals to contain $\theta$. For $n=100$, the intervals are more accurate: $83$ normal (top right) and $86$ pivotal (bottom right) contain $\theta$. }
	\label{fig:9fig5}%
\end{figure}}

 \section{Further Readings}
 \begin{enumerate}
 	\item The original bootstrap paper B. Efron (1979) ``\href{http://projecteuclid.org/euclid.aos/1176344552}{Bootstrap methods: another look at the jackknife},'' \textit{The Annals of Statistics}, 7:	1-26. is classical. It is very readable and highly recommended. 
 	\item A clear textbook-length treatment of the bootstrap is by Efron and his former PhD student Tibshirani: B.~Efron \& R.J.~Tibshirani (1993) \textit{\href{https://books.google.com/books?id=gLlpIUxRntoC}{An Introduction to the Bootstrap}}. It  focuses more on theory.
 	\item Another good textbook that focuses more on applications is A.C.~Davison \& D.V.~Hinkley  (1997)
 	\textit{\href{https://books.google.com/books/about/Bootstrap_Methods_and_Their_Application.html?id=4aCDbm_t8jUC}{Bootstrap Methods and their Applications}}.
 	\item If you encounter a serious application, the review A.J.~Canty et al (2006) ``\href{http://onlinelibrary.wiley.com/doi/10.1002/cjs.5550340103/abstract}{Bootstrap diagnostics
 	and remedies},'' \textit{Canadian Journal of Statistics},
 	34:5-27  might be useful. They describe typical problems with bootstrap, provide workable diagnostics, and discuss efficient ways to implement the necessary computations.
 	\item  For a more conceptual and somewhat philosophical discussion of the bootstrap, see the beautiful essay by C.~ Shalizi (2010) ``\href{http://www.americanscientist.org/issues/pub/2010/3/the-bootstrap/1}{The bootstrap},'' \textit{American Scientist,} 98: 186-190. For a much more complete list of references on the bootstrap, go to [\href{http://bactra.org//notebooks/bootstrap.html}{web}].
 \end{enumerate}
 
  \section{What is Next?}
  Nonparametric inference is made under minimal assumptions on the underlaying statistical model for the data in hand. In general, the more assumptions we make, the more powerful methods are available for data analysis, and, as a result, the more we can infer from the data\footnote{If the assumptions are correct!}. In the next Lecture, we will start discussing the \textit{parametric inference}, which makes stronger assumptions about the~data.
  
\chapter{The Method of Moments}\label{ch:MOM}

\newthought{Let} us turn our attention to \textit{parametric inference}, where the data $X_1,\ldots,X_n$ is modeled as a random sample from a finitely parametrized distribution:
\begin{equation}\label{eq:setup}
X_1\ldots,X_n\sim f, \hspace{5mm} f\in\mathcal{F}=\{f(x;\theta): \theta\in\Theta\},
\end{equation}
where $f$ is a probability density function (PDF)\footnote{Or, a probability mass function (PMF) if the data is discrete.}, $\theta$ is the model parameter, and  $\Theta$ is the parameter space. In general $\Theta\subset\mathbb{R}^p$ and $\theta=(\theta_1,\ldots,\theta_p)$ is a vector of parameters. Recall that whenever $p=\dim\Theta<\infty$, we call the corresponding model $\mathcal{F}$ in (\ref{eq:setup}) a parametric model. In this framework, the problem of inference reduces to estimating $\theta$ from the data. But before we start talking about different parametric methods, let us discuss the following conceptual question:

	\section{How could we possibly know that $X_1,\ldots,X_n\sim f(x;\theta)$ ?}
	In other words, how would we ever know that there exists a  family of distribution $\mathcal{F}=\{f(x;\theta)\}$ and a specific value of the parameter $\theta$\footnote{This value is often called the ``true'' value of the parameter.} such that our data $X_1\ldots,X_n$ is generated exactly from $f(x;\theta)$. 
	
	First of all, when we assume the parametric model (\ref{eq:setup}), we don't really believe that  the data is exactly generated from $f(x;\theta)$ for some value of $\theta$. Rather, we  believe that there exists some value of $\theta$ in $\Theta$ such that the distribution  $f(x;\theta)$ does well (for all practical purposes) in describing the randomness in the data. That is
	\begin{equation}
	X_1,\ldots,X_n\approxdist f(x;\theta).
	\end{equation} 
	
	Ok, but still, how do we know what parametric model to chose?  Indeed, in many applications we would not have such knowledge\footnote{In such cases, nonparametric inference is preferable.}. But there are cases where background knowledge and prior experience suggest that a certain parametric model provides a reasonable approximation\footnote{For example, it is known that counts of traffic accidents approximately follow a Poisson model.}.
	
	Finally, whenever we assume a parametric model (\ref{eq:setup}), we can always check this assumption. One possibility is to check (\ref{eq:setup}) informally by inspecting plots of the data\footnote{For example, if a histogram of the data looks bimodal, then the assumption of normality $\mathcal{F}=\{\mathcal{N}(\mu,\sigma^2)\}$ would be at least questionable.}. A formal way to check a parametric model is to use some formal \textit{test}, for instance \textit{permutation test}, which will consider later in Lecture~\ref{ch:Permutation}.
	
	\section{Parameters in Parametric Models}
	Recall that in the nonparametric setup, a parameter $\theta$ of a distribution $F$ is the value of a certain (known) functional $t$ on $F$, $\theta=t(F)$. In the parametric setup (\ref{eq:setup}), a parameter of interest can be $\theta$, or a component of $\theta$, or, more generally\footnote{And quite often in applications.}, any function of $\theta$.  
	
	\paragraph{Example:} Suppose our data is $X_1,\ldots,X_n$, where $X_i$ is the outcome of a blood test of subject $i$. And suppose we are interested in the fraction $\tau$ of the entire population whose test score is larger than a certain threshold $\alpha$. Since many measurements taken on humans approximately follow normal distribution, it is reasonable to model the data as a sample $X_1,\ldots,X_n\sim\mathcal{N}(\mu,\sigma^2)$. The parameter of interest is then 
	\begin{equation}
	\tau=1-\Phi\left(\frac{\alpha-\mu}{\sigma}\right),
	\end{equation}
	which can be estimated by $\hat{\tau}=1-\Phi\left(\frac{\alpha-\hat{\mu}}{\hat{\sigma}}\right)$, where $\hat{\mu}$ and $\hat{\sigma}$ are the estimates of $\mu$ and $\sigma$ obtained from the data\footnote{If we ignore the normal model for the data, we can estimate $\tau$ simply by the fraction of sample whose score is large than $\alpha$, $\check{\tau}=|\{X_i: X_i>\alpha\}|/n$.}.  %What estimate --- parametric $\hat{\tau}$ or nonparametric $\check{\tau}$ is more accurate?
	\hfill $\square$
	
	\section{The Method of Moments}
	The first method for constructing parametric estimates  that we will consider is called the method of moments (MOM). MOM is perhaps the oldest general method for finding point estimates, dating back at least to Karl Pearson in the late 1800s. Its main advantage is that MOM estimates are usually easy to compute for ``standard'' models\footnote{Normal, Bernoulli, Poisson, etc.}. The main drawback is that they often not optimal and better estimates exist\footnote{For example, \textit{maximum likelihood estimates}, which we will discuss in the next Lecture.}. Nevertheless, MOM estimates are often useful as starting values for other methods that require
	iterative numerical routines.
	
	So, suppose that we model the data parametrically 
	\begin{equation}
	X_1,\ldots,X_n\sim f(x;\theta),
	\end{equation}
	and that $\dim\theta=k$, that is $\theta=(\theta_1,\ldots,\theta_k)$.  Recall that the $q^{\mathrm{th}}$ \textit{moment} of a distribution $f$ is the expected value $\mathbb{E}_f[X^q]$. Since, $f$ depends on $\theta$, so do the moments\footnote{If the data is discrete, then $f$ is a probability mas function and the integral in (\ref{eq:moment}) is replaced with the sum.}:
	\begin{equation}\label{eq:moment}
	m_q(\theta)=\mathbb{E}_f[X^q]=\int x^qf(x;\theta)dx.
	\end{equation}
	Can we estimate these ``theoretical'' moments from the data? Yes, of course. Let is define the $q^{\mathrm{th}}$ \textit{sample moment} as follows:
	\begin{equation}
	\hat{m}_q=\frac{1}{n}\sum_{i=1}^nX_i^q.
	\end{equation} 
	By the law of large numbers, $m_q(\theta)\approx\hat{m}_q$ when $n$ is large\footnote{Note that the sample moment $\hat{m}_q$ is simply the plug-in estimate of the theoretical moment $m_q$: $\hat{m}_q=\int x^qd\hat{F}_n(x)$.}.
	The method of moments exploits this approximation. The MOM estimate $\hat{\theta}_{\mathrm{MOM}}$ of $\theta$ is the solution of the following system of $k$ equations with $k$ unknowns:
	\begin{equation}\label{eq:equations}
	m_q(\theta)=\hat{m}_q, \hspace{3mm} q=1,\ldots,k.
	\end{equation}
	Let us consider a couple of classical examples. 
	
	\paragraph{Example:} Let $X_1,\ldots,X_n\sim\mathrm{Bernoulli}(p)$, then $\hat{p}_{\mathrm{MOM}}=\overline{X}_n$.
	\hfill $\square$ %\bigskip
	
	\paragraph{Example:} Let $X_1,\ldots,X_n\sim\mathcal{N}(\mu,\sigma^2)$, then 
	\begin{equation}
	\hat{\mu}_{\mathrm{MOM}}=\overline{X}_n \hspace{5mm}and\hspace{5mm}
	\hat{\sigma}^2_{\mathrm{MOM}}=\frac{1}{n}\sum_{i=1}^n(X_i-\overline{X}_n)^2.
	\end{equation}
	\hfill $\square$
	
	These examples show that, at least in these two specific cases, MOM produces very reasonable estimates, which, in fact, coincide with the corresponding plug-in estimates. This leads to a natural question: are there examples where the MOM and plug-in estimates are different? As expected, the answer is yes. 
	
	\paragraph{Example:} Let $X_1,\ldots,X_n\sim U[0,\theta]$. Recall %\footnote{Homework 3, problem 4.}  
	that the plug-in estimate is $\hat{\theta}_n=X_{(n)}$. The MOM estimate is $\hat{\theta}_{\mathrm{MOM}}=2\overline{X}_n$. Note that this estimate, although unbiased, can give impossible results. For example, if $n=3$, $X_1=X_2=1$ and $X_3=7$, then $\hat{\theta}_{\mathrm{MOM}}=6$, which makes $X_3=7$ impossible.  \hfill $\square$\bigskip
	
	This example serves as a ``red flag:'' we should be careful when using MOM, it may give unreasonable estimates. On the other hand, MOM estimates have a nice property. 
	
\section{Consistency of MOM Estimates}
Let $X_1,\ldots,X_n\sim f(x;\theta)$ and let $\hat{\theta}_n$ denote the MOM estimate. Under certain regularity conditions on the model $f$, $\hat{\theta}_n$ is consistent:
\begin{equation}
\hat{\theta}_n\stackrel{\mathbb{P}}{\longrightarrow}\theta.
\end{equation}

A couple of remarks are in order:
\begin{enumerate}
	%\item[a)] The first property roughly means the following. If you work hard enough, you will find examples of a parametric model $\mathcal{F}=\{f(x;\theta)\}$ for which the system (\ref{eq:equations}) does not have a solution. But if you don't have this specific goal in mind, and take any reasonable statistical model, then for large enough $n$, the MOM estimate will always exist. 
	\item[a)] Consistency says roughly that $\hat{\theta}_n$ gives the right answer in the long run, as the sample size $n$ goes to infinity. This is a very basic test for a quality of an estimate: it is rare to use an estimate which is not consistent\footnote{Indeed, if we can't get the true value of the parameter even with infinite data, the estimator we are using is rubbish!}. Note that if $\mathbb{E}[\hat{\theta}_n]\rightarrow\theta$ and $\mathbb{V}[\hat{\theta}_n]\rightarrow0$, then $\hat{\theta}_n$ is consistent\footnote{See, \eg [CH, Proposition 6.6, p. 158].}. For all considered examples, this helps to check the consistency of the corresponding MOM estimates. The general proof of consistency is beyond our scope. 
	\item[b)] Finally, it us useful to keep in mind a scenario when MOM can be preferable to other approaches. Namely: the chosen statistical model $\mathcal{F}=\{f(x;\theta)\}$  is in doubt\footnote{If it is not, use MLE (Lecture~\ref{ch:MLE})!} and one wishes to make sure $\mathcal{F}$ accurately fits certain aspects of the data that can be expressed in term of moments. 
\end{enumerate}

\section{Further Reading}
\begin{enumerate}
	\item In complicated statistical models, theoretical moments $m_q(\theta)$ are generally expressed as intractable integrals, in which case matching theoretical and sample moments requires solving a system of integral equations. A. Gelman (1995) ``\href{http://stat.columbia.edu/~gelman/research/published/moments.pdf}{Method of moments using Monte Carlo simulation},'' \textit{J. Comp and Graph. Statistics}, 4(1), 36-54 presents a computation approach to MOM that efficiently resolves this technical problem. 
\end{enumerate}

\section{What is Next?} 
We will discuss one of the most fundamental and iconic methods of parametric inference: \textit{maximum likelihood estimation}.
	
\chapter{Maximum Likelihood: Intuition, Definitions, Examples}\label{ch:MLE}

\newthought{Maximum likelihood} is the most popular method for estimating parameters in parametric models. It was strongly recommended by Ronald Fisher, one of the greatest statisticians of all times. The maximum likelihood estimates (MLEs) are known to be very powerful, especially with large samples. We start from describing the intuition behind this method, then define the likelihood function and MLE, and consider several classical examples.
\begin{marginfigure}
		\vspace{-20mm}
	
	\centerline{\includegraphics[width=.6\linewidth]{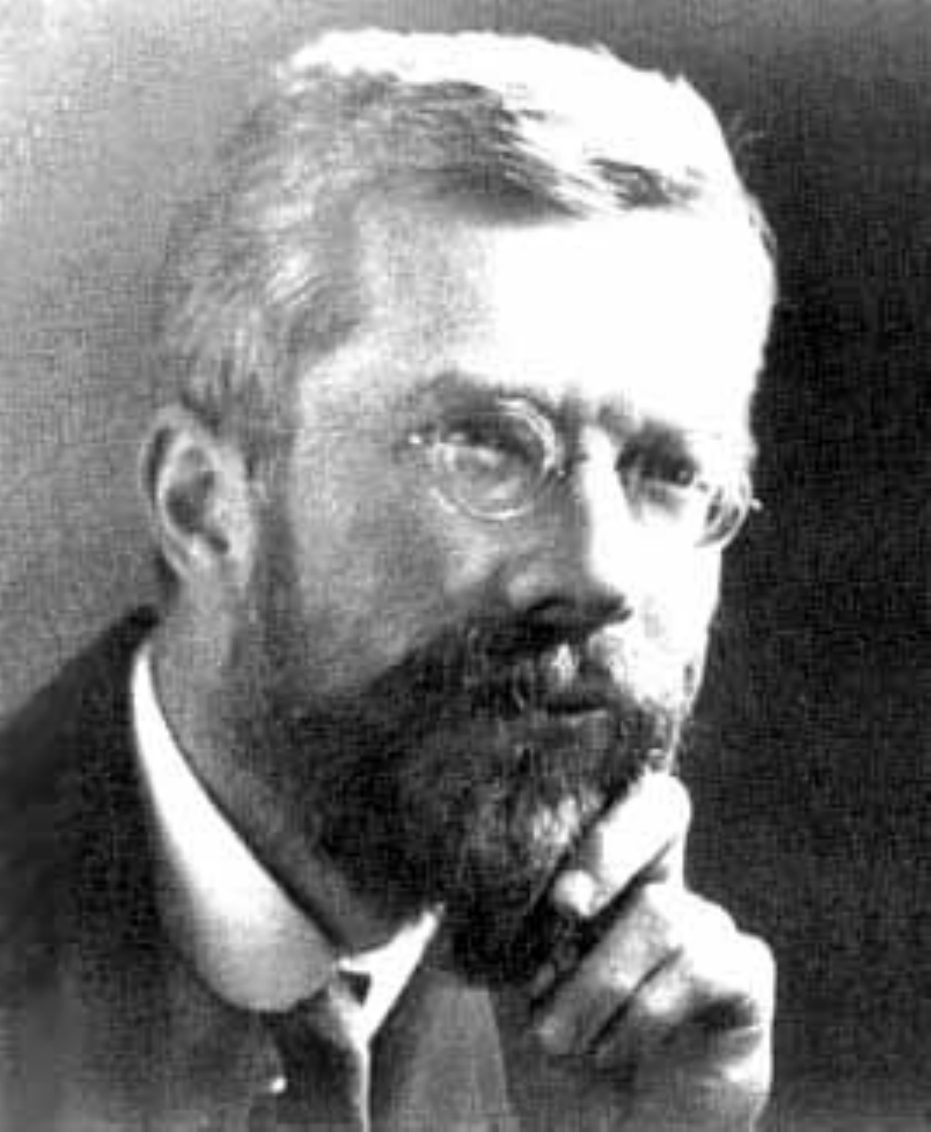}}
	\caption{Sir Ronald Fisher. Photo source: \href{https://en.wikipedia.org/wiki/Ronald_Fisher}{wikipedia.org}.}
	\label{fig:Fisher}
\end{marginfigure}

\section{Intuition}
Suppose I tell you I have $100$ cookies in my backpack.  The cookies are of two types: chocolate chip cookies and fortune cookies. Moreover, I tell you that the number of fortune cookies is either $10$ or $90$. You draw a cookie out of my backpack at random and see that it is a fortune cookie. Based on this ``data,'' what is more likely: there are  
\begin{itemize}
	\item[a)] $10$  fortune cookies and $90$ chocolate chip cookies, or
	\item[b)] $90$ fortune cookies  and $10$ chocolate chip cookies?
\end{itemize}
%a) $10$  fortune cookies and $90$ chocolate chip cookies, or \\
%b) $90$ fortune cookies  and $10$ chocolate chip cookies? %\footnote{In this example, the data is \{fortune cookie\}, the parametric model for the data is the to cases a) and b), and the parameter is $\theta\in\{10,90\}$, the number of fortune cookies.} 
%\\
%\hspace{-5mm} 
Based solely on one sample (fortune cookie), b) is more likely.   

This is exactly the idea behind maximum likelihood estimation. The method asks: what value of the a parameter is most consistent with the data. In other words, \textit{what value of a parameter makes the data most likely}?
\begin{marginfigure}
	\vspace{-60mm}	
	\centerline{\includegraphics[width=.5\linewidth]{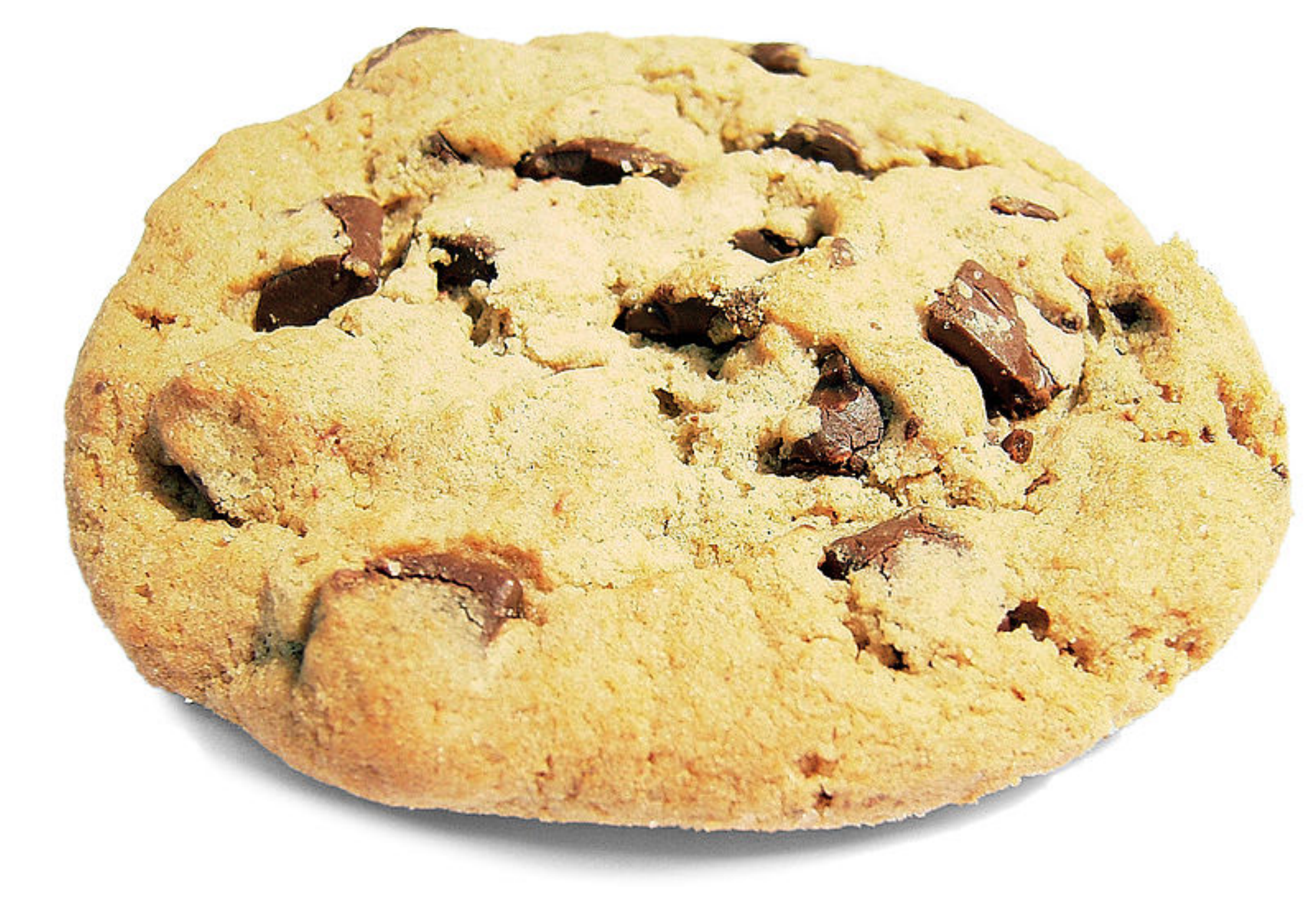}\includegraphics[width=.5\linewidth]{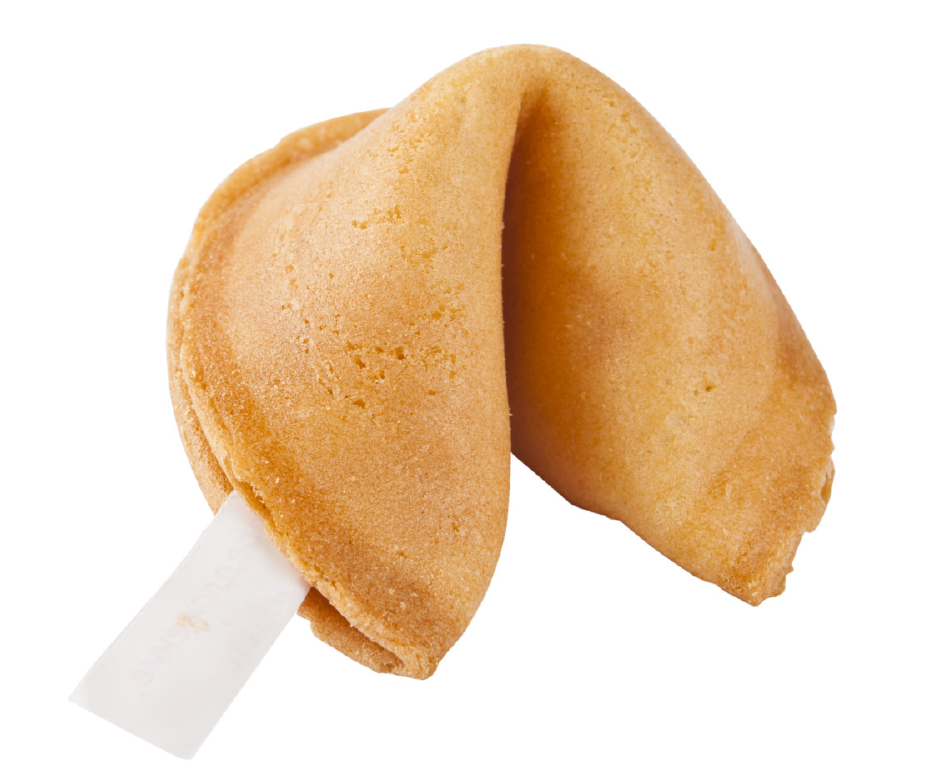}}
	\caption{A chocolate chip cookie (left) was invented in 1938 in Massachusetts and a fortune cookie (right). The exact origin of fortune cookies is unclear, though various immigrant groups in California claim to have popularized them in the early 20th century. Photo source: \href{https://commons.wikimedia.org/wiki/File:Choco_chip_cookie.png}{wikipedia.org} and \href{http://i.huffpost.com/gen/1117772/images/h-FORTUNE-COOKIE-628x314.jpg}{huffingtonpost.com}.}
	\label{fig:cookies}
\end{marginfigure}

\section{Likelihood Function and MLE}
Let us consider the discrete and continuous cases separately. The discrete case is somewhat more intuitive, but at the end, we will see that there is no much difference between the two cases. 

\subsection{Discrete Models}

Let $X_1\ldots,X_n$ be data modeled as a sample from a discrete distribution with the probability mass function (PMF) $f(x;\theta)$.  What is the probability of observing the data? Given the value of the parameter, we can write it as follows\footnote{We are using independence of $X_i$ in the first equality in (\ref{eq:1}).}:
\begin{equation}\label{eq:1}
\mathbb{P}(X_1,\ldots,X_n|\theta)=\prod_{i=1}^n\mathbb{P}(X_i|\theta)=\prod_{i=1}^n f(X_i;\theta).
\end{equation}
The \textit{likelihood function} is the joint probability of the data \textit{viewed as a function of the parameter} $\theta$:
\begin{equation}
\mathcal{L}(\theta|X_1,\ldots,X_n)=\prod_{i=1}^n f(X_i;\theta).
\end{equation}
In spite the fact that the likelihood $\mathcal{L}(\theta|X_1,\ldots,X_n)$ is expressed in therms of $f(X_i;\theta)$, the two functions are conceptually different. When we consider the probability mass function $f(x;\theta)$, we consider $x$ to be variable and the parameter $\theta$ is fixed. When we consider the likelihood $\mathcal{L}(\theta|X_1,\ldots,X_n)$, we consider the data to be fixed (observed) and $\theta$ to be variable: $\mathcal{L}(\theta_1|X_1,\ldots,X_n)$ is the probability of observing the data if $\theta=\theta_1$, $\mathcal{L}(\theta_2|X_1,\ldots,X_n)$ is the probability of observing the data if $\theta=\theta_2$, etc. Often, the likelihood function is denoted simply by $\mathcal{L}_n(\theta)$. This notation emphasizes the fact that likelihood is a function of a parameter.

The maximum likelihood method looks for the value of $\theta$ that makes the data as likely as possible\footnote{Assuming the model is correct!}. Technically, that means looking for $\theta$ which maximizes the likelihood $\mathcal{L}_n(\theta)\equiv\mathcal{L}_n(\theta|X_1,\ldots,X_n)$. Thus, a \textit{maximum likelihood estimate} (MLE) is a value $\hat{\theta}_{\mathrm{MLE}}$ such that
\begin{equation}
\mathcal{L}_n(\hat{\theta}_{\mathrm{MLE}})\geq\mathcal{L}_n(\theta)\hspace{5mm}\mbox{for all } \theta\in\Theta,
\end{equation}
or, equivalently,
\begin{equation}
\hat{\theta}_{\mathrm{MLE}}=\arg\max\limits_{\theta\in\Theta}\mathcal{L}_n(\theta).
\end{equation}
Notice that by construction, the range of the MLE coincides with the range of the parameter $\Theta$. Let us consider an example.

\begin{figure}
	%	\vspace{-5mm}	
	\centerline{\includegraphics[width=.8\linewidth]{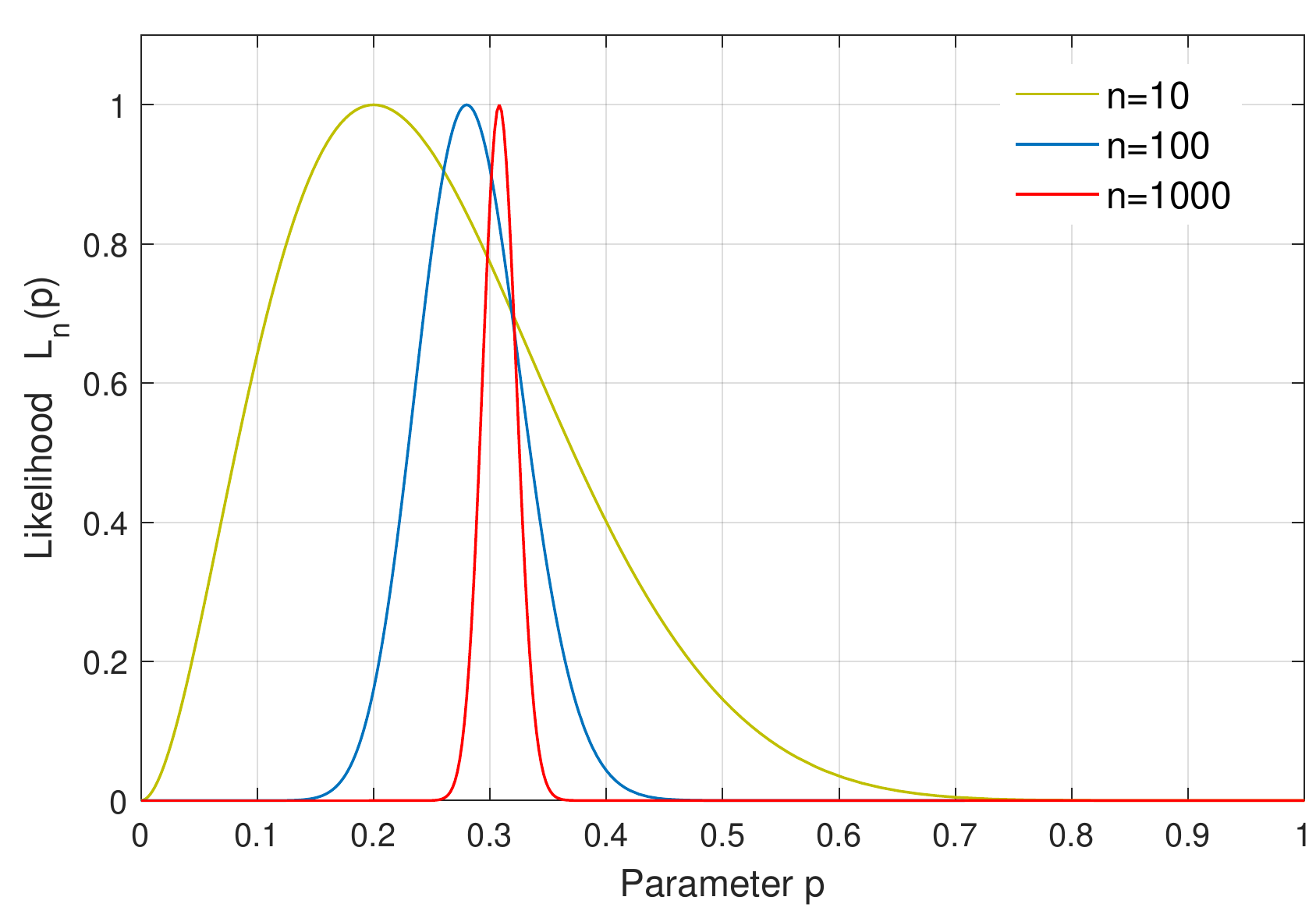}}
	\caption{Three normalized (so that $\max\mathcal{L}_n(p)=1$) likelihood functions for the data $X_1,\ldots,X_n\sim \mathrm{Bernoulli}(1/3)$ with $n=10$, $n=100$, and $n=1000$. Notice how the likelihood function becomes more and more concentrated around the true value $p=1/3$ as the sample size $n$ increases. }
	\label{fig:bernoulli}
\end{figure}

\paragraph{Example:} Let $X_1\ldots,X_n\sim\mathrm{Bernoulli}(p)$. Let us find the MLE of the model parameter $p\in[0,1]$.  Since the PMF is $f(x;p)=p^x(1-p)^{1-x}$, $x=0,1,$ the likelihood function is 
\begin{equation}\label{eq:likBern}
\begin{split}
\mathcal{L}_n(p|X_1,\ldots,X_n)&=\prod_{i=1}^np^{X_i}(1-p)^{1-X_i}\\
&=
p^{\sum_{i=1}^nX_i}(1-p)^{n-\sum_{i=1}^nX_i}=p^{S}(1-p)^{n-S},
\end{split}
\end{equation}
where $S=\sum_{i=1}^nX_i$. Figure~\ref{fig:bernoulli} shows the likelihood function (\ref{eq:likBern}) (up to a multiplicative constant) for the data generated from $\mathrm{Bernoulli}(p)$ with $p=1/3$ with $n=10, 100,$ and $1000$. 
%\vspace{-5mm}	

Now we need to find the value of $p$ that maximizes the likelihood. To this end, we need to solve $\mathcal{L}'_n(p)=0$ and check that $\mathcal{L}''_n(p)<0$ at the solution. This leads to a very familiar estimate\footnote{The same estimate is given by the plug-in princimple and the method of moments.}: 
\begin{equation}
\hat{p}_{\mathrm{MLE}}=\frac{S}{n}=\overline{X}_n.
\end{equation}
\hfill $\square$

\vspace{-5mm}

\subsection{Continuous Models}
Now let us turn to the continuous parametric case, where the data is modeled as a sample $X_1,\ldots,X_n\sim f(x;\theta)$ from a continuous  distribution with PDF $f(x;\theta)$. If we were to mimic the discrete case exactly, we would fail since the probability of observing the data $\mathbb{P}(X_1,\ldots,X_n|\theta)=\prod_{i=1}^n\mathbb{P}(X_i|\theta)=0$ in continuous settings. But there is a walk around this technical problem.

Recall that if $X\sim f(x;\theta)$, then for small $\epsilon\ll1$
\begin{equation}
\mathbb{P}(x-\epsilon<X<x+\epsilon|\theta)=\int_{x-\epsilon}^{x+\epsilon}f(t;\theta)dt\approx 2\epsilon f(x;\theta).
\end{equation}
Therefore,
\begin{equation}\label{eq:2}
\begin{split}
&\mathbb{P}(x_1-\epsilon<X_1<x_1+\epsilon,\ldots,x_n-\epsilon<X_n<x_n+\epsilon|\theta)\\
&=\prod_{i=1}^n\mathbb{P}(x_i-\epsilon<X_i<x_i+\epsilon|\theta)\\
&=\prod_{i=1}^n2\epsilon f(x_i;\theta)=(2\epsilon)^n\prod_{i=1}^nf(x_i;\theta).
\end{split}
\end{equation}
Of course if $\epsilon\rightarrow0$, both sides of (\ref{eq:2}) quickly converge to zero. But for small non zero $\epsilon$ the probability on the left hand side that we want to maximize, is proportional to $\prod_{i=1}^nf(x_i;\theta)$. This leads to the following natural definition of the likelihood function:
\begin{equation}
\mathcal{L}_n(\theta)=\mathcal{L}(\theta|X_1,\ldots,X_n)=\prod_{i=1}^nf(X_i;\theta).
\end{equation}
In words, the likelihood function is the joint density of the data, excepts that we treat it is a \textit{function of the model parameter} $\theta$\footnote{It is important, so once again: $\mathcal{L}_n(\theta)$ is not a density, in particular, $\int_\Theta \mathcal{L}_n(\theta)d\theta\neq 1$.}. As in the discrete case, the MLE is the value of $\theta$ that maximizes $\mathcal{L}_n(\theta)$.

\begin{figure}
	%\vspace{-5mm}	
	\includegraphics[width=\linewidth]{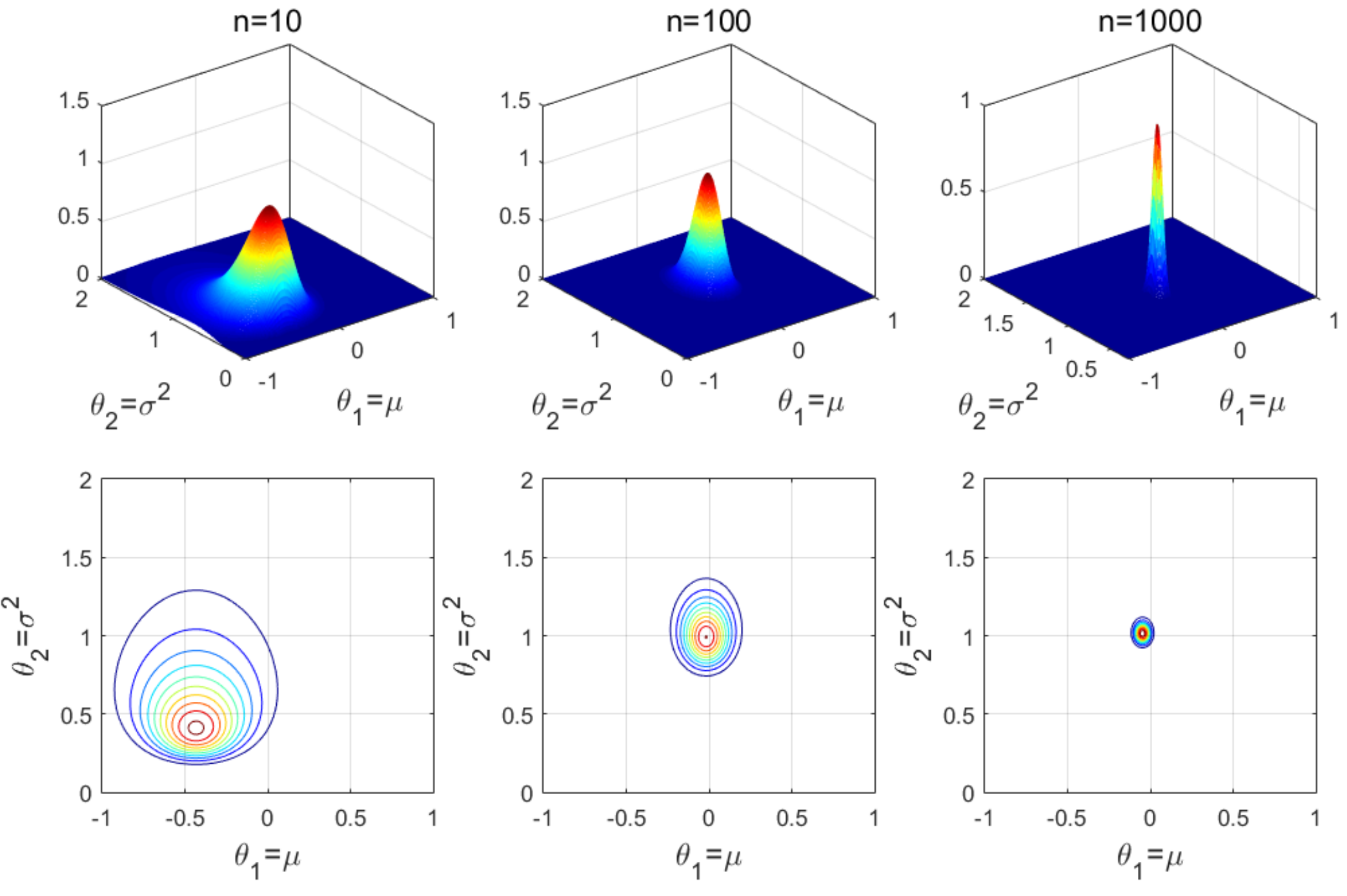}
	\caption{Top row: three likelihood functions for the data $X_1,\ldots,X_n\sim \mathcal{N}(0,1)$ with $n=10$, $n=100$, and $n=1000$. Bottom row: the corresponding contour plots. Red color corresponds to higher values of the likelihood function. Notice how the likelihood function becomes more and more concentrated around the true values  $\theta_1=\mu=0$ and $\theta_2=\sigma^2=1$ as the sample size $n$ increases. }
	\label{fig:normal}
\end{figure}

\paragraph{Example:} Let $X_1\ldots,X_n\sim\mathcal{N}(\mu,\sigma^2)$. Let us find the MLEs of the model parameters $\theta=(\theta_1,\theta_2)$, where $\theta_1=\mu$ and $\theta_2=\sigma^2$. The likelihood function (ignoring some multiplicative constants\footnote{Multiplication of $\mathcal{L}_n(\theta)$ by some positive constant $c$ (not depending on $\theta$) does notchange the MLE. Hence, for convenience, we will often drop some irrelevant constants in the likelihood function.}) is
\begin{equation}\label{eq:liknorm}
\begin{split}
\mathcal{L}(\theta|X_1,\ldots,X_n)&=\prod_{i=1}^n\frac{1}{\sigma}\exp\left(-\frac{(X_i-\mu)^2}{2\sigma^2}\right)\\
&=\theta_2^{-\frac{n}{2}}\exp\left(-\frac{1}{2\theta_2}\sum_{i=1}^n(X_i-\theta_1)^2\right).
\end{split}
\end{equation}
Figure~\ref{fig:normal} shows the likelihood function (\ref{eq:liknorm}) for the data generated from $\mathcal{N}(0,1)$ with $n=10, 100$ and $1000$.
\vspace{-5mm}

The next step is to find the MLEs by solving the system of two equations $\frac{\partial\mathcal{L}_n(\theta)}{\partial\theta_1}=0$ and $\frac{\partial\mathcal{L}_n(\theta)}{\partial\theta_2}=0$ for $\theta_1$ and $\theta_2$\footnote{And then checking that this is indeed the maximum, not the minimum!}. But differentiating (\ref{eq:liknorm}) is a daunting task\footnote{At least for me :)}. Technically, it is much easier to differentiate its logarithm $l_n(\theta)=\log\mathcal{L}_n(\theta)$. Note that since $\log$ is an increasing function, maximizing $\mathcal{L}_n(\theta)$ is equivalent to maximizing $l_n(\theta)$. 
\begin{equation}
l_n(\theta|X_1,\ldots,X_n)=-\frac{n}{2}\log\theta_2-\frac{1}{2\theta_2}\sum_{i=1}^n(X_i-\theta_1)^2.
\end{equation}
Solving $\frac{\partial l_n(\theta)}{\partial\theta_1}=0$ and $\frac{\partial l_n(\theta)}{\partial\theta_2}=0$ for $\theta_1=\mu$ and $\theta_2=\sigma^2$ gives a familiar result:
\begin{equation}
\hat{\mu}_{\mathrm{MLE}}=\overline{X}_n \hspace{3mm}\mbox{ and } \hspace{3mm} \hat{\sigma}^2_{\mathrm{MLE}}=\frac{1}{n}\sum_{i=1}^n(X_i-\overline{X}_n)^2.
\end{equation}
It can be verified\footnote{\eg [CB], Example 7.2.11 on p.321.} that these values indeed define the global maximum of the likelihood. \hfill $\square$

\section{Log-Likelihood}
In the last example, we maximized the logarithm of the likelihood function because it was theoretically equivalent\footnote{Maximizing $h(x)$ is equivalent to maximizing $\log h(x).$}, but technically much easier. This is often the case for many parametric models. The log of the likelihood function is called, es expected, the \textit{log-likelihood}:
\begin{equation}
l_n(\theta)=\log\mathcal{L}_n(\theta)=\sum_{i=1}^n \log f(X_i;\theta).
\end{equation}
Sums are easier to differentiate than products.

\section{Plug-In, MOM, and  MLE}

So far we consider two examples of maximum likelihood estimation --- Bernoulli and normal models --- and in both cases MLE agrees with the corresponding MOM and plug-in estimates. Recall (lecture~8) that plug-in and MOM disagree in estimating the upper bound of the uniform model $U[0,\theta]$: $\hat{\theta}_{\mathrm{plug\mbox{-}in}}=X_{(n)}$ and $\hat{\theta}_{\mathrm{MOM}}=2\overline{X}_{n}$. What about MLE?

\begin{figure}
	\vspace{-4mm}	
	\centerline{\includegraphics[width=.8\linewidth]{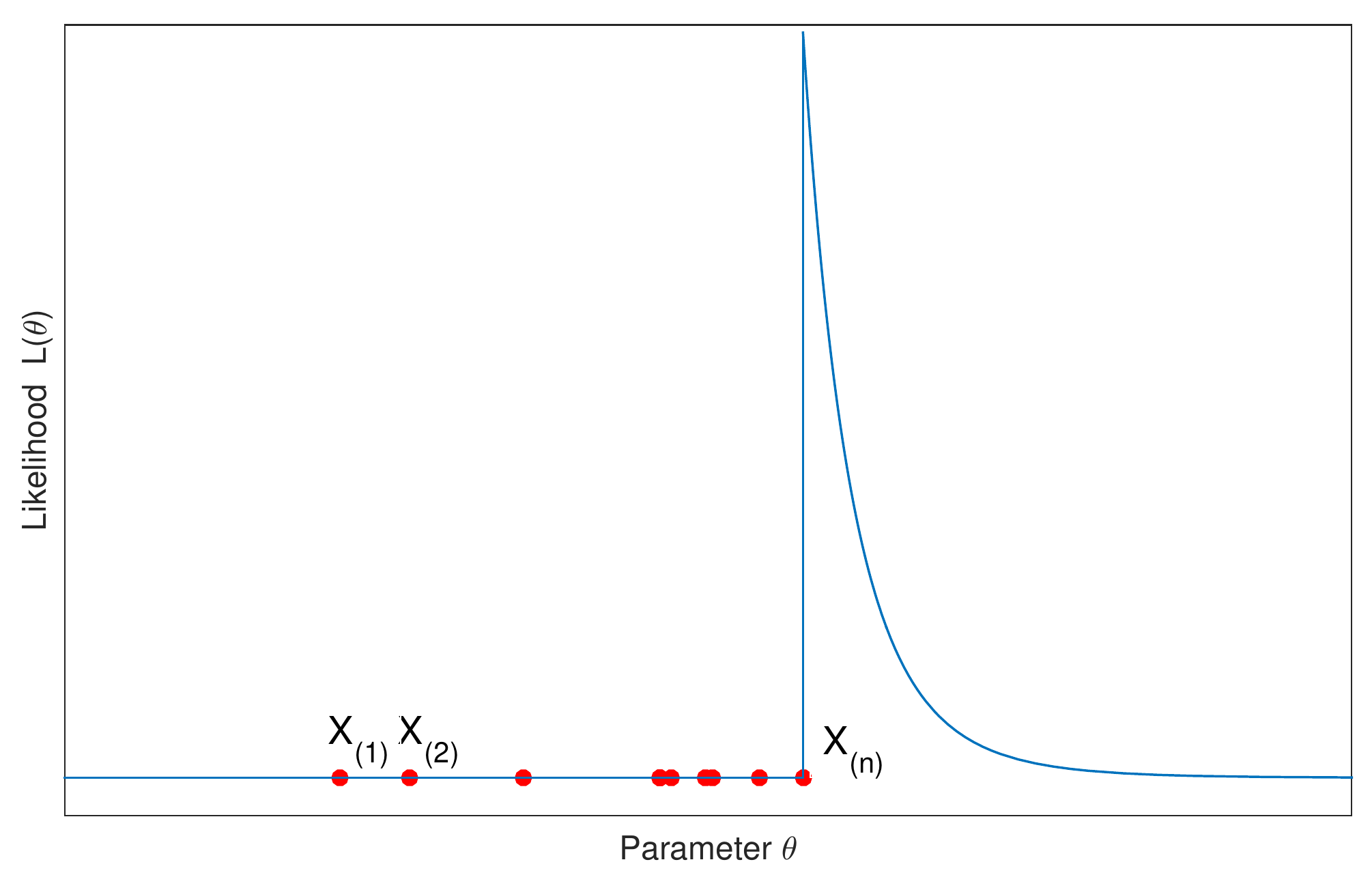}}
	\caption{The likelihood function (\ref{eq:likuni}) for the uniform model.}
	\label{fig:uniform}
\end{figure}

\paragraph{Example:} Let $X_1\ldots,X_n\sim U[0,\theta]$. Let us find the MLE of the model parameter $\theta$. Given $\theta$, the PDF is 
\begin{equation}
f(x;\theta)=\begin{cases}
\frac{1}{\theta}, & \mbox{if } x\in[0,\theta], \\
0, & \mbox{if } x\notin[0,\theta].
\end{cases}
\end{equation}
The likelihood function is then (keep in mind that in (\ref{eq:likuni}), $X_1,\ldots,X_n$ are fixed, and $\theta$ is a variable)
\begin{equation}\label{eq:likuni}
\mathcal{L}(\theta|X_1,\ldots,X_n)=\prod_{i=1}^n f(X_i;\theta)=
\begin{cases}
0, & \mbox{if } \theta<X_{(n)},\\
\frac{1}{\theta^n}, & \mbox{if } \theta\geq X_{(n)}. 
\end{cases}
\end{equation}
The likelihood function (\ref{eq:likuni}) is shown schematically in Fig.~\ref{fig:uniform}.

The MLE of $\theta$ is therefore $\hat{\theta}_{\mathrm{MLE}}=X_{(n)}$, the same as the plug-in estimate. \hfill $\square$\bigskip

Thus, MLE agrees with the plug-in estimate in all three examples. It turns out however, that the above example can be slightly modified in such a way that all three methods give different results. Namely, let us consider the model $U(0,\theta)$, where the support of the uniform distribution is the open interval $(0,\theta)$. The plug-in and MOM estimates remain the same in this case, but it is easy to see that the MLE simply does not exist, since the new likelihood function
\begin{equation}\label{eq:likuninew}
\mathcal{L}(\theta|X_1,\ldots,X_n)=
\begin{cases}
0, & \mbox{if } \theta \leq X_{(n)},\\
\frac{1}{\theta^n}, & \mbox{if } \theta> X_{(n)},
\end{cases}
\end{equation}
which does not have a maximum. But of course, this example is quite  artificial: for all practical purposes $\hat{\theta}=X_{(n)}+\epsilon$, where $\epsilon\ll1$ will be a good estimate for $\theta$.  Can you come up with a better example where all three methods give substantially  different estimates?

\section{Final Remarks}
In the next Lecture, we will discuss good properties of MLEs. So let us list a few bad ones here. There are a few  issues associated with the general problem of finding the global maximum of a function, and, therefore, with maximum likelihood estimation.
\begin{enumerate}
	\item  The MLE may not exist.     
	\item The MLE may not be unique. 
	\item Finding the global maximum can be a nontrivial task. In some cases, \eg for the Bernoulli and normal models, this problem reduces to a simple calculus problem. But in most applications (even using common statistical models) MLEs can't be found analytically and some numerical optimization methods must be used. 
	\item The likelihood function may have several local maxima. In this case, a maximum found by a numerical method may not be the global maximum, and, therefore, some additional checks are required.
	\item Sensitivity to small changes in data: sometimes a slightly different sample will produce a vastly different MLE\footnote{For example, I. Otkin et al (1981) ``\href{http://www.jstor.org/stable/2287523}{A comparison of $n$ estimators for the binomial distribution},'' \textit{Journal of the American Statistical Association} 76(375), 637-642 demonstrate this effect with the binomial model $\mathrm{Binomial}(k,p)$. If the data is $X_1=16, X_2=18, X_3=22, X_4=25, X_5=27$, then $\hat{k}_{\mathrm{MLE}}=99$, but if $X_5=28$, then  $\hat{k}_{\mathrm{MLE}}=190$!}, making the use of the method at least questionable. This happens when the likelihood function is very flat in the neighborhood of its maximum.  
\end{enumerate}

\section{Further Reading}
\begin{enumerate}
	\item The MLEs for the multivariate normal model $\mathcal{N}(\mu,\Sigma)$, where $\mu\in\mathbb{R}^k$ and $\Sigma$ is a $k\times k$ symmetric, positive definite covariance matrix are a straightforward generalization of the considered univariate model. The expressions can be found in many textbooks, for example, in [Wa], Sec. 14.3.
	\item A tutorial exposition of maximum	likelihood estimation  is provided in J. Myung (2003) ``\href{http://www.sciencedirect.com/science/article/pii/S0022249602000287}{Tutorial on maximum likelihood estimation},'' \textit{Journal of Mathematical Psychology} 47: 90-100.
\end{enumerate}

\section{What is Next?} 
The popularity of MLEs stems from their attractive properties for large sample sizes. We will discuss these properties in the next~\mbox{Lecture}.

\chapter{Properties of Maximum Likelihood Estimates}

\newthought{The method} of maximum likelihood is, by far, the most popular method of parametric inference. Its popularity stems from the nice asymptotic properties of MLEs: if the parametric model $\mathcal{F}=\{f(x;\theta)\}$ satisfies certain regularity conditions\footnote{These are essentially smoothness conditions.}, then the MLE is consistent, asymptotically normal, asymptotically unbiased, asymptotically efficient, and  equivariant. Let us  discuss  these properties in detail. So, assume that
\begin{equation}\label{eq:model}
X_1,\ldots,X_n\sim f(x;\theta),
\end{equation} 
let $\theta_0$ denote the true value of $\theta\in\Theta$, and let $\hat{\theta}_n$ be the MLE of $\theta$.

\section{Consistency}
Consistency is a basic ``must have'' property for any reasonable estimate. As MOM, the MLE is consistent: 
\begin{equation}\label{eq:consistentcy}
\hat{\theta}_n \stackrel{\mathbb{P}}{\longrightarrow}\theta_0.
\end{equation}
In words, as we get more and more data, the MLE $\hat{\theta}_n $ becomes more and more accurate, and gives the correct answer in the long run.

I will use the urge to give you a sketch of a proof of consistency as an opportunity to introduce another important notion which is frequently used in probability and information theory: \textit{Kullback-Leibler (KL) distance} which measures  the difference between two probability distributions $f$ and $g$\footnote{As usual, in the discrete settings, where $f$ and $g$ are PMFs, the integral sign in (\ref{eq:KL}) is replaced by  a sum.},
\begin{equation}\label{eq:KL}
D(f,g)=\int f(x)\log\frac{f(x)}{g(x)}dx. 
\end{equation}   
It can be shown that $D(f,g)\geq0$ and $D(f,g)=0$ if and only if $f=g$. However, it is not symmetric: $D(f,g)\neq D(g,f)$. Despite the name, the KL distance is not really a distance in the formal sense\footnote{That is why it is often called the \textit{KL  divergence.}}.

\paragraph{Sketch of Proof} of (\ref{eq:consistentcy}). To find the MLE, we need to maximize the log-likelihood:
\begin{equation}
l_n(\theta)=\sum_{i=1}^n\log f(X_i;\theta) \rightarrow \max.
\end{equation}
The expression of the log-likelihood as a sum of iid quantities calls for the use of the law of large numbers: 
\begin{equation}\label{12eq:1}
\begin{split}
\frac{l_n(\theta)}{n}&=\frac{1}{n}\sum_{i=1}^n\log f(X_i;\theta)\stackrel{\mathbb{P}}{\longrightarrow}\mathbb{E}\left[\log f(X;\theta)\right]\\
&=\int f(x;\theta_0)\log f(x;\theta)dx.
\end{split}
\end{equation}
The right hand side of (\ref{12eq:1}) reminds us a bit the KL distance between $f(x;\theta_0)$ and $f(x;\theta)$. It is not quite the distance, but it is straightforward to cook it up: 
\begin{equation}
\begin{split}
\frac{l_n(\theta)}{n}&\stackrel{\mathbb{P}}{\longrightarrow}\int f(x;\theta_0)\log\left( \frac{f(x;\theta)}{f(x;\theta_0)}f(x;\theta_0)\right)dx\\
&=\int f(x;\theta_0)\log\frac{f(x;\theta)}{f(x;\theta_0)}dx+
\int f(x;\theta_0)\log  f(x;\theta_0)dx\\
&=-D(\theta_0,\theta)+\xi(\theta_0),
\end{split}
\end{equation}
where $D(\theta_0,\theta)=D(f(x;\theta_0),f(x;\theta))$ and $\xi(\theta_0)=\int f(x;\theta_0)\log  f(x;\theta_0)dx$ is a function of $\theta_0$. Thus, for large $n$,
\begin{equation}
l_n(\theta)\approx -nD(\theta_0,\theta) + n\xi(\theta_0).
\end{equation}
Since $D(\theta_0,\theta)\geq0$ and $D(\theta_0,\theta)=0$ if and only if $\theta=\theta_0$\footnote{Strictly speaking, we must assume that different values of the parameter $\theta$ correspond to different distributions: $if \theta_1\neq\theta_2 \Rightarrow D(\theta_1,\theta_2)>0$. Such models $\mathcal{F}$ are called \textit{identifiable}.}, the log-likelihood is maximized at $\hat{\theta}_n\approx\theta_0$, where the approximation becomes the exact equality in the limit $n\rightarrow\infty$. \hfill$\square$

\section{Asymptotic Normality}
It turns out that for large $n$,  the distribution of the MLE $\hat{\theta}_{n}$ is approximately normal. Namely, under appropriate regularity conditions\footnote{More precisely, $\frac{\hat{\theta}_{n}-\theta_0}{\mathrm{se}}$ converges to the standard normal variable in distribution.}, 
\begin{equation}\label{eq:AN}
\hat{\theta}_{n}\approxdist\mathcal{N}(\theta_0,\mathrm{se}^2),
\end{equation}
where $\mathrm{se}$ is the standard error of MLE, $\mathrm{se}=\mathrm{se}[\hat{\theta}_{n}]=\sqrt{\mathbb{V}[\hat{\theta}_{n}]}$. Moreover, the standard error can be approximated analytically: 
\begin{equation}\label{12eq:se}
\mathrm{se}\approx\frac{1}{\sqrt{nI(\theta_0)}}, 
\end{equation}
where $I(\theta_0)$ is the \textit{Fisher information} of a random variable $X$ with distribution $f(x;\theta_0)$ from the family $\mathcal{F}=\{f(x;\theta), \theta\in\Theta\}$:
\begin{equation}\label{eq:FI1}
I(\theta_0)=\mathbb{E}\left[\left(\left.\frac{\partial\log f(X;\theta)}{\partial\theta}\right|_{\theta=\theta_0}\right)^2\right].
\end{equation}

Let us give an informal interpretation of the Fisher information. The derivative\footnote{It is called the \textit{score function}. In terms of the score function, $I(\theta_0)=\mathbb{E}[s^2(X;\theta_0)]$.}
\begin{equation}\label{eq:derivative}
s(x;\theta_0)=\left.\frac{\partial\log f(x;\theta)}{\partial\theta}\right|_{\theta=\theta_0}=\frac{\left.\frac{\partial f(x;\theta)}{\partial\theta}\right|_{\theta=\theta_0}}{f(x;\theta_0)}
\end{equation}
can be interpreted as a measure of how quickly the distribution $f$ will change at $X=x$ when we change the parameter $\theta$ near $\theta_0$.  To get the measure of the magnitude  of the change (we don't care about the sign\footnote{In fact, the expected value of the score function (\ref{eq:derivative}) is zero. To see this, differentiate $\int f(x;\theta)dx=1$ with respect to $\theta$ at $\theta_0$.}), we square the derivative (\ref{eq:derivative}). To get the average value of the measure across different values of $X$, we take the expectation. Thus, if $I(\theta_0)$ is large, the distribution $f(x;\theta)$ will change quickly when  $\theta$ moves near $\theta_0$. This means that $f(x;\theta_0)$ is quite different from ``neighboring'' distributions, and we should be able to pin it down well from the data. So, large $I(\theta_0)$ is good\footnote{This also follows from (\ref{12eq:se}): the larger $I(\theta_0)$, the smaller the standard error.}: $\theta_0$ is easier to estimate. If $I(\theta_0)$ is small, we have the opposite story: distribution $f(x;\theta)$ are similar to $f(x;\theta_0)$, and therefore the estimation of the true value is troublesome\footnote{If $I(\theta_0)$ is small, (\ref{12eq:se}) tells us that we need a lot of data, to get small standard error.}. 

While (\ref{eq:FI1}) provides an intuitive interpretation of the Fischer information, it is not convenient for actual computations. It can be shown that\footnote{The proof is straightforward computation of the integral in (\ref{eq:FI2}). } 
\begin{equation}\label{eq:FI2}
\begin{split}
I(\theta_0)&=-\mathbb{E}\left[\left.\frac{\partial^2\log f(X;\theta)}{\partial\theta^2}\right|_{\theta=\theta_0}\right]\\
&=-\int\left.\frac{\partial^2\log f(x;\theta)}{\partial\theta^2}\right|_{\theta=\theta_0}f(x;\theta_0)dx.
\end{split}
\end{equation}

To get a better feel for it, let us compute the Fischer information for some particular models. 

\paragraph{Example:} For the $\mathrm{Bernoulli}(p)$, the Fisher information is 
\begin{equation}
I(p)=\frac{1}{p(1-p)}.
\end{equation} 
It agrees with our intuitive interpretation: the closer $p$ to $0$ or $1$, the larger the Fisher information, and the easier to infer $p$ from the data. The fair coin $p=\frac{1}{2}$ provides the global minimum of the Fisher information. \hfill $\square$

\paragraph{Example:} For the normal model $\mathcal{N}(\mu,\sigma^2)$ with known $\sigma^2$, the Fisher information is constant:
\begin{equation}
I(\mu)=\frac{1}{\sigma^2}.
\end{equation}
Indeed, it is equally difficult (or easy) to infer different values of $\mu$. \hfill $\square$\bigskip

Let us now come back to the MLE standard error approximation (\ref{12eq:se}). We now know how to compute and (most importantly) how to think of $I(\theta_0)$. How does the factor $n$ appear in the denominator? Since the expected value of the score function is zero\footnote{See footnote $^7$.}, the Fisher information (\ref{eq:FI1}) is simply the variance of the score function:
\begin{equation}
I(\theta_0)=\mathbb{V}[s(X;\theta_0)].
\end{equation}
This is the Fisher information of a \textit{single} random variable distributed according to $f(x;\theta_0)$. If we have an iid sample of size $n$, $X_1\ldots,X_n\sim f(x;\theta_0)$, then the Fisher information of this sample is defined as the variance of the sum of score function:
\begin{equation}
I_n(\theta_0)=\mathbb{V}\left[\sum_{i=1}^ns(X_i,\theta_0)\right]=
\sum_{i=1}^n\mathbb{V}[s(X_i,\theta_0)]=nI(\theta_0).
\end{equation} 
Thus, the denominator of (\ref{12eq:se}) is the square root of the Fisher information of the sample $X_1,\ldots,X_n$.

\paragraph{Sketch of Proof}  of asymptotic normality of the MLE:
\begin{equation}
\hat{\theta}_n\approxdist\mathcal{N}\left(\theta_0,\frac{1}{nI(\theta_0)}\right).
\end{equation}
Recall that by definition,
\begin{equation}
\hat{\theta}_n=\arg\max_{\theta\in\Theta}l_n(\theta),
\end{equation}
where $l_n(\theta)$ is the log-likelihood. Let us Taylor-expand the derivative of $l_n(\theta)$ at $\theta=\theta_0$:
\begin{equation}
l'_n(\theta)= l'_n(\theta_0)+(\theta-\theta_0)l''_n(\theta_0) + \mbox{higher order terms} 
\end{equation}
Setting $\theta=\hat{\theta}_n$, noting that $l'_n(\hat{\theta}_n)=0$, and dropping the higher order terms, we obtain:
\begin{equation}\label{eq:20}
\hat{\theta}_n-\theta_0=-\frac{l'_n(\theta_0)}{l''_n(\theta_0)}.
\end{equation}
\begin{itemize}
	\item The nominator (using the central limit theorem): 
	\begin{equation}\label{eq:21}
	\begin{split}
	\frac{1}{n}l'_n(\theta_0)&=\frac{1}{n}\sum_{i=1}^n \left.\frac{\partial\log f(X_i;\theta)}{\partial\theta}\right|_{\theta=\theta_0}=\frac{1}{n}\sum_{i=1}^n s(X_i;\theta_0)\\
	&\approxdist\mathcal{N}\left(\mathbb{E}[s(X;\theta_0)],\frac{\mathbb{V}[s(X;\theta_0)]}{n}\right)=\mathcal{N}\left(0,\frac{I(\theta_0)}{n}\right).
	\end{split}
	\end{equation}
	\item The denominator (using the law of large numbers):
	\begin{equation}\label{eq:22}
	\begin{split}
	\frac{1}{n}l''_n(\theta_0)&=\frac{1}{n}\sum_{i=1}^n \left.\frac{\partial^2\log f(X_i;\theta)}{\partial\theta^2}\right|_{\theta=\theta_0}\\
	&\approx \mathbb{E}\left[\left.\frac{\partial^2\log f(X;\theta)}{\partial\theta^2}\right|_{\theta=\theta_0}\right]=-I(\theta_0).
	\end{split}
	\end{equation}
\end{itemize}
Combining\footnote{Slutzky's theorem allows to to that: if $X_n\stackrel{D}{\longrightarrow}X$ and $Y_n\stackrel{\mathbb{P}}{\longrightarrow}a$, then $\frac{X_n}{Y_n}\stackrel{D}{\longrightarrow}\frac{X}{a}$.} (\ref{eq:20}), (\ref{eq:21}), and (\ref{eq:22}), we get what we need:
\begin{equation}
\hat{\theta}_n-\theta_0\approxdist \mathcal{N}\left(0,\frac{1}{nI(\theta_0)}\right).
\end{equation}
\hfill$\square$

\paragraph{Remark.} Why is this a sketch, not a proof? Because we need to make an appropriate regularity conditions on the statistical model $\mathcal{F}=\{f(x;\theta)\}$ to make sure that all considered derivatives exist, the Fisher information is well defined, the higher order terms in the Taylor series go to zero, the conditions for the Law of Large Numbers and Central Limit Theorem are satisfied,~etc. \hfill$\square$ 

\subsection{Asymptotic Confidence Intervals}

The asymptotic normality (\ref{eq:AN}) is a nice theoretical result, but how to use it in practice if the standard error (\ref{eq:se})  is unknown since $\theta_0$ is unknown. It can be shown that the standard error of the MLE can be estimated by 
\begin{equation}
\widehat{\mathrm{se}}=\frac{1}{\sqrt{nI(\hat{\theta}_n)}}, 
\end{equation}
and the asymptotic normality result will still hold: for large $n$,
\begin{equation}\label{eq:AN2}
\frac{\hat{\theta}_{n}-\theta_0}{\widehat{\mathrm{se}}}\approxdist\mathcal{N}(0,1).
\end{equation}

We can use (\ref{eq:AN2}) for construction an asymptotic confidence interval for $\theta_0$. Indeed let
\begin{equation}
\mathcal{I}_n=\hat{\theta}_n\pm z_{\alpha/2}\widehat{\mathrm{se}},
\end{equation}
then the usual computation shows that
\begin{equation}
\mathbb{P}(\theta_0\in\mathcal{I}_n)\rightarrow 1-\alpha, \hspace{3mm}\mbox{ as } n\rightarrow\infty. 
\end{equation} 
For example, if $\alpha=0.05$, then $z_{\frac{\alpha}{2}}\approx -2$, and $\hat{\theta}_n\pm 2\widehat{\mathrm{se}}$ is an approximate $95\%$ confidence interval for $\theta_0.$

\paragraph{Example:} Supposed that  $X_1,\ldots,X_n\sim\mathrm{Bernoulli}(p)$, then $\hat{p}_{\mathrm{MLE}}=\overline{X}_n$, $\widehat{\mathrm{se}}=\sqrt{\frac{\overline{X}_n(1-\overline{X}_n)}{n}}$, and an approximate $95\%$ confidence interval for $p$ is
\begin{equation}
\mathcal{I}_n=\overline{X}_n\pm2\sqrt{\frac{\overline{X}_n(1-\overline{X}_n)}{n}}.
\end{equation}\hfill $\square$

\vspace{-5mm}
\section{Asymptotic Unbiasedness}
As a byproduct of asymptotic normality with vanishing variance, we have that the MLE is asymptotically unbiased\footnote{By the way, this is not a by-product of consistency! Somewhat counterintuitive, $X_n \stackrel{\mathbb{P}}{\longrightarrow}a$ does not imply that $\mathbb{E}[X_n]\rightarrow a$. Can you give a counterexample?}:
\begin{equation}
\lim_{n\rightarrow\infty}\mathbb{E}[\hat{\theta}_n]=\theta_0.
\end{equation} 

\paragraph{Example:} Recall that for the normal $\mathcal{N}(\mu,\sigma^2)$ and uniform $U[0,\theta]$ models the MLEs are:
\begin{equation}
\hat{\mu}_{\mathrm{MLE}}=\overline{X}_n, \hspace{3mm} \hat{\sigma}^2_{\mathrm{MLE}}=\frac{1}{n}\sum_{i=1}^n(X_i-\overline{X}_n)^2, \hspace{3mm} \hat{\theta}_{\mathrm{MLE}}=X_{(n)}.
\end{equation}
The first estimate is unbiased for any $n$, but the last two are biased:
\begin{equation}
\mathbb{E}[\hat{\sigma}^2_{\mathrm{MLE}}]=\frac{n-1}{n}\sigma^2, \hspace{3mm} \mathbb{E}[\hat{\theta}_{\mathrm{MLE}}]=\frac{n}{n+1}\theta.
\end{equation}
In both cases, the bias disappears as the sample size increases. \hfill $\square$

\section{Asymptotic Efficiency}
We see that, when $n$ is large, the MLE is approximately unbiased. This leads to a natural question: what is the smallest possible value of the variance of an unbiased estimate? 

The answer is given by the \textit{Cramer-Rao inequality}. Let $\tilde{\theta}_n$ be any \textit{unbiased} estimate of the parameter $\theta$ whose true value is $\theta_0$, then\footnote{Again, (\ref{eq:CR}) is true if the underlying statistical model satisfies certain regularity conditions.} 
\begin{equation}\label{eq:CR}
\mathbb{V}[\tilde{\theta}_n]\geq\frac{1}{nI(\theta_0)}.
\end{equation}

An unbiased estimate whose variance achieves this lower bound is said to be \textit{efficient}. In some sense, it is the best estimate\footnote{An efficient estimate may not exist.}.  Note that the right-hand-side of (\ref{eq:CR}) is exactly the asymptotic variance of the MLE. Therefore, the MLE is \textit{asymptotically efficient}. 
Roughly speaking, this means that among all well-behaved estimates, the MLE has the smallest variance, at least for large samples. 

\paragraph{Example:} Assume that $X_1\ldots,X_n\sim\mathcal{N}(\mu,\sigma)$ with known $\sigma^2$. Then the MLE $\hat{\mu}_{\mathrm{MLE}}=\overline{X}_n$ and its variance is $\frac{\sigma^2}{n}$. This is exactly the Cramer-Rao lower bound. Another reasonable estimate of the mean is the sample median $\tilde{\mu}_n$. It can be shown that its asymptotically unbiased and its variance is $\frac{\pi}{2}\frac{\sigma^2}{n}$. Thus, $\tilde{\mu}_n$ converges to the right value, but it has a larger variance than the MLE. \hfill $\square$\bigskip

Note, however, that 
\begin{itemize}
	\item[a)] For a finite sample size $n$, MLE may not be efficient,
	\item[b)] If the MLE is efficient\footnote{\eg for the normal or Poisson models.}, there may still exist a biased estimate with a smaller mean squared error, and 
	\item[c)] MLEs are not the only asymptotically efficient estimates.
\end{itemize}

\section{Equivariance}
This is a non-asymptotic\footnote{Valid for all $n$.} ``bonus'' to the nice asymptotic properties of MLEs. 
Suppose we are interested in estimating a parameter $\tau$, which is a function of $\theta$ which parametrizes the model (\ref{eq:model})\footnote{Recall a blood test example from Lecture~\ref{ch:MOM}.}, $\tau=g(\theta)$. It turns out that if we know $\hat{\theta}_{\mathrm{MLE}}$, then the MLE of $\tau$ is simply
\begin{equation}\label{eq:equi}
\hat{\tau}_{\mathrm{MLE}}=g(\hat{\theta}_{\mathrm{MLE}}).
\end{equation}   
This property is called \textit{equivariance} or \textit{transformation invariance}.  

\paragraph{Example:} Recall that in Lecture~\ref{ch:MLE}, we found that the MLE of variance under the normal model $\mathcal{N}(\mu,\sigma)$ is 
\begin{equation}
\hat{\sigma}^2_{\mathrm{MLE}}=\frac{1}{n}\sum_{i=1}^n(X_i-\overline{X}_n)^2.
\end{equation}
Thanks to the equivariance of maximum likelihood estimation, we can readily find the MLE of the standard deviation\footnote{We don't need to solve the calculus problem again with respect to $\theta_2=\sigma$.}:
\begin{equation}
\hat{\sigma}_{\mathrm{MLE}}=\sqrt{\frac{1}{n}\sum_{i=1}^n(X_i-\overline{X}_n)^2}.
\end{equation}\vspace{-6mm}

\hfill $\square$\bigskip

Equivariance holds for arbitrary functions $g$\footnote{Theorem 7.2.10 in [CB].}. If $g$ is a one-to-one correspondence, then it is easy to understand why. In this case, there exists the inverse function $\theta=g^{-1}(\tau)$. This means that we can parametrize the model using $\theta$, as in (\ref{eq:model}), or using $\tau$:
\begin{equation}
\begin{split}
& f(x;\theta)=f(x;g^{-1}(\tau))=:\tilde{f}(x;\tau),  \\
&{\mathcal{F}}=\{\tilde{f}(x;\tau), \tau\in T\}, \hspace{2mm} T=g(\Theta).
\end{split}
\end{equation}
Let $\widetilde{\mathcal{L}}$ denote the likelihood in the $\tau$-parametrization. Then 
\begin{equation}
\widetilde{\mathcal{L}}(\tau)=\prod \tilde{f}(X_i;\tau)=\prod {f}(X_i;\theta)={\mathcal{L}}(\theta).
\end{equation}
Therefore, for any $\tau$:
\begin{equation}
\widetilde{\mathcal{L}}(\tau)={\mathcal{L}}(\theta)\leq {\mathcal{L}}(\hat{\theta}_{\mathrm{MLE}})=\widetilde{\mathcal{L}}(g(\hat{\theta}_{\mathrm{MLE}})),
\end{equation}
which means exactly (\ref{eq:equi}).

\section{Final Remark}
The considered nice properties explain the popularity of MLEs. But we should always keep in mind, that if the statistical model $\mathcal{F}=\{f(x;\theta), \theta\in\Theta\}$ is wrong, meaning that there is no $\theta$ in $\Theta$ that model the data adequately, then the inference based on $f(x;\hat{\theta}_{\mathrm{MLE}})$ may be very poor. Moreover, even if the model is correct, it may not satisfy the regularity conditions (which is often difficult to check) required for the MLE to have the above asymptotic properties. Finally, even if the model is correct and satisfies the regularity conditions, finding the MLE could be very challenging\footnote{Here we focused on a one parameter case. Everything could be generalized to the arbitrary number of parameters. High-dimensional optimization is, in general, a non-trivial task.}: the (log) likelihood may not be analytically tractable, may have many local maxima, be sensitive to data, etc. 

\section{Further Reading}
\begin{enumerate}
	\item The regularity conditions, so often mentioned in these notes, are discussed in detail in many advanced (and often unreadable) texts on mathematical statistics. This [\href{http://www2.econ.iastate.edu/classes/econ671/hallam/documents/Asymptotic_Dist.pdf}{note}]
	provides a good trade-off between rigor and readability. 
	\item Computing the MLE using two standard numerical methods, Newton-Raphson and the expectation-maximization algorithm, are discussed in Sec.~9.13.4 of [Wa].
	\item For extension to multiparameter models, see Sec.~9.10 of [Wa].
\end{enumerate}

\section{What is Next?} 
``To be, or not to be...'' In the next Lecture, we will start discussing move to hypothesis testing. 

\chapter{Hypothesis Testing: General Framework}\label{ch:HypoTesting}

\newthought{In previous lectures}, we discussed how to estimate parameters in parametric and nonparametric settings. Quite often, however, researchers are interesting in checking a certain statement about a parameter, not its exact value. Suppose, for instance, that someone developed a new drug for reducing blood pressure. Let $\theta$ denote the average change in a patient's blood pressure after taking a drug. The big question is to test 
\begin{equation}
H_0:\hspace{1mm} \theta=0 \hspace{2mm}\mbox{ versus }\hspace{2mm} H_1:\hspace{1mm}\theta\neq0.
\end{equation}
The hypothesis $H_0$ is called the \textit{null hypothesis}. It states that, on average, the new treatment has zero effect\footnote{Hence the name ``null.''} on blood pressure. The \textit{alternative hypothesis}\footnote{Also sometimes called the \textit{research hypothesis}.} states that there is some effect. In this context, testing $H_0$ against $H_1$ is a primary problem. Even if we find out that $\theta\neq0$\footnote{Hopefully, $\theta<0$!}, estimating the value of $\theta$ is important, yet a secondary problem.  A part of statics that deals with this sort of ``yes/no'' problems is called \textit{hypothesis testing}. 

In this lecture, we discus a general framework of hypothesis testing. To get started let us consider the following toy example, that will help us to illustrate all main notions and ideas.

\section{Two Coins Example}
Suppose that Alice has two coins: fair and unfair, with the probabilities of heads $p_0=0.5$ and $p_1=0.7$ respectively. Alice chooses one of the coins, tosses it $n = 10$ times and tells Bob the number of heads, but does not tell him what coin she tossed. Based of the number of heads $k$, Bob has to decide which coin it was.
\begin{marginfigure}
	%\vspace{20mm}
	\centerline{\includegraphics[width=.8\linewidth]{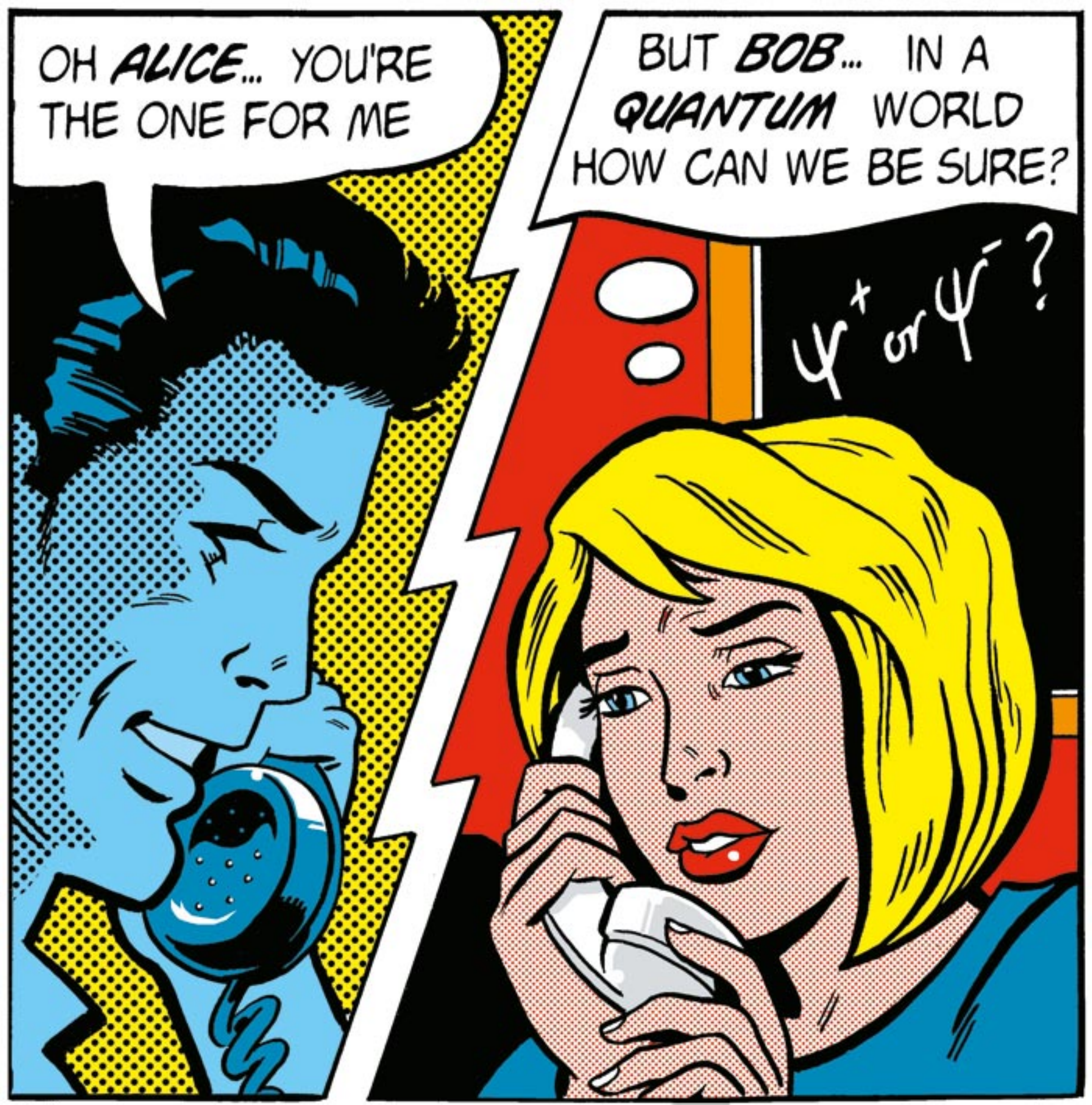}}\caption{Alice and Bob are two archetypal characters commonly used in  cryptography, game theory, physics, and now.... in statistics. Comics source: \href{http://physicsworld.com/cws/article/news/2013/apr/16/alice-and-bob-communicate-without-transferring-a-single-photon}{physicsworld.com}.}\label{fig:AB}
\end{marginfigure}

\begin{figure}
	\centerline{\includegraphics[width=.75\linewidth]{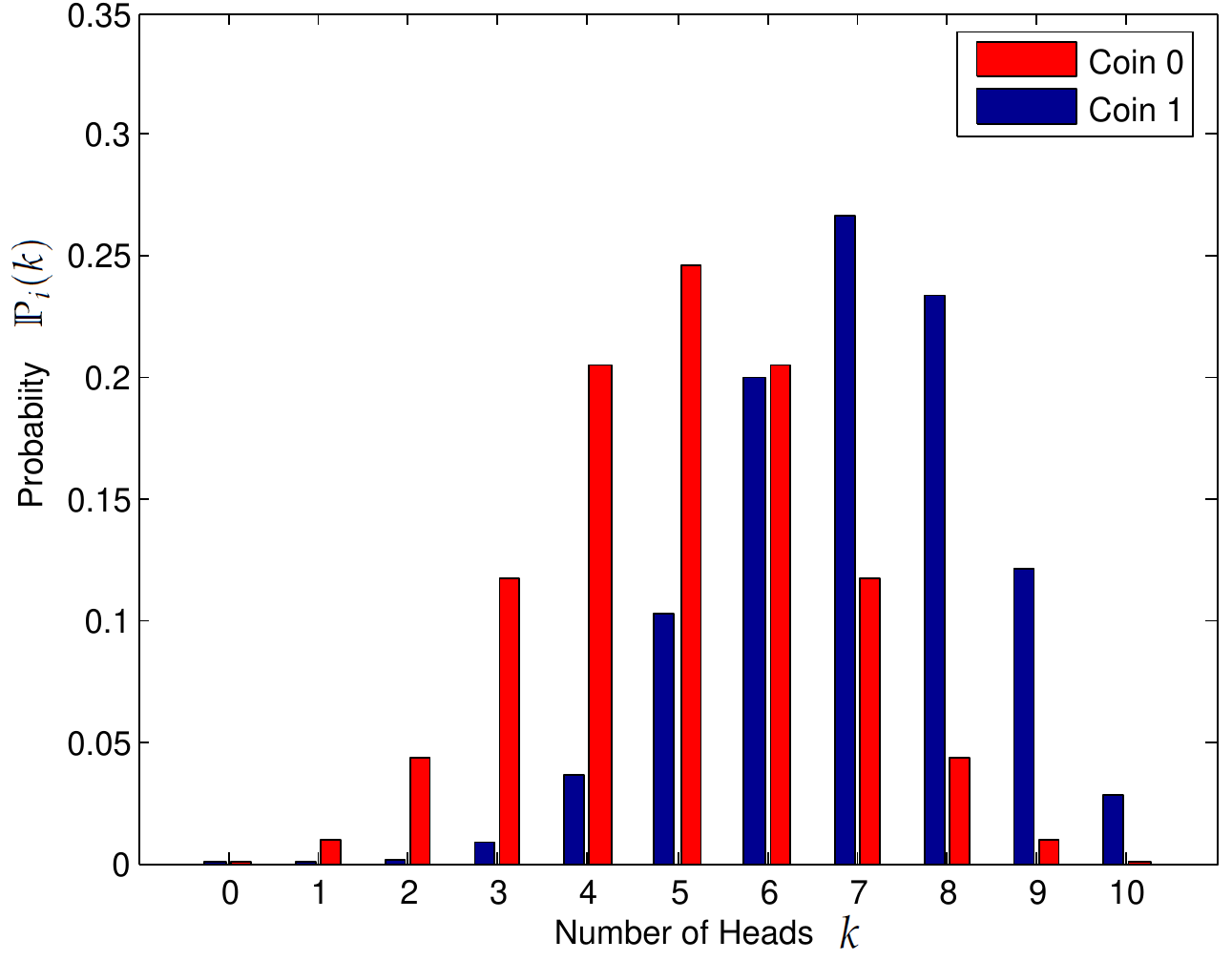}}\caption{Probabilities (\ref{eq:13prob}).}\label{fig:coins}
\end{figure}

Intuitively, it is clear that the larger $k=0,1,\ldots,n$, the more likely it was an unfair coin. If Alice tossed coin $i$ ($i=0$ is fair and $i=1$ is unfair), then the probability of getting exactly $k$ heads is given by the Binomial distribution $\mathrm{Bin}(n,p_i)$:
\begin{equation}\label{eq:13prob}
\mathbb{P}_i(k)={n \choose k} p_i^k(1-p_i)^{n-k}, \hspace{5mm} i=0,1.
\end{equation}
Figure~{\ref{fig:coins}} shows the values of these probabilities for different $k$.
Suppose that Bob observed only $k=2$ heads. Then 
\begin{equation}
\frac{\mathbb{P}_0(k=2)}{\mathbb{P}_1(k=2)}\approx 30,
\end{equation}
and, therefore, the fair coin is about 30 times more likely to produce this result than the unfair one. On the other hand, if there were $k=8$ heads, then
\begin{equation}
\frac{\mathbb{P}_0(k=8)}{\mathbb{P}_1(k=8)}\approx 0.19,
\end{equation}
which would favor the unfair coin. So, based on Fig.~(\ref{fig:coins}), Bob should guess that the coin is unfair if 
\begin{equation}\label{eq:78910}
k\in\{7,8,9,10\},
\end{equation}
and unfair otherwise. This is the simplest example of testing.

\section{General Framework}

Suppose that data $X_1,\ldots,X_n$ is modeled as a sample from a distribution $f\in\mathcal{F}$\footnote{The statistical model $\mathcal{F}$ can be either parametric or nonparametric.}. Let $\theta$ be the parameter of interest, and $\Theta$ be the set of all its possible values, called the \textit{parameter space}. Let $\Theta=\Theta_0\sqcup\Theta_1$ be a partition of the parameter space into two disjoint sets\footnote{Recall that $A=B\sqcup C$ means that $A=B\cup C$ and $B\cap C=\varnothing$.}. Suppose we wish to test
\begin{equation}
H_0:\hspace{1mm} \theta\in \Theta_0 \hspace{2mm}\mbox{ versus }\hspace{2mm} H_1:\hspace{1mm}\theta\in\Theta_1.
\end{equation}
We call $H_0$ the null hypothesis and $H_1$ the alternative hypothesis. 

Let $\Omega$ be the \textit{samples space}, \ie the range of data, $X=(X_1,\ldots,X_n)\in\Omega$. We test a hypothesis by finding an appropriate subset of outcomes $\mathcal{R}\subset\Omega$, called the \textit{rejection region}: 
\begin{equation}
\begin{split}
&\mbox{If } X\in\mathcal{R} \hspace{3mm}\Rightarrow \hspace{3mm} \mbox{reject } H_0,\\
&\mbox{If } X\notin\mathcal{R} \hspace{3mm}\Rightarrow \hspace{3mm} \mbox{accept } H_0.
\end{split}
\end{equation}

Usually, the rejection region has the following form:
\begin{equation}
\mathcal{R}=\{X\in\Omega: s(X)>c\},
\end{equation}
where $s$ is a \textit{test statistic} and $c$ is a \textit{critical value}. The problem of testing is then boils down to finding 
%\vspace{-2mm}
\begin{itemize}
	\item an appropriate statistic $s$ and
%	\vspace{-2mm}
	\item an appropriate critical value $c$.
\end{itemize}

In the two coin example, the data is the total number of heads $X=k$, which is modeled as a sample from the binomial distribution $\mathrm{Bin}(n,\theta)$, where $n=10$ and $\theta\in\Theta=\{0.5, 0.7\}$. The hull hypothesis is that the coin is fair: $H_0: \theta\in\Theta_0=\{0.5\}$, and the alternative is that the coin is not fair, $H_1: \theta\in\Theta_1=\{0.7\}$. The sample space is $\Omega=\{0,\ldots,10\}$. Bob tested the hypothesis using the rejection region $\mathcal{R}$ given by (\ref{eq:78910})\footnote{What is the test statistic and the critical value in this example?}. 

\section{The Null and Alternative}
Mathematically, the null and alternative hypotheses seem to play symmetric roles. Traditionally, however, the null hypothesis $H_0$ says that ``nothing interesting'' is going on\footnote{Recall the drag example from the beginning. $H_0$ says the new drag no effect on the blood pressure.}, the current theory is correct, no new effects, etc. The null hypothesis is a ``status quo.'' The alternative hypothesis, on the other hand, says that something interesting, something unexpected is happening: the old theory needs to be updated, new previously unseen effects are present, etc\footnote{This explains why we focus on the rejection region and not the acceptance region. The rejection region is where the surprise is living.}.

It is useful to think of hypothesis testing is a legal trial.  By default, we assume that someone is innocent\footnote{Presumption of innocence.} (the null hypothesis)  unless there is strong evidence that s/he is guilty (alternative). 

\begin{marginfigure}
	%\vspace{20mm}
	\centerline{\includegraphics[width=.6\linewidth]{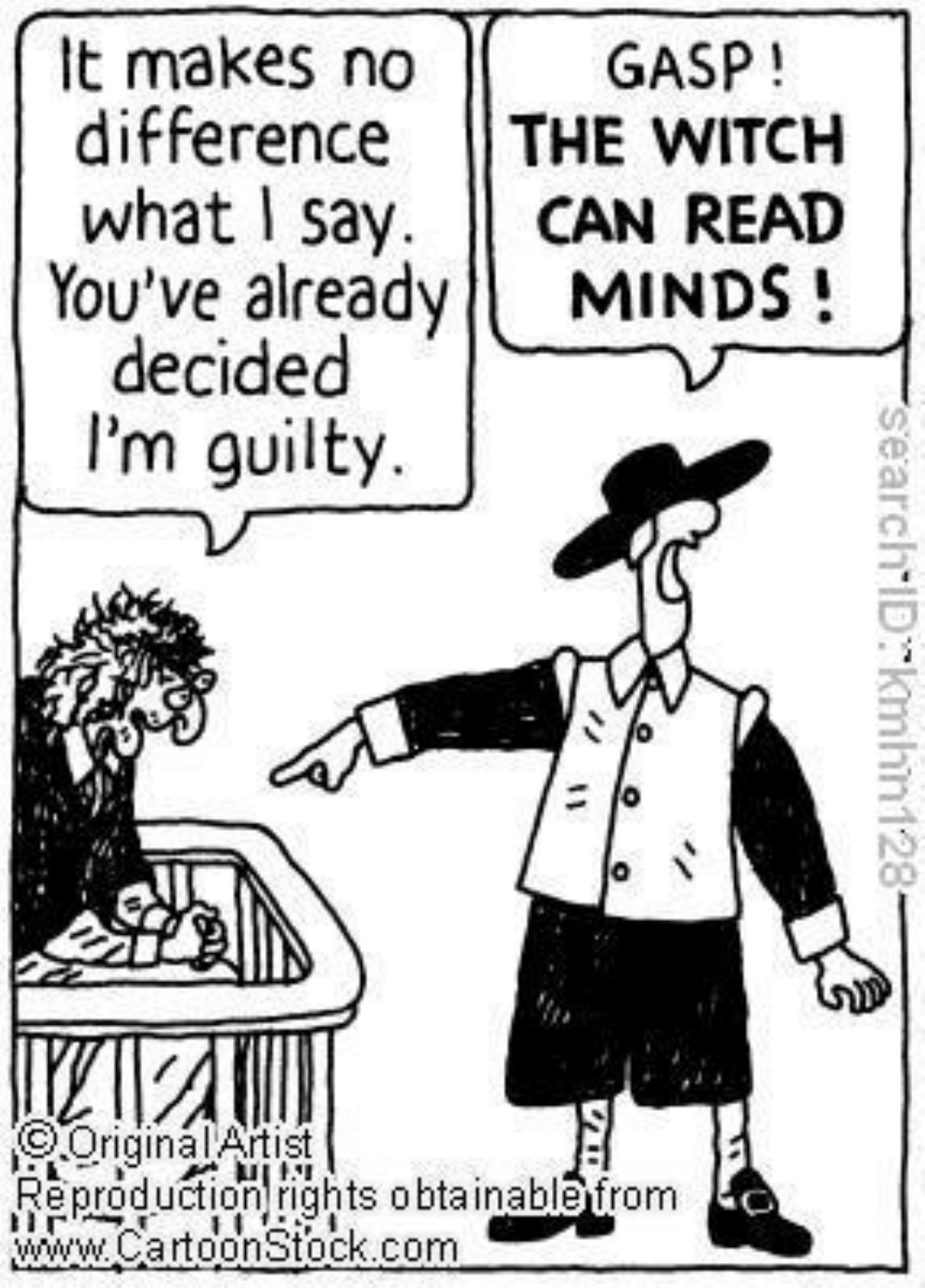}}\caption{Unfortunately, the presumption of innocence does on always work in real life. }\label{fig:guilty}
\end{marginfigure}

\paragraph{Question:} Suppose an engineer designed a new earthquake-resistant building. Let $p_F$ be the failure probability of the building under earthquake excitation. How would you formulate the null and alternative hypotheses if you wish to test whether or not the failure probability is smaller than a certain acceptable threshold $p_F^*$?

\section{Errors in Testing}
Can we guarantee that we make no errors when making conclusions from data? Of course, not. Data provides some, but not full, information about the unknown quantity of interest and helps to reduce the uncertainty, but not completely illuminate it. The errors are thus unavoidable\footnote{The unfair coin may produce 5 heads in which case Bob will make in error by accepting the hypothesis that the coin is fair.}. 

There are two types of errors in hypothesis testing with very boring names: \textit{type I error} and \textit{type II error}:
\begin{itemize}
	\item Type I error: rejecting $H_0$ when it is true.
	\item Type II error: accepting $H_0$ when it is not true.
\end{itemize} 
Purely mathematically\footnote{That is when we focus on equations and forget about the context. }, making both errors are equally bad. But, given the context discussed in the previous section, making a type I error is much worse than making a type II error: declaring an innocent person guilty is much worse than declaring a guilty person innocent. Probabilities of both errors can be computed using the so-called power function.

\section{Power Function}

\begin{marginfigure}
	\centerline{\includegraphics[width=\linewidth]{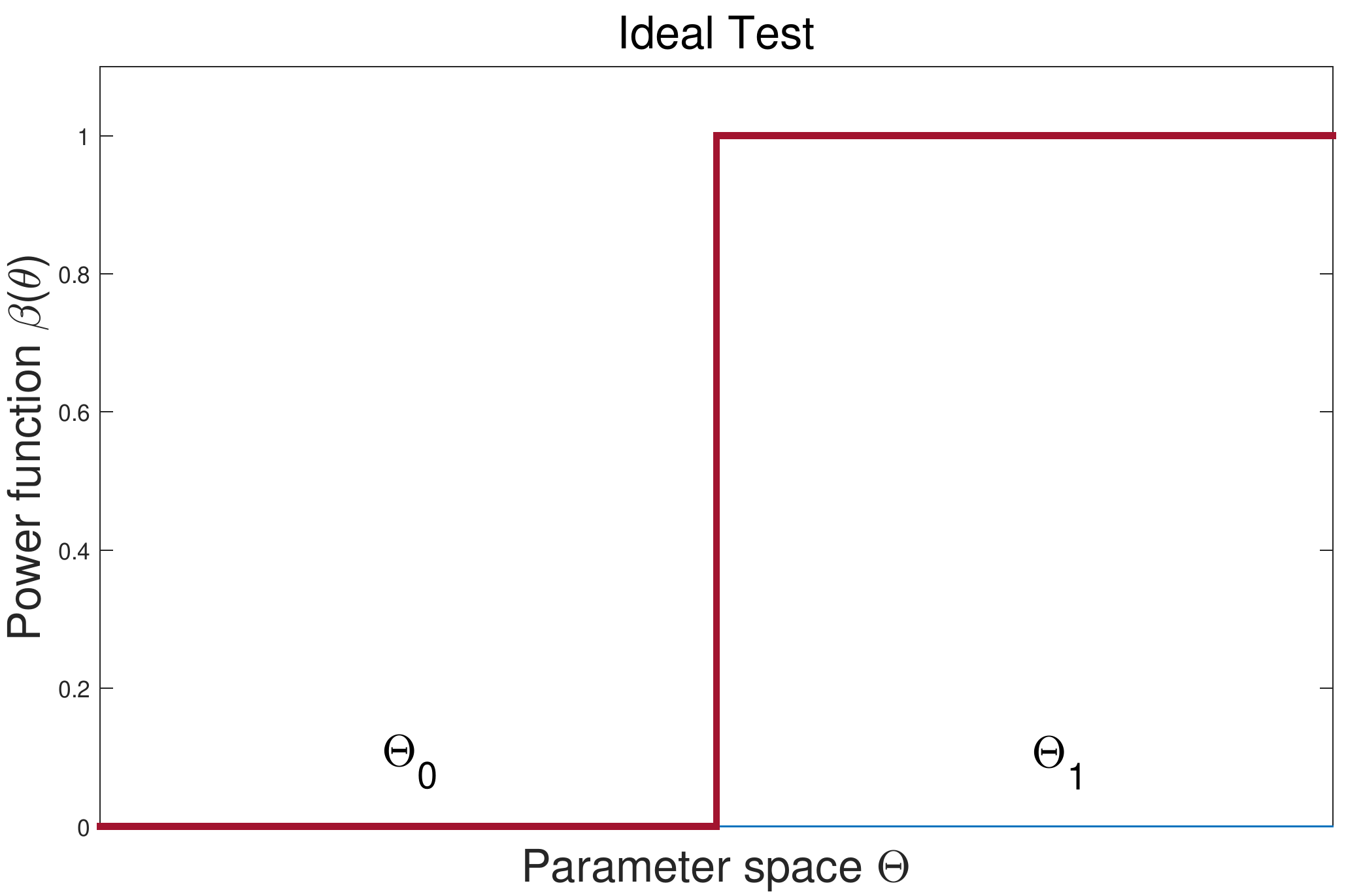}}\caption{The ideal power function.}\label{fig:ideal_power}
\end{marginfigure}

If $\mathcal{R}$ is the rejection region, then the probability of a type I error is
\begin{equation} \label{eq:error1}
\mathbb{P}(\mbox{Type I error})=\mathbb{P}(X\in\mathcal{R}|\theta\in\Theta_0).
\end{equation}
The probability of a type II error is
\begin{equation}\label{eq:error2}
\begin{split}
\mathbb{P}(\mbox{Type II error})&=\mathbb{P}(X\notin\mathcal{R}|\theta\in\Theta_1)\\
&=1-\mathbb{P}(X\in\mathcal{R}|\theta\in\Theta_1).
\end{split}
\end{equation}
From (\ref{eq:error1}) and (\ref{eq:error2}), we see that probabilities of both error are determined by function on the parameter space $\mathbb{P}(X\in\mathcal{R}|\theta)$ . This leads to the following definition. 
\begin{definition}
	The \textit{power function} of a hypothesis test with rejection region $\mathcal{R}$ is the function of $\theta$ defined by
	\begin{equation}
	\beta(\theta)=\mathbb{P}(X\in\mathcal{R}|\theta).
	\end{equation}
\end{definition}
In term of error probabilities:
\begin{equation}
\beta(\theta)=\begin{cases}
\mathbb{P}(\mbox{Type I error}), & \mbox{if } \theta\in\Theta_0,\\
1-\mathbb{P}(\mbox{Type II error}), & \mbox{if } \theta\in\Theta_1.
\end{cases}
\end{equation}
The ideal test will thus have the power function which is zero on $\Theta_0$ and one on $\Theta_1$, see Fig.~\ref{fig:ideal_power}. This ideal is rarely (never) achieved in practice. 

\paragraph{Example:} In the two coin example, the parameter space is a two point set $\Theta=\{0.5, 0.7\}$, $\Theta_0=\{0.5\}$, $\Theta_1=\{0.7\}$, and the power function is 
\begin{equation}
\begin{split}
\beta(\theta)&=\mathbb{P}(k\in\{7,8,9,10\}|\theta)\\
&=\sum_{k=7}^{10}{10 \choose k} \theta^k(1-\theta)^{10-k}
\approx\begin{cases}
0.17, & \mbox{if } \theta=0.5,\\
0.65, & \mbox{if } \theta=0.7.
\end{cases}
\end{split}
\end{equation}
This power function is not exactly what Bob would like to have, but in some sense (will discuss later) this is the best possible test. \hfill $\square$\bigskip

\begin{marginfigure}
	\centerline{\includegraphics[width=\linewidth]{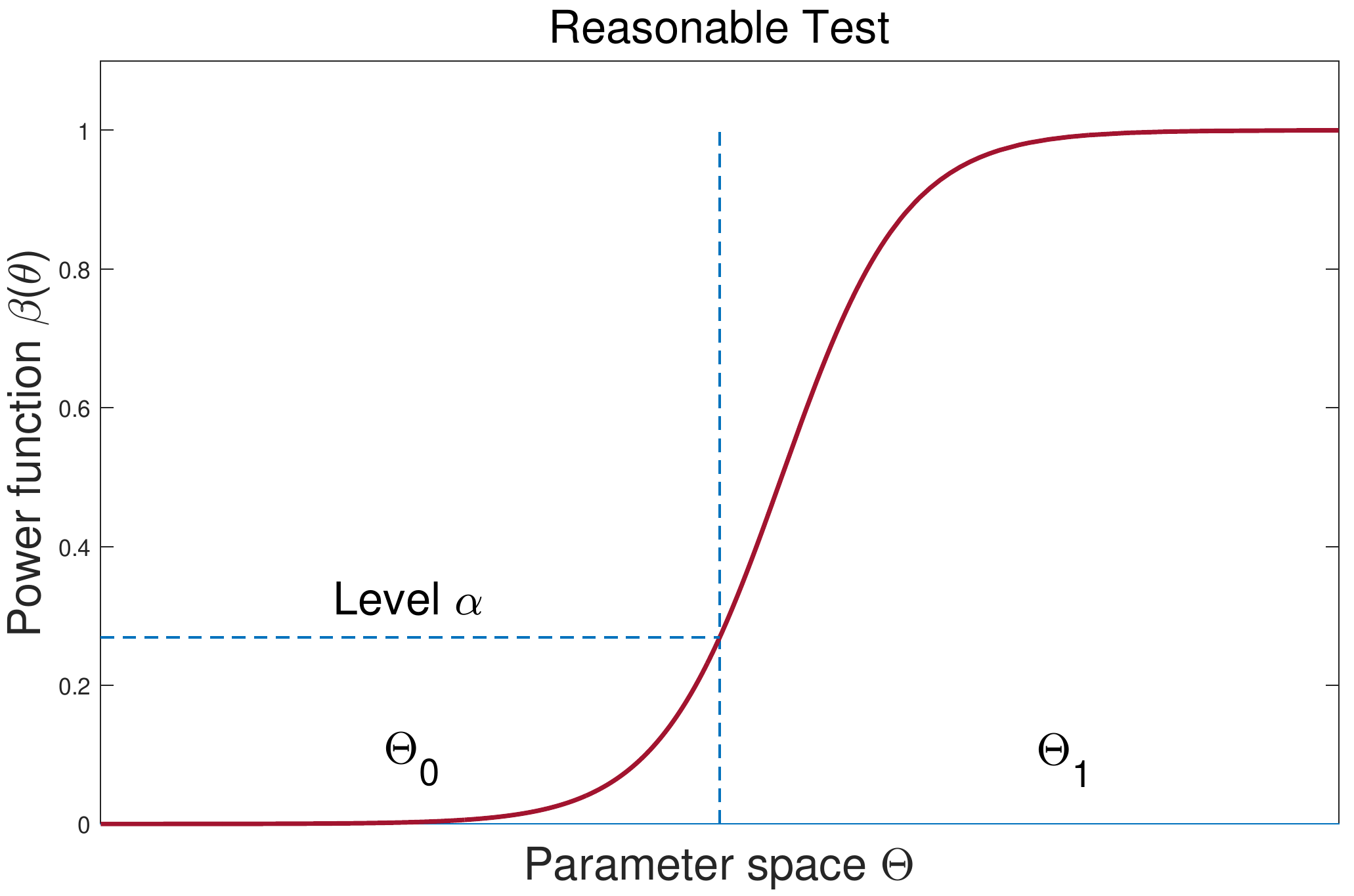}}\caption{The power function of a reasonably good test of size $\alpha$.}\label{fig:good_power}
\end{marginfigure}

In reality, a reasonable test has power function near zero on $\Theta_0$ and near one on $\Theta_1$. So, qualitatively, the power function of a good test looks like the one in Fig.~\ref{fig:good_power}.

\section{Controlling Errors}
Usually it is impossible to control both types of errors and make their probabilities arbitrary small. Roughly, the reason behind this is the following. Choosing a test is choosing the rejection region $\mathcal{R}\subset\Omega$. If we want to make the  type I error probability (\ref{eq:error1}) smaller, we need to shrink $\mathcal{R}$. In the extreme case, we can completely exclude the type I error by taking $\mathcal{R}=\varnothing$. On the other hand, to make the type II error probability (\ref{eq:error2}) smaller, we need to inflate $\mathcal{R}$. By taking $\mathcal{R}=\Omega$, we can guaranty that the type II error will not be made. So, typically, decrease in the probability of one error leads to the increase of the probability of the other error\footnote{The provided intuition is ``rough'' because instead of shrinking and inflating $\mathcal{R}$ we could move it around. }. 

As we discussed previously, type I error is more dangerous, and therefore, controlling its probability is more important. This leads to the following definition. 

\begin{definition}
	The \textit{size} of a test with power function $\beta(\theta)$ is \begin{equation}\label{eq:size}
	\alpha=\sup_{\theta\in\Theta_0}\beta(\theta).
	\end{equation}
	A test is said to have \textit{level} $\alpha$ if its size is $\leq\alpha$\footnote{In practice, the terms ``size'' and ``level'' are often used interchangeably because both are upper-bounds for the type I error probability.}. 
\end{definition}
In words, the size of the test is the largest possible probability of the type I error (rejecting $H_0$ when it is true). See Fig.~\ref{fig:good_power}. Researchers usually specify the size of the test they wish to use\footnote{With typical choice being $\alpha=0.01, 0.05,$ and $0.1$.} (to make sure that the type I error is under control), and then search for the test with the highest power under $H_1$ (\ie on $\Theta_1$) among all test with level $\alpha$. Such a test, if it exists, is called \textit{most powerful}. Finding most powerful tests is hard and, in many cases, they don't even exist. So in practice, researchers use a test with power which is high enough.

\paragraph{Example:} Let $X_1,\ldots,X_n\sim\mathcal{N}(\mu,\sigma^2)$, where $\sigma^2$ is known\footnote{\ie estimated.}.  We want to test 
\begin{equation}
H_0:\hspace{1mm} \mu\leq0 \hspace{2mm}\mbox{ versus }\hspace{2mm} H_1:\hspace{1mm}\mu>0.
\end{equation}
So, here $\Theta=\mathbb{R}$, $\Theta_0=(-\infty,0]$, and $\Theta_1=(0,\infty)$. It seems reasonable to use the sample mean $\overline{X}_n$ as a test statistic and reject $H_0$ if $\overline{X}_n$ is large enough. The rejection region is thus
\begin{equation}\label{eq:rr}
\mathcal{R}=\{(X_1,\ldots,X_n) : \overline{X}_n>c\}\subset\Omega=\mathbb{R}^n,
\end{equation}
where $c$ is the critical value. Let us find the power function of this test. 
\begin{equation}
\beta(\mu)=\mathbb{P}(\overline{X}_n>c|\mu).
\end{equation}
Since $\overline{X}_n\sim\mathcal{N}\left(\mu,\frac{\sigma^2}{n}\right)$, we have that $\frac{\sqrt{n}(\overline{X}_n-\mu)}{\sigma}\sim\mathcal{N}(0,1)$. Therefore,
\begin{equation}\label{eq:normal}
\begin{split}
\beta(\mu)&=\mathbb{P}\left(\frac{\sqrt{n}(\overline{X}_n-\mu)}{\sigma}>\frac{\sqrt{n}(c-\mu)}{\sigma}\right)\\
&=1-\Phi\left(\frac{\sqrt{n}(c-\mu)}{\sigma}\right).
\end{split}
\end{equation}
\vspace{-4mm}
\begin{figure}
	\centerline{\includegraphics[width=.8\linewidth]{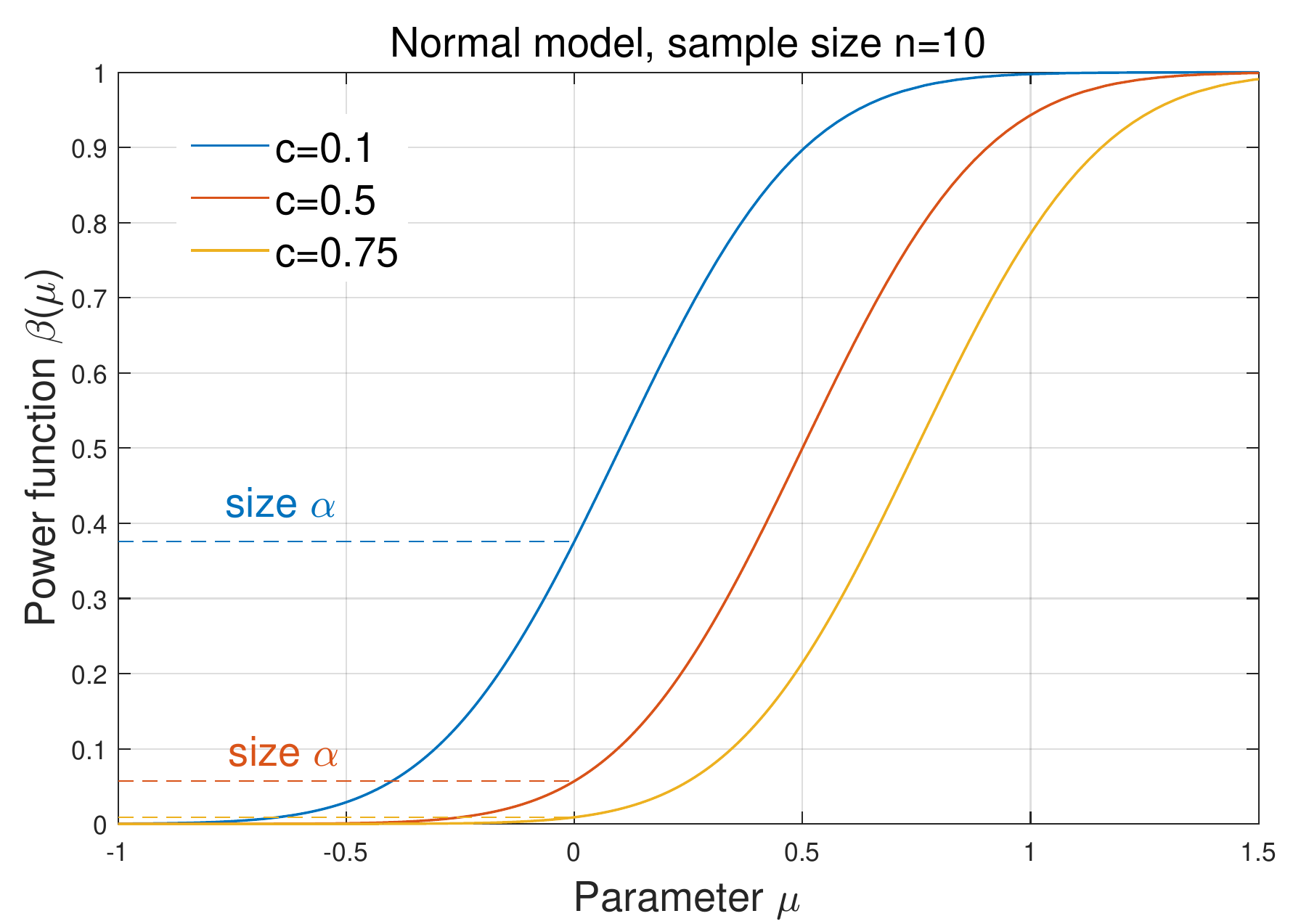}}\caption{The normal power function (\ref{eq:normal}) for $n=10$, $\sigma=1$, and different values of $c$. Notice that as $c$ increases (rejection region shrinks), the size of the test decreases, as expected.}\label{fig:normal_power}
\end{figure}
\vspace{-4mm}
The power function is an increasing function of $\mu$. It is shown in Fig.~\ref{fig:normal_power} for $n=10$, $\sigma=1$, and different values of $c$. As expected, when the rejection region (\ref{eq:rr}) shrinks (the critical value $c$ increases), the size of the test $\alpha$ decreases meaning that it becomes less and less likely to make the type I error. On the other hand, the type II error probability increases.  
To  make a test with a \textit{specific size} $\alpha$, we need to find the corresponding critical value $c$. Thanks to monotonicity of $\beta$, $\alpha=\beta(0)$. Together with (\ref{eq:normal}), this give an equation for $c$, whose solution is 
\vspace{-2mm}
\begin{equation}\label{eq:c}
c=\frac{\sigma\Phi^{-1}(1-\alpha)}{\sqrt{n}}.
\end{equation}
A halfway summary: the test which rejects $H_0$ whenever $\overline{X}_n>c$, where $c$ is given by (\ref{eq:c}), has size $\alpha$. 

Suppose now that we can also control the sample size $n$\footnote{For example, we are designing an experiment, and trying to determine what sample size is appropriate.}. Note that the power function does depend on the sample size, and by choosing $n$ large enough we can hope to reduce the type II error probability. Since $\beta$ is continuous and $\beta(0)=\alpha\ll1$, $\beta(\mu)\ll1$ in the neighborhood of zero, and, therefore, the type II error probability is large in this neighborhood. However, we may step apart from zero by $\delta>0$, $\delta\ll1$ and ask the power function to be large at $\delta$:
\begin{equation}\label{eq:delta}
\beta(\delta)=1-\epsilon,
\end{equation}
where $\epsilon>0$, $\epsilon\ll1$ plays similar role to type II error as $\alpha$ plays for the type I error. Combining (\ref{eq:normal}), (\ref{eq:c}), and $(\ref{eq:delta})$, gives an equation for the sample size, whose solution is
\begin{equation}\label{eq:n}
\sqrt{n}=\frac{\sigma}{\delta}\left(\Phi^{-1}(1-\alpha)-\Phi^{-1}(\epsilon))\right).
\end{equation}
\begin{marginfigure}
	\centerline{\includegraphics[width=\linewidth]{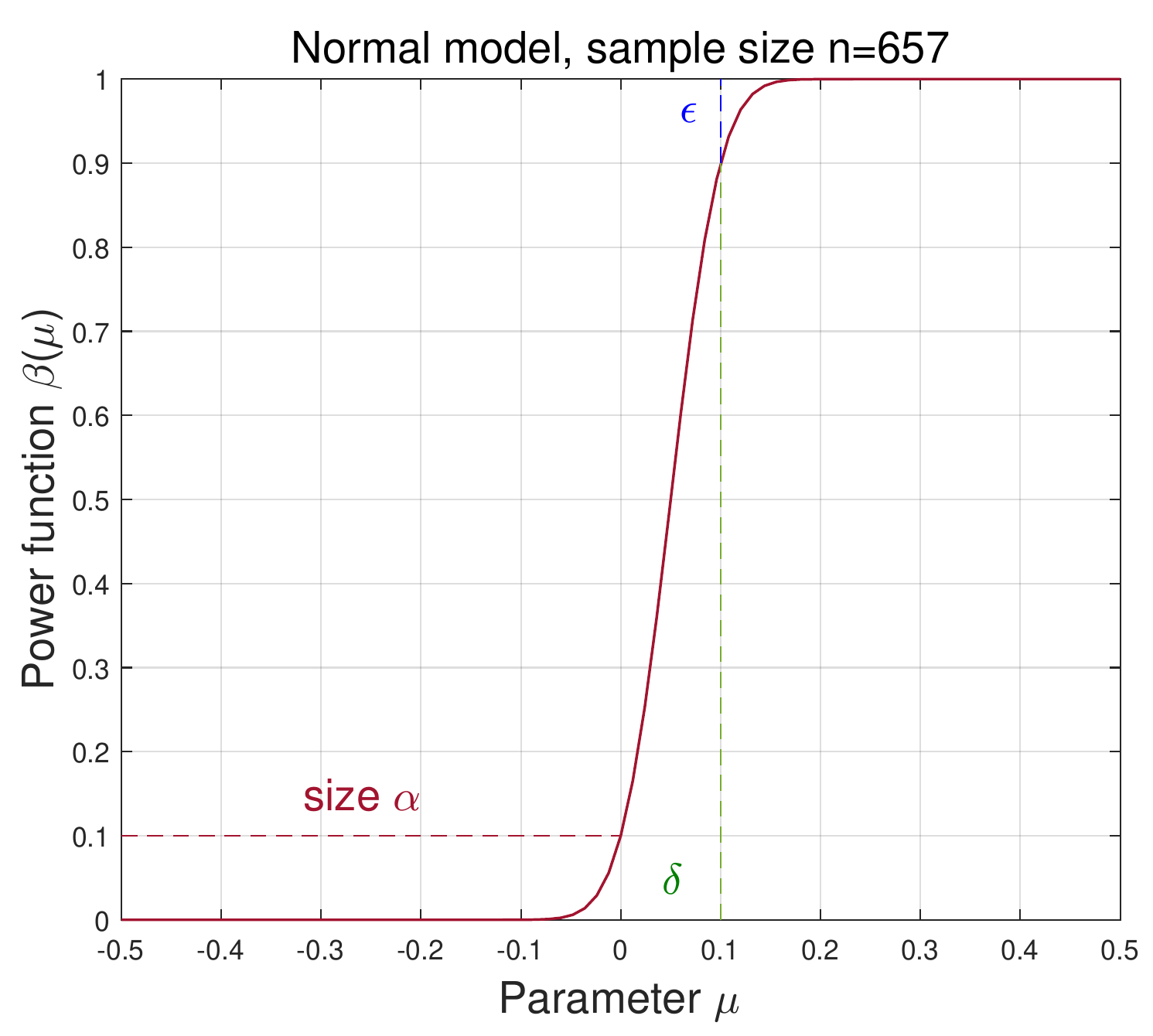}}\caption{The normal power function (\ref{eq:normal}) with $n$ and $c$ defined from (\ref{eq:n}) and (\ref{eq:c}) with $\alpha=\delta=\epsilon=0.1$. Notice the sample size increase!}\label{fig:deltaeps}
\end{marginfigure}
Thus, the test which rejects $H_0$ whenever $\overline{X}_n>c$, where $n$ and $c$ are given by (\ref{eq:n}) and (\ref{eq:c}), has size $\alpha$ and, moreover, the type II error probability is at most $\epsilon$ if $\mu\in[\delta,\infty]\subset\Theta_1$. If $\mu\in[0,\delta]$, this probability is, unfortunately, larger. Figure~\ref{fig:deltaeps} shows the power function for $\alpha=\delta=\epsilon=0.1$.\hfill $\square$\bigskip

The strategy described in this example is often employed in other cases. Namely, to design a test, we need to specify the rejection region $\mathcal{R}=\{X\in\Omega: s(X)>c\}$ by choosing a test statistic $s$ and its critical value $c$. Choosing the test statistic is an art, but often reasonable candidates are rather obvious\footnote{Sometimes after long analysis :)}. After choosing $s$, the rejection region is parametrized by $c$. We chose $c$ to get the desired size $\alpha$. To this end, we need to solve\footnote{Analytically if you are lucky, but most likely numericlly.} for $c$ the following equation:
\begin{equation}
\sup_{\theta\in\Theta_0}\mathbb{P}(X\in\mathcal{R}_c|\theta)=\alpha,
\end{equation}
where $X=(X_1,\ldots,X_n)$. If  we can't control $n$\footnote{The data comes from an observational study, or experiments are too expensive.}, then we are done. If we can control $n$, then we may try to reduce the type II error probability by exploiting  the fact that the power function depends on $n$.
\vspace{-3mm}

\section{Further Reading}
\vspace{-3mm}
\begin{enumerate}
	\item The most complete book on testing is E.~Lehmann \& J.~Romano (2005) \textit{\href{http://www.springer.com/us/book/9780387988641}{Testing Statistical Hypotheses}}.
\end{enumerate}
\vspace{-3mm}

\section{What is Next?} 
\vspace{-3mm}
In real applications, finding most powerful tests is a very hard problem which often does not have a solution. So, instead of focusing on the theory of most powerful tests, we will consider several widely used tests that often perform reasonably well.

\chapter{The Wald test and $t$-test}\label{ch:Wald}
	\begin{marginfigure}
		\centerline{\includegraphics[width=.6\linewidth]{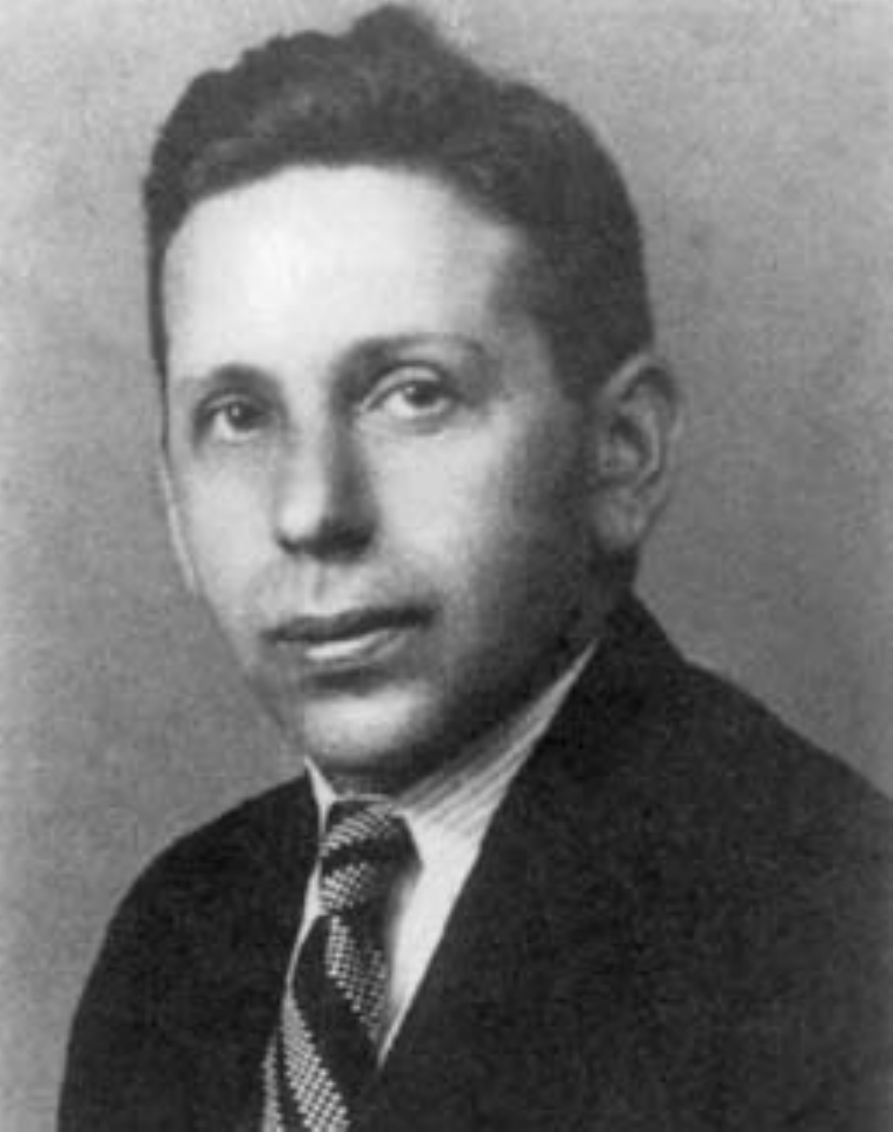}}\caption{Abraham Wald, a Hungarian statistician. Photo source: \href{https://en.wikipedia.org/wiki/Abraham_Wald}{wikipedia.org}.}\label{fig:Wald}
	\end{marginfigure}
	\newthought{In this lecture}, we will discuss two straightforward and often used parametric tests: the Wald test and the $t$-test. %, and introduce an important, yet often misunderstood, concept of the $p$-value.
	
\section{The Wald Test}
The Wald test bridges the gap between three statistical inference methods: estimation, confidence sets, and hypothesis testing. 

Let $\theta$ be the parameter of interest, and suppose we want to test\footnote{A hypothesis of the form $\theta=\theta_0$ is called a \textit{simple hypothesis}, and a test of the form (\ref{eq:two-sided}) is called a \textit{two-sided test}.} 
\begin{equation}\label{eq:two-sided}
H_0:\hspace{1mm} \theta=\theta_0\hspace{2mm}\mbox{ versus }\hspace{2mm} H_1:\hspace{1mm}\theta\neq\theta_0.
\end{equation}
Let $\hat{\theta}$ be an estimate of $\theta$\footnote{For example, $\hat{\theta}=\hat{\theta}_{\mathrm{MLE}}$.}, and let $\widehat{\mathrm{se}}$ be the estimated standard error of $\hat{\theta}$\footnote{For example, we can estimate $\mathrm{se}$ using the bootstrap; or, if $\hat{\theta}$ is the MLE, then $\widehat{\mathrm{se}}=1/\sqrt{nI(\hat{\theta})}$.}. Assume that $\hat{\theta}$ is approximately normally distributed: 
\begin{equation}\label{eq:normality}
\frac{\hat{\theta}-\theta}{\widehat{\mathrm{se}}}\approxdist\mathcal{N}(0,1).
\end{equation}
%where $\theta^*$ is the true value of the parameter\footnote{A remark on notation: $\theta$ is the ``name'' of the parameter, $\theta^*$ is its true value (unknown), and $\theta_0$ is the value we wish to test (known). So the null hypothesis essentially saying that $\theta^*=\theta_0$.}. 
Note that this assumption is not very strong: it holds for many reasonable estimates\footnote{For example, if $\hat{\theta}$ is the MLE, then (\ref{eq:normality}) holds, since the MLE is asymptotically normal. }. 

In this settings, if $H_0$ is true, then  $\left|\frac{\hat{\theta}-\theta_0}{\widehat{\mathrm{se}}}\right|$ is likely to be small. Therefore,
it seems rational to reject the null hypothesis if 
\begin{equation}
W=\left|\frac{\hat{\theta}-\theta_0}{\widehat{\mathrm{se}}}\right|>c.
\end{equation}
As usual, we find the critical value $c$ from the upper-bound for the probability of the type I error, \ie the size of the test. Since $H_0$ is a simple hypothesis, the size
\begin{equation}
\alpha=\sup_{\theta\in\Theta_0}\beta(\theta)=\beta(\theta_0).
\end{equation}
Under $H_0$, $\frac{\hat{\theta}-\theta_0}{\widehat{\mathrm{se}}}\approxdist\mathcal{N}(0,1)$, and therefore, 
\begin{equation}
%\begin{split}
\beta(\theta_0)=\mathbb{P}\left(W>c|\theta=\theta_0\right)=\mathbb{P}\left(\left|\frac{\hat{\theta}-\theta_0}{\widehat{\mathrm{se}}}\right|>c\right)
\approx 2\Phi(-c).
%\end{split}
\end{equation}
The critical value $c$ is thus\footnote{Recall that $\Phi(z_{\alpha})=\alpha$.}
\begin{equation}
c=-\Phi^{-1}\left(\frac{\alpha}{2}\right)=-z_{\frac{\alpha}{2}}.
\end{equation}
To summarize, the size $\alpha$ \textit{Wald test} rejects $H_0$ when
\begin{equation}\label{eq:W}
W=\left|\frac{\hat{\theta}-\theta_0}{\widehat{\mathrm{se}}}\right|>-z_{\frac{\alpha}{2}}=z_{1-\frac{\alpha}{2}}.
\end{equation}
\begin{marginfigure}
	\vspace{45mm}
	\centerline{\includegraphics[width=1.1\linewidth]{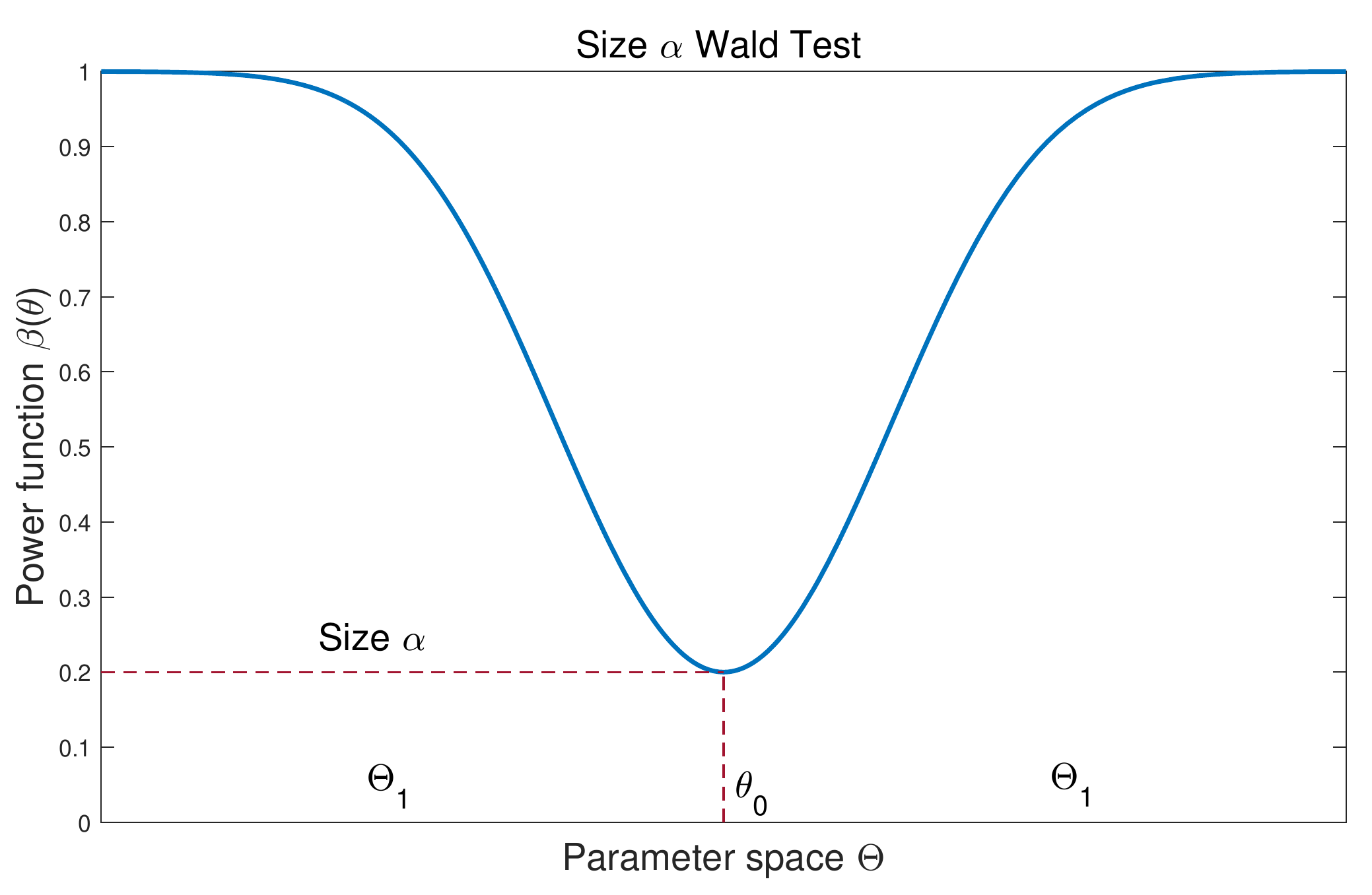}}\caption{The power function of the Wald test with size $\alpha$.}\label{fig:WaldPower}
\end{marginfigure}

How does the power function look?
\begin{equation}
\begin{split}
\beta(\theta)&=\mathbb{P}(W>-z_{\frac{\alpha}{2}}|\theta)=\mathbb{P}\left(\left.\left|\frac{\hat{\theta}-\theta_0}{\widehat{\mathrm{se}}}\right|>-z_{\frac{\alpha}{2}}\hspace{1mm}\right|\theta\right)\\
&=\mathbb{P}\left(\left.\frac{\hat{\theta}-\theta_0}{\widehat{\mathrm{se}}}>-z_{\frac{\alpha}{2}}\hspace{1mm}\right|\theta\right)+\mathbb{P}\left(\left.\frac{\hat{\theta}-\theta_0}{\widehat{\mathrm{se}}}<z_{\frac{\alpha}{2}}\hspace{1mm}\right|\theta\right)\\
&=\mathbb{P}\left(\left.\frac{\hat{\theta}-\theta}{\widehat{\mathrm{se}}}>-z_{\frac{\alpha}{2}}+\frac{\theta_0-\theta}{\widehat{\mathrm{se}}}\hspace{1mm}\right|\theta\right)+\mathbb{P}\left(\left.\frac{\hat{\theta}-\theta}{\widehat{\mathrm{se}}}<z_{\frac{\alpha}{2}}+\frac{\theta_0-\theta}{\widehat{\mathrm{se}}}\hspace{1mm}\right|\theta\right)\\
&=1-\Phi\left(\frac{\theta_0-\theta}{\widehat{\mathrm{se}}}-z_{\frac{\alpha}{2}}\right)+\Phi\left(\frac{\theta_0-\theta}{\widehat{\mathrm{se}}}+z_{\frac{\alpha}{2}}\right).
\end{split}
\end{equation}
The power function of the Wald test is shown schematically in Fig.~\ref{fig:WaldPower}. As expected, $\beta(\theta_0)=\alpha$. Also, $\beta(\theta)\rightarrow1$ as $|\theta-\theta_0|\rightarrow\infty$.  Recall that $\widehat{\mathrm{se}}$ often tends to zero as the sample size $n$ increases. As a result, $\beta(\theta)\rightarrow1$ as $n\rightarrow\infty$ for all $\theta\neq\theta_0$. We can therefore reduce the probability of the type II error outside of a certain neighborhood of $\theta_0$ by choosing $n$ sufficiently  large. 

In a nutshell, given a point estimate of the parameter of interest, which is approximately normally distributed (\ref{eq:normality}), the Wald test allows to test simple hypothesis (\ref{eq:two-sided}) essentially with zero intellectual effort. 

\subsection{Connection to Confidence Intervals}
Given the approximate normality (\ref{eq:normality}), we can immediately construct an approximate $(1-\alpha)100\%$ confidence interval for $\theta$:
\begin{equation}
\mathcal{I}=\hat{\theta}\pm  z_{\frac{\alpha}{2}}\widehat{\mathrm{se}}.
\end{equation}
The size $\alpha$ Wald test
\begin{equation}
\mbox{Rejects } H_0: \theta=\theta_0 \hspace{2mm}\Leftrightarrow\hspace{2mm} \theta_0\notin\mathcal{I}.
\end{equation}
In words, testing the hypothesis is equivalent to checking whether the null value is in the confidence interval. 

\subsection{Comparing Means of Two Populations}
Let us finish the discussion of the Wald test with a nonparametric example. Suppose we are interesting in comparing the unknown means $\mu_1$ and $\mu_2$ of two populations\footnote{For example, mean income of males and females.}. In particular, we want to test 
\begin{equation}\label{eq:14means}
H_0:\hspace{1mm} \Delta\mu=0\hspace{2mm}\mbox{ versus }\hspace{2mm} H_1:\hspace{1mm}\Delta\mu\neq0,
\end{equation}
where $\Delta\mu=\mu_1-\mu_2$. Let $X_1\ldots,X_n$ and $Y_1,\ldots,Y_m$ be two independent samples from the populations. The plug-in estimates of the means are: $\hat{\mu}_1=\overline{X}_n$ and $\hat{\mu}_2=\overline{Y}_m$. A nonparametric estimates of $\Delta\mu$ is thus
\begin{equation}
\widehat{\Delta\mu}=\overline{X}_n-\overline{Y}_m.
\end{equation}
The standard error of $\widehat{\Delta\mu}$ is
\begin{equation}
\mathrm{se}^2=\mathrm{se}^2[\overline{X}_n]+\mathrm{se}^2[\overline{Y}_m]=\frac{\sigma_1^2}{n}+\frac{\sigma_2^2}{m},
\end{equation}
where $\sigma_1^2$ and $\sigma_2^2$ are the population variances. It can be estimated by
\begin{equation}
\widehat{\mathrm{se}}=\frac{s_1^2}{n}+\frac{s_2^2}{m},
\end{equation}
where $s_i^2$ are the sample variances. Thus, the size $\alpha$ Wald test rejects $H_0$ when 
\begin{equation}
\left|\frac{\overline{X}_n-\overline{Y}_m}{\sqrt{\frac{s_1^2}{n}+\frac{s_2^2}{m}}}\right|>z_{1-\frac{\alpha}{2}}.
\end{equation}

\subsection{The $t$-Test}
When testing simple hypothesis (\ref{eq:two-sided}), it is common to use the t-test\footnote{The t-test was introduced in 1908 by William Gosset, an English statistician. At that time he was an employee of Guinness in Dublin. To prevent disclosure of confidential information that could potentially be used by other competitors, the brewery prohibited its employees from publishing any papers.  Gosset published his results under a pseudonym ``Student.''} instead of the Wald test if 
\begin{enumerate}
	\item The data is modeled as a sample from the normal distribution,
	\item The sample size is small. 
\end{enumerate}
\begin{marginfigure}
%	\vspace{0.5mm}
	\centerline{\includegraphics[width=.6\linewidth]{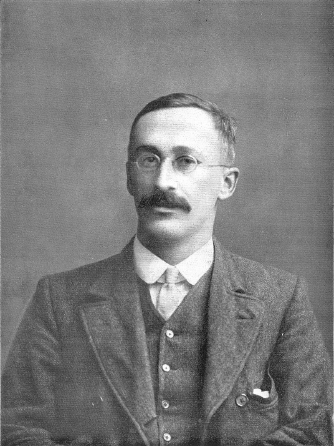}}\caption{William Sealy Gosset (aka ``Student''). Photo source: \href{https://en.wikipedia.org/wiki/William_Sealy_Gosset}{wikipedia.org}.}\label{fig:Gosset}
\end{marginfigure}

\begin{marginfigure}
	%	\vspace{-10mm}
	\centerline{\includegraphics[width=\linewidth]{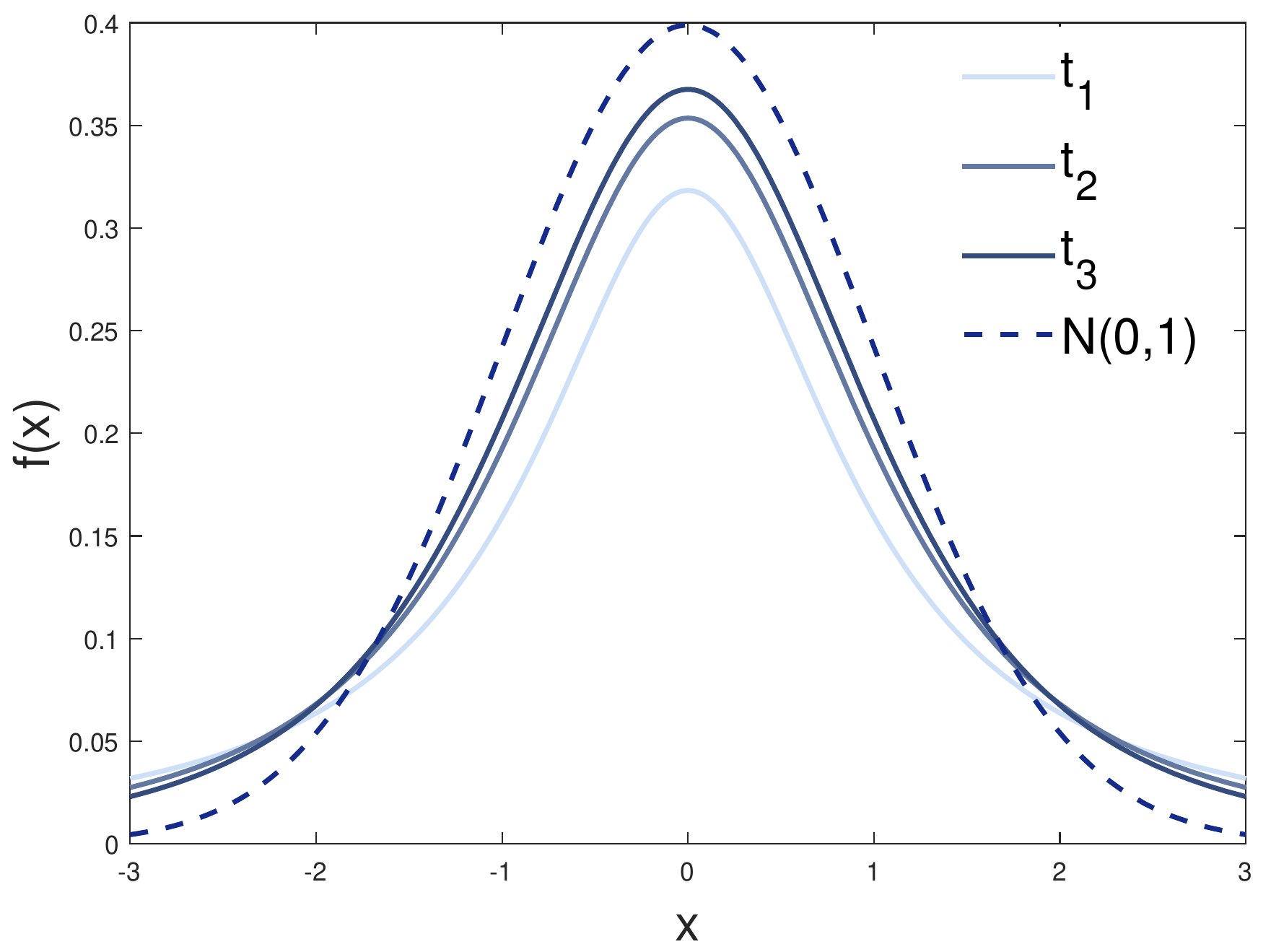}}\caption{The PDF of t-distribution with $k=1,2,$ and $3$ degrees of freedom.}\label{fig:t}
\end{marginfigure}

Let $X_1,\ldots,X_n\sim\mathcal{N}(\mu,\sigma^2)$, where both $\mu$ and $\sigma$ are unknown. Suppose we want to test
\begin{equation}\label{eq:two-sided2}
H_0:\hspace{1mm} \mu=\mu_0\hspace{2mm}\mbox{ versus }\hspace{2mm} H_1:\hspace{1mm}\mu\neq\mu_0.
\end{equation}
Let us estimate $\mu$ by the sample mean $\hat{\mu}=\overline{X}_n$ and the standard error of $\hat{\mu}$ by $\widehat{\mathrm{se}}=s_n/\sqrt{n}$, where $s_n^2$ is the sample variance. If $n$ is large,  then, under $H_0$, the random variable 
\begin{equation}
T=\frac{\hat{\mu}-\mu_0}{\widehat{\mathrm{se}}}=\frac{\overline{X}_n-\mu_0}{s_n/\sqrt{n}}\approxdist\mathcal{N}(0,1),
\end{equation}
and we can use the Wald test. It turns out however, that for \textit{any} $n$, the \textit{exact} distribution of $T$ under $H_0$ is  \textit{Student's t-distribution} with $(n-1)$ degrees of freedom:
\begin{equation}
T\sim t_{n-1}.
\end{equation}
The formula for the PDF of this distribution looks rather complicated:
\begin{equation}
f_{t_k}(x)=\frac{\Gamma\left(\frac{k+1}{2}\right)}{\sqrt{k\pi}\Gamma\left(\frac{k}{2}\right)\left(1+\frac{x^2}{k}\right)^{\frac{k+1}{2}}}.
\end{equation}
But it has a couple of nice properties: a) the t-distribution is symmetric about zero and, b) as $k\rightarrow\infty$, it tends to the standard normal distribution (as expected)\footnote{In fact, if $k<30$, the two distributions are very close.}.  Figure~\ref{fig:t} shows the PDF of t-distribution for several values of $k$. 
By analogy with the Wald test, the size $\alpha$ t-test rejects $H_0$ when 
\begin{equation}
\left|\frac{\overline{X}_n-\mu_0}{s_n/\sqrt{n}}\right|>t_{n-1,1-\frac{\alpha}{2}},
\end{equation}
where $t_{n-1,{\alpha}}$ plays the same role as  $z_{{\alpha}}$ plays in the Wald test, \ie controls the size $\alpha$. More precisely, $t_{k,\alpha}$ is such point that the probability that the t random variable with $k$ degrees of freedom is less than $t_{k,\alpha}$ is exactly $\alpha$:
\begin{equation}
\int_{-\infty}^{t_{k,\alpha}} f_{t_k}(x)dx=\alpha.
\end{equation}
When $n$ is moderately large (say, $n\approx30$), the t-test is essentially identical  to the Wald test.

\section{Further Reading}
\begin{enumerate}
	\item ``Sometimes the most important step in creative work is simply to ask the right question.'' 
	J.F.~Box (1987) ``\href{https://projecteuclid.org/euclid.ss/1177013437}{Guinness, Gosset, Fisher, and small samples},'' \textit{Statistical Science}, 2(1), 45-52 is a nice story about the two men, one of whom invented the $t$-test and the other generalized it so greatly.  
\end{enumerate}

\section{What is Next?} 
Can we be more informative than simply reporting ``reject'' or ``accept'' when testing a hypothesis? Yes, we can. This leads to the cornerstone concept of statistical inference, the $p$-value. 

\chapter{P-values}\label{ch:p-values}
\newthought{The $p$-value} is an iconic concept in statistics.  First computations of $p$-values date back to the 1770s, where they were used by Laplace. The modern use of $p$-values was popularized by Fisher in the 1920s. Nowadays, most of the research papers that use statistical analysis of data report $p$-values. Yet quite shamefully, many researchers using the $p$-value can't even explain what exactly the $p$-value means\footnote{C.~Aschwanden (2016) ``\href{http://fivethirtyeight.com/features/statisticians-found-one-thing-they-can-agree-on-its-time-to-stop-misusing-p-values/}{Statisticians found one thing they can agree on: it's time to stop misusing p-values}'' http://fivethirtyeight.com/.}.  

In this lecture, we will provide a rigorous statistical definition of the $p$-value, its intuitive meaning, geometrical and probabilistic interpretations, and an analytical recipe for computing $p$-values. 

\section{Definition of the $p$-value}

\begin{marginfigure}
	%\vspace{35mm}
	\centerline{\includegraphics[width=1\linewidth]{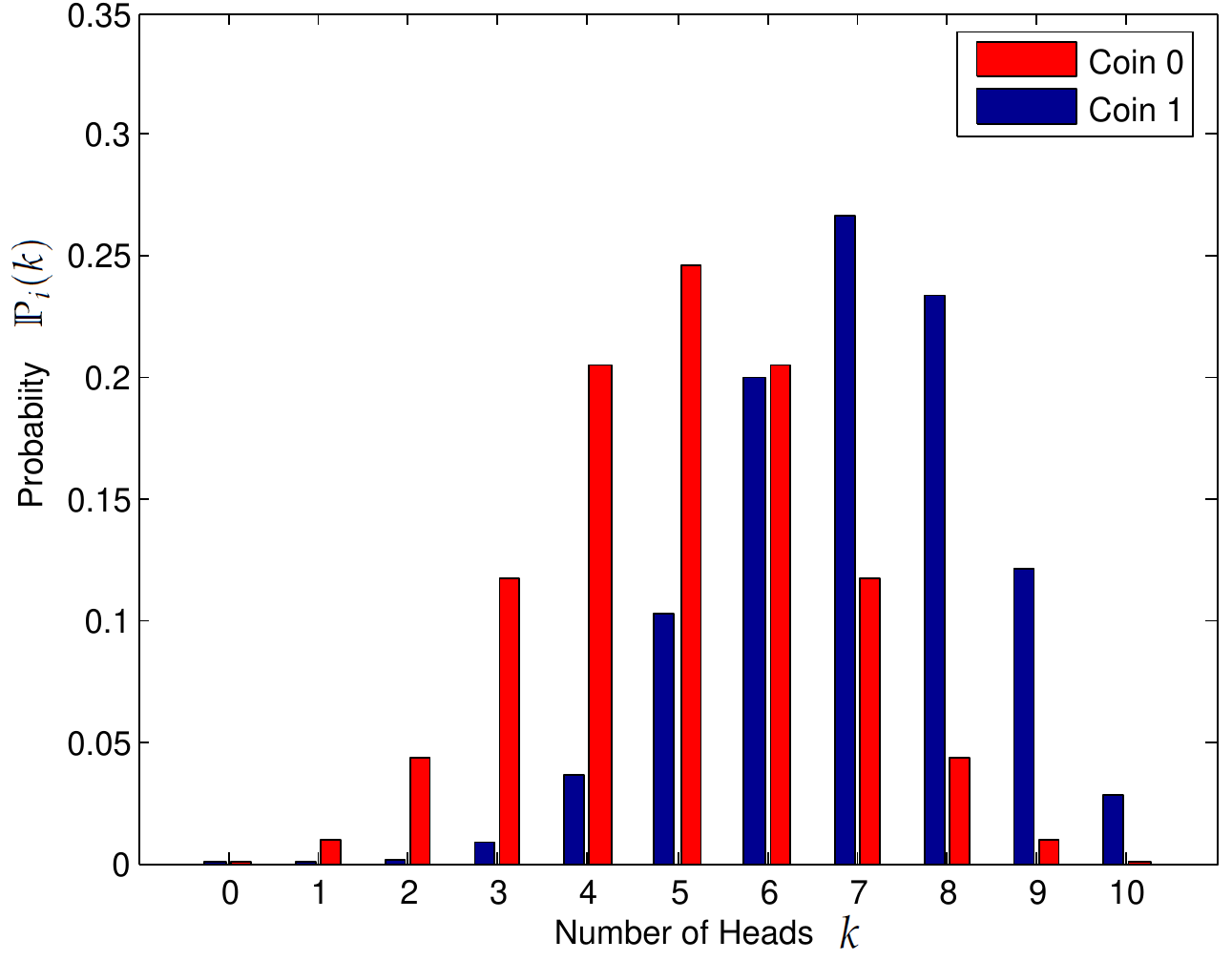}}\caption{Binomial probabilities $$\mathbb{P}_i(k)={n \choose k} p_i^k(1-p_i)^{n-k}, $$ where $n=10$, $p_0=0.5$, and $p_1=0.7$.}\label{fig:15coins}
\end{marginfigure}

Reporting ``reject'' or ``accept'' a hypothesis is not very informative.
Recall the two coin example from the previous lecture, Fig.~\ref{fig:15coins}. Accepting the hypothesis that the coin is fair observing $k=3$ heads is much more comfortable than accepting it observing $k=6$ heads. 

To start with, we can always report the size $\alpha$ of the test used and the decision the reject $H_0$ or accept $H_0$. Can we be more informative? It turns out we can, and this leads to the concept of $p$-value.

Recall that the rejection region $\mathcal{R}$ has the following form:
\begin{equation}
\mathcal{R}=\{X\in\Omega: s(X)>c\},
\end{equation}
where $\Omega$ is the sample space\footnote{ 
	Set of all possible  outcomes of data $X=(X_1,\ldots,X_n)$.}, $s$ is the test statistic, and $c$ is its critical value. The critical value $c$ is determined by the test size $\alpha$, $c=c_\alpha$. By varying $\alpha\in(0,1)$, we generally  obtain a one-parameter family of nested  rejection regions $\mathcal{R}_\alpha$\footnote{The intuition behind this is the following. The size $\alpha$ is the largest possible probability to reject $H_0$ when it is true: $\alpha=\mathbb{P}(X\in\mathcal{R}_\alpha|H_0)$. Increasing $\alpha$ leads to inflating $\mathcal{R}_{\alpha}$: it becomes easier to reject. The size $\alpha$, therefore, controls the size of the rejection region.}:
\begin{equation}
\begin{split}
&\mathcal{R}_{\alpha}\subset\mathcal{R}_{\alpha'}\hspace{2mm}\mbox{for } \alpha'>\alpha,\\
&\mathcal{R}_{\alpha}\rightarrow\varnothing, \hspace{3mm}\mbox{ as } \alpha\rightarrow0 \hspace{2mm}\mbox{(never reject)},\\
&\mathcal{R}_{\alpha}\rightarrow\Omega, \hspace{3mm}\mbox{ as } \alpha\rightarrow1 \hspace{2mm}\mbox{(always reject)}.
\end{split}
\end{equation}
This means that if a test rejects at size $\alpha$, it will also rejects at size $\alpha'>\alpha$. Therefore, given the observed data $X\in\Omega$, there exits the smallest $\alpha$ at which the test rejects. This number is called the $p$\textit{-value}:
\begin{equation}
p(X)=\inf_{\alpha\in(0,1)}\{\alpha: X\in\mathcal{R}_\alpha\}.
\end{equation}

\section{Geometric Interpretation and Intuitive Meaning}

Schematically, the picture looks as follows:
\begin{figure}
	\vspace{-5mm}
	\centerline{\includegraphics[width=0.9\linewidth]{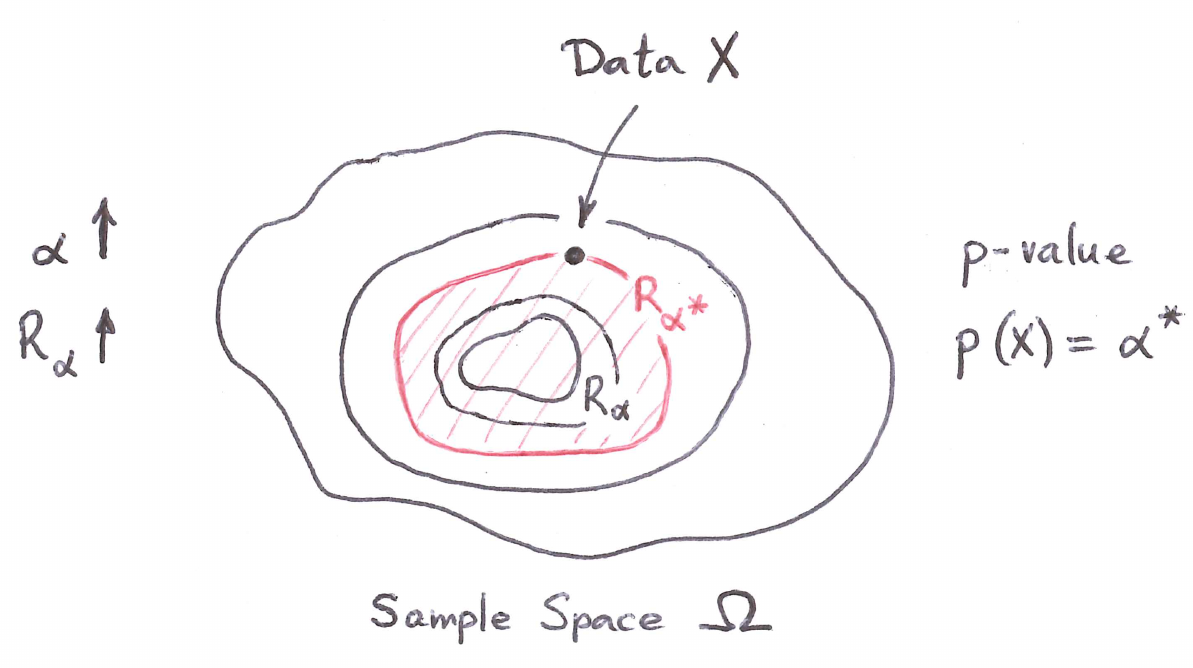}}\vspace{5mm}
	\centerline{\includegraphics[width=0.5\linewidth]{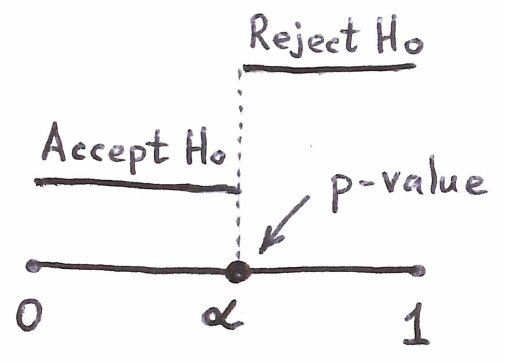}}\caption{The concept of $p$-value. Here $X$ is the observed data. If the size $\alpha$ is too small, $X\notin\mathcal{R}_\alpha$, and we accept $H_0$. Gradually increasing $\alpha$, we reach the moment where we reject $H_0$. The corresponding value of $\alpha$ is the $p$-value. Reporting simply a particular size and the corresponding decision (reject/accept), gives only a point on the bottom graph. Reporting the $p$-value, determines the graph completely.}\label{fig:pvalue}
\end{figure}

Intuitively, the $p$-value is a \textit{measure of the evidence against $H_0$} provided by the data: the smaller the $p$-value, the stronger the evidence against $H_0$\footnote{Indeed, if the $p$-value $\alpha^*$ is small, than the test of size $\alpha^*$ is a) very conservative about rejecting $H_0$, and yet b) it rejects $H_0$ based on the obrained data.}. Typically, researchers use the following evidence scale:
\begin{equation}\label{eq:classification}
\begin{split}
p(X)<0.01&\hspace{5mm} \mbox{very strong evidence agianst } H_0,\\
p(X)\in(0.01,0.05)&\hspace{5mm} \mbox{strong evidence agianst } H_0,\\
p(X)\in(0.05,0.1)&\hspace{5mm} \mbox{weak evidence agianst } H_0,\\
p(X)>0.1&\hspace{5mm} \mbox{little or no evidence agianst } H_0.\\
\end{split}
\end{equation}

To get accustomed to the new notion, let us compute the $p$-values for several examples. 

\section{Examples}

\paragraph{Two coins.} Here we have the data $k\sim\mathrm{Bin}(n,\theta)$, where $k$ is the number of heads in $n=10$ trials and $\theta\in\Theta=\{0.5, 0.7\}$. We wish to test
\begin{equation}
H_0:\hspace{1mm} \theta=\theta_0=0.5\hspace{2mm}\mbox{ versus }\hspace{2mm} H_1:\hspace{1mm}\theta=\theta_1=0.7.
\end{equation}
The test statistic that we used last time is the ratio of likelihoods 
\begin{equation}
s(k)=\frac{\mathbb{P}_1(k)}{\mathbb{P}_0(k)}, \hspace{5mm}\mbox{where }\hspace{2mm} \mathbb{P}_i(k)={n \choose k} \theta_i^k(1-\theta_i)^{n-k},
\end{equation}
and the rejection region is\footnote{We used $c=1$, which led to $\mathcal{R}=\{7,\ldots,10\}$ and size $\alpha=0.17$. To compute $p$-value, we need to consider a family of rejection regions and find the smallest that contain our data.} 
\begin{equation}
\mathcal{R}=\{k: s(k)>c\}.
\end{equation}
Since $s(k+1)>s(k)$ (see Fig.~\ref{fig:coins}),  if $k\in\mathcal{R}$, then $k+1\in\mathcal{R}$. This means the all rejection regions have the following intuitive form:
\begin{equation}\label{eq:RR}
\mathcal{R}=\{k_{\min},\ldots,n\}, \hspace{5mm}\mbox{where }\hspace{2mm} k_{\min}=0,\ldots,n,n+1.
\end{equation}
Here $k_{\min}=n+1$ corresponds to the empty rejection region. Let us compute the size of the test with rejection region (\ref{eq:RR}).
\begin{equation}
\alpha=\mathbb{P}(k\in\{k_{\min},\ldots,n\}|\theta=\theta_0)=\sum_{i=k_{\min}}^n\mathbb{P}_0(i).
\end{equation}
\begin{marginfigure}
	\vspace{40mm}
	\centerline{\includegraphics[width=1\linewidth]{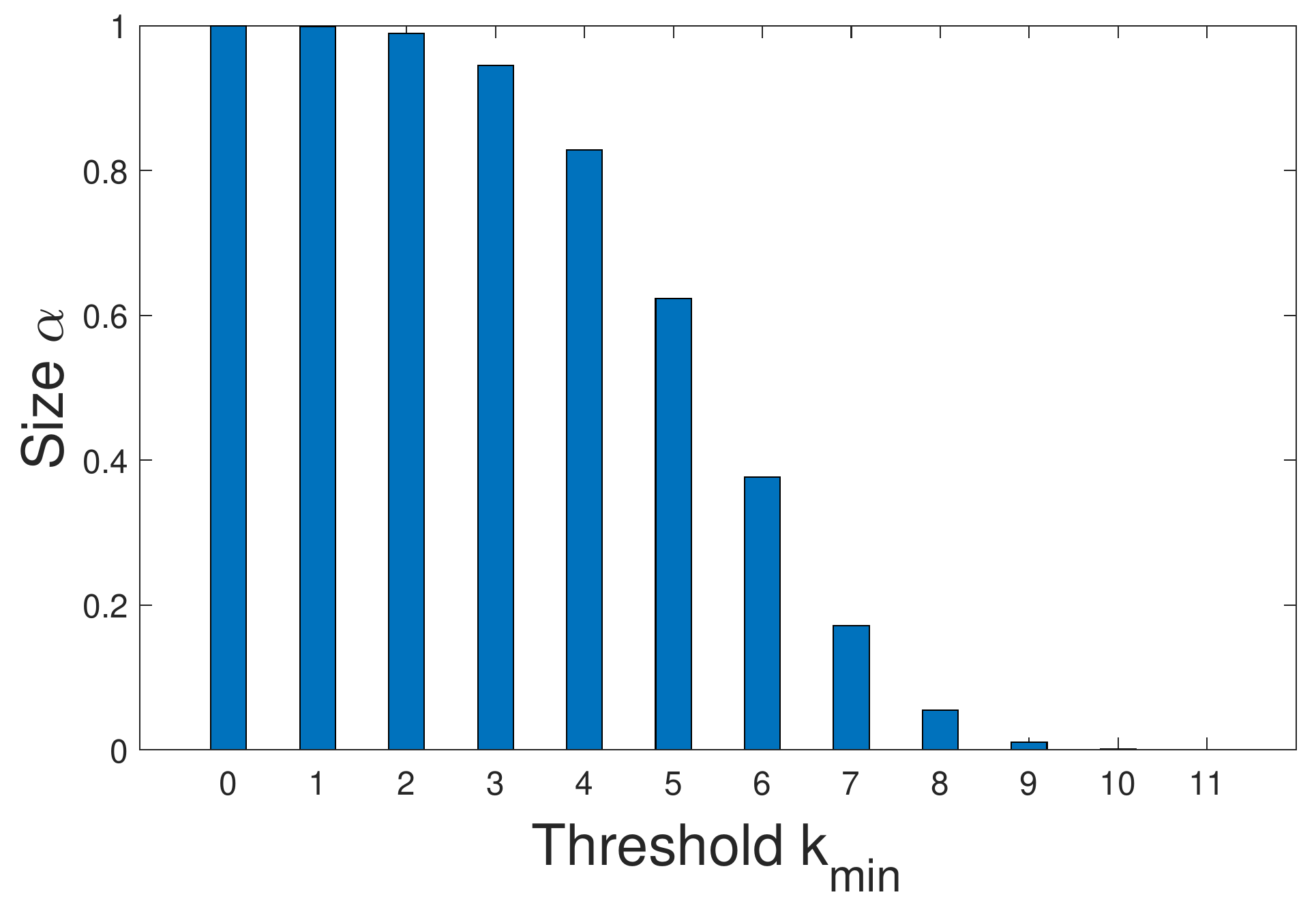}}\caption{The size $\alpha$ as a function of the rejection region boundary $k_{\min}$. As expected, $\alpha=1$ corresponds to $\mathcal{R}=\Omega$ ($k_{\min}=0$) and $\alpha=0$ corresponds to $\mathcal{R}=\varnothing$ ($k_{\min}=11$). }\label{fig:kminalpha}
\end{marginfigure}
Figure~\ref{fig:kminalpha} shows the dependence of $\alpha$ on $k_{\min}$. Notice that because of the discreteness, we can't construct a test of arbitrary size: only sizes appeared on the $y$-axes of Fig.~\ref{fig:kminalpha} are available. 

Given $k$, to find the $p$-value, we need to find the smallest size at which the test rejects $H_0$. This smallest size corresponds to the smallest rejection region (\ref{eq:RR}) that contains $k$. This smallest rejection region is $\mathcal{R}=\{k,\ldots,n\}$. And, thus, the $p$-value is
\begin{equation}
p(k)=\sum_{i=k}^n\mathbb{P}_0(i).
\end{equation} 
Figure~\ref{fig:kminalpha} shows the $p$-value as a function of $k$. According to the classification (\ref{eq:classification}), $k=10$ provides very strong evidence against $H_0$; $k=9$ provides strong evidence; $k=8$ corresponds to weak evidence; and $k\leq7$ corresponds to little or no evidence. 
\hfill $\square$ %\bigskip

\paragraph{Normal model} Last time we constructed a test of size $\alpha$ for testing 
\begin{equation}
H_0:\hspace{1mm} \mu\leq0 \hspace{2mm}\mbox{ versus }\hspace{2mm} H_1:\hspace{1mm}\mu>0,
\end{equation}
where  $X_1,\ldots,X_n\sim\mathcal{N}(\mu,\sigma^2)$, where $\sigma^2$ is known. The rejection region was 
\begin{equation}
\mathcal{R}_\alpha=\left\{X: \overline{X}_n>\frac{\sigma z_{1-\alpha}}{\sqrt{n}}\right\}.
\end{equation}
Geometrically,  this region is a half-space in $\Omega=\mathbb{R}^n$. To find the smallest rejection region that contains the data  $X\in\mathbb{R}^n$, we must require that $X$ lies on the boundary of that rejection region. That is the equation for $p$-value $\alpha^*=p(X)$ is 
\begin{equation}
\overline{X}_n=\frac{\sigma z_{1-\alpha^*}}{\sqrt{n}},
\end{equation}
which leads to 
\begin{equation}
p(X)=1-\Phi\left(\frac{\sqrt{n}\overline{X}_n}{\sigma}\right).
\end{equation}
Qualitatively, large positive values of $\overline{X}_n$ (strong evidence against $H_0$) corresponds to small $p$-values. 
\hfill $\square$ %\bigskip

\paragraph{The Wald Test} As we know, the rejection region of the size $\alpha$ \text{Wald test} is
\begin{equation}\label{eq:W2}
\mathcal{R}_\alpha=\left\{X\in\Omega: W(X)>z_{1-\frac{\alpha}{2}}\right\},
\end{equation}
where $W(X)=\left|\frac{\hat{\theta}(X)-\theta_0}{\widehat{\mathrm{se}}(X)}\right|$ is the Wald statistic. Given the data $X$, the $p$-value is that value of $\alpha$ for which $X$ lies exactly on the boundary of $\mathcal{R}_\alpha$. So, to find the $p$-value, we need to solve $W(X)=z_{1-\frac{\alpha}{2}}$ for $\alpha$. This leads to
\begin{equation}
p(X)=2\Phi(-W(X)).
\end{equation}
\hfill $\square$\vspace{-5mm}

\section{Computing $p$-values}

In general, if the rejection region of a test with size $\alpha$ has the form
\begin{equation}
\mathcal{R}_\alpha=\{X: s(X)>c_\alpha\},
\end{equation}
where $c_\alpha$ is a decreasing function of $\alpha$\footnote{Which means that $\mathcal{R}_\alpha$ inflates  with $\alpha$.}, then to find the $p$-value $\alpha^*$, we need to solve $s(x)=c_{\alpha^*}$ for $\alpha^*$, where $x$ is actually observed data.

\subsection{Probabilistic Interpretation}
The $p$-value is a certain value $\alpha^*$ of the test size. Recall that by definition
\begin{equation}
\begin{split}
\alpha^*&=\sup_{\theta\in\Theta_0}\beta(\theta)=\sup_{\theta\in\Theta_0}\mathbb{P}(X\in\mathcal{R}_{\alpha^*}|\theta)\\
&=\sup_{\theta\in\Theta_0}\mathbb{P}(s(X)>c_{\alpha^*}|\theta)=\sup_{\theta\in\Theta_0}\mathbb{P}(s(X)>s(x)|\theta).
\end{split}
\end{equation}
Hence, the $p$-value is the probability (under $H_0$) of observing a value of the test statistic more extreme that was actually observed.

\subsection{Misinterpretations}

Let us finish with two main \textit{misinterpretation }of the $p$-values, which often appear even in published research papers in respected journals. 
\begin{itemize}
	\item \textit{A  large $p$-value is not  a strong evidence in favor of $H_0$.}
	
	A large $p$-value can occur for two reasons. First, indeed $H_0$ is true. Second, $H_0$ is false, but the probability of the type II error (accept $H_0$ when it is false) is high (\ie the power of the test is low). 
	\item \textit{The $p$-value is not the probability that the null hypothesis is true.}
	
	The $p$-value is merely a measure of the evidence against $H_0$. It is not meant to be the measure of whether or not $H_0$ is true. Rather it is a measure of whether or not the data should be taken seriously. 
\end{itemize}

\section{Further Reading}
\begin{enumerate}
	\item In 2015 the journal \textit{Basic and Applied Social Psychology} has banned the use of $p$-values. The editors argued that, in practice, the use of $p$-values is more misleading than informative. In particular, $p$-values encourage lazy thinking:  if you reach the magical $p$-value$<0.05$ then the null is false.  Steven Novella discusses this issue in his recent post ``\href{https://www.sciencebasedmedicine.org/psychology-journal-bans-significance-testing/}{Psychology journal bans significance testing}'' on www.sciencebasedmedicine.org.
	\item  Recent critical literature on $p$-values is reviewed in B.~Vidgen \& T.~Yasseri (2016) ``\href{http://arxiv.org/abs/1601.06805}{P-values: misunderstood and misused},'' arXiv:1601.06805. In particular, the difference between the $p$-value and the False Discovery Rate (Lecture~\ref{ch:multipleTesting}) is discussed.  
\end{enumerate}

\section{What is Next?} 
We will discuss the permutation test, which is a nonparametric method for testing whether two samples were generated by the same data generation process.

\chapter{The Permutation Test}\label{ch:Permutation}

\newthought{The Wald test} for simple hypothesis $H_0: \theta=\theta_0$  assumes that the estimate $\hat{\theta}$ of the parameter of interest is approximately normally distributed. The $t$-test assumes that the data itself comes from a normal distribution. But what if these \textit{normality assumptions do not hold}?  

For example, recall comparing the means $\mu_1$ and $\mu_2$ of two populations: given two independent samples $X_1\ldots,X_n$ and $Y_1,\ldots,Y_m$ from the populations, we want to test
\begin{equation}\label{eq:16means}
H_0:\hspace{1mm} \Delta\mu=\mu_1-\mu_2=0\hspace{2mm}\mbox{ versus }\hspace{2mm} H_1:\hspace{1mm}\Delta\mu\neq0.
\end{equation}
If $n$ and $m$ are small, then $\widehat{\Delta\mu}=\overline{X}_n-\overline{Y}_m$ may not be  normal. 

The permutation test is a general nonparametric method for testing whether two distributions are the same. It is completely free of any normality assumptions, or any other distributional assumptions.  In the spirit, it is very similar to the bootstrap. Like bootstrap, it has been known for awhile\footnote{``the statistician does not carry out this very simple and very tedious process, but his conclusions
	have no justification beyond the fact that they agree
	with those which could have been arrived at by this
	elementary method,'' \\ \hspace{3cm}Fisher (1936).}, but became popular only with availability of cheap computing power.
	
Let $X_1\ldots,X_n\sim F_X$ and $Y_1\ldots,Y_m\sim F_Y$ be two independent samples from two populations, and $H_0$ is the hypothesis that two populations are identical\footnote{This is the type of hypothesis we would consider when testing wheter a treatment differes from a placebo.}. Namely, we want to test
\begin{equation}\label{eq:F}
H_0:\hspace{1mm} F_X=F_Y\hspace{2mm}\mbox{ versus }\hspace{2mm} H_1:\hspace{1mm}F_X\neq F_Y.
\end{equation}
Let $s(X;Y)=s(X_1\ldots,X_n;Y_1\ldots,Y_m)$ be some test statistic that discriminates between the null and alternative\footnote{For example, $s(X;Y)=|\overline{X}_n-\overline{Y}_m|$, or the Kolmogorov-Smirnov statistic $s(X;Y)=\sup_x|\hat{F}_{X,n}(x)-\hat{F}_{Y,m}(x)|$.}, and, as usual, we  reject $H_0$ if $s(X;Y)$ is large enough. 

If $H_0$ is true, then all $N=n+m$ random variables that constitute the data $X_1\ldots,X_n, Y_1\ldots,Y_m$ come from essentially one population:
\begin{equation}
\mbox{Under } H_0: \hspace{2mm}X_1\ldots,X_n,Y_1\ldots,Y_m\sim F=F_X=F_Y.
\end{equation} 
This means that, conditional on the observed values, any of the $N!$ permutations of the data has the same probability of being observed. Let $s_1,\ldots,s_{N!}$ denote the values of the test statistic computed for all permutations of the data\footnote{One of these values is what we actually observe, $s_{\mathrm{obs}}$.}. Then, under the null hypothesis, all these values are equally likely. The distribution $\mathbb{P}_0$ that puts mass $\frac{1}{N!}$ on each $s_i$ is called the \textit{permutation distribution} of $s$. 

Recall that the $p$-value is the probability (under $H_0$) of observing a value of the test statistic more extreme that was actually observed. The $p$-value of the permutation test is then
\begin{equation}\label{eq:p-value}
p\mbox{-value}=\mathbb{P}_0(s>s_{\mathrm{obs}})=\frac{1}{N!}\sum_{i=1}^{N!}I(s_i>s_{\mathrm{obs}}).
\end{equation}

In most cases, $N=n+m$ is large enough so that summing over all permutations in (\ref{eq:p-value}) is infeasible. In this case we can simply approximate the $p$-value using the Monte Carlo method, that is by using a random sample of permutations. This leads to the following algorithm for testing (\ref{eq:F}):
\begin{enumerate}
	\item Compute the observed value of the test statistic 
	\begin{equation}
	s_{\mathrm{obs}}=s(X_1\ldots,X_n;Y_1\ldots,Y_m).
	\end{equation}
	\item Randomly permute the data. That is pick a permutation\footnote{Racall that a permutation of $N$ elements is a one-to-one map from $\{1,\ldots,N\}$ to itself.} $\pi$ at random and define 
	\begin{equation}
	\begin{split}
	Z_\pi&=(Z_{\pi(1)},\ldots,Z_{\pi(N)}), \hspace{3mm}\mbox{where}\\
	Z&=(Z_1,\ldots,Z_N)=(X_1,\ldots,X_n;Y_1,\ldots,Y_m).
	\end{split}
	\end{equation}
	\item Compute the statistic for the permuted data: 
	\begin{equation}
	s_\pi=s(Z_\pi).
	\end{equation}
	\item Repeat the last two steps $K$ times and let $s_1,\ldots,s_K$ denote the resulting statistic values.
	\item The estimated $p$-value is\footnote{The number of permutations $K$ is a trade-off between the accuracy and computer time. The more permutation the better. It is suggested to use $K$ of the order $K\sim10^4,10^5$.} 
	\begin{equation}
	p\mbox{-value}\approx \frac{1}{K}\sum_{i=1}^{K}I(s_i>s_{\mathrm{obs}}).
	\end{equation}
\end{enumerate}

The permutation test is especially useful for small samples, since for large samples, the normality assumptions used in parametric tests usually hold and the tests give similar results. 
\vspace{-3mm}

\section{Example: Hot Wings} \vspace{-3mm}
\begin{marginfigure}
	%\vspace{35mm}
	\centerline{\includegraphics[width=.7\linewidth]{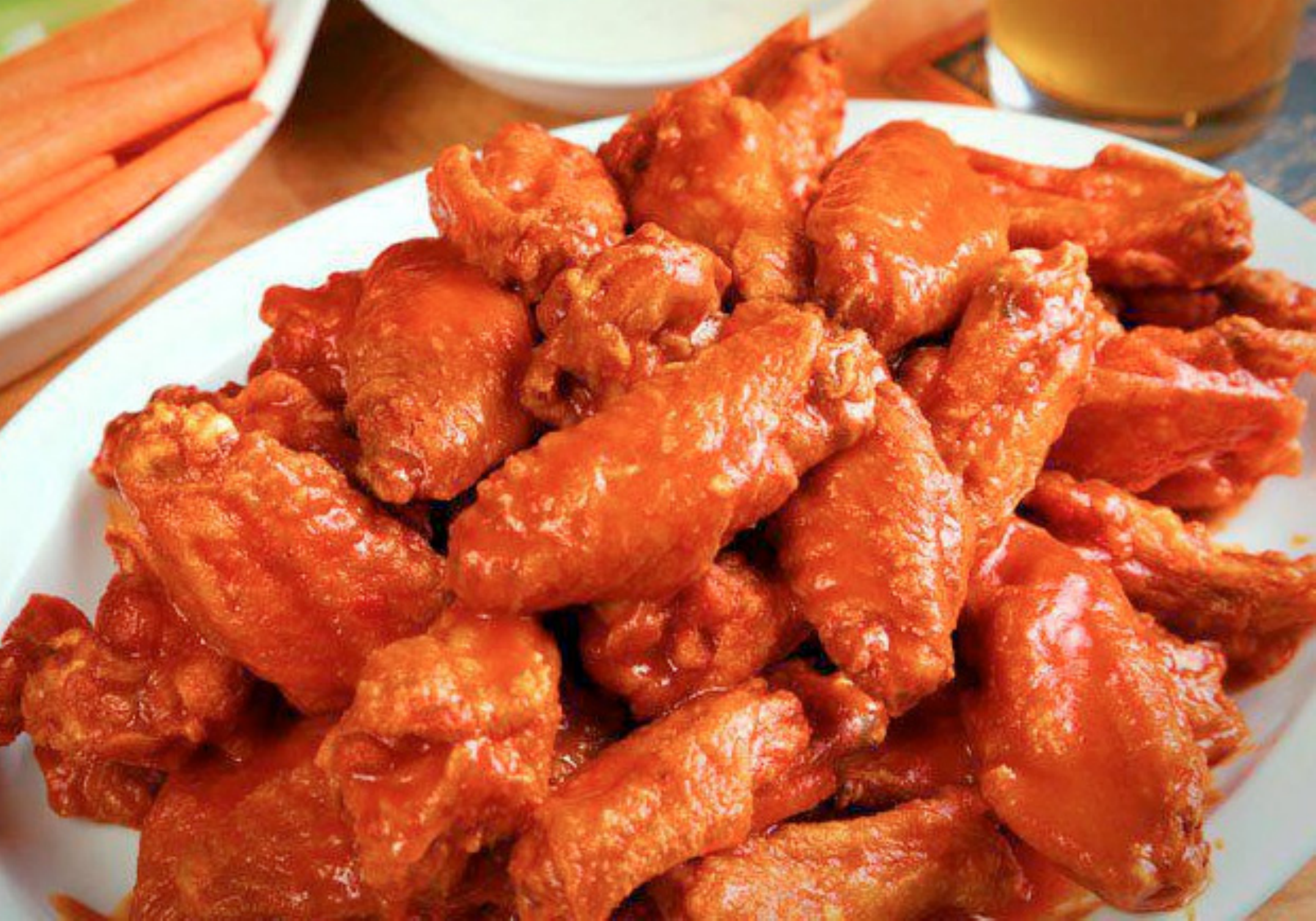}}\caption{Hot Chicken Wings. Photo source: \href{http://www.losangeles.com/articles/la-s-best-chicken-wings.html}{losangeles.com}.}\label{fig:wings}
\end{marginfigure}
Carleton student Nicki Catchpole conducted a study of hot wings consumption at the Williams bar in the Uptown area of Minneapolis. She asked patrons at the bar to record the consumption of hot wings during several hours. One of the questions she wanted to investigate is \textit{whether or not gender had an impact on hot wings consumption}.

\begin{marginfigure}
	\vspace{5mm}
	\centerline{\includegraphics[width=\linewidth]{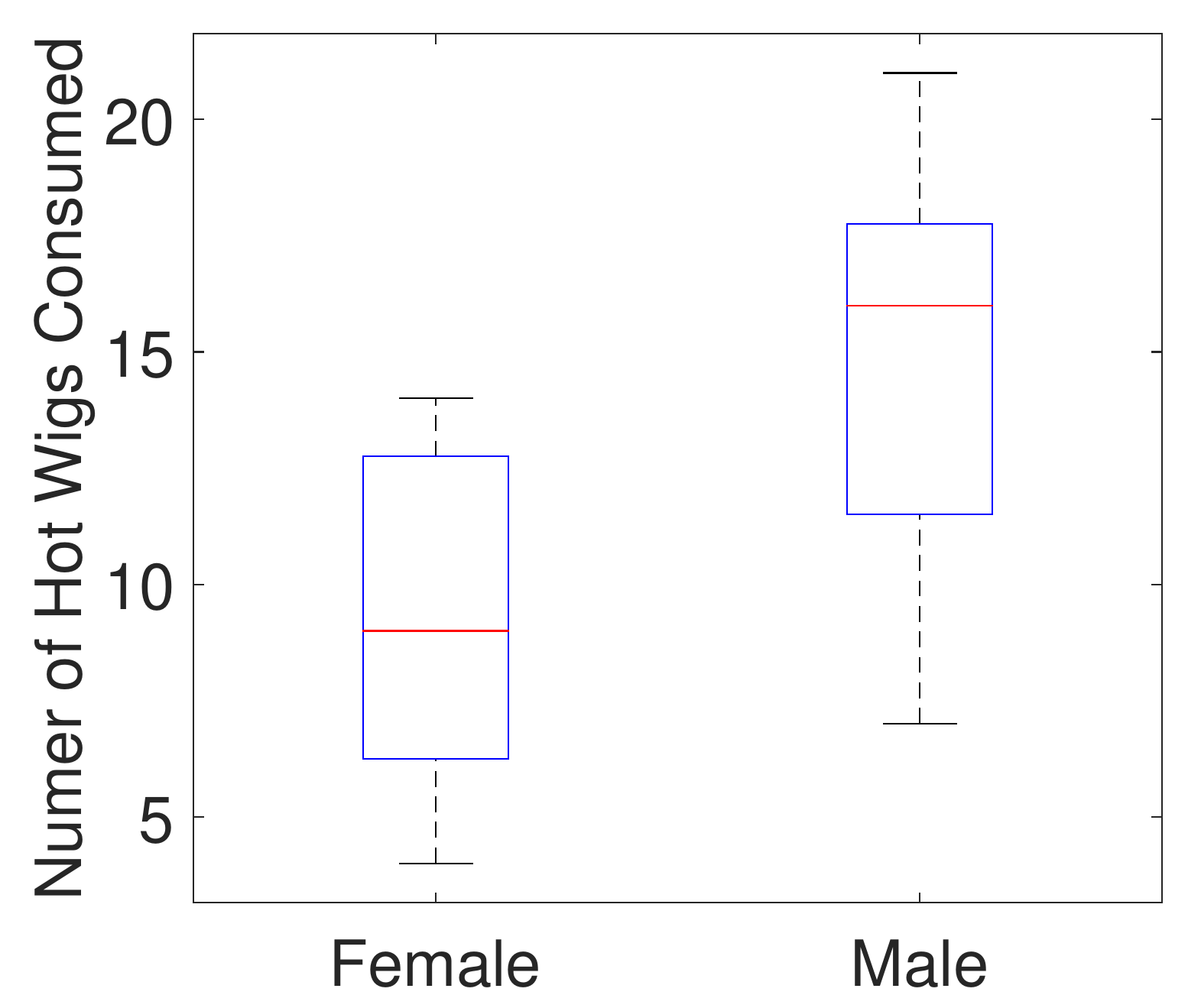}}\caption{Boxplots of hot wings consumption by Females and Males at the Williams bar, MN.}\label{fig:boxplots}
\end{marginfigure}

The data\footnote{Available at \href{http://www.
		its.caltech.edu/~zuev/teaching/
		2016Winter/wings.xlsx}{wings.xlsx}.} obtained in the course of this study consists of $N=30$ observations: $X_1,\ldots,X_n$ are $Y_1,\ldots,Y_m$, where $X$'s and $Y$'s correspond to males and females, and $n=m=30$. The boxplots of the data are shown in Fig.~\ref{fig:boxplots}. 
The sample means for males and females are clearly different: $\mu_M=14.53$ and $\mu_F=9.33$. But could  this difference arise by chance? In other words, could it be the case that the male and female consumptions are the same, and the difference that we observe is simply due to high variability of the consumption? Let us test this using the permutation test. 

So, we assume that $X_1,\ldots,X_n\sim W_M$, $Y_1,\ldots,Y_m\sim W_F$, and our null hypothesis is that the two populations (male and female consumptions of hot wings) are identical:
\begin{equation}\label{eq:G}
H_0:\hspace{1mm} W_M=W_F\hspace{2mm}\mbox{ versus }\hspace{2mm} H_1:\hspace{1mm}W_M\neq W_F.
\end{equation}
Let us use the absolute value of the sample means
\begin{equation}\label{eq:statistic}
s(X;Y)=|\overline{X}_n-\overline{Y}_m|,
\end{equation}
as a test statistic. The observed value is $s_{\mathrm{obs}}=5.2$ 
\begin{marginfigure}
	%\vspace{35mm}
	\centerline{\includegraphics[width=\linewidth]{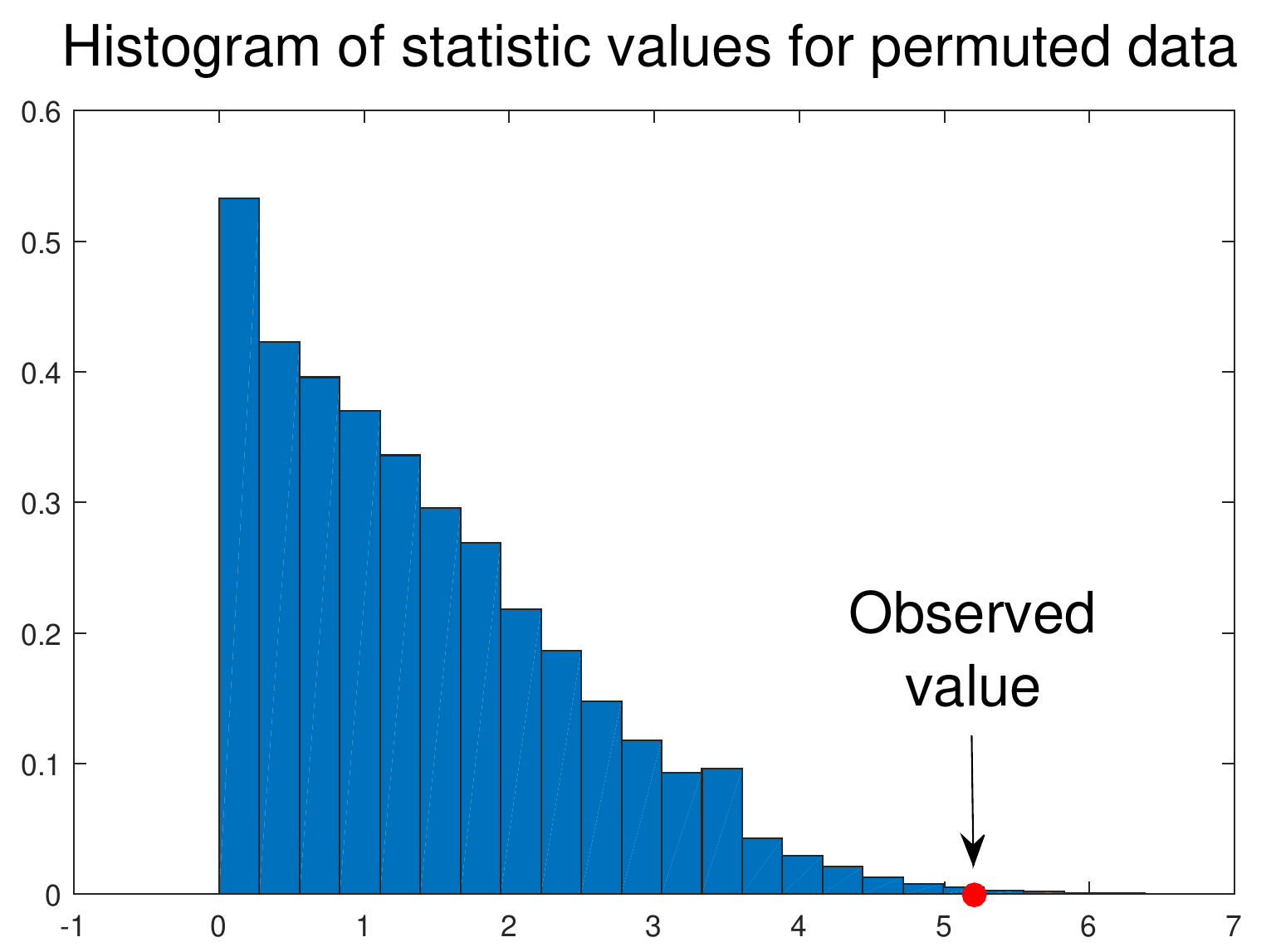}}\caption{Histogram of the obtained values of the test static (\ref{eq:statistic}) for $K=10^5$ data permutations.}\label{fig:histogram}
\end{marginfigure}

Figure~\ref{fig:histogram} shows the histogram of the statistic values  (\ref{eq:statistic}) obtained from $K=10^5$ random permutations of the original data. As we can see, the observed value of the statistic is quite extreme. The corresponding estimated $p$-value is $0.0017$, which means that the data provided quite strong evidence against the hypothesis that males and females consume hot wigs in equal amounts. 

What if instead of (\ref{eq:statistic}), we will use the Kolmogorov-Smirnov statistic:
\begin{equation}\label{eq:KS}
s(X;Y)=\sup_x|\hat{F}_{X,n}(x)-\hat{F}_{Y,m}(x)|.
\end{equation}
\begin{marginfigure}
	%\vspace{35mm}
	\centerline{\includegraphics[width=\linewidth]{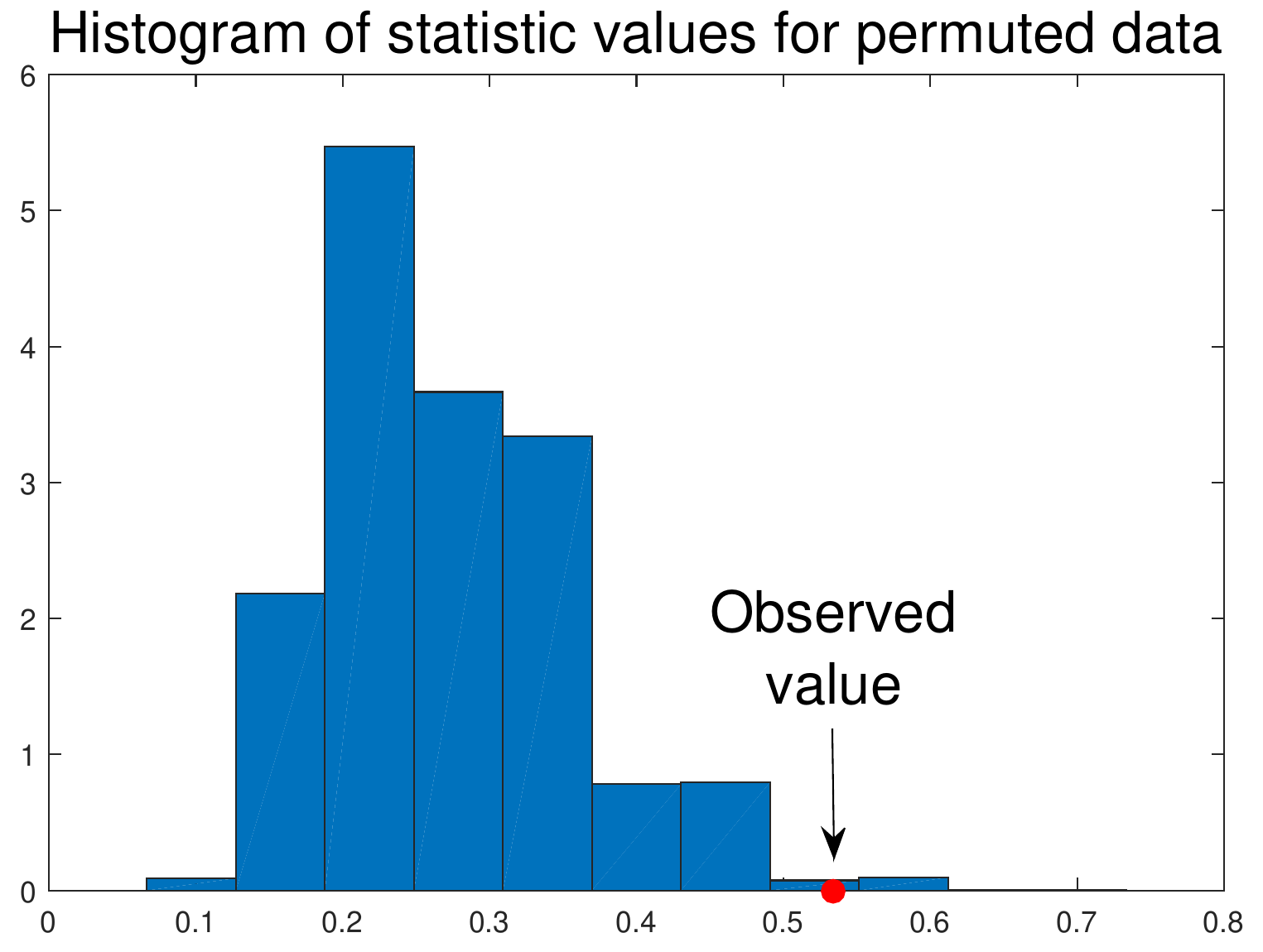}}\caption{Histogram of the obtained values of the test static (\ref{eq:KS}) for $K=10^5$ data permutations.}\label{fig:histogram2}
\end{marginfigure}
Figure~\ref{fig:histogram2} shows the distribution of the statistic values in this case. Again, the observed value $s_{\mathrm{obs}}=0.53$ is extreme. The estimated $p$-value $0.0062$ is a bit larger than in the previous case, but still small enough to reject the null.

\vspace{-4mm}
\section{Further Reading}
\vspace{-3mm}
\begin{enumerate}
	\item Introductory texts on statistical inference rarely cover permutation tests. And yet permutation procedures are the primary methods for testing hypothesis in many application areas, especially in biostatistics and genetics. The book P.~Good (2005) \textit{\href{https://books.google.com/books/about/Permutation_Parametric_and_Bootstrap_Tes.html?id=tQtedCBEgeAC}{Permutation, Parametric, and Bootstrap Tests of Hypothesis}} is highly recommended. 
\end{enumerate}
\vspace{-6mm}

\section{What is Next?} 
\vspace{-3mm}
We will discuss the maximum likelihood ratio test which plays the same role in testing as the MLE plays in estimation.

\chapter{The Likelihood Ratio Test}\label{ch:LRT}

\newthought{In this lecture}, we will discuss the likelihood ratio method for testing hypotheses, which plays the same role as the maximum likelihood estimates play in point estimation. The likelihood ratio tests are as widely applicable as MLEs and are one of the most popular methods of testing in parametric settings. 

\section{Likelihood Ratio Test for Simple Hypotheses}
Let us first consider a simple case where both the null and alternative hypotheses are simple. Namely, suppose that $X_1,\ldots,X_n$ is modeled as a sample from $f(x;\theta)$, where $\theta\in\Theta=\{\theta_0,\theta_1\}$, and we wish to test
\begin{equation}\label{eq:simpleHYPO}
H_0:\hspace{1mm} \theta=\theta_0\hspace{2mm}\mbox{ versus }\hspace{2mm} H_1:\hspace{1mm}\theta=\theta_1.
\end{equation}
Recall that the likelihood function, which is the join probability/density of the data viewed as a function of the parameter, 
\begin{equation}
\mathcal{L}(\theta|X)=\prod_{i=1}^nf(X_i;\theta),
\end{equation}
measures the consistency of the parameter $\theta$ and the observed data.
If $\mathcal{L}(\theta_1|X)>\mathcal{L}(\theta_0|X)$, then it is more likely that the data $X=\{X_1,\ldots,X_n\}$ was generated by $f(x;\theta_1)$ and vice versa. This motivates the likelihood ratio test (LRT): 
\begin{equation}\label{eq:LRT}
\mbox{Reject } H_0 \hspace{2mm}\Leftrightarrow\hspace{2mm} \lambda(X)=\frac{\mathcal{L}(\theta_1|X)}{\mathcal{L}(\theta_0|X)}>c,
\end{equation}
where $c$ is some critical value\footnote{Which is, as usual, found from the requirement for the size $\alpha$ of the test.}. The statistic $\lambda$ is called the \textit{likelihood ratio statistic}. 

Note that the LRT is exactly the test Bob used in the two coin example (Lecture~\ref{ch:HypoTesting}). Let us consider one more classical example. 

\paragraph{Example: Normal LRT.}  Let $X_1,\ldots,X_n\sim N(\mu,\sigma^2)$, where $\sigma^2$ is known\footnote{If the variance is unknown, then (\ref{eq:17mu}) are no longer simple hypotheses.}, and let us test 
\begin{equation}\label{eq:17mu}
H_0:\hspace{1mm} \mu=\mu_0\hspace{2mm}\mbox{ versus }\hspace{2mm} H_1:\hspace{1mm}\mu=\mu_1,
\end{equation}
where $\mu_1>\mu_0$. The likelihood function is
\begin{equation}\label{eq:lik}
\begin{split}
\mathcal{L}(\mu|X)&=\prod_{i=1}^n\frac{1}{\sqrt{2\pi}\sigma}\exp\left(\frac{-(X_i-\mu)^2}{2\sigma^2}\right)\\
&=\left(\frac{1}{\sqrt{2\pi}\sigma}\right)^n\exp\left(-\frac{\sum_{i=1}^n(X_i-\mu)^2}{2\sigma^2}\right).
\end{split}
\end{equation}
The likelihood ratio statistic is then
\begin{equation}
\lambda(X)=\frac{\mathcal{L}(\mu_1|X)}{\mathcal{L}(\mu_0|X)}=\exp\left(\frac{\sum_{i=1}^n(X_i-\mu_0)^2-\sum_{i=1}^n(X_i-\mu_1)^2}{2\sigma^2}\right).
\end{equation}
After some algebra, the rejection region of the LRT $\{X: \lambda(X)>c\}$ reduces to 
\begin{equation}
\mathcal{R}=\left\{X: \overline{X}_n>c'=\frac{2\sigma^2\log c +n(\mu_1^2-\mu_0^2)}{2n(\mu_1-\mu_0)}\right\},
\end{equation}
which looks intuitive: the test rejects $H_0$ when $\overline{X}_n$ (which is supposed to be around $\mu_0$ under $H_0$) is large enough\footnote{If $\mu_1<\mu_0$, the rejection region would be of the form $\{\overline{X}_n<c''\}$.}. The critical value $c'$ is determined from the condition on the size $\alpha$:
\begin{equation}
\begin{split}
\alpha=&\mathbb{P}(\overline{X}_n>c'|\mu=\mu_0)=\mathbb{P}\left(\left.\frac{\overline{X}_n-\mu_0}{\sigma/\sqrt{n}}>\frac{c'-\mu_0}{\sigma/\sqrt{n}}\right|\mu=\mu_0\right)\\
&=1-\Phi\left(\frac{c'-\mu_0}{\sigma/\sqrt{n}}\right),
\end{split}
\end{equation} 
where we used that $\overline{X}_n\sim\mathcal{N}\left(\mu_0,\frac{\sigma^2}{n}\right)$.
The previous equations which leads to 
\begin{equation}
c'=\mu_0+\frac{\sigma z_{1-\alpha}}{\sqrt{n}}.
\end{equation}
To sum up, the size $\alpha$ LRT\footnote{What is the $p$-value of this test?} 
\begin{equation}\label{eq:test}
\mbox{Rejects } H_0 \hspace{2mm}\Leftrightarrow\hspace{2mm}\overline{X}_n>\mu_0+\frac{\sigma z_{1-\alpha}}{\sqrt{n}}.
\end{equation} \vspace{-7mm}

\hfill $\square$\bigskip

We know that in general, finding the most powerful test is a daunting task. It turns out however, that in the special of simple null and simple alternative (\ref{eq:simpleHYPO}), the LRT is the most powerful test\footnote{J. Neyman \& E.S. Pearson (1933) ``\href{http://www.jstor.org/stable/91247}{On the problem of the most efficient tests of statistical hypotheses},'' \textit{Philosophical Transactions of the Royal Society A: Mathematical, Physical and Engineering Sciences}, 231 (694-706): 289-337.}. 
\begin{marginfigure}
	%\vspace{35mm}
	\centerline{\includegraphics[width=.5\linewidth]{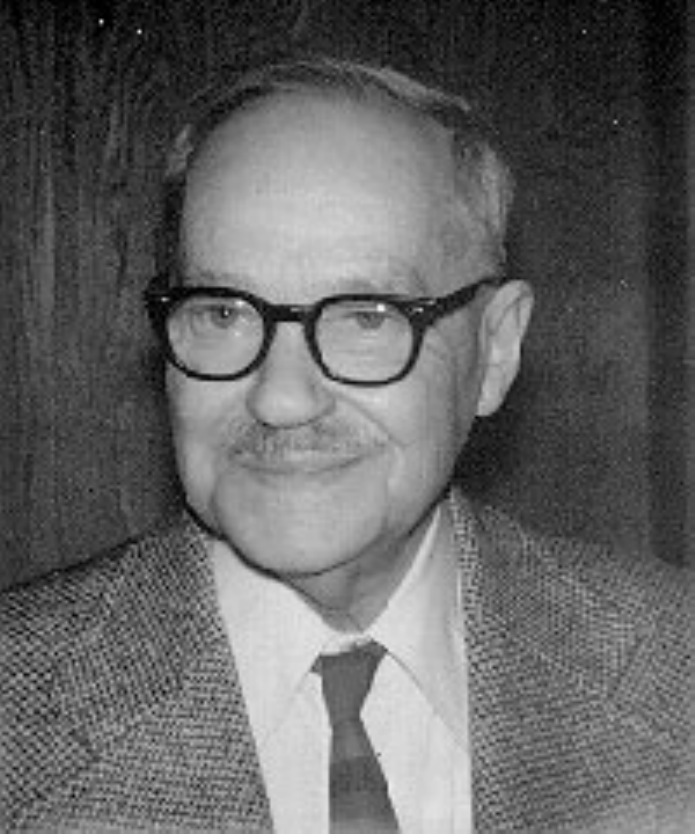}
		\includegraphics[width=.5\linewidth]{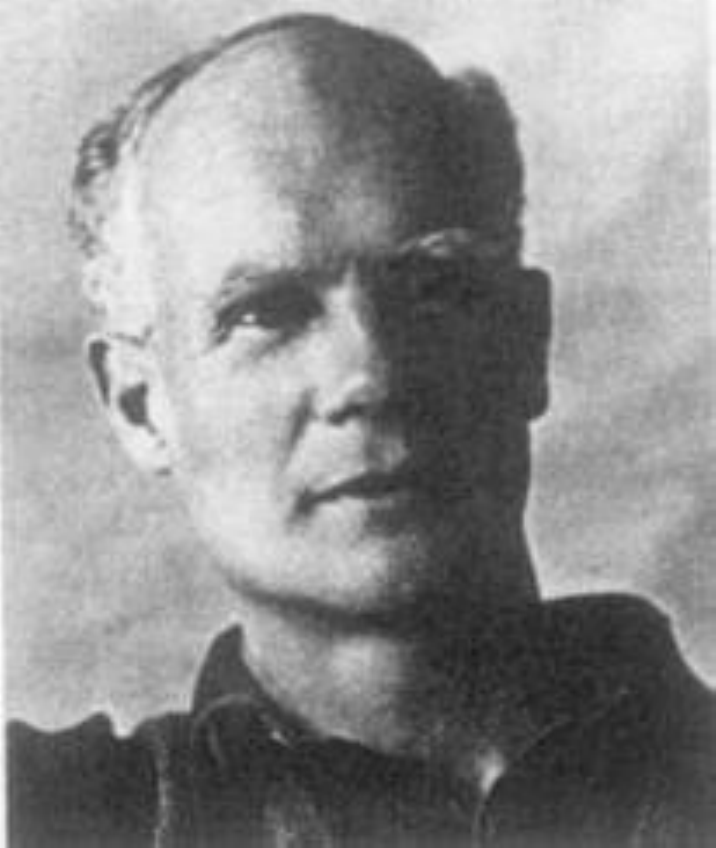}}\caption{Jerzy Neyman, Polish-American mathematician, and Egon Pearson, leading British statistician.}\label{fig:stats}
\end{marginfigure}
\begin{theorem}[Neyman-Pearson Lemma] Let $X_1\ldots,X_n$ is modeled as a random sample from a distribution with parameter $\theta$. Suppose we wish to test $H_0: \theta=\theta_0$ vesus $H_1: \theta=\theta_1$. The size $\alpha$ LRT is the most powerful test of size $\alpha$. That is, among all tests with size $\alpha$, the LRT has the largest power $\beta(\theta_1)$  (\ie~the smallest probability of the type II error). 
\end{theorem}

In particular, the test (\ref{eq:test}) is the most powerful for testing (\ref{eq:17mu}), and  Bob did his best in testing the hypothesis that the coin is fair. 

\section{Likelihood Ratio Test: General Case}
Let us now consider a general case, where the hypothesis are not necessarily simple\footnote{In this case, they are often called \textit{composite}.}. That is, suppose that $X_1,\ldots,X_n$ is modeled as a sample from $f(x;\theta)$, where $\theta\in\Theta=\Theta_0\sqcup\Theta_1$, and we wish to test
\begin{equation}\label{eq:compositeHYPO}
H_0:\hspace{1mm} \theta\in\Theta_0\hspace{2mm}\mbox{ versus }\hspace{2mm} H_1:\hspace{1mm}\theta\in\Theta_1.
\end{equation}

We need to generalize the definition of the likelihood ratio statistic  (\ref{eq:LRT}) because now it does not makes sense: we have sets $\Theta_0$ and $\Theta_1$ instead of points $\theta_0$ and $\theta_1$. In general case, the likelihood ratio statistic is defined as follows:
\begin{equation}\label{eq:GLRS}
\lambda(X)=\frac{\sup_{\theta\in\Theta}\mathcal{L}(\theta|X)}{\sup_{\theta\in\Theta_0}\mathcal{L}(\theta|X)}.
\end{equation} 

Looking at (\ref{eq:LRT}), you might have expected to see $\tilde{\lambda}(X)=\frac{\sup_{\theta\in\Theta_1}\mathcal{L}(\theta|X)}{\sup_{\theta\in\Theta_0}\mathcal{L}(\theta|X)}$. In practice, these two statistics often have similar values\footnote{Notice that $\lambda=\max\{1,\tilde{\lambda}\}$.}. But theoretical properties of the statistic (\ref{eq:GLRS}) are much simpler and nicer.

Large values of $\lambda(X)$ provide evidence against the null hypothesis. Indeed, if $\lambda(X)$ is large, then the value of parameter $\theta$ most consistent with the observed data does not lie in $\Theta_0$. So, the LRT\footnote{Sometimes this test is called the generalized likelihood ratio test.}:
\begin{equation}\label{eq:GLRT}
\mbox{Rejects } H_0 \hspace{2mm}\Leftrightarrow\hspace{2mm} \lambda(X)>c.
\end{equation}
Since $\lambda(X)\geq1$, the critical value $c$ should be also $c\geq1$.

If we think of maximization over $\Theta$ and $\Theta_0$, then the close relationship between LRTs and MLEs become clear. Let $\hat{\theta}$ be the MLE of $\theta$, and $\hat{\theta}_0$ be the MLE when $\theta$ is required to lie in $\Theta_0$ (\ie when we consider $\Theta_0$ as the full parameter space). Then $\lambda$ can be written as follows:
\begin{equation}
\lambda(X)=\frac{\mathcal{L}(\hat{\theta}|X)}{\mathcal{L}(\hat{\theta}_0|X)}. 
\end{equation}

The Neyman-Pearson Lemma says that the likelihood ratio tests are 
optimal for simple hypotheses. For composite hypothesis, the LRTs are  generally not optimal\footnote{But often the most powerful test simply do not exist.}, but perform reasonable well\footnote{Just like MLEs.}. This explains the popularity of LRTs.

\paragraph{Example: Normal GLRT.}  Let $X_1,\ldots,X_n\sim\mathcal{N}(\mu,\sigma^2)$, where variance $\sigma^2$ is known. Consider testing the following hypothesis:
\begin{equation}
H_0:\hspace{1mm} \mu=\mu_0\hspace{2mm}\mbox{ versus }\hspace{2mm} H_1:\hspace{1mm}\mu\neq\mu_0.
\end{equation}
Here, $\Theta=\mathbb{R}$, $\Theta_0=\{\mu_0\},$ and $\Theta_1=(-\infty,\mu_0)\cup(\mu_0,\infty)$. As in the previous example, the likelihood is given by (\ref{eq:lik}). The likelihood ratio statistic is
\begin{equation}
\begin{split}
\lambda(X)&=\frac{\sup_{\mu\in\Theta}\mathcal{L}(\mu|X)}{\sup_{\mu\in\Theta_0}\mathcal{L}(\mu|X)}=\frac{\mathcal{L}(\hat{\mu}_{\mathrm{MLE}}|X)}{\mathcal{L}(\mu_0|X)}=\frac{\mathcal{L}(\overline{X}_n|X)}{\mathcal{L}(\mu_0|X)}\\
&=\exp\left(\frac{\sum_{i=1}^n\left((X_i-\mu_0)^2-(X_i-\overline{X}_n)^2\right)}{2\sigma^2}\right)\\
&=\exp\left(\frac{n(\overline{X}_n-\mu_0)^2}{2\sigma^2}\right).
\end{split}
\end{equation}
Rejecting when $\lambda(X)>c$ is equivalent to rejecting when 
\begin{equation}
\left|\frac{\overline{X}_n-\mu_0}{\sigma/\sqrt{n}}\right|>c'=\sqrt{2\log c}.
\end{equation}
Since under the null $\overline{X}_n\sim\mathcal{N}\left(\mu_0,\frac{\sigma^2}{n}\right)$, to construct the size $\alpha$ test, we need to set $c'=z_{1-\frac{\alpha}{2}}$. So, in this example, the LRT  essentially coincides with the Wald test\footnote{How would you construct the size $\alpha$ LRT test if $\sigma$ is unknown? Does it remind you any other test?}.
\hfill $\square$

\subsection{Null Distribution of $\lambda(X)$}
In order to construct the LRT of size $\alpha$, we need to find the critical value $c$ from the following equation:
\begin{equation}
\alpha=\sup_{\theta\in\Theta_0}\mathbb{P}(\lambda(X)>c).
\end{equation}
To compute the probability of the type I error on the right-hand side, we need to know the null distribution\footnote{That is, the distribution under $H_0$.} of the LRT statistic $\lambda(X)$.  Let us look at the previous example: under $H_0$
\begin{equation}
2\log\lambda(X)=\left(\frac{\overline{X}_n-\mu_0}{\sigma/\sqrt{n}}\right)^2\sim\chi^2_1,
\end{equation}
where $\chi^2_1$ is the $\chi^2$-distributions with $1$ degree of freedom\footnote{Recall, that if $Z_1,\ldots,Z_q$ are i.i.d. standard normal variables, then the the distribution of $Q=||Z||^2=Z_1^2+\ldots Z_q^2$ is called the $\chi^2$-distribution with $q$ degrees of freedom.}.
\begin{marginfigure}
	%\vspace{35mm}
	\centerline{\includegraphics[width=\linewidth]{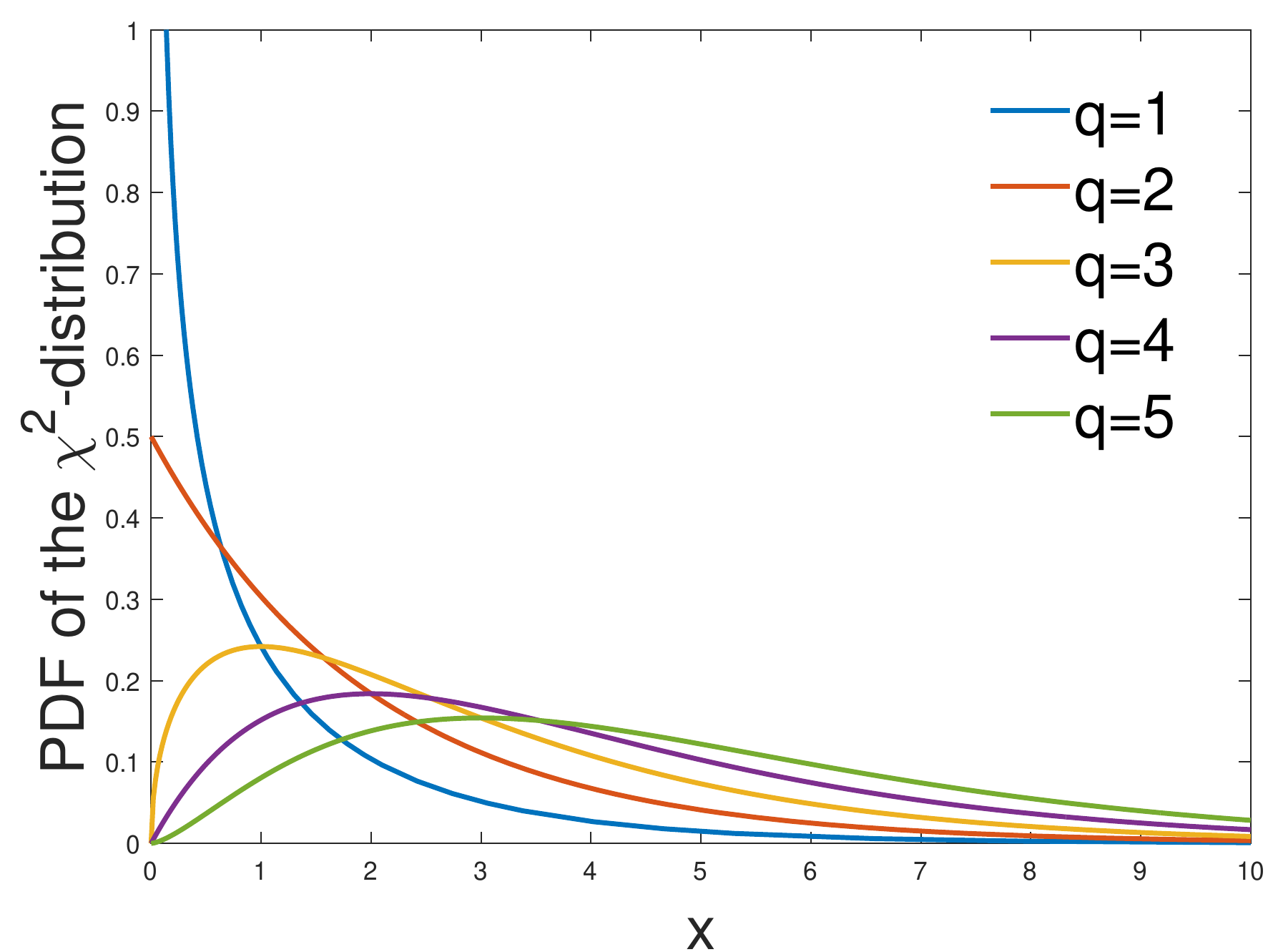}}\caption{The PDF of the $\chi^2$-distribution with $q$ degrees of freedom.}\label{fig:chi}
\end{marginfigure}
It turns out that similar result holds in a more general case. Assuming that the probability model $f(x;\theta)$ satisfies certain regularity conditions, the null distribution of $2\log\lambda(X)$ tends to  $\chi^2$-distribution with $q$ degrees of freedom as the sample size $n\rightarrow\infty$:
\begin{equation}\label{eq:limitingpdf}
2\log\lambda(X)\stackrel{D}{\longrightarrow}\chi^2_q, \hspace{3mm}\mbox{where }q=\dim\Theta-\dim\Theta_0,
\end{equation}
where $\dim\Theta$ and $\dim\Theta_0$ are the numbers of free parameters in $\Theta$ and $\Theta_0$. For instance, in the previous example, $\Theta=\mathbb{R}$, $\dim\Theta=1$, $\Theta_0=\{\mu_0\}, \dim\Theta_0=0$.

The result (\ref{eq:limitingpdf}) may appear counter intuitive at the first glance: indeed, how come that regardless of the probability model for data $X$ (as long as it is smooth enough), the LRT statistic $2\log\lambda(X)$ converges to the same $\chi^2$-distribution. So let us give 

\paragraph{Sketch of Proof of (\ref{eq:limitingpdf})}  for a special case:  
$
\Theta=\mathbb{R}, \Theta_0=\{\theta_0\}, q=1.
$
\begin{equation}
\begin{split}
2\log\lambda(X)=2\log\mathcal{L}(\hat{\theta}_{\mathrm{MLE}})-2\log\mathcal{L}(\theta_0)=2l(\hat{\theta}_{\mathrm{MLE}})-2l(\theta_0),
\end{split}
\end{equation}
where $l(\theta)$ is the log-likelihood. Using the Taylor expansion of $l(\theta)$ at $\theta=\hat{\theta}_{\mathrm{MLE}}$, we have:
\begin{equation}
\begin{split}
l(\theta)&\approx l(\hat{\theta}_{\mathrm{MLE}})+(\theta-\hat{\theta}_{\mathrm{MLE}})l'(\hat{\theta}_{\mathrm{MLE}})+\frac{(\theta-\hat{\theta}_{\mathrm{MLE}})^2}{2}l''(\hat{\theta}_{\mathrm{MLE}})\\
&=l(\hat{\theta}_{\mathrm{MLE}})+\frac{(\theta-\hat{\theta}_{\mathrm{MLE}})^2}{2}l''(\hat{\theta}_{\mathrm{MLE}}).
\end{split}
\end{equation}
Therefore,
\begin{equation}
\begin{split}
2\log\lambda(X)&\approx 2l(\hat{\theta}_{\mathrm{MLE}})-2\left(l(\hat{\theta}_{\mathrm{MLE}})+\frac{(\theta_0-\hat{\theta}_{\mathrm{MLE}})^2}{2}l''(\hat{\theta}_{\mathrm{MLE}})\right)\\
&=-l''(\hat{\theta}_{\mathrm{MLE}})(\theta_0-\hat{\theta}_{\mathrm{MLE}})^2.
\end{split}
\end{equation}
Under $H_0$, the true value of the parameter is $\theta_0$. Recall that the MLE is asymptotically normal, $\hat{\theta}_{\mathrm{MLE}}\stackrel{D}{\longrightarrow}\mathcal{N}\left(\theta_0,\frac{1}{nI(\theta_0)}\right)$. Therefore,
\begin{equation}
\begin{split}
2\log\lambda(X)&\approx-\frac{l''(\hat{\theta}_{\mathrm{MLE}})}{nI(\theta_0)}\left((\theta_0-\hat{\theta}_{\mathrm{MLE}})\sqrt{nI(\theta_0)}\right)^2\\
&=-\frac{l''(\hat{\theta}_{\mathrm{MLE}})}{nI(\theta_0)}Z_n^2, 
\end{split}
\end{equation}
where $Z_n\stackrel{D}{\longrightarrow}\mathcal{N}(0,1)$. We are almost there. Since the MLE is consistent, it converges to the true value of the parameter, with is $\theta_0$ under $H_0$, \ie $\hat{\theta}_{\mathrm{MLE}}\stackrel{\mathbb{P}}{\longrightarrow}\theta_0$. Therefore, $l''(\hat{\theta}_{\mathrm{MLE}})\stackrel{\mathbb{P}}{\longrightarrow}l''(\theta_0)$. In lecture 10, while discussing the asymptotic normality of the MLE, we obtained that $-\frac{l''(\theta_0)}{nI(\theta_0)}\approx1$. Combining these results, we finally have that 
\begin{equation}
2\log\lambda(X)\stackrel{D}{\longrightarrow}(\mathcal{N}(0,1))^2=\chi^2_1.
\end{equation}
Thus, in essence, the null distribution of the LRT statistic is a consequence of the nice analytical properties of the MLE.  \hfill$\square$

\subsection{Approximate $p$-value of the LRT}

Using the asymptotic result (\ref{eq:limitingpdf}), it is straightforward to derive the approximate $p$-value of the LRT. Indeed, if the sample size is sufficiently large, then $2\log\lambda(X)\approxdist\chi^2_q$, and the test size is therefore
\begin{equation}
\alpha=\mathbb{P}(\lambda(X)>c)=\mathbb{P}(2\log\lambda(X)>2\log c),
\end{equation}
which means that 
\begin{equation}
2\log c\approx \chi^2_{q,1-\alpha},
\end{equation}
where, $\chi^2_{q,\alpha}$ is such point that the probability that the $\chi^2$-random variable with $q$ degrees of freedom is lass than $\chi^2_{q,\alpha}$ is $\alpha$\footnote{$\chi^2_{q,\alpha}$ is the $\chi^2$ analog of $z_\alpha$. \\ Check that $\sqrt{\chi^2_{1,\alpha}}=-z_{\frac{1-\alpha}{2}}$.}. Thus, the LRT with the rejection region 
\begin{equation}
\mathcal{R}_\alpha=\left\{X: \lambda(X)>\exp\left(\frac{\chi^2_{q,1-\alpha}}{2}\right)\right\}
\end{equation}
has approximate size $\alpha$. 

The $p$-value is smallest (infimum) size $\alpha^*$ at which the test rejects. The approximate $p$-value is thus the solution of
\begin{equation}
\lambda(X)=\exp\left(\frac{\chi^2_{q,1-\alpha^*}}{2}\right),
\end{equation} 
which is 
\begin{equation}
\alpha^*=\mathbb{P}(Y>2\log\lambda(X)), \hspace{5mm}Y\sim\chi^2_q,
\end{equation}
where $X$ is the actually observed data.

\section{Further Reading}
\begin{enumerate}
	\item I encourage you to read (or at least to look through) the original paper J. Neyman \& E.S. Pearson (1933) ``\href{http://www.jstor.org/stable/91247}{On the problem of the most efficient tests of statistical hypotheses},'' \textit{Philosophical Transactions of the Royal Society A: Mathematical, Physical and Engineering Sciences}, 231 (694-706): 289-337. This is a very good literature with beautiful illustrations.\footnote{Notice the vintage terminology and notation. They call ``character'' and ``elementary probability'' what we now call ``test statistic'' and ``probability density function'', and they spell coordinates as ``co-ordinates.''}. 
	
\end{enumerate}

\section{What is Next?} 
We will use the LRT to test Mendel's theory of inheritance.

\chapter{Testing Mendel's Theory}\label{ch:Mendel}

\begin{marginfigure}
%	\vspace{10mm}
	\centerline{\includegraphics[width=.5\linewidth]{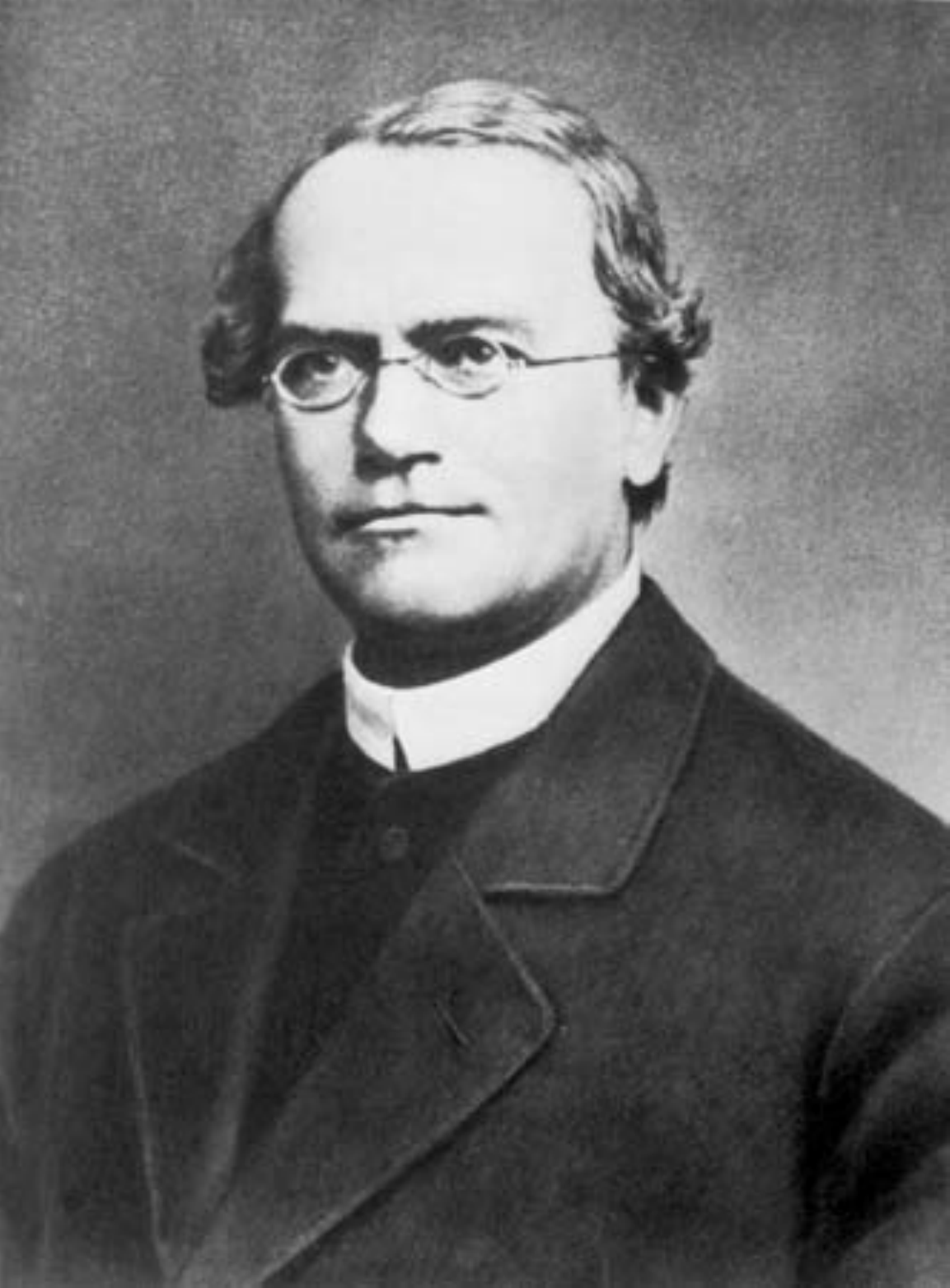}
	}\caption{Gregor Mendel, a scientist and a monk, the father of modern genetics. Photo source: \href{http://www.britannica.com/biography/Gregor-Mendel}{britannica.com}}\label{fig:Mendel}
\end{marginfigure}
\newthought{Here} we will statistically  test the theory of inheritance proposed by Gregor Mendel, after he experimented with pea plants \footnote{A brief interactive introduction to Mendelian inheritance is available at  \href{http://www.wiley.com/college/test/0471787159/biology\_basics/animations/mendelianInheritance.swf}{wiley.com}.}. We focus on his third law:  the \textit{principle of independent assortment}, which states that alleles for different traits are distributed uniformly at random to the offspring. Using Mendel's data, we will construct the likelihood ratio test and compute the corresponding $p$-value. \vspace{-2mm}

\section{Mendel's Peas}
\begin{figure}
%	\vspace{-5mm}
	\centerline{\includegraphics[width=.9\linewidth]{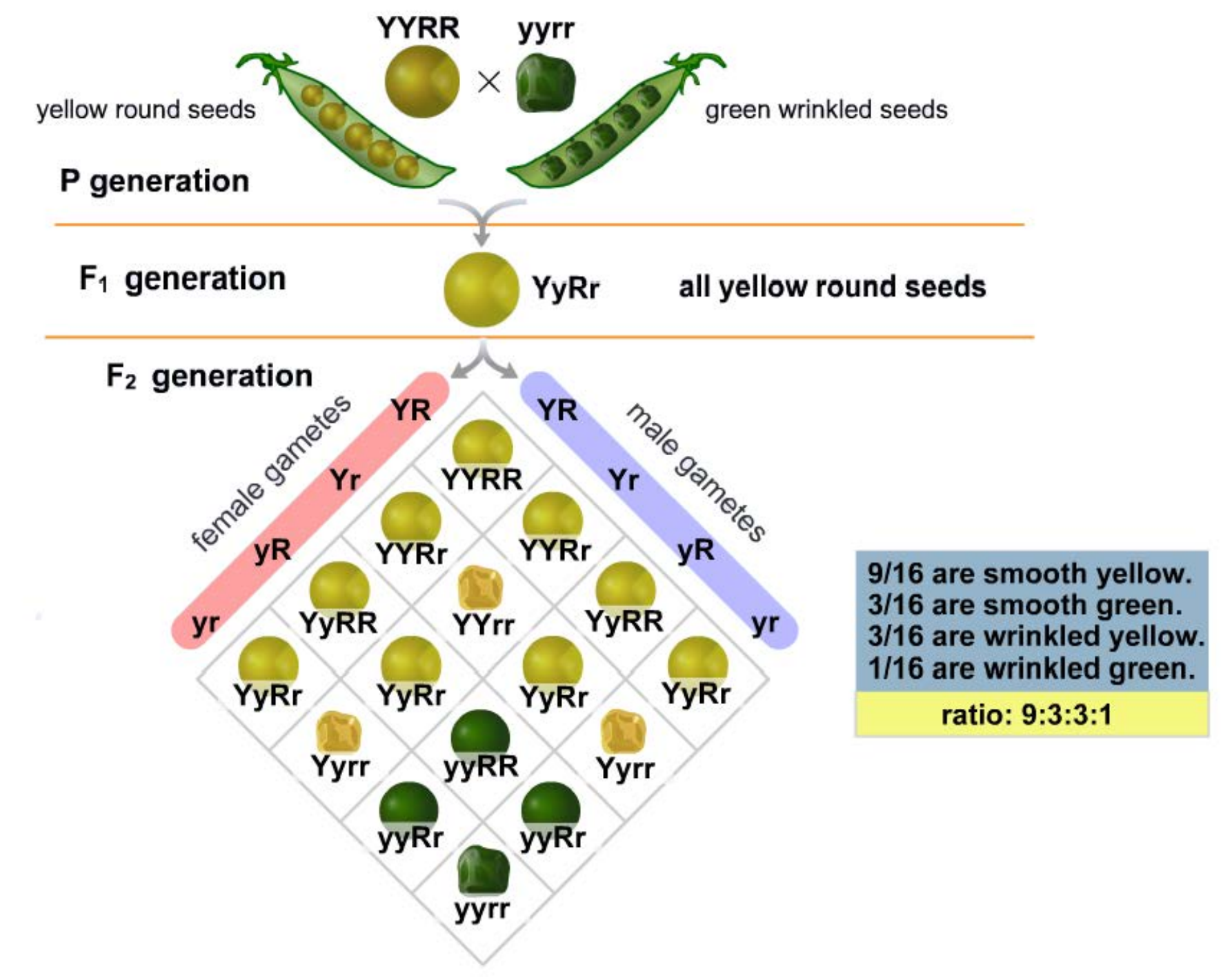}
	}\caption{Mendel's Principle of Independent Assortment. Picture source: \href{http://www.wiley.com/college/test/0471787159/biology\_basics/animations/mendelianInheritance.swf}{wiley.com}. }\label{fig:peas}
\end{figure}

\vspace{-1mm}
Suppose we are going to breed peas with \textit{round yellow} seeds and \textit{wrinkled green} seeds. Then, according to the Mendel's principle of dominance, all of the offspring in the first generation will be round and yellow, since yellow trait is dominant to green and round trait is dominant to wrinkled. If we now allow the offspring of the first generation to self-fertilize, then we will get all four types of progeny: round yellow, round green, wrinkled yellow, and wrinkled green. Moreover, Mendel's principle of independent assortment predicts the proportions of each type. Namely 
\begin{equation}\label{eq:theory}
\frac{9}{16}, 
\hspace{2mm} \frac{3}{16}, \hspace{2mm} \frac{3}{16}, \hspace{2mm}\frac{1}{16},
\end{equation} respectively. The breeding process is schematically shown in Fig.~\ref{fig:peas}. 
\vspace{-3mm}

\section{The Data}\vspace{-2mm}
In his original paper\footnote{G. Mendel (1866), ``\href{http://www.esp.org/foundations/genetics/classical/gm-65-f.pdf}{Versuche \"{u}ber Pflanzen-Hybriden},'' Verh. Naturforsch. Ver. Br\"{u}nn, 4: 3-47. For the English translation, see: W.~Bateson (1901). ``\href{http://www.esp.org/foundations/genetics/classical/gm-65.pdf}{Experiments in plant hybridization}.''}, Mendel described the results of his experiments, and, in particular, he reported that in $N=556$ trials he observed
\begin{equation}\label{eq:data}
\begin{split}
&n_1=315 \hspace{5mm}\mbox{round and yellow},\\
&n_2=108 \hspace{5mm}\mbox{round and green},\\
&n_3=101 \hspace{5mm}\mbox{wrinkled and yellow},\\
&n_4=32 \hspace{6.5mm}\mbox{wrinkled and green}.
\end{split}
\end{equation}
Let us test whether or not Mendel's theoretical prediction (\ref{eq:theory}) is consistent with the observed data (\ref{eq:data}). Based on the data, should we accept his theory or reject it?

\section{Probability Model for the Data}
In order to quantitatively answer the question on how plausible or unlikely the observed data under Mendel's theory, we need to inject some stochasticity in the picture. Namely, we need to assume some probability model for the data\footnote{Without probability model, the data are just numbers. With probability model, these numbers are a sample from probability distribution, which allows to us to use the machinery of probability theory to make quantitative statements.}.
What is a natural model for the observed numbers of peas of different types?

\subsection{Multinomial Distribution}
As the name suggests, the multinomial distribution is a straightforward  generalization of the binomial distribution. Recall, that the binomial distribution $\mathrm{Bin}(n,p)$ is the discrete probability distribution of the number of successes in a sequence of $n$ independent success/failure  experiments, each of which has success with probability $p$\footnote{That is, $\mathrm{Bin}(n,p)$ is the distribution if the sum on $n$ i.i.d. Bernoulli trials $\mathrm{Bernoulli}(p)$.}.   Multinomial distributions is a generalization for the case where there are more than two possible outcomes and a ``success-failure'' description is insufficient to understand the underlying system or phenomenon\footnote{For example, temperature can be ``high,'' ``medium,'' or ``low,'' or, as in the Mendel experiments, the seeds can be round yellow, round green, wrinkled yellow, and wrinkled green.}. 

Consider drawing a ball from a box which has balls with $k$ different colors labeled $1,2\ldots,k$. Let $p=(p_1,\ldots,p_k)$, where $p_i$ is is the probability of drawing a ball of color $i$,
\begin{equation}
p_i\geq0, \hspace{5mm}\sum_{i=1}^kp_i=1.
\end{equation}
Let us draw $N$ times\footnote{Independent draws with replacement.}, and let $n=(n_1,\ldots,n_k)$, where $n_i$ is the number of times that color $i$ appeared,
\begin{equation}
\sum_{i=1}^kn_i=N.
\end{equation}
We say that $n$ has a multinomial distribution, $n\sim\mathrm{Mult}(N,p)$, with parameters $N$, number of trials, and $p=(p_1,\ldots,p_k)$, vector of probabilities of $k$ different outcomes.

Here is a couple of properties of the multinomial distribution:
\begin{itemize}
	\item The probability mass function of $\mathrm{Mult}(N,p)$ is 
	\begin{equation}
	f(n|N,p)=\frac{N!}{n_1!\ldots n_k!}p_1^{n_1}\ldots p_k^{n_k}.
	\end{equation}
	\item The marginal distribution of $n_i$ is $\mathrm{Bin}(N,p_i)$.
\end{itemize}

\section{Problem Formulation}

It seems very natural to model the observed data (\ref{eq:data}) as a sample from the multinomial distribution:
\begin{equation}
n=(n_1,\ldots,n_k)\sim\mathrm{Mult}(N,p), \hspace{5mm}\mbox{where } k=4 \mbox{ and } N=556. 
\end{equation}
The Mendel theory, the null hypothesis, is then
\begin{equation}
H_0:\hspace{1mm} p=p^*=\left(\frac{9}{16}, 
\frac{3}{16},  \frac{3}{16}, \frac{1}{16}\right),
\end{equation}
and we want to test versus the alternative
\begin{equation}
H_1:\hspace{1mm} p\neq p^*.
\end{equation}
The full parameter space is $\Theta=\{(p_1,\ldots,p_k): p_i\geq0, \sum p_i=1\}$, which is geometrically a $3$-simplex (\ie tetrahedron), and $\Theta_0=\{p^*\}$. 

\section{Constructing the LRT}
The first step in constructing the LRT is to find the likelihood function, which is in this case is simply
\begin{equation}
\mathcal{L}(p)=f(n|N,p)=\frac{N!}{n_1!\ldots n_k!}p_1^{n_1}\ldots p_k^{n_k}.
\end{equation}
The LRT statistic is then 
\begin{equation}
\lambda(n)=\frac{\sup_{p\in\Theta}\mathcal{L}(p)}{\sup_{p\in\Theta_0}\mathcal{L}(p)}=\frac{\mathcal{L}(\hat{p}_{\mathrm{MLE}})}{\mathcal{L}(p^*)}.
\end{equation}
To proceed, we need to find the MLE of $p$.

\subsection{The MLE of $p$}
The log-likelihood is
\begin{equation}
l(p)=\log\mathcal{L}(p)=\log N! - \sum_{i=1}^k\log n_i! +\sum_{i=1}^k n_i\log p_i.
\end{equation}
The MLE $\hat{p}_{\mathrm{MLE}}$ is thus the solution of the following constrained optimization problem:
\begin{equation}
\begin{split}
&\sum_{i=1}^k n_i\log p_i \longrightarrow \max,\\
& \mbox{subject to } \sum_{i=1}^kp_i=1.
\end{split}
\end{equation}
The solution is readily obtained by the method of Lagrange multipliers: 
\begin{equation}\label{eq:mle}
\hat{p}_{\mathrm{MLE}}=\left(\frac{n_1}{N},\ldots,\frac{n_k}{N}\right). 
\end{equation}

\subsection{Computing the $p$-value}
Using (\ref{eq:mle}), we can now compute the LRT statistic:
\begin{equation}
\lambda(n)=\frac{\frac{N!}{n_1!\ldots n_k!}\left(\frac{n_1}{N}\right)^{n_1}\ldots \left(\frac{n_k}{N}\right)^{n_k}}{\frac{N!}{n_1!\ldots n_k!}(p_1^*)^{n_1}\ldots (p_k^*)^{n_k}}=\prod_{i=1}^k\left(\frac{n_i}{Np_i^*}\right)^{n_i}.
\end{equation}
Recall\footnote{See Lecture~\ref{ch:LRT}.} that for large $N$, the null distribution of $2\log\lambda(n)$ is approximately the $\chi^2$-distribution with $q=\dim\Theta-\dim\Theta_0$ degrees of freedom. In our case, $\dim\Theta=3$, $\dim\Theta_0=0$, and, thus $q=3$. 

The $p$-value is then\footnote{See Lecture~\ref{ch:LRT}.}
\begin{equation}
p(n)=\mathbb{P}(Y>2\log\lambda(n))=\mathbb{P}\left(Y>2\sum_{i=1}^kn_i\log\frac{n_i}{Np^*_i}\right), \hspace{1mm}Y\sim\chi^2_3.
\end{equation}
For Mendel's data (\ref{eq:data}), we have:
\begin{equation}
p(n)=\mathbb{P}(Y>0.475)=0.92.
\end{equation}
This is a huge $p$-value\footnote{In fact, this $p$-value is so large that there is some controversy about whether Mendel's results are ``too good'' to be true. See the reference in the next section ``Further Reading.''}. This means that the Mendel data does not provide evidence for rejecting Mendel's theory. As expected.

\section{Further Reading}
\begin{enumerate}
	\item In 1866 Mendel published his seminal paper containing the
	foundations of modern genetics, where he reported the data that we analyzed in this lecture. In 1936 Fisher published a statistical
	analysis of Mendel's data concluding that ``the data of most, if not all, of the experiments have been falsified so as to agree closely with Mendel's expectations.'' A recent paper A.M.~Pires \& J.A.~Branco (2010) ``\href{https://projecteuclid.org/euclid.ss/1300108237}{A statistical model to explain the Mendel-Fisher controversy},'' \textit{Statistical Science}, 25(4): 545-565 provides a brief history of the controversy and offers a possible resolution, which suggests that perhaps Mendel performed several experiments, but reported only the results that best fit his theory.
\end{enumerate}

\section{What is Next?} 
There are applications where we need to to test thousands or even millions of hypotheses. For any one test, the chance of a false rejection\footnote{\ie the probability of the type I error.} may be small $\alpha\ll1$, but the chance of at least one false rejection may still be large. This is called the multiple testing problem. In the next lecture,  we will discuss how to deal with it.

\chapter{Multiple Testing}\label{ch:multipleTesting}

\newthought{There are applications} where one needs to perform multiple testing, that is, to conduct several hypotheses tests simultaneously. The basic paradigm for single-hypothesis testing dictates: fix the maximum acceptable value $\alpha$ for the the type I error probability\footnote{That is, fix the size of the test.}, and then search for the test with the lowest type II error probability\footnote{That is, the most powerful test.}.
When testing multiple hypotheses,  the situation in more subtle since each test has type I and type II errors, and it becomes unclear how to measure the overall error rate and how to control it\footnote{In other words, it is not clear what is the analog of $\alpha$ in multiple testing.}. In this lecture we will introduce two popular measures: the \textit{family-wise error rate} (FWER) and the \textit{false discovery rate} (FDR), and two methods for controlling these measures: the Bonferroni correction and the Benjamini-Hochberg algorithm. 

\section{Multiple Testing Problem}
Let us first fully appreciate the importance of the multiple testing problem. Suppose a pharmaceutical company is testing a new drug for efficacy:
\begin{equation}
H_0:\hspace{1mm} \mbox{no effect}  \hspace{2mm}\mbox{ versus }\hspace{2mm} H_1:\hspace{1mm} \mbox{effect}
\end{equation}
They performed a test with type I error probabiity $\alpha(\ll1)$ on a whole population and the data forced them to accepted the null: the benefit of the drug was not\footnote{Unfortunately for the company.} statistically significant. Regardless of this failure, they could repeat the test for several subpopulation\footnote{For example, males, females, children, students, etc.}. The probability of making \textit{at least one} type I error among the family of hypotheses tests is called the  \textit{family-wise error rate} (FWER). Let us compute it assuming there are $m$ tests, all tests are independent, and have the same type I error probability $\alpha$:
\begin{equation}
\begin{split}
&\mathrm{FWER}=\mathbb{P}(\mbox{at least one type I error})=1-\mathbb{P}(\mbox{no type I errors})\\&=1-\mathbb{P}(\mbox{no type I error in test } 1, \ldots, \mbox{no type I error in test } m)\\
&=1-\prod_{i=1}^m\mathbb{P}(\mbox{no type I error in test } i)\\&=
1-\prod_{i=1}^m(1-\mathbb{P}(\mbox{type I error in test } i))=1-(1-\alpha)^m.
\end{split}
\end{equation}

Thus, even if $\alpha$ is small for each individual test, considering sufficiently large number of tests $m$, it is possible to make FWER very large. For example, if $\alpha=0.05$ and $m=100$, then FWER $\approx0.99$, meaning that almost certainly the company will obtain at least one false rejection\footnote{This effect can be formulated in coin-tossing language: if we toss a coin, no matter how strongly biased against heads, long enough, sooner a later we will observe heads.}. Purely by chance. The corresponding subpopulation may be reported as the one for which the drug produces the desired effect... This situation does not look good.

\section{Bonferroni Correction}

The first idea that comes in mind is that instead of fixing $\alpha$ for each individual test, we need to fix the overall FWER. Since $\alpha\ll1$,
\begin{equation}
\mathrm{FWER}\approx m\alpha.
\end{equation}
Therefore, if we wish to get $\mathrm{FWER}=\alpha$, the new value of $\alpha$ for each individual test must be 
\begin{equation}
\alpha \dashrightarrow \tilde{\alpha}=\frac{\alpha}{m}.
\end{equation}
This method of controlling the overall error rate is called the \textit{Bonferroni correction}: if we run $m$ tests and want the FWER to be $\alpha$, then the type I error for each test should be set to $\frac{\alpha}{m}$. 

The Bonferroni method of controlling the FWER is historically the first attempt to deal with the multiple testing problem. It has two main drawbacks. 
\begin{enumerate}
	\item \textit{Technical:} In practice, it is often too conservative: the corrected sizes $\tilde{\alpha}$ are much smaller than they need to be\footnote{Especially if the number of tests $m$ is large.}. Let us explain why. In practice, tests are rarely independent and 
	\begin{equation}
	\begin{split}
	&\mathbb{P}(\mbox{no type I error in test } 1, \ldots, \mbox{no type I error in test } m)\\
	&\gg\prod_{i=1}^m\mathbb{P}(\mbox{no type I error in test } i).
	\end{split}
	\end{equation}
	This results into
	\begin{equation}
	\mathrm{FWER}\ll m\tilde{\alpha}=\alpha.
	\end{equation}
	Thus, the true FWER will be significantly less then the prescribed value $\alpha$. The tests will  be  unwilling to reject raising the number of false acceptances (type II errors).
	\item \textit{Conceptual:} In many applications, especially in exploratory analysis, one is more interested in finding potentially interesting effects, \ie having mostly true rejections and maybe a few false ones, rather than guarding against one or more false rejections\footnote{For example, DNA microarrays measure the expression levels of thousands of genes simultaneously. An important problem is to identify genes that are differently expressed in different biological conditions (\eg different types of cancer). In this context, failing to identify truely differentially expressed genes is a major concern.}. This led to a new measure, called \textit{false discovery rate} (FDR), which is designed to for this kind of applications and allows to maintain the overall rate of false rejections (type I errors) without inflating the rate of false acceptances (type II errors). 
\end{enumerate}

\section{False Discovery Rate}

Consider the problem of testing simultaneously $m$ null hypotheses: $H_0^{(1)},\ldots,H_0^{(m)}$. Let $p_1,\ldots,p_m$ denote the $p$-values for the corresponding tests. Suppose that we reject $H_0^{(i)}$ if $p_i$ is below some threshold. The question is how to chose the threshold?

Let us introduce some notation:
\begin{itemize}
	\item $m_0$ is the number of true null hypotheses (unknown).
	\item $m_1$ is the number of false null hypotheses\footnote{$m_1=m-m_0$.} (unknown).
	\item $R_f$ is the number of false rejections, \ie the number of type I errors (unobservable random variable).
	\item $R_t$ is the number of true rejections (unobservable random variable).
	\item $R$ is the total number of rejections\footnote{$R=R_f+R_t$.} (observable random variable).
	\item $A_f$ is the number of false acceptances, \ie the number of type II errors (unobservable random variable).
	\item $A_t$ is the number of true acceptances (unobservable random variable).
	\item $A$ is the total number of acceptances\footnote{$A=A_f+A_t$, $A+R=m$.} (observable random variable)
\end{itemize}
The following table summarizes the error outcomes.

\begin{center}
	\begin{tabular}{|c|c|c|c|}
		\hline \rule[-2ex]{0pt}{5.5ex}  & Accepted  & Rejected & Total \\ 
		\hline \rule[-2ex]{0pt}{5.5ex}  True nulls& $A_t$  & $R_f$  & $m_0$  \\ 
		\hline \rule[-2ex]{0pt}{5.5ex}  False nulls &  $A_f$ & $R_t$  & $m_1$ \\ 
		\hline \rule[-2ex]{0pt}{5.5ex}  Total&  $A$& $R$  & $m$ \\ 
		\hline 
	\end{tabular}  
\end{center}

Note that in this notation, the FWER is simply 
\begin{equation}
\mbox{FWER}=\mathbb{P}(R_f\geq1).
\end{equation}
In the language of $p$-values, the Bonferroni method can be formulated as follows: for all $i=1,\ldots,m$,
\begin{equation}
\mbox{Reject } H_0^{(i)} \hspace{2mm}\Leftrightarrow\hspace{2mm} p_i<\frac{\alpha}{m}. 
\end{equation}
It can be shown that this guarantees  $\mbox{FWER}\leq\alpha$\footnote{The value $\frac{\alpha}{m}$ is sometimes called the Bonferroni threshold.}. 

In their seminal paper\footnote{Y. Benjamini \& Y.Hochberg (1995) ``\href{http://www.jstor.org/stable/2346101}{Controlling the false discovery rate: a practical and powerful approach to multiple testing},'' \textit{Journal of the Royal Statistical Society. Series B (Methodological)}, 57(1): 289-300.}, Benjamini and Hochberg defined the false discovery rate as the expected value of the proportion of false rejection among all rejections:
\begin{equation}
\mathrm{FDR}=\mathbb{E}\left[\frac{R_f}{R}\right].
\end{equation} 
This formula assumes $R>0$. If $R=0$, then obviously $\mathrm{FDR}=0$. It should be clear why ``false'' and why ``rate.'' Why ``discovery''? A true rejection of the null hypothesis, which represents a current theory or belief, is considered as a discovery.

To keep the $\mathrm{FDR}$ below a certain acceptable value $\alpha$, the following algorithm can be used. 

\begin{center}
	\textit{The Benjamini-Hochberg Algorithm for controlling FDR}
\end{center}
\begin{enumerate}
	\item Let $p_{(1)}\leq\ldots\leq p_{(m)}$ be the ordered $p$-values, and denote $H_0^{((i))}$ the null hypotheses corresponding to $p_{(i)}$.
	\item Let $i^*$ be the largest $i$ for which $p_{(i)}\leq\frac{\alpha}{m}\frac{i}{\beta_m}$, where $\beta_m=1$ if the $p$-values are independent and $\beta_m=\sum_{i=1}^m\frac{1}{i}$ otherwise.
	\begin{equation}
	i^*=\max\left\{i=1,\ldots,m:\hspace{1mm} p_{(i)}\leq\frac{\alpha}{m}\frac{i}{\beta_m}\right\}.
	\end{equation}
	\item  Reject all $H_0^{((1))},\ldots,H_0^{((i^*))}$. In other words, reject all $H_0^{(i)}$ for which $p_i<p_{(i^*)}$. The $p$-value $p_{(i^*)}$ is called the BH threshold. 
\end{enumerate}
It can be shown\footnote{The proof is out of our scope and can be found in the original paper.} that in this case, 
\begin{equation}
\mathrm{FDR}\leq\alpha.
\end{equation}

\section{Example}
\begin{figure}
%	\vspace{-5mm}
	\centerline{\includegraphics[width=\linewidth]{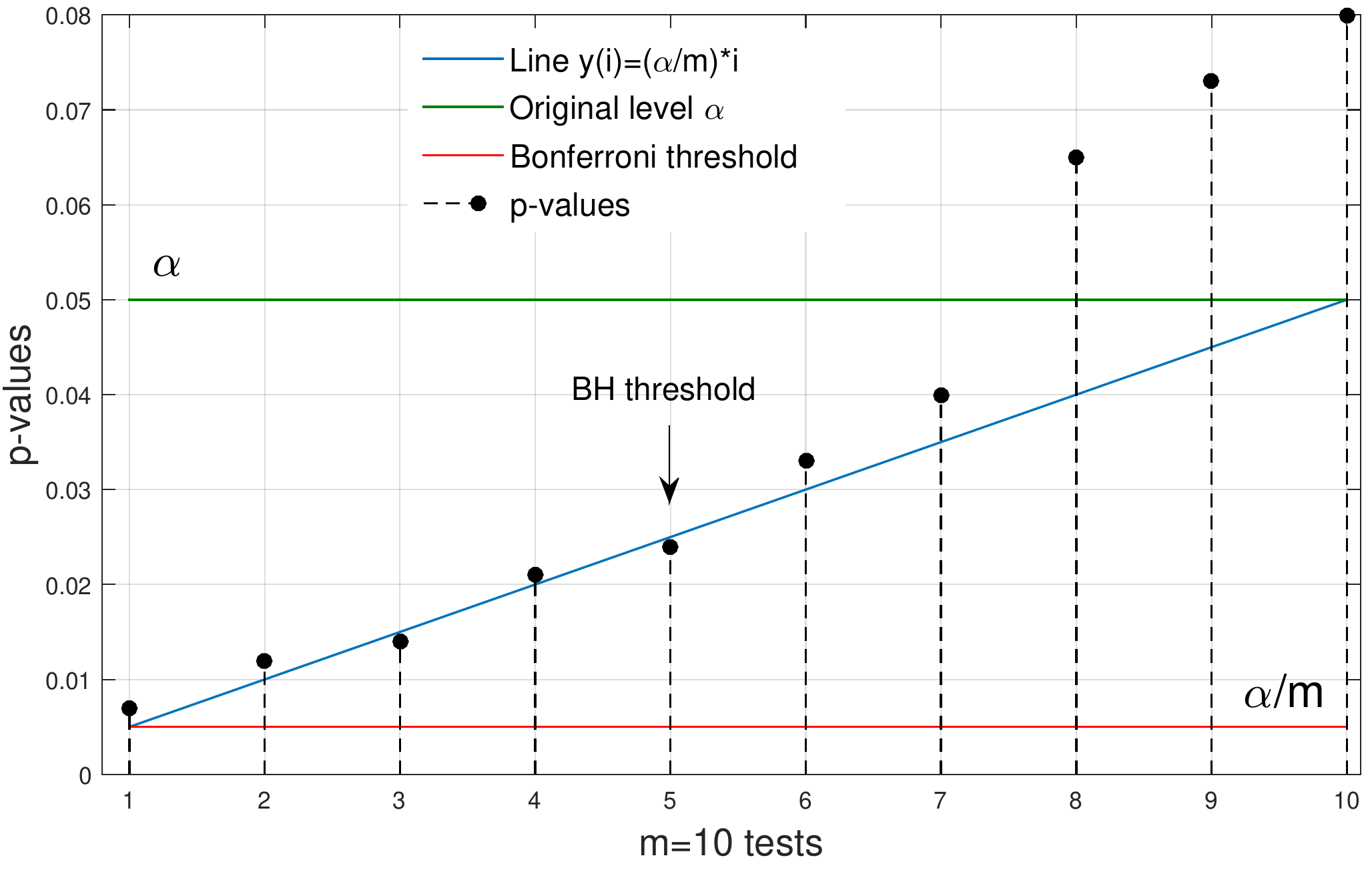}
	}\caption{Uncorrected testing vs. Bonferroni vs Benjamini-Hochberg.}\label{fig:bh}
\end{figure}

Suppose that we performed $m=10$ independent hypothesis tests and obtained the following (ordered) $p$-values:
\begin{equation}
\begin{split}
&p_{(1)}=0.007, p_{(2)}=0.012, p_{(3)}=0.014, p_{(4)}=0.021, p_{(5)}=0.024,\\
&p_{(6)}=0.033, p_{(7)}=0.04, p_{(8)}=0.065, p_{(9)}=0.073, p_{(10)}=0.08.\\
\end{split}
\end{equation}
These $p$-values as well as the Benferroni and Benjamini-Hochberg rejection thresholds  are shown in Fig.~\ref{fig:bh}.\vspace{2mm}
If we tested at level $\alpha$ without doing any corrections for multiple testing, we would reject all tests whose $p$-values are less then $\alpha$. In this example, $\alpha=0.05$, so we would reject seven null hypotheses with the smallest $p$-values. The Bonferroni method rejects all nulls whose $p$-values are less than $\alpha/m$. In this example, $\alpha/m=0.005$ and none hypotheses are rejected. The BH threshold corresponds to the last $p$-value that falls under the line with slope $\alpha/m$. Here, it is $p_{(5)}$. This leads to five hypothesis being rejected. 

\section{Bottom line}
When testing multiple hypothesis, uncorrected testing is simply unacceptable. The Bonferroni correction, which controls the FWER, provides a simple solution, but it may be too conservative for certain applications. The Benjamini-Hochberg algorithm controls the FDR. The main advantage of controlling the FDR instead of FWER is that the former is better  detecting true effects. The FDR control is especially popular in genomics and neuroscience.

\section{Further Reading}
\begin{enumerate}
	\item Y.~Benjamini (2010) ``\href{http://onlinelibrary.wiley.com/doi/10.1111/j.1467-9868.2010.00746.x/abstract}{Discovering the false discovery rate},'' \textit{Journal of the Royal Statistical Society}, 72(4): 405-416 describes the background for the original paper Y. Benjamini \& Y.Hochberg (1995) ``\href{http://www.jstor.org/stable/2346101}{Controlling the false discovery rate: a practical and powerful approach to multiple testing},'' \textit{Journal of the Royal Statistical Society. Series B (Methodological)}, 57(1): 289-300, and  reviews the progress made on the false discovery rate.
\end{enumerate}

\section{What is Next?} 
We'll turn to  \textit{regression}, one of the most popular statistical techniques.

\chapter{Regression Function and General Regression Model} \label{ch:Regression}

	\newthought{Regression} is the study of \textit{dependence}. It is one of the most important and perhaps the most popular statistical technique\footnote{In recent years many other methods have been developed: neural networks, support vector machines, tree-based methods, no name but a few. These methods often outperforms the old good regression. This leads to a natural question: why do we need to study regression? The main reason is that most of these fancy new methods are really just modifications of regression. So, understanding, say SVMs, is impossible without understanding of regression. Going for neural networks without understanding regression is like studying string theory without knowing calculus.}. 
	
	Recall the schematic picture of a certain phenomenon of interest from Lecture~\ref{ch:BigPicture}:
	\vspace{-3mm}
	\begin{figure}
		\centerline{\includegraphics[width=0.5\linewidth]{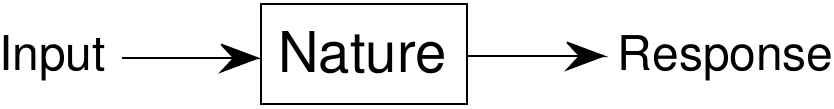}}
	\end{figure}
	
	\vspace{-3mm}\hspace{-5mm}
	So far, we have been ignoring the inputs and discussed the classical methods of statistical inference tailored for analyzing responses. In many applications, however, data comes in the form $(X_1, Y_1),\ldots, (X_n,Y_n)$, where $X_i$ is an input and $Y_i$ is the corresponding response. Moreover, inputs and responses are often depended, and ignoring inputs, when trying to understand the phenomenon, is not wise. 
	
	Regression analysis explores the dependence of responses on inputs with the following two major goals: 
	\begin{enumerate}
		\item \textit{Understanding.} How does Nature associate response $Y$ to input $X$?\footnote{In general, the question ``how $Y$ depends on $X$'' is one of the most fundamental in Science.} 
		\item \textit{Prediction.} Given a future input $X$ what will be the response $Y$?\footnote{Prediction is one of the main goals of Applied Science and Engineering.}
	\end{enumerate} 
	Besides it is direct ``mercantile'' purpose, being able to make predictions also tests our understanding of the phenomena: if we misunderstand, we might still be able to predict, but we are not able to predict, then it is hard to claim that we understand.
	
	Let us see how the attempt to predict the response naturally leads to the regression function, the key element of the regression methodology.  
	
	\section{Regression Function}
	
	Suppose that, given the data $(X_1, Y_1),\ldots, (X_n,Y_n)$, we want to predict the value of the response $Y$ to future input $X$. For the moment, let us again forget about inputs\footnote{Assuming, for example, that don't actually affect responses, or, simply that we don't have access to the input data.}, and focus on the response data $Y_1,\ldots, Y_n$. Let $r$ denote our prediction. What is the optimal value for $r$? The answer, of course, depends on what we mean by ``optimal.'' Suppose that we want to minimize the mean squared error:
	\begin{equation}
	\mathrm{MSE}[r]=\mathbb{E}[(Y-r)^2]\rightarrow\min.
	\end{equation}
	This is a well-defined calculus problem that has an expected solution. Using the bias-variance decomposition for MSE\footnote{See Lecture~\ref{ch:BigPicture}.}, we have:
	\begin{equation}
	\mathrm{MSE}[r]=(\mathbb{E}[Y]-r)^2+\mathbb{V}[Y],
	\end{equation}
	and, therefore, the MSE is minimized when 
	\begin{equation}\label{eq:toy}
	r=\mathbb{E}[Y].
	\end{equation}
	Given the response data, we can estimate $r$ by $\hat{r}=\overline{Y}_n=\frac{1}{n}\sum_{i=1}^nY_i$.
	
	But if we have the input data and we believe that inputs and responses are depended, then it is natural to bring $X_i$s in to the game. Let $r(X)$ denote our prediction of the response $Y$ to input $X$, which now explicitly depends on the input.  What function should we use? As above, let us use the MSE as a measure of goodness:
	\begin{equation}
	\mathrm{MSE}[r]=\mathbb{E}[(Y-r(X))^2]\rightarrow\min.
	\end{equation}
	Using the law of total expectation\footnote{$\mathbb{E}[Y]=\mathbb{E}[\mathbb{E}[Y|X]]$. Here the inner expectation is wrt $Y$ and the outer expectation is wrt $X$.}, we have:
	\begin{equation}
	\begin{split}
	\mathrm{MSE}[r]=\mathbb{E}\left[\mathbb{E}\left[\left.(Y-r(X))^2\right|X\right]\right].
	\end{split}
	\end{equation}
	Since $r(X)$ is a constant when conditioned on $X$, we can work with the inner expectation as about, that is we can use the bias-variance decomposition:
	\begin{equation}
	\begin{split}
	\mathrm{MSE}[r]&=\mathbb{E}\left[\left(\mathbb{E}\left[\left.Y-r(X)\right|X\right]\right)^2+\mathbb{V}[Y|X]\right]\\
	&=\mathbb{E}\left[\left(\mathbb{E}[Y|X]-r(X)\right)^2+\mathbb{V}[Y|X]\right].
	\end{split}
	\end{equation}
	And, thus, the MSE is minimized when 
	\begin{equation}
	r(X)=\mathbb{E}[Y|X].
	\end{equation}
	In other words, if we observe that input $X=x$, then our optimal\footnote{In the MSE sense.} prediction for the response should be 
	\begin{equation} \label{eq:regression_function}
	r(x)=\mathbb{E}[Y|X=x].
	\end{equation}
	Note that if the response does not depend on the input, then $\mathbb{E}[Y|X]=\mathbb{E}[Y]$, and (\ref{eq:regression_function}) reduces to (\ref{eq:toy}). 
	
	The function $r(x)$ in (\ref{eq:regression_function}) is called the \textit{regression function}. This is what we want to know when we want to predict the response. 
	
\section{General Regression Model}

Suppose that the regression function is known. It is important to realize that the true response $Y$ to input $X=x$ typically will not be exactly equal to our prediction $r(x)$. Simply because there are measurement errors and, most importantly, because often $Y$ can take a range of values for given $x$\footnote{For example, for different patients, the improvement in blood cholesterol ($Y$) due to the same dose of drug ($x$) is different.}. In other words, the observed response is a sample from the conditional distribution  $Y|X=x$ and generally does not equal to the expected value $r(x)=\mathbb{E}[Y|X=x]$,
\begin{equation}
Y\neq r(x).
\end{equation} 
But we hope, especially if the variability of $Y$ for a given $X$ is small\footnote{For example, if we hang weight $x$ on a spring, then, according to Hooke's law, the length of the elongated spring is $Y=a+bx$, where $a$ and $b$ are constants that depend on the spring. If, however, we repeat this experiment $n$ times with the same weight $x$, we will get slightly different values $Y_1,\ldots,Y_n$ because of the measurement error.}, that our prediction is not too bad and that approximately
\begin{equation}
Y\approx r(x).
\end{equation}
To account for this discrepancy between the observed data and the expected value, we introduce a quantity called a \textit{statistical error}\sidenote[][5mm]{In engineering fileds, it is often called a \textit{prediction error} or \textit{noise}.} %\footnote{In engineering fileds, it is often called a \textit{prediction error} or \textit{noise}.}
:
\begin{equation}
e=Y-r(x).
\end{equation}
Note that, in general, the distribution of $e$ depends on $X$ (since $Y$ depends on $X$), but the mean is zero:
\begin{equation}\label{eq:0}
\mathbb{E}[e|X=x]=\mathbb{E}[Y|X=x]-r(x)=0.
\end{equation}

The response $Y$ is thus a sum of a deterministic prediction term, which is simply the conditional mean value of $Y$, and a random statistical error\footnote{In some texts, the statistical error is denoted by $\epsilon$, which unconsciously makes us to think about the error as a small quantity, which, although desirable, may not be the case.}:
\begin{equation}\label{eq:GRM}
Y=\underbrace{\mathbb{E}[Y|X=x]}_{r(x)}+e.
\end{equation}
This equation constitutes a \textit{general regression model} of $Y$ on $X$. Note that so far we did not make any assumptions whatsoever and (\ref{eq:GRM}) is \textit{always} true\footnote{We simply say: here is our prediction, if we are off, we call the difference an error.}. Different specific regression models are obtained when we start  making certain assumption about the regression function and statistical error. 
In nonparametric regression, one tries to estimate the regression function directly from the data, without making any specific assumptions.
In classical parametric regression\footnote{The main focus of these notes.}, we assume a particular functional form of the regression function  $r(x)\in\mathcal{F}=\{f(x;\theta), \theta\in\Theta\}$\footnote{For example, $f(x;\theta)=\theta_1+\sin(\theta_2x)+\exp(-\theta_3x^2)$. You can pick your favorite.} and then try to obtain a good estimate for $\theta$. 

\paragraph{Remark:} In regression analysis, the input variable $X$ is often called \textit{predictor}, and the parameters are traditionally denotes by $\beta$s\footnote{Rather than $\theta$s.} 

\section{Further Reading}
\begin{enumerate}
	\item Cosma Shalizi writes \href{http://bactra.org/notebooks/}{notebooks} on various topics. Highly recommended. One of the notebooks is on \href{http://bactra.org/notebooks/regression.html}{regression}, mostly on nonparametric regression though. Highly recommended.
\end{enumerate}

\section{What is Next?}
 How to guess and choose a good functional form for the regression function and a reasonable assumption about the statistical error? A good approach for answering this question is examining the \textit{scatter plot }of the data, which is a starting point of any regression analysis. In the next lecture, we will examine in detail a scatter plot of the data collected by Karl Pearson and see how our observations will naturally lead to the \textit{simple linear regression model}.

\chapter{Scatter Plots and Simple Linear Regression Model} \label{ch:SLR}

\newthought{Recall} that in Lecture~\ref{ch:Summarizing Data}, we discussed how to summarize data $X_1,\ldots,X_n$ using graphical tools such as histograms, boxplots, and Q-Q plots. A \textit{scatter plot}, which is simply a plot of the response $Y$ versus the predictor $X$, is a fundamental graphic tool for looking at the regression data $(X_1, Y_1),\ldots, (X_n,Y_n)$. 

\section{Example: Inheritance of Height}
\begin{figure}
	%\vspace{35mm}
	\centerline{\includegraphics[width=\linewidth]{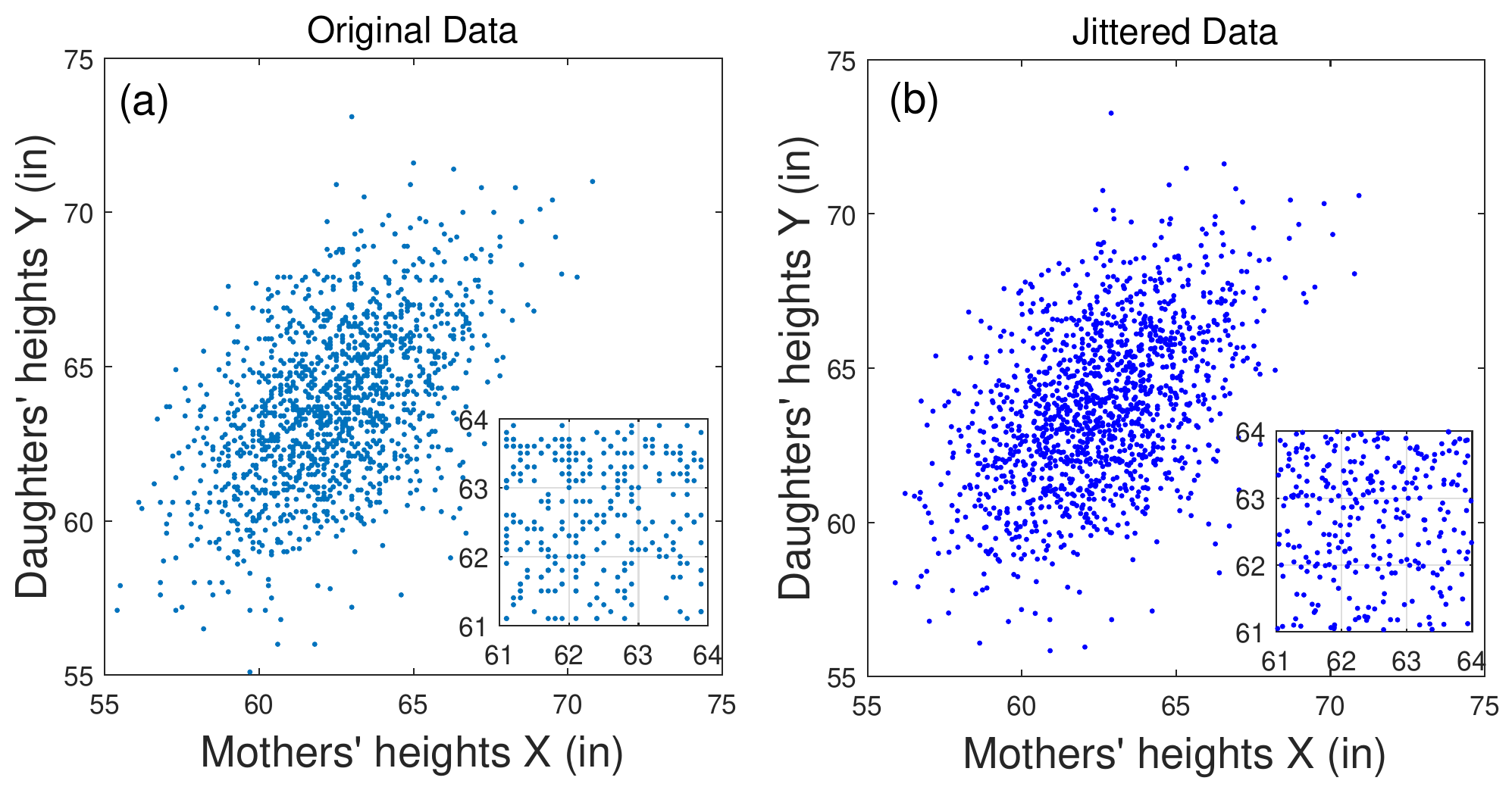}}\caption{Scatterplots of (a) the original Pearson's data  and (b) jittered data with added small random noise.}\label{fig:heights}
\end{figure}

Karl Pearson studied inheritance of different traits from generation to generation. In 1893-1898, he collected $n=1375$ heights of mothers  and their adult daughters in the UK\footnote{K. Pearson \& A. Lee (1903) ``\href{http://www.jstor.org/stable/2331507}{On the laws of inheritance in man},'' \textit{Biometrika}, 2, 357-463, Table 31.}. Figure~\ref{fig:heights}(a) shows the the scatter plot of the original data, where we consider the mother's height $X_i$ as a predictor, and her daughter's height $Y_i$ as the response.
\begin{marginfigure}
	\vspace{-10mm}
	\centerline{\includegraphics[width=\linewidth]{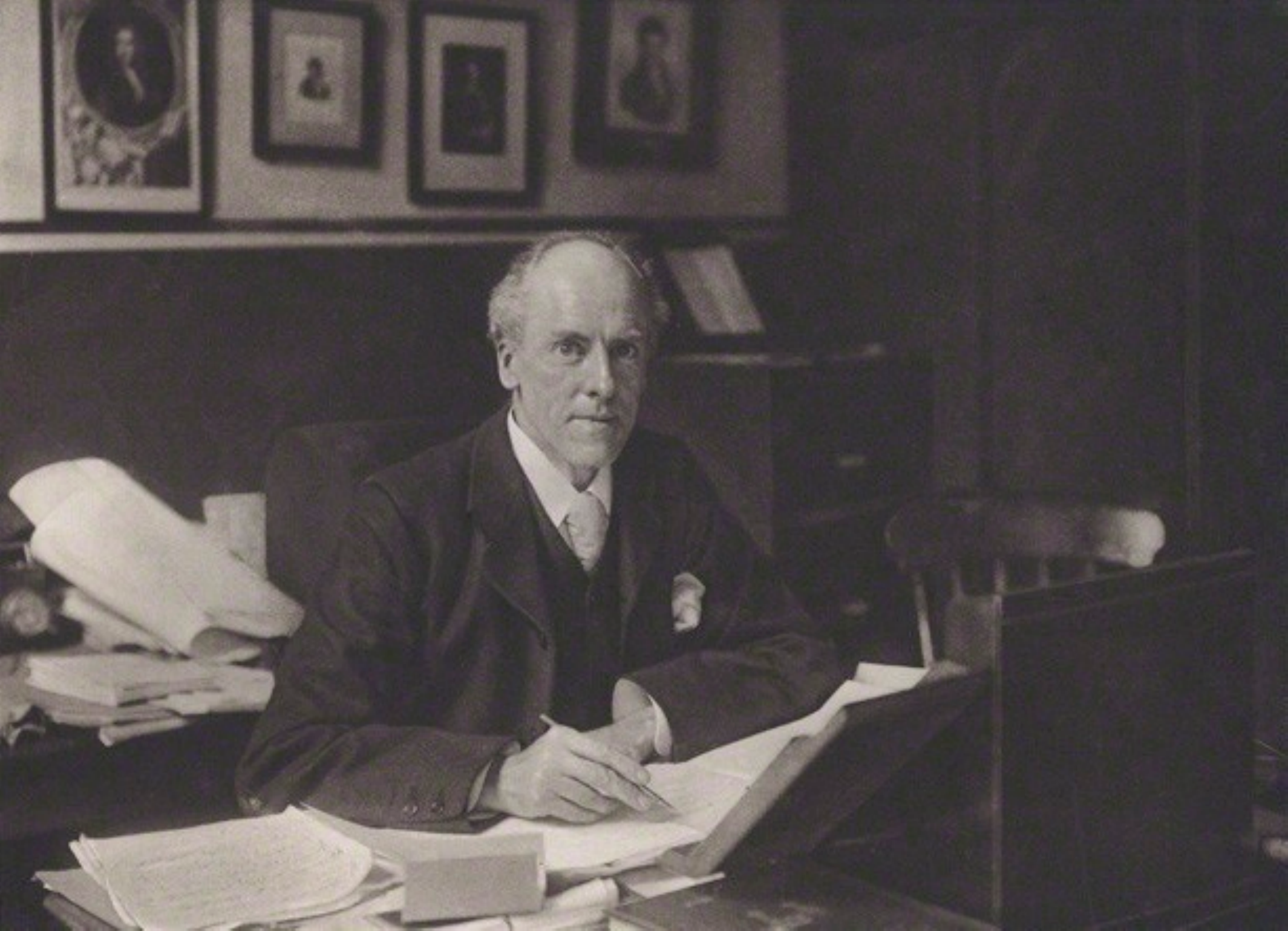}}\caption{Karl Pearson, English mathematician, one of the fathers of mathematical statistics and the father of Egon Pearson. Photo source: \href{https://en.wikipedia.org/wiki/Karl_Pearson}{wikipedia.org}.}\label{fig:Pearson}
\end{marginfigure}

Let us discuss several important features of this scatter plot. 
\begin{enumerate}
	\item \textit{Size \& Scale}. The range of heights appears to be about the same for mother and for daughters: between 55 and 74 inches. That is why we make the lengths and the scale of the $x$- and $y$-axis the same. In general, it is useful to play with the scatter plot by resizing and changing scales, and see how the visual appearance of the data changes.  
	\item \textit{Jittering}. The original Pearson's data were rounded: each height was given to the nearest tenth of an inch. This is very well seen in the zoomed portion of the scatter plot in Fig.~\ref{fig:heights}(a). This may lead to substantial \textit{overplotting}: having many data points $(X_i,Y_i)$ at exactly the same location. This is undesirable since by looking at the scatter plot we will not know if one point represents one case or many cases. This can be very misleading. The easiest solution is \textit{jittering}: add a small uniform random number to each $X_i$ and $Y_i$\footnote{The original data is likely to be noisy anyway!}:
	\begin{equation}
	X_i \rightarrow X_i+u_i, \hspace{3mm} Y_i\rightarrow Y_i + v_i, \hspace{3mm} u_i,v_i\sim U[-\delta,\delta].
	\end{equation}
	In our case, $\delta=0.05$ seems to be a good choice: the jittered values would round to the original numbers. The scatter plot of the jittered data is shown in Fig.~\ref{fig:heights}(b). In what follows, we work with the jittered data\footnote{Available at \href{http://www.its.caltech.edu/~zuev/teaching/2016Winter/heights.xlsx}{heights.xlsx}.}.
	
	\item \textit{Dependence}. One important function of the scatter plot is to decide if we can reasonably assume that the response $Y$ indeed depends on the predictor $X$. This assumption is clearly reasonable for the heights data: when $X$ increases, the scatter of $Y$s shifts upwards\footnote{This is expected of course: higher mothers tend to have higher daughters.}.  This effect is illustrated in Fig.~\ref{fig:slices}(a).
		\begin{figure}
			%\vspace{35mm}
			\centerline{\includegraphics[width=\linewidth]{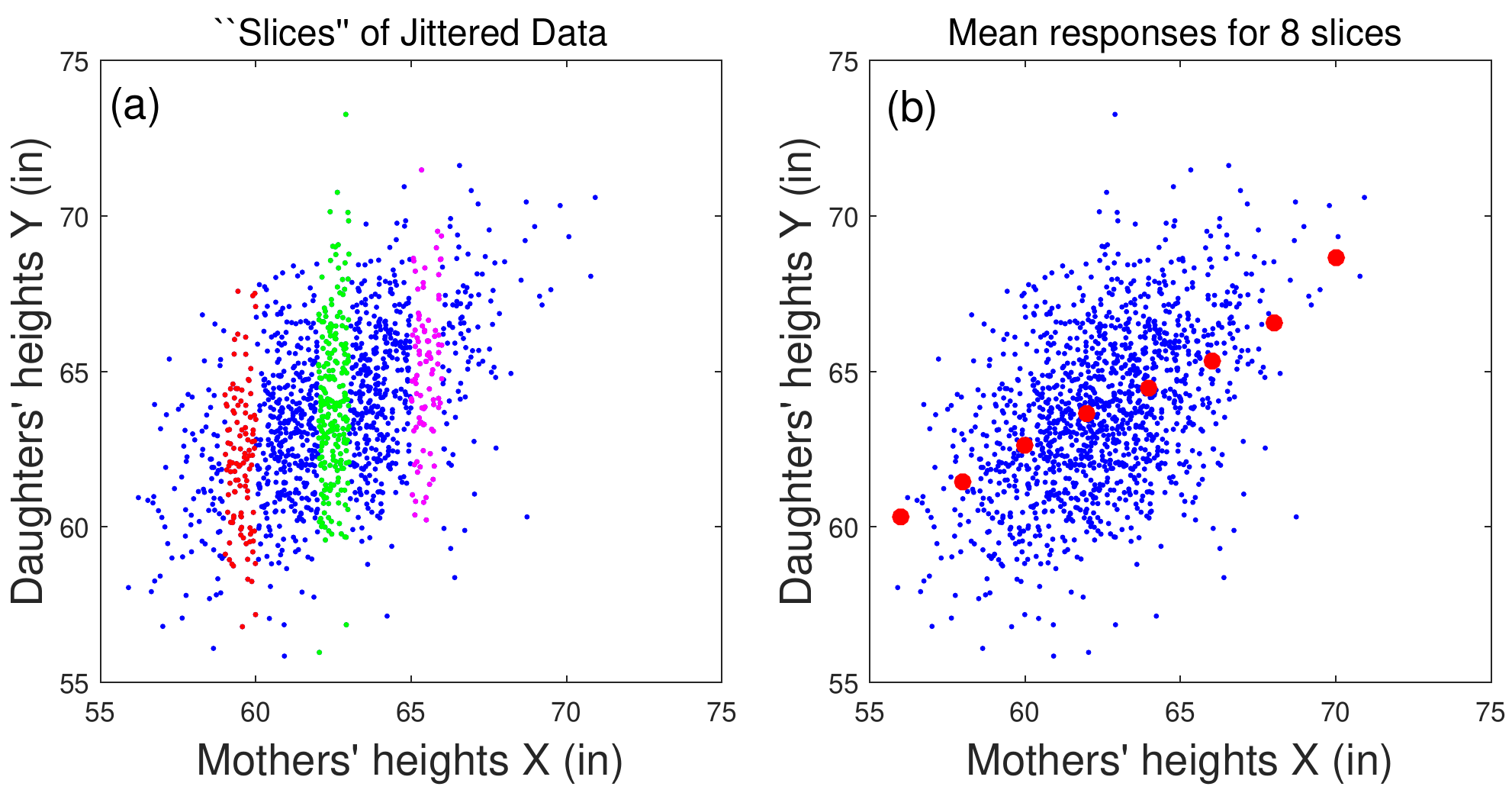}}\caption{Panel (a): Examining three vertical slices of the data suggest that the response $Y$ indeed depends on the predictor $X$. Panel (b): Nonparametric smoother suggests that the regression function can be reasonably well modeled by a liner function.}\label{fig:slices}
		\end{figure}
	\item \textit{Regression Function}. It appears form in Fig.~\ref{fig:slices}(a) that the mean of $Y$ increases when $X$ increases, \ie that the regression function $r(x)$ is an increasing function. Let us look into this in more detail. Let us consider 8 vertical slices of data:
	\begin{equation}
	S_1=\{(X,Y): X\in(55,57)\},  \ldots, S_8=\{(X,Y): X\in(69,71)\}.
	\end{equation}
	For each slice, we compute the mean response 
	\begin{equation}
	\overline{Y}_k=\frac{1}{|S_k|}\sum_{Y_i\in S_k}Y_i, \hspace{5mm} k=1,\ldots,8,
	\end{equation}
	and plot points $(56,\overline{Y}_1), \ldots (70,\overline{Y}_8)$ with big red dots in Fig.~\ref{fig:slices}(b)\footnote{Values $56,\ldots,70$ are simply horizontal ``centers'' of the slices. }. The points almost perfectly lie on a straight line. This\footnote{This method of approximation of the regression function --- averaging the observed responses for all values of $X$ close to $x$ ---  is called \textit{nonparametric smoother}. It is at the core of may nonparametric regression methods.} suggests that a linear function is a very reasonable parametric model for the regression function:
	\begin{equation}\label{21eq:1}
	r(x)=\mathbb{E}[Y|X=x]=\beta_0+\beta_1x.
	\end{equation}
	It has two parameters: an intercept $\beta_0$ and a slope $\beta_1$ that can be estimated from the data\footnote{We will learn how to do this in the next lecture.}.
	
	\item \textit{Statistical Errors}. Let us look again at the slices in Fig.~\ref{fig:slices}(a). While the mean value $\mathbb{E}[Y|X=x]$ increases with $x$, the conditional variance $\mathbb{V}[Y|X=x]$ seems to be constant: the spread of all three slices looks the same\footnote{This appears not to be true for the farmost left and farmost right slices where $X\approx56$ and $X\approx70$. But most likely these regions are simply undersampled: very few very short mothers and very few very tall mothers in the data.}. Thus, it is reasonable to assume that
	\begin{equation}
	\mathbb{V}[Y|X=x]=\sigma^2,
	\end{equation}
	where $\sigma^2$ is some positive constant. In view of (\ref{eq:GRM}), this assumption can be rewritten in terms of the statistical error: 
	\begin{equation}\label{21eq:2}
	\mathbb{V}[e|X=x]=\sigma^2.
	\end{equation}
\end{enumerate}

The general regression equation (\ref{eq:GRM}) together with linear model for the regression function (\ref{21eq:1}) and properties of the statistical error (\ref{eq:0}) and (\ref{21eq:2}) gives the \textit{simple liner regression model}, arguably the most popular and widely used statistical model.

\section{Simple Linear Regression Model}

To sum up, given the data $(X_1, Y_1),\ldots, (X_n,Y_n)$, where $X$ is viewed as the predictor (input) variable that affects the response variable $Y$, the simple linear regression model is 
\begin{equation}\label{eq:SLR}
Y_i=\beta_0+\beta_1X_i+e_i,
\end{equation}
where the errors $e_i$ are \textit{independent}\footnote{We assume that the knowing the error $e_i$ made in case $i$, does not affect the error $e_j$ made in case $j$.} random variables with 
\begin{equation}\label{eq:error}
\mathbb{E}[e_i|X_i]=0, \hspace{3mm}\mbox{ and }\hspace{3mm} \mathbb{V}[e_i|X_i]=\sigma^2.
\end{equation}
The predictor variable can be either fully deterministic if we can control and chose its values $X_1,\ldots,X_n$ as we wish\footnote{For example, $X_i$ can be the $i^{\mathrm{th}}$ time we measure a certain quantity, or $X_i$ can be a weight we attach to a spring in the $i^{\mathrm{th}}$ experiment.}, or $X_1,\ldots,X_n$ can be viewed as a sample from a certain distribution\footnote{For example, a simple random sample from the populations of mothers in UK.}. In either case,  (\ref{eq:SLR}) \& (\ref{eq:error}) tell us that if $X_i$ is known, then $Y_i$ is simply $\beta_0+\beta_1X_i$ plus zero-mean ``noise'' with constant variance\footnote{The property of constant variance is called \textit{ homoscedasticity}.}. The model has three parameters: $\beta_0, \beta_1$, and $\sigma^2$. 

\begin{itemize}
	\item Why ``simple''? The model is called simple because the predictor is one-dimensional. In general, $X_i$ can be a vector, $X_i=(X_{i,1},\ldots,X_{i,p})$. In this case, the model is called  \textit{multiple} linear regression\footnote{This should not be confused with the \textit{multivariate} linear regression, where there are several response variables.}. 
	\item Why ``linear''? Because the regression function is assumed to be linear in \textit{parameters}:
	\begin{equation}
	r=\beta_0+\beta_1\wasylozenge+\beta_2\kreuz+\ldots+\beta_p\sun.
	\end{equation} 
	Whatever $\wasylozenge, \kreuz,\ldots, \sun$ are, we can call them predictors. For example, $Y_i=\beta_0+\beta_1\exp(X_i)+e_i$ is also a simple linear regression model: in this case we could just redefine $\widetilde{X}_i=\exp(X_i)$ to obtain a more familiar form (\ref{eq:SLR}); $Y_i=\beta_0+\beta_1\exp(X_{i,1})+\beta_2\sin(X_{i,2})+e_i$ is multiple linear regression; but $Y_i=\beta_0+\exp(\beta_1X_i)+e_i$ is simple \textit{non-linear} regression.
	\item Why ``regression''\footnote{The term was coined by Sir Francis Galton.}? Let us look at Fig.~\ref{fig:regression}, where the estimated values of the conditional expectation function $r(x)=\mathbb{E}[Y|X=x]$ are plotted together with the the line $y=x$, where all data points would like if all daughters would have exactly the same hight as their mothers. Notice that the slope of $r(x)$ is less than one. This means that tall mothers tend to have tall daughters (the slope of $r(x)$ is positive), but not as tall as themselves (the slope is less than one).  Likewise, the short mothers tend to have short daughters, but not as  as short as themselves. This effect is observed in relationships between many attributes of parents and children and was called ``regression towards the mean.'' The line relating the mean attribute of children to that of their parents was called ``'the regression line.''
	\begin{marginfigure}
		\vspace{-45mm}
		\centerline{\includegraphics[width=\linewidth]{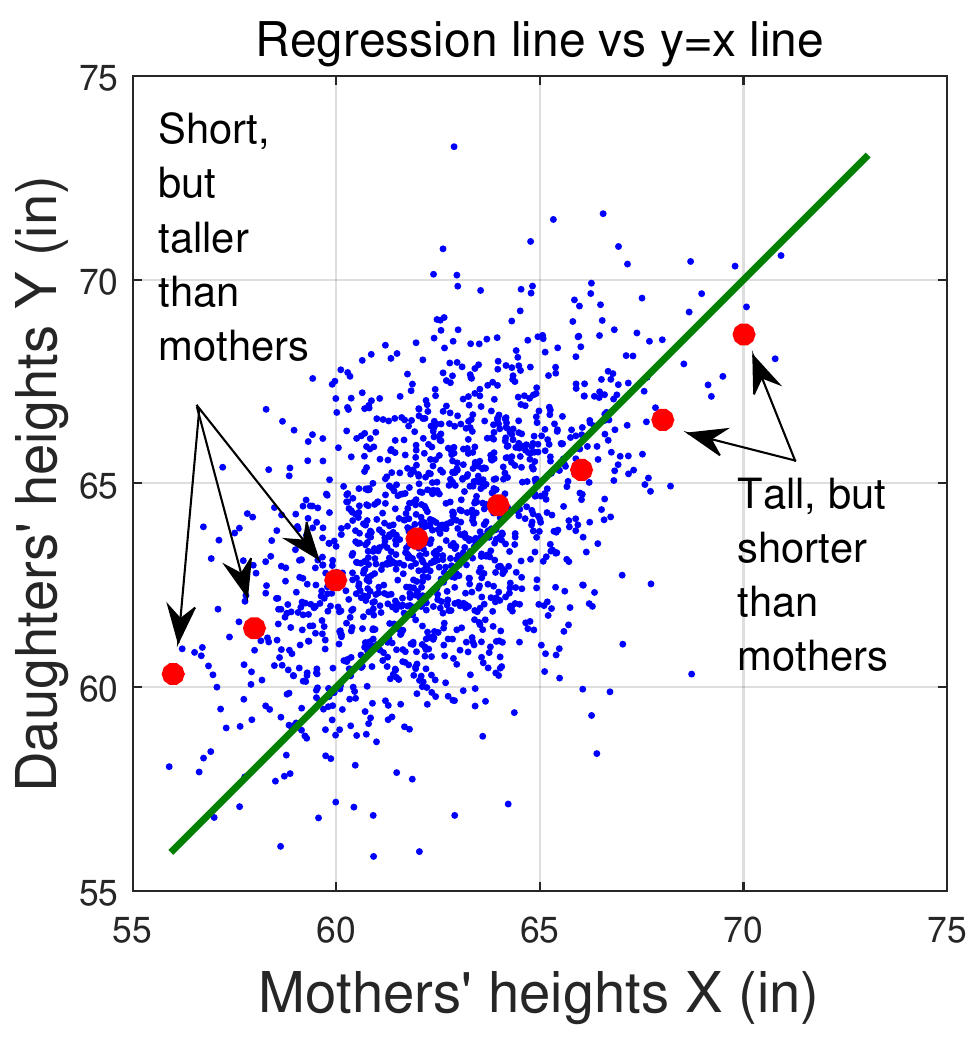}}\caption{``Regression towards the mean.''}\label{fig:regression}
	\end{marginfigure}
	\item 
	Why ``model''?... I leave this to you as a question for reflection :o)  
\end{itemize}

\section{Further Reading}
\begin{enumerate}
	\item Simple linear regression is just the first little (yet very important) step towards a major area of statistical inference, called regression analysis. S. Weisberg (2014) \textit{\href{https://books.google.com/books/about/Applied_Linear_Regression.html?id=FHt-AwAAQBAJ}{Applied Linear Regression}} is recommended as a gentle introduction.
\end{enumerate}

\section{What is Next?} 
We will  discuss how to estimate the parameters of the simple linear regression model using the method of ordinary least squares.

\chapter{Ordinary Least Squares} \label{ch:OLS}

	\newthought{The simple linear regression} (SLR) model is
	\begin{equation}\label{eq:22SLR}
	Y_i=\beta_0+\beta_1X_i+e_i,  \hspace{5mm} i=1,\ldots,n,
	\end{equation}
	where $X_i$ is the predictor variable, $Y_i$ is the corresponding response, and $e_i$ is the random statistical error\footnote{Note that (\ref{eq:22SLR}) is always true as long as we do not make any assumptions of the errors.}.  The model assumptions are:
	\begin{equation}\label{eq:SLR_Assumptions}
	\begin{split}
	1.\hspace{1mm}& e_i \mbox{ are independent,}\\
	2.\hspace{1mm} & \mathbb{E}[e_i|X_i]=0,\\
	3.\hspace{1mm} & \mathbb{V}[e_i|X_i]=\sigma^2.
	\end{split}
	\end{equation}
	The parameters $\beta_0$ and $\beta_1$ are called the \textit{regression coefficients}. There are  many methods  for estimating the regression coefficients, since there are many reasonable ways to fit a line to a cloud of points. In this lecture, we will discus the most common method: ordinary least squares (OLS). 

\section{Ordinary Least Squares}
\begin{figure}
	%	\vspace{0mm}
	\centerline{\includegraphics[width=.9\linewidth]{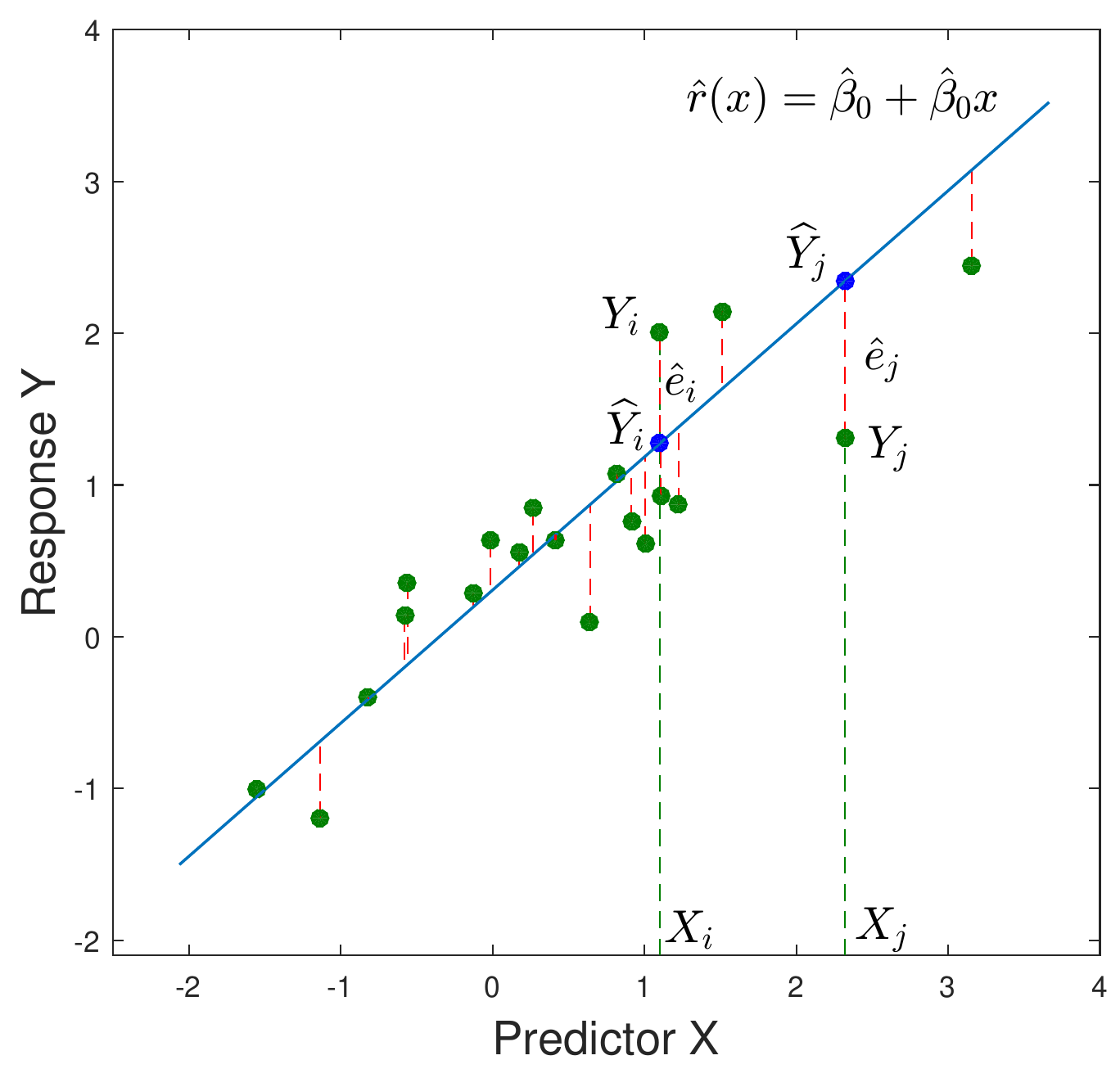}}\caption{A cloud of green points (data) and a fitted blue line $\hat{r}(x)=\hat{\beta}_0+\hat{\beta}_1x$. A couple of predicted values ($\widehat{Y}_i$ and $\widehat{Y}_j$) are shown in blue. The residuals are depicted by red dashed lines. }\label{fig:OLS}
\end{figure}

Let $\hat{\beta}_0$ and $\hat{\beta}_1$ denote estimates of $\beta_0$ and $\beta_1$. The line 
\begin{equation}
\hat{r}(x)=\hat{\beta}_0+\hat{\beta}_1x
\end{equation}
is then called the \textit{fitted line}, and 
\begin{equation}
\widehat{Y}_i=\hat{r}(X_i)=\hat{\beta}_0+\hat{\beta}_1X_i
\end{equation}
are called the \textit{fitted} or \textit{predicted} values. The difference between the actually observed data point $Y_i$ and the predicted value $\widehat{Y}_i$ is called the \textit{residual}:
\begin{equation}
\hat{e}_i=Y_i-\widehat{Y}_i=Y_i-\hat{\beta}_0-\hat{\beta}_1X_i.
\end{equation} 
Residuals $\hat{e}_i$ can be viewed as realizations of random errors $e_i$ and they play an important role in checking the model assumptions (\ref{eq:SLR_Assumptions}). Geometrically, residuals are simply the (signed) vertical distances between the fitted line and the actual $Y$-values. See Fig.~\ref{fig:OLS}.

The OLS method choses the estimates  $\hat{\beta}_0$ and $\hat{\beta}_1$
to minimize the quantity called \textit{residual sum of squares}:
\begin{equation}
\mathrm{RSS}=\sum_{i=1}^n\hat{e}_i^2=\sum_{i=1}^n(Y_i-\hat{\beta}_0-\hat{\beta}_1X_i)^2 \longrightarrow\min.
\end{equation}
This minimization criterion is very natural: RSS is a measure of the overall prediction error, and we want to minimize it by choosing $\hat{\beta}_0$ and $\hat{\beta}_1$ appropriately. In the Appendex, we show that the solution, \ie the OLS estimates of $\beta_0$ and $\beta_1$ are
\begin{equation}\label{eq:OLS}
\hat{\beta}_0=\overline{Y}-\hat{\beta}_1\overline{X} \hspace{3mm}\mbox{and}\hspace{3mm}\hat{\beta}_1=\frac{S_{XY}}{S_{XX}},
\end{equation} 
where $\overline{X}$ and $\overline{Y}$ are the sample means\footnote{To make the notation simpler, we drop the usual subscript $n$, $\overline{X}\equiv\overline{X}_n$,  $\overline{Y}\equiv\overline{Y}_n$.}, $S_{XX}$ is the \textit{sum of squares}, and $S_{XY}$ is the \textit{sum of cross-products}:
\begin{equation}
S_{XX}=\sum_{i=1}^n(X_i-\overline{X})^2\hspace{3mm}\mbox{and}\hspace{3mm} S_{XY}=\sum_{i=1}^n(X_i-\overline{X})(Y_i-\overline{Y}).
\end{equation}
\begin{marginfigure}
	\vspace{-7mm}
	\centerline{\includegraphics[width=\linewidth]{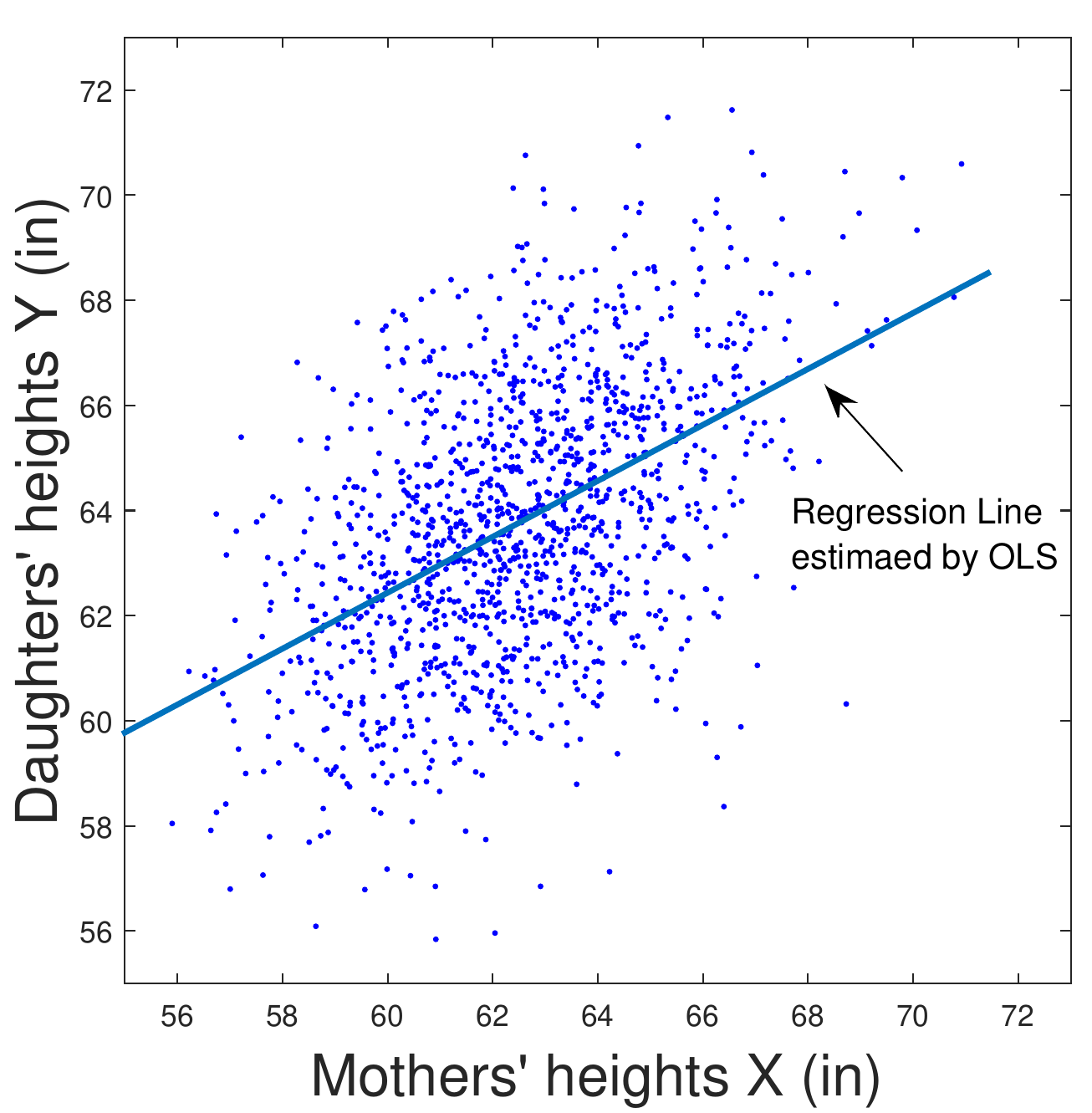}}\caption{OLS in action: the regression line for the heights data (Lecture~\ref{ch:SLR}), \href{http://www.its.caltech.edu/~zuev/teaching/2016Winter/heights.xlsx}{heights.xlsx}.}\label{fig:OLS_heights}
\end{marginfigure}

As an example, Fig.~\ref{fig:OLS_heights} shows the regression line estimated by the OLS for the Pearson's heights data. The estimated values are $\hat{\beta}_0=30.5$ and $\hat{\beta}_1=0.53$.

\section{Anscombe's Quartet}
Notice that the OLS estimates (\ref{eq:OLS}) depend on data only through the statistics $\overline{X}$, $\overline{Y}$, $S_{XX}$, and $S_{XY}$. This means that any two data sets for which these statistics are the same will have identical fitted regression lines,
\begin{equation}
\hat{r}(x)=\overline{Y}+(x-\overline{X})\frac{S_{XY}}{S_{XX}},
\end{equation}
even if a straight-line model is appropriate for one but not the other.  This effect was beautifully demonstrated by Frank Anscombe\sidenote[][-4mm]{F.J.~Anscombe (1973) ``\href{http://www.jstor.org/stable/2682899}{Graphs in statistical analysis},'' \textit{The American Statistician}, 27(1): 17-21.}.

\begin{figure}
	%	\vspace{0mm}
	\centerline{\includegraphics[width=.65\linewidth]{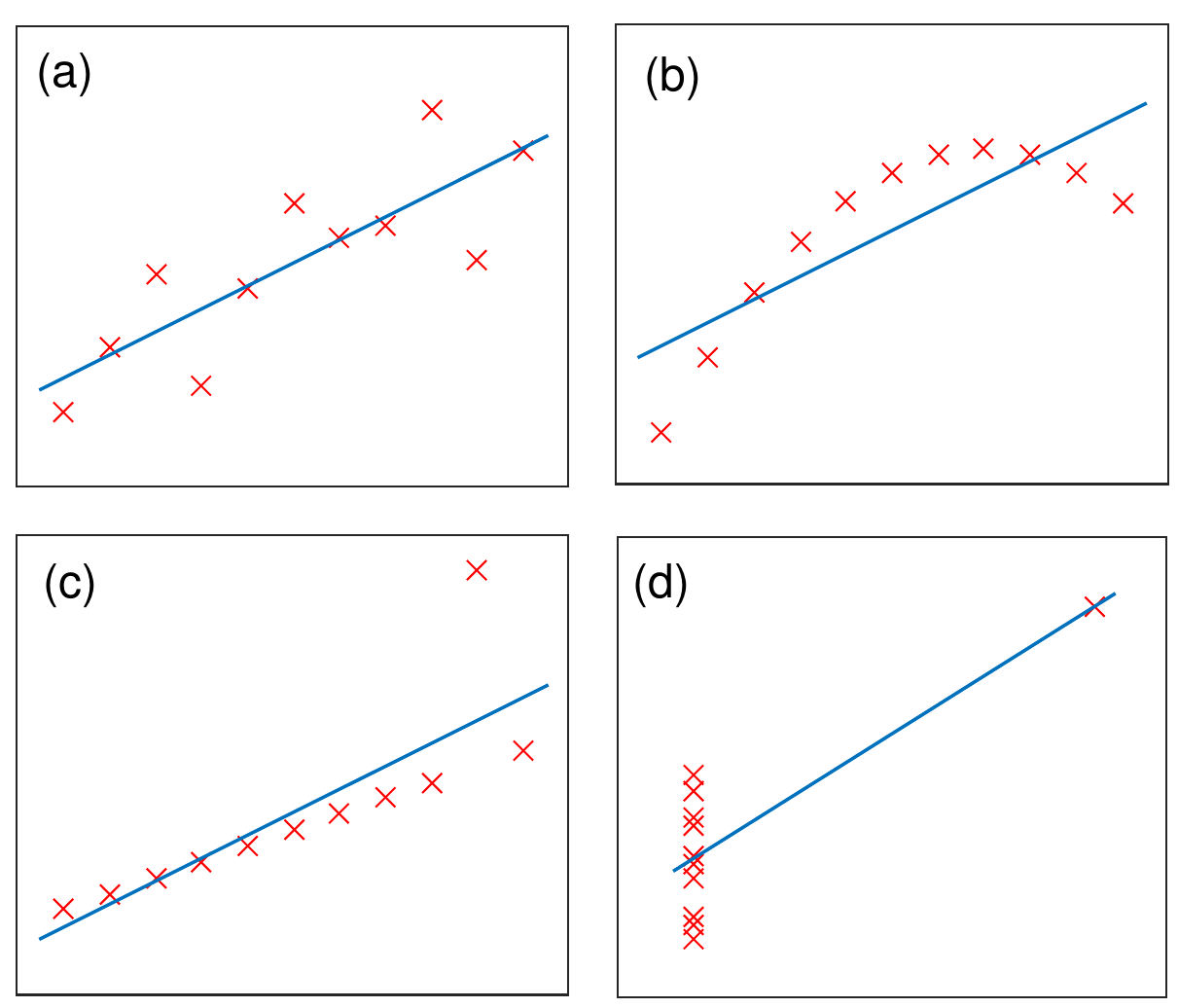}}\caption{Anscombe's quartet.}\label{fig:anscombe}
\end{figure}

Anscombe came up with four artificial data sets for which the OLS method fits the same regression line, but the visual impression of the scatter plots is very different. Anscombe's ``quartet'' is shown in Fig.~\ref{fig:anscombe}. For the first data set in Fig.~\ref{fig:anscombe}(a) the SLR model is appropriate. The scatter plot of the second data set in Fig.~\ref{fig:anscombe}(b) suggests that fitting a line is incorrect, fitting a quadratic polynomial would be more natural. In Fig.~\ref{fig:anscombe}(c), we see that the SLR model maybe correct for most of the data, but one of the cases is too far away from the fitted regression line. This is called the \textit{outlier problem}\footnote{In this case, the outlier is often removed from the data and the regression line is refit from the remaining data. This of course assumes that the outlier is not a true response to the corresponding input: it occurred because of the noise in the measurement or some other error. }. Finally, the scatter plot in Fig.~\ref{fig:anscombe}(d) is different from the previous three in that there is not enough data to make a judgment regarding the regression function $r(x)=\mathbb{E}[Y|X=x]$. Essentially, we have information about $r(x)$ only at two points. Moreover, there is only one response value for the larger input. Without that point, we would not be able even estimate the slope. Even if the SRL model is correct here, we can't trust the regression line whose slope is so heavily dependent on a single case. More data is needed. 

\paragraph{The moral is clear:} don't fit the regression line blindly, check the scatter plot first.  

\section{OLS and MLE}
The OLS method is very intuitive. It turns out that the OLS estimates can be (at first glance unexpectedly) justified purely statistically if we assume that the errors $e_i$ are normally distributed\footnote{The model (\ref{eq:22SLR}) \& (\ref{eq:22normal}) is called the \textit{conditional normal model}. It is the most common SLR model and the most straightforward to analyze.}: 
\begin{equation}\label{eq:22normal}
e_i|X_i\hspace{1mm}\sim\hspace{1mm}\mathcal{N}(0,\sigma^2).
\end{equation}
Note that this assumption is much stronger than (\ref{eq:SLR_Assumptions}): whereas (\ref{eq:SLR_Assumptions}) specifies only the first two moments, $\mathbb{E}[Y_i|X_i]=\beta_0+\beta_1X_i$ and $\mathbb{V}[Y_i|X_i]=\sigma^2$, the normality assumption (\ref{eq:22normal}) specifies the exact form of the distribution of the response variable:
\begin{equation}
Y_i|X_i\hspace{1mm} \sim \hspace{1mm}\mathcal{N}(\beta_0+\beta_1X_i,\sigma^2).
\end{equation}

Under the conditional normal SLR model, the likelihood function of model parameters is the joint density of the data: 
\begin{equation}\label{eq:likelihood}
\begin{split}
\mathcal{L}&(\beta_0,\beta_1,\sigma^2|\{(X_i,Y_i)\})=\prod_{i=1}^n\frac{1}{\sqrt{2\pi}\sigma}\exp\left(-\frac{(Y_i-\beta_0-\beta_1X_i)^2}{2\sigma^2}\right)\\
&=\left(2\pi\sigma^2\right)^{-\frac{n}{2}}\exp\left(-\frac{\sum_{i=1}^n(Y_i-\beta_0-\beta_1X_i)^2}{2\sigma^2}\right)\\
&=\left(2\pi\sigma^2\right)^{-\frac{n}{2}}\exp\left(-\frac{\mathrm{RSS}(\beta_0,\beta_1)}{2\sigma^2}\right).
\end{split}
\end{equation}
This means that the MLEs of $\beta_0$ and $\beta_1$ are exactly those values that minimize $\mathrm{RSS}(\beta_0,\beta_1)$. Thus, under the assumption of normality (\ref{eq:22normal}), the OLS are also the MLEs:
\begin{equation}
\hat{\beta}_0=\hat{\beta}_{0,\mathrm{MLE}} \hspace{3mm}\mbox{and}\hspace{3mm}\hat{\beta}_1=\hat{\beta}_{1,\mathrm{MLE}}
\end{equation}

To find the MLE of $\sigma^2$, we need to substitute $\hat{\beta}_0$ and $\hat{\beta}_1$ in (\ref{eq:likelihood}) and maximize the likelihood over $\sigma^2$. Maximizing the log-likelihood is more convenient. Dropping non relevant terms, 
\begin{equation}
l(\hat{\beta}_0,\hat{\beta}_1,\sigma^2)=-n\log\sigma-\frac{1}{2\sigma^2}\mathrm{RSS}(\hat{\beta}_0,\hat{\beta}_1).
\end{equation}
Differentiating with respect to $\sigma$, setting the derivative to zero, and solving the corresponding equations gives:
\begin{equation}
\hat{\sigma}^2_{\mathrm{MLE}}=\frac{1}{n}\mathrm{RSS}(\hat{\beta}_0,\hat{\beta}_1)=\frac{1}{n}\sum_{i=1}^n \hat{e}_i^2,
\end{equation}
which is a natural estimate if you think about it. It is natural, but biased. An unbiased\footnote{Even under the original weaker assumptions (\ref{eq:SLR_Assumptions}).} estimate of $\sigma^2$ is\footnote{The proof is a long calculation. See, for example, Appendix C.3 in D.C.~Montgomery et al (2006) \textit{\href{https://books.google.com/books?id=0yR4KUL4VDkC}{Introduction to Linear Regression Analysis}}.}
\begin{equation}
\hat{\sigma}^2=\frac{1}{n-2}\sum_{i=1}^n \hat{e}_i^2,
\end{equation}
You might expect to see $\frac{1}{n-1}$ as in the case of the sample variance, but the residuals are not independent\footnote{In the next lecture, we will show (you are welcome to check this now) that $\sum_{i=1}^n\hat{e}_i$ is zero, hence the dependence.}. The dependence is low though, and the factor $\frac{1}{n-2}$ takes care of it. 

\section{Appendix: Proof of (\ref{eq:OLS})}
The least squares estimates $\hat{\beta}_0$ and $\hat{\beta}_1$  are defined as those values that minimize the RRS
\begin{equation}
\mathrm{RSS}(\hat{\beta}_0,\hat{\beta}_1)=\sum_{i=1}^n(Y_i-\hat{\beta}_0-\hat{\beta}_1X_i)^2.
\end{equation}
This function of two variables can be minimized in the following way. For any fixed $\hat{\beta}_1$, the value of $\hat{\beta}_0$ that minimizes 
\begin{equation}
\mathrm{RSS}(\hat{\beta}_0,\hat{\beta}_1)=\sum_{i=1}^n((Y_i-\hat{\beta}_1X_i)-\hat{\beta}_0)^2
\end{equation}
is given by\footnote{We use elementary fact that $$\arg\min_a\sum_{i=1}^n(x_i-a)^2=\bar{x}.$$ To show this, simply add and subtract $\bar{x}$ inside the brackets, and then expand.}
\begin{equation}
\hat{\beta}_0=\overline{Y-\hat{\beta}_1X}=\overline{Y}-\hat{\beta}_1\overline{X}.
\end{equation}
Thus, for a given value of $\hat{\beta}_1$, the minimum value of RSS is 
\begin{equation}
\begin{split}
&\mathrm{RSS}(\overline{Y}-\hat{\beta}_1\overline{X},\hat{\beta}_1)=\sum_{i=1}^n(Y_i-\overline{Y}+\hat{\beta}_1\overline{X}-\hat{\beta}_1X_i)^2\\
&=\sum_{i=1}^n((Y_i-\overline{Y})-\hat{\beta}_1(X_i-\overline{X}))^2=S_{YY}-2\hat{\beta}_1S_{XY}+\hat{\beta}_1^2S_{XX}.
\end{split}
\end{equation}
Since $S_{XX}>0$, the value of $\hat{\beta}_1$ that gives the overall minimum of RSS is 
\begin{equation}
\hat{\beta}_1=\frac{S_{XY}}{S_{XX}}.
\end{equation}

Note that the OLS method, strictly speaking, is not a method of statistical inference. It does not use any model assumptions (\ref{eq:SLR_Assumptions}). It simply fits the line to the data using  using the $\mathrm{RSS}\rightarrow\min$ criterion, and RSS is one of many reasonable ways of measuring the distance from the line $\hat{r}(x)=\hat{\beta}_0+\hat{\beta}_1x$ to the data points. But we will see that under (\ref{eq:SLR_Assumptions}), the OLS estimates have nice optimality properties. 

\section{Further Reading}
\begin{enumerate}
	\item The original F.J.~Anscombe (1973) ``\href{http://www.jstor.org/stable/2682899}{Graphs in statistical analysis},'' \textit{The American Statistician}, 27(1): 17-21 is worth reading. 
\end{enumerate}

\section{What is Next?} 
We will discuss several important properties of the OLS estimates, in particular, their color :)

\chapter{Properties of the OLS Estimates}\label{ch:OLSproperties}

	\newthought{Recall} that the simple linear regression (SLR) model is:
	\begin{equation}\label{eq:23SLR}
	Y_i=\beta_0+\beta_1X_i+e_i,  \hspace{5mm} i=1,\ldots,n,
	\end{equation}
	where $X_i$ is the predictor variable, $Y_i$ is the corresponding response, and $e_i$ is the random statistical error.  The model assumptions are:
	\begin{equation}\label{eq:23SLR_Assumptions}
	\begin{split}
	1.\hspace{1mm}& e_i \mbox{ are independent,}\\
	2.\hspace{1mm} & \mathbb{E}[e_i|X_i]=0,\\
	3.\hspace{1mm} & \mathbb{V}[e_i|X_i]=\sigma^2
	\end{split}
	\end{equation}
	Last time we discussed the ordinary least squares (OLS) estimates of the regression coefficients $\beta_0$ and $\beta_1$:
	\begin{equation}\label{eq:23OLS}
	\hat{\beta}_0=\overline{Y}-\hat{\beta}_1\overline{X} \hspace{3mm}\mbox{and}\hspace{3mm}\hat{\beta}_1=\frac{S_{XY}}{S_{XX}}.
	\end{equation} 
	The OLS estimates have several important properties, some of which we will derive in this lecture.
	
\section{OLS and Data Centroid}
The point $(\overline{X},\overline{Y})$ is called the \textit{centroid} of the data. It is straightforward to check that the least-squares regression line always passes through the centroid. Indeed:
\begin{equation}
\hat{r}(\overline{X})=\hat{\beta}_0+\hat{\beta}_1\overline{X}=
\overline{Y}-\frac{S_{XY}}{S_{XX}}\overline{X}+\frac{S_{XY}}{S_{XX}}\overline{X}=\overline{Y}.
\end{equation} 
This is somewhat expected: our prediction for the average input is the average response. 

\section{OLS and the Sum of the Residuals}
The sum of the residuals is always zero:
\begin{equation}
\begin{split}
\sum_{i=1}^n\hat{e}_i&=\sum_{i=1}^n(Y_i-\widehat{Y}_i)=\sum_{i=1}^n(Y_i-\hat{\beta}_0-\hat{\beta}_1X_i)\\
&=\sum_{i=1}^n\left(Y_i-\overline{Y}+\frac{S_{XY}}{S_{XX}}\overline{X}-\frac{S_{XY}}{S_{XX}}X_i\right)\\&=\sum_{i=1}^n(Y_i-\overline{Y})+\frac{S_{XY}}{S_{XX}}\sum_{i=1}^n(\overline{X}-X_i)=0.
\end{split}
\end{equation}
This property is also natural: on average the fitted value $\widehat{Y}_i$ neither overestimates nor underestimates the true response $Y_i$.

\section{OLS is Linear} 
There many possible estimates of the regression coefficients\footnote{For example, one may chose to minimize (instead of RSS) the sum of squared Euclidean distances from data points to the fitted line, or some other measure of the overall fit.}. Let us restrict our attention to the class of liner estimates. An estimate $\hat{\hat{\beta}}$  of a regression coefficient $\beta$ is called \textit{linear} if it as a linear combination of the responses: 
\begin{equation}
\hat{\hat{\beta}}=\sum_{i=1}^n \alpha_iY_i, \hspace{5mm} \alpha_i\in\mathbb{R}.
\end{equation}
The OLS estimates are linear: 
\begin{equation}\label{eq:linear}
\begin{split}
\hat{\beta}_1&=\frac{S_{XY}}{S_{XX}}=\frac{\sum_{i=1}^n(X_i-\overline{X})(Y_i-\overline{Y})}{S_{XX}}\\
&=\sum_{i=1}^n\frac{X_i-\overline{X}}{S_{XX}}Y_i-\frac{\overline{Y}}{S_{XX}}\sum_{i=1}^n(X_i-\overline{X})=\sum_{i=1}^n\underbrace{\frac{X_i-\overline{X}}{S_{XX}}}_{\alpha_i}Y_i.
\end{split}
\end{equation}
The estimate $\hat{\beta}_0$ is also linear since both terms $\overline{Y}$ and $\hat{\beta}_1$ are linear. 

\section{OLS is Unbiased}
In regression problems, we always focus on properties conditional on the values $\{X_i\}$ of predictor variable\footnote{Recall that we think of $\{X_i\}$ as either being fully deterministic or being an observed sample from a certain distribution. The context is: observing $X$ we want to predict $Y$.}. Assuming the SLR model is correct and using representation (\ref{eq:linear}) and model assumption 2 in (\ref{eq:23SLR_Assumptions}), we have:
\begin{equation}
%\hspace{-10mm}
\begin{split}
&\mathbb{E}[\hat{\beta}_1|\{X_i\}]=\mathbb{E}\left[\left.\sum_{i=1}^n\frac{X_i-\overline{X}}{S_{XX}}Y_i\hspace{1mm}\right|\{X_i\}\right]=\sum_{i=1}^n\frac{X_i-\overline{X}}{S_{XX}}\mathbb{E}[Y_i|X_i]\\&=
\sum_{i=1}^n\frac{X_i-\overline{X}}{S_{XX}}(\beta_0+\beta_1X_i)=\frac{\beta_0}{S_{XX}}\sum_{i=1}^n(X_i-\overline{X})+\frac{\beta_1}{S_{XX}}\sum_{i=1}^nX_i(X_i-\overline{X})\\
&=\frac{\beta_1}{S_{XX}}\sum_{i=1}^n(X_i-\overline{X})(X_i-\overline{X})+\frac{\beta_1\overline{X}}{S_{XX}}\sum_{i=1}^n(X_i-\overline{X})=\beta_1.
\end{split}
\end{equation}

Using the unbiasedness of $\hat{\beta}_1$, 
\begin{equation}
%\hspace{-10mm}
\begin{split}
\mathbb{E}[\hat{\beta}_0|\{X_i\}]&=\mathbb{E}[\overline{Y}-\hat{\beta}_1\overline{X}|\{X_i\}]=\frac{1}{n}\sum_{i=1}^n\mathbb{E}[Y_i|X_i]-\beta_1\overline{X}\\
&=\frac{1}{n}\sum_{i=1}^n(\beta_0+\beta_1X_i)-\beta_1\overline{X}=\beta_0.
\end{split}
\end{equation}

\section{Variance of OLS}
To quantify the variability of the OLS estimates, let us compute their variances. We will also need this result later on when we will discuss the prediction based on the OLS regression line. Using model assumptions 1 and 3 in (\ref{eq:23SLR_Assumptions}), we have:
\begin{equation}\label{eq:V1}
\begin{split}
\mathbb{V}[\hat{\beta}_1|\{X_i\}]&=\mathbb{V}\left[\left.\sum_{i=1}^n\frac{X_i-\overline{X}}{S_{XX}}Y_i\hspace{1mm}\right|\{X_i\}\right]\\
&=\sum_{i=1}^n\left(\frac{X_i-\overline{X}}{S_{XX}}\right)^2\mathbb{V}[Y_i|X_i]=\frac{\sigma^2}{S_{XX}}.
\end{split}
\end{equation}

Let's now look at this expression and let's suppose that we can control the inputs $X_1,\ldots,X_n$, that is we can choose them as we wish. Then (\ref{eq:V1}) suggests that we must chose them such that $S_{XX}$ is as large as possible. This would make the variance small. For example, if all $X_i$ must be in an interval $[x_0,x_1]$, then the choice of $X_i$ that maximizes $S_{XX}$ is to take half\footnote{Assume for simplicity that $n$ is even.} of them equal to $x_0$ and the other half equal to $x_1$. This would be the best\footnote{``Best'' in the sense that it would give the most precise estimate of the slope $\beta_1$ of the regression line.} design \textit{if} we are certain that the SLR model is correct. In practice, however, this \textit{two-point design} is almost never used, since researchers are rarely certain of the model. If the regression function $r(x)=\mathbb{E}[Y|X=x]$ is, in fact, non-linear, it could never be detected from data obtained using the two-point design\footnote{Recall also the 4th example in Anscombe's quartet. }. 

Computing the variance of $\hat{\beta}_0$ is a bit more involved. 
\begin{equation}
\begin{split}
\mathbb{V}[\hat{\beta}_0|\{X_i\}]&=\mathbb{V}[\overline{Y}-\hat{\beta}_1\overline{X}|\{X_i\}]\\&=
\mathbb{V}[\overline{Y}|\{X_i\}]+\overline{X}^2\mathbb{V}[\hat{\beta}_1|\{X_i\}]-2\overline{X}\mathrm{Cov}[\overline{Y},\hat{\beta}_1|\{X_i\}].
\end{split}
\end{equation}
Since $Y_1,\ldots,Y_n$ are independent\footnote{Assumption 1 in (\ref{eq:23SLR_Assumptions}).}, the first term is simply $\frac{\sigma^2}{n}$. The second term has been just computed in (\ref{eq:V1}). The sample mean response and the OLS estimate $\hat{\beta}_1$ constitute an example of two dependent but uncorrelated  random variables:
\begin{equation}
\begin{split}
\mathrm{Cov}&[\overline{Y},\hat{\beta}_1|\{X_i\}]=\mathrm{Cov}\left[\left.\frac{1}{n}\sum_{i=1}^nY_i,\sum_{j=1}^n\frac{X_j-\overline{X}}{S_{XX}}Y_j\hspace{1mm}\right|\{X_i\}\right]\\
&=\frac{1}{n}\sum_{i=1}^n\frac{X_i-\overline{X}}{S_{XX}}\mathrm{Cov}[Y_i,Y_i|X_i]=\frac{\sigma^2}{nS_{XX}}\sum_{i=1}^n(X_i-\overline{X})=0.
\end{split}
\end{equation}
Thus, 
\begin{equation}
\mathbb{V}[\hat{\beta}_0|\{X_i\}]=\sigma^2\left(\frac{1}{n}+\frac{\overline{X}^2}{S_{XX}}\right).
\end{equation}

\section{OLS is {\color{blue}BLUE}}
So, we have shown that the OLS estimates are linear, unbiased, and computed their variances. It turns out that $\hat{\beta}_0$ and $\hat{\beta}_1$ are the \textit{best liner unbiased estimates} (BLUE). Here ``best'' means that the estimate has the smallest variance among all linear and unbiased estimates. This result, called the \textit{Gauss-Markov theorem}, is valid not only for the SLR model, but also for a more general multiple regression model. 

Let us show that $\hat{\beta}_1$ is BLUE. To start, let us describe in more detail the class of estimates we consider. First, the estimate must be linear:
\begin{equation}\label{eq:linear1}
\hat{\hat{\beta}}_1=\sum_{i=1}^n \alpha_iY_i, \hspace{5mm} \alpha_i\in\mathbb{R}.
\end{equation}
Second, it must be unbiased. That is $\mathbb{E}[\hat{\hat{\beta}}_1|\{X_i\}]=\beta_1$. This condition induces the following requirement\footnote{We use assumption 2 in (\ref{eq:23SLR_Assumptions}).}: 
\begin{equation}
\begin{split}
\beta_1=\mathbb{E}[\hat{\hat{\beta}}_1|\{X_i\}]&=\mathbb{E}\left[\left.\sum_{i=1}^n \alpha_iY_i\hspace{1mm}\right|\{X_i\}\right]=\sum_{i=1}^n\alpha_i\mathbb{E}[Y_i|X_i]\\
&=\sum_{i=1}^n\alpha_i(\beta_0+\beta_1X_i)=\beta_0\sum_{i=1}^n\alpha_i+\beta_1\sum_{i=1}^n\alpha_iX_i.
\end{split}
\end{equation}
The RHS must be equal to the LHS for any $\beta_0$ and $\beta_1$. This is possible if and only if 
\begin{equation}\label{eq:constraints}
\sum_{i=1}^n\alpha_i=0 \hspace{3mm}\mbox{and}\hspace{3mm} \sum_{i=1}^n\alpha_iX_i=1.
\end{equation}
So, we consider the estimates of the form (\ref{eq:linear1}) with coefficients satisfying (\ref{eq:constraints}).

\begin{figure}
	%	\vspace{0mm}
	\centerline{\includegraphics[width=.8\linewidth]{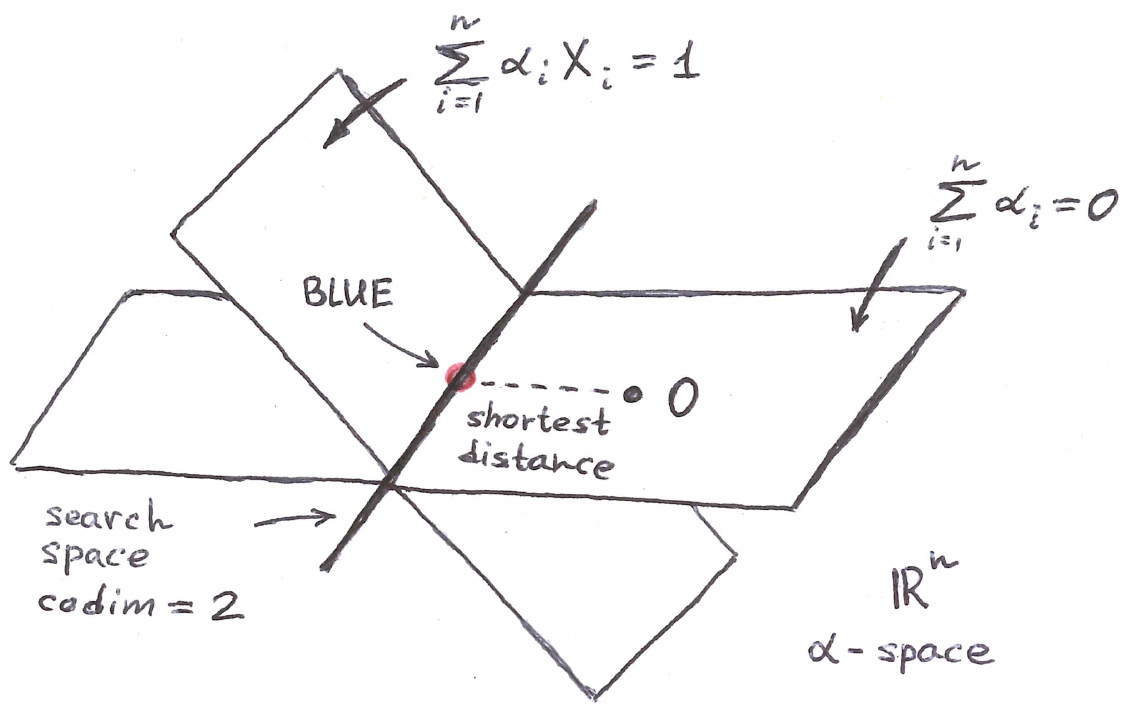}}\caption{The constrain minimization (\ref{eq:problem}) is equivalent to finding the closest to the origin point (red dot) in the search space of codimension $2$, which is the intersection of two hyperplanes defined by the constraints.   }\label{fig:visualization}
\end{figure}

The variance of $\hat{\hat{\beta}}_1$ is\footnote{We use assumptions 1 and 3 in (\ref{eq:23SLR_Assumptions}). } 
\begin{equation}
\mathbb{V}[\hat{\hat{\beta}}_1|\{X_i\}]=\sigma^2\sum_{i=1}^n\alpha_i^2.
\end{equation}
To find the BLUE, we thus need to minimize
\begin{equation}\label{eq:problem}
\begin{split}
&\sum_{i=1}^n\alpha_i^2\longrightarrow\min\\
&\mbox{subject to } \sum_{i=1}^n\alpha_i=0 \hspace{3mm}\mbox{and}\hspace{3mm} \sum_{i=1}^n\alpha_iX_i=1.
\end{split}
\end{equation}
This constrained minimization problem can be solved, for instance, by the method of Lagrange multipliers. But before we delve into computations, let us see what is going on geometrically. Figure~\ref{fig:visualization} shows what we are trying to find. 

From this visualization it is clear that there exists the unique critical point which is the global minimum. Let us now go back to work. The Lagrangian is
\begin{equation}
L(\alpha_1,\ldots,\alpha_n;\lambda_1,\lambda_2)=\sum_{i=1}^n\alpha_i^2+\lambda_1\sum_{i=1}^n\alpha_i+\lambda_2\left(\sum_{i=1}^n\alpha_iX_i-1\right).
\end{equation} 
To find the critical points of the Lagrangian we need to set all its partial derivatives to zero:
\begin{equation}
\begin{cases} 2\alpha_i+\lambda_1+\lambda_2X_i=0, \hspace{2mm}i=1,\ldots,n, \\ 
\sum_{i=1}^n\alpha_i=0,\\ \sum_{i=1}^n\alpha_iX_i-1=0. \end{cases}
\end{equation}
It is readily verifiable that this system has the unique solution:
\begin{equation}\label{eq:solution}
\lambda_1=\frac{2\overline{X}}{S_{XX}}, \hspace{3mm} \lambda_2=-\frac{2}{S_{XX}}, \hspace{3mm} \alpha_i=\frac{X_i-\overline{X}}{S_{XX}}.
\end{equation} 
It remains to observe that $\alpha_i$ in (\ref{fig:visualization}) are exactly the same as in (\ref{eq:linear}), which proves that $\hat{\beta}_1$ is the BLUE of the slope $\beta_1$. A similar analysis shows that $\hat{\beta}_0$ is also the BLUE of the intercept $\beta_0$.

Thus, if you believe the SLR model correctly describes your data and want to use linear unbiased estimates for $\beta_0$ and $\beta_1$, the OLS estimates are the ones to use. 

\section{Further Reading}
\begin{enumerate}
	\item But what if we don't care about the unbiasedness and linearity, and simply want (possibly biased) estimates of $\beta_0$ and $\beta_1$ with small MSE? Can we beat OLS? A short answer is ``yes.'' See T.L.~Burr \& H.A.~Fry (2005) ``\href{http://www.jstor.org/stable/25471022}{Biased regression: the case for cautious application},'' \textit{Technometrics}, 47(3), 284-296 for a review of biased estimates that have lower MSE than OLS. 
	\item See also a discussion on \href{http://stats.stackexchange.com/questions/2957/ols-is-blue-but-what-if-i-dont-care-about-unbiasedness-and-linearity}{stats.stackexchange.com}.
\end{enumerate}

\section{What is Next?} 
We will see how hypothesis testing and interval estimation work in the context of simple linear regression.

\chapter{Hypothesis Testing \& Interval Estimation}\label{ch:Hypo and Intervals}

\newthought{The OLS method} allows us to construct point estimates for the regression parameters. In many applications, however, we are often interested in testing hypothesis about the parameters and constructing confidence intervals for them. In previous lectures, we discussed hypothesis testing and intervals in general settings. Here, we will see how these methods of statistical inference work for the SLR model.

To both test hypothesis and construct confidence intervals, we need to make a parametric assumption about the statistical errors. Namely, we assume that $e_i$ are normally distributed, and thus we will work with the conditional normal model\footnote{We introduced in Lecture~\ref{ch:OLS}, where we have shown that, in this model, the OLS estimates are simply the MLEs.}:
\begin{equation}\label{eq:NormalSLR}
\begin{split}
&Y_i=\beta_0+\beta_1X_i+e_i,  \\
&e_i|X_i\hspace{1mm}\sim\hspace{1mm}\mathcal{N}(0,\sigma^2), \hspace{5mm} i=1,\ldots,n.
\end{split}
\end{equation}
In the next lecture, we will discuss how this assumption can be checked using the residual analysis. 

\section{The t-Test for the Regression Parameters}
Suppose we want to test the hypothesis (``current theory'') that the slope $\beta_1$ equals to some constant $\beta_1^*$:
\begin{equation}\label{eq:24two-sided}
H_0:\hspace{1mm} \beta_1=\beta_1^*\hspace{2mm}\mbox{ versus }\hspace{2mm} H_1:\hspace{1mm}\beta_1\neq\beta_1^*.
\end{equation} 
Thanks to the normality assumption in (\ref{eq:NormalSLR}), the responses are independently and normally\footnote{But, of course, not identically.} distributed:
\begin{equation}
Y_i|X_i\hspace{1mm}\sim\hspace{1mm}\mathcal{N}(\beta_0+\beta_1X_i,\sigma^2).
\end{equation}
Recall that the OLS estimate $\hat{\beta}_1$ is a liner combination of responses, 
\begin{equation}
\hat{\beta}_1=\sum_{i=1}^n\frac{X_i-\overline{X}}{S_{XX}}Y_i,
\end{equation}
and, therefore, it is also normal. We also found its mean ($\mathbb{E}[\hat{\beta}_1|\{X_i\}]=\beta_1$) and variance ($\mathbb{V}[\hat{\beta}_1|\{X_i\}]=\frac{\sigma^2}{S_{XX}}$) last time. Thus,
\begin{equation}
\hat{\beta}_1|\{X_i\}\hspace{1mm}\sim\hspace{1mm}\mathcal{N}\left(\beta_1,\frac{\sigma^2}{S_{XX}}\right).
\end{equation}
This means that, under $H_0$, 
\begin{equation}
\left.\frac{\hat{\beta}_1-\beta_1^*}{\sqrt{\frac{\sigma^2}{S_{XX}}}}\right|\hspace{-1mm}\{X_i\}\hspace{1mm}\sim\hspace{1mm}\mathcal{N}(0,1).
\end{equation}
The parameter $\sigma^2$ is unknown, but recall\footnote{Lecture~\ref{ch:OLS}.} that we know its unbiased estimate:
\begin{equation}\label{eq:unbiased}
\hat{\sigma}^2=\frac{1}{n-2}\sum_{i=1}^n \hat{e}_i^2.
\end{equation}
The random variable 
\begin{equation}
T_1=\left.\frac{\hat{\beta}_1-\beta_1^*}{\sqrt{\frac{\hat{\sigma}^2}{S_{XX}}}}\right|\hspace{-1mm}\{X_i\}\hspace{1mm}\approxdist\hspace{1mm}\mathcal{N}(0,1).
\end{equation}
is then approximately normally distributed\footnote{So, in principle, we can use the Wald test.}. It can be shown\footnote{For example, see Appendix C.3 in D.C.~Montgomery et al (2006) \textit{\href{https://books.google.com/books?id=0yR4KUL4VDkC}{Introduction to Linear Regression Analysis}}.} however, that the exact distribution of $T_1$ under $H_0$ is $t$-distribution with $(n-2)$ degrees of freedom:
\begin{equation}
T_1|\{X_i\}\hspace{1mm}\sim\hspace{1mm}t_{n-2}.
\end{equation}
So, the size $\alpha$ $t$-test rejects $H_0$ when 
\begin{equation}\label{eq:ttest}
\left|\frac{\hat{\beta}_1-\beta_1^*}{\hat{\sigma}/\sqrt{S_{XX}}}\right|>t_{n-2,1-\frac{\alpha}{2}}.
\end{equation}
To find the $p$-value of the test, we need to solve $|T_1|=t_{n-2,\frac{\alpha}{2}}$ for $\alpha$. 

Similarly, we can construct a $t$-test for the intercept:
\begin{equation}\label{eq:24two-sided2}
H_0:\hspace{1mm} \beta_0=\beta_0^*\hspace{2mm}\mbox{ versus }\hspace{2mm} H_1:\hspace{1mm}\beta_0\neq\beta_0^*.
\end{equation} 
The test statistic in this case is\footnote{In (\ref{eq:T_0}), as usual, $\widehat{\mathrm{se}}(\hat{\beta}_0)$ denotes the estimated standard error.} 
\begin{equation}\label{eq:T_0}
T_0=\frac{\hat{\beta}_0-\beta_0^*}{\widehat{\mathrm{se}}(\hat{\beta}_0)}=\frac{\hat{\beta}_0-\beta_0^*}{\hat{\sigma}\sqrt{\frac{1}{n}+\frac{\overline{X}^2}{S_{XX}}}},
\end{equation}
and, as before, the size $\alpha$ $t$-test rejects $H_0$ when 
\begin{equation}
|T_0|>t_{n-2,1-\frac{\alpha}{2}},
\end{equation}
and the $p$-value is $p=2F_{n-2}(-|T_0|)$, where  $F_{n-2}$ is the CDF of the $t$-distribution with $(n-2)$ degrees of freedom. 

\section{Testing Significance of Linear Regression}

\begin{figure}
	\centerline{\includegraphics[width=\linewidth]{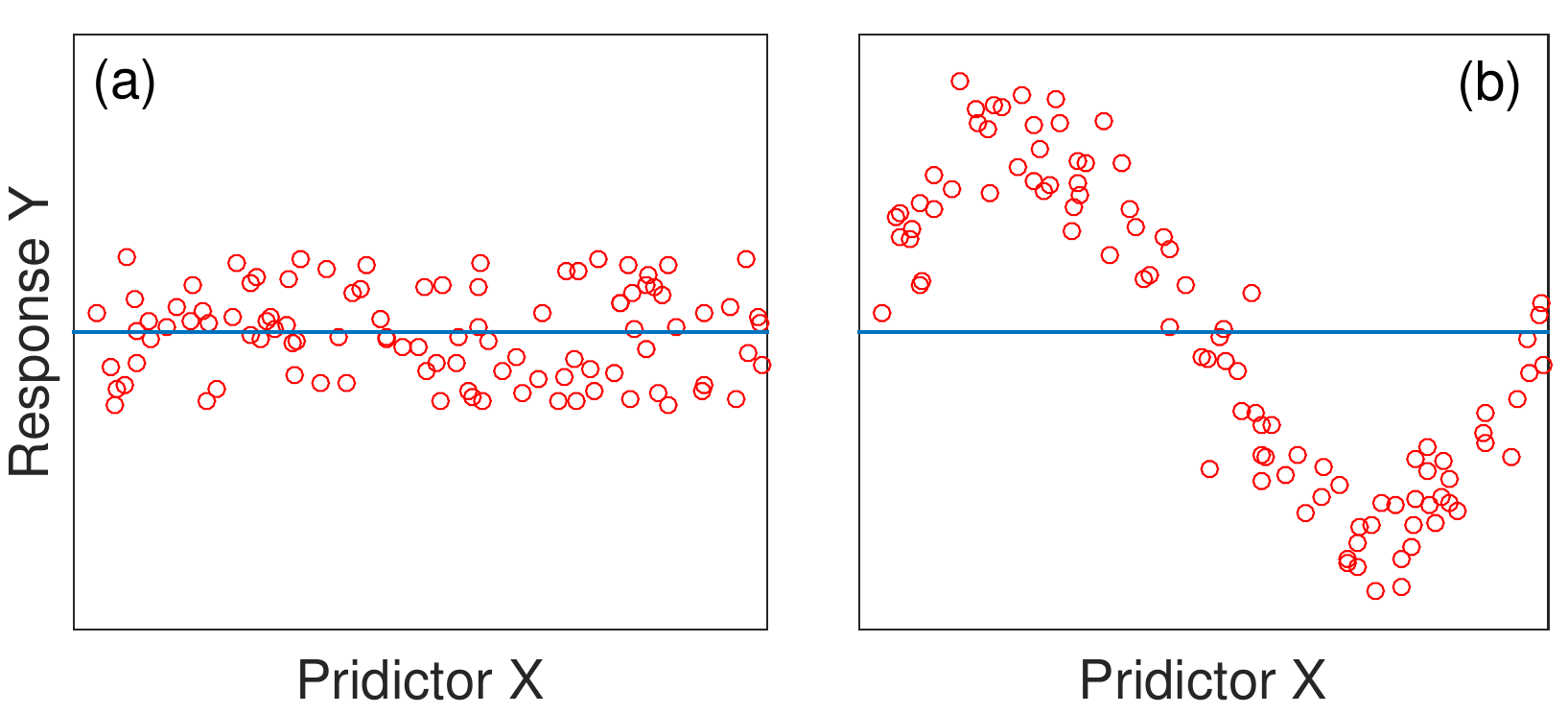}}\caption{Two cases where the null hypothesis $H_0: \beta_1=0$ is accepted.}\label{fig:zeroslope}
\end{figure}

A very important special case of (\ref{eq:24two-sided}) is 
\begin{equation}\label{eq:significance}
H_0:\hspace{1mm} \beta_1=0\hspace{2mm}\mbox{ versus }\hspace{2mm} H_1:\hspace{1mm}\beta_1\neq0.
\end{equation} 
Here we test the existence of linear relationship between the predictor  and response. Accepting the null hypothesis implies that we have one of the following two scenarios:
\begin{enumerate}
	\item The response $Y$  does not really depend on the input $X$, and the best prediction $\widehat{Y}$ of the response to any future input $X$ is simply the sample mean, $\widehat{Y}=\overline{Y}$. This situation is illustrated in Fig.~\ref{fig:zeroslope}(a).
	\item The response $Y$ does depend on $X$, but the true relationship is not linear, Fig.~\ref{fig:zeroslope}(b).
\end{enumerate}
Thus, accepting $H_0$ is equivalent to saying that there is \textit{no linear relationship} between $Y$ and $X$.

\begin{figure}
	\centerline{\includegraphics[width=\linewidth]{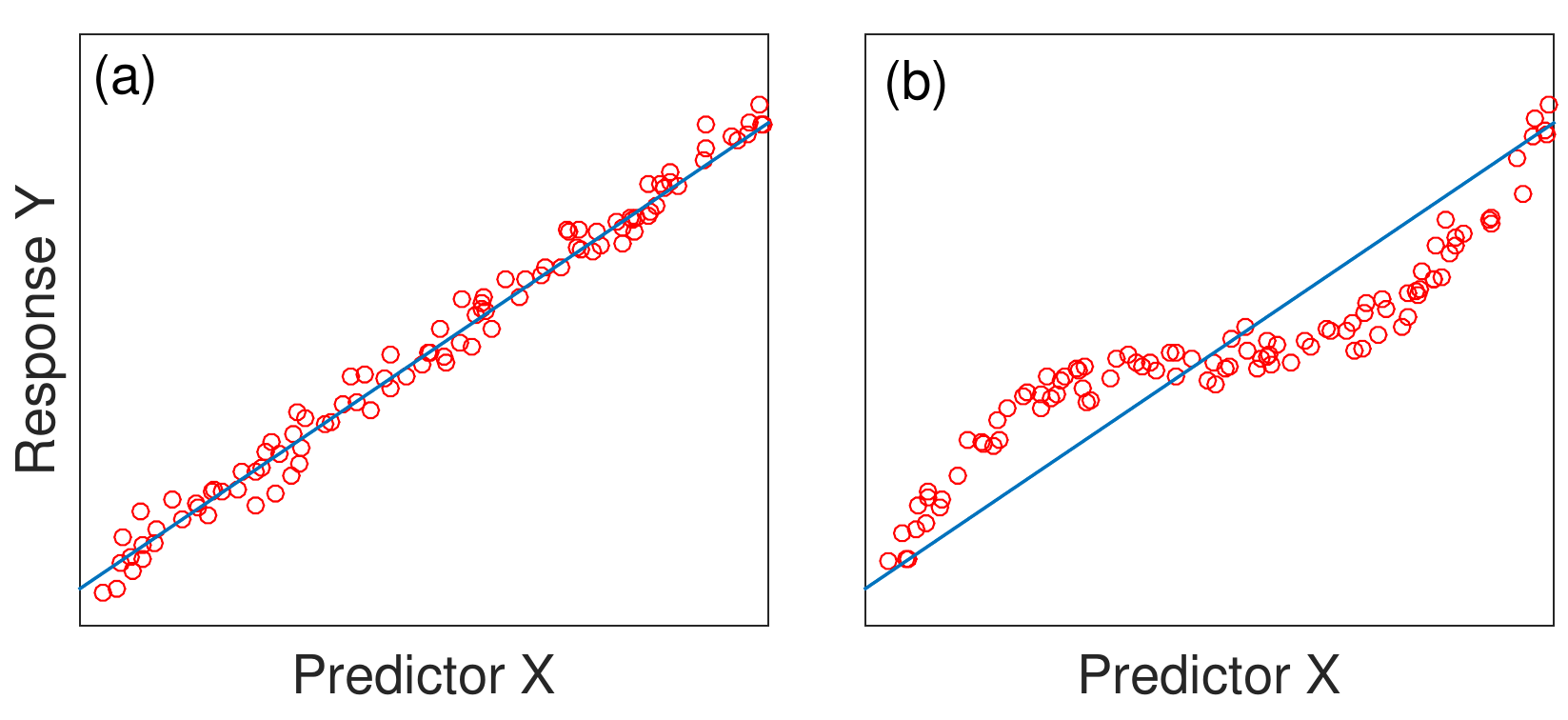}}\caption{Two cases where the null hypothesis $H_0: \beta_1=0$ is rejected.}\label{fig:nonzeroslope}
\end{figure}

On the other hand, rejecting $H_0$ means that the predictor indeed influences the response, but the relationship is not necessarily linear. Namely,  two cases are possible:
\begin{enumerate}
	\item The SLR model (\ref{eq:NormalSLR}) is an accurate model for the data, Fig.~\ref{fig:nonzeroslope}(a).
	\item There is a linear  trend $Y=\beta_0+\beta_1X$, but the data is more accurately modeled with the addition of  higher order terms $Y=\beta_0+\beta_1X+\beta_2X^2+\ldots$, Fig.\ref{fig:nonzeroslope}(b).
\end{enumerate}

The procedure  for testing significance of linear regression is obtained from (\ref{eq:ttest})  by setting $\beta_1^*$ to zero. Namely, we claim the liner regression significant at level $\alpha$ if 
\begin{equation}\label{eq:ttest2}
\left|\frac{\hat{\beta}_1}{\hat{\sigma}/\sqrt{S_{XX}}}\right|>t_{n-2,\frac{\alpha}{2}}.
\end{equation}
The corresponding $p$-value is $p(X,Y)=2F_{n-2}\left(-\left|\frac{\hat{\beta}_1}{\hat{\sigma}/\sqrt{S_{XX}}}\right|\right)$.
\subsection{Example: Heights of Mothers and Daughters}
\begin{marginfigure}
	\vspace{-30mm}
	\centerline{\includegraphics[width=\linewidth]{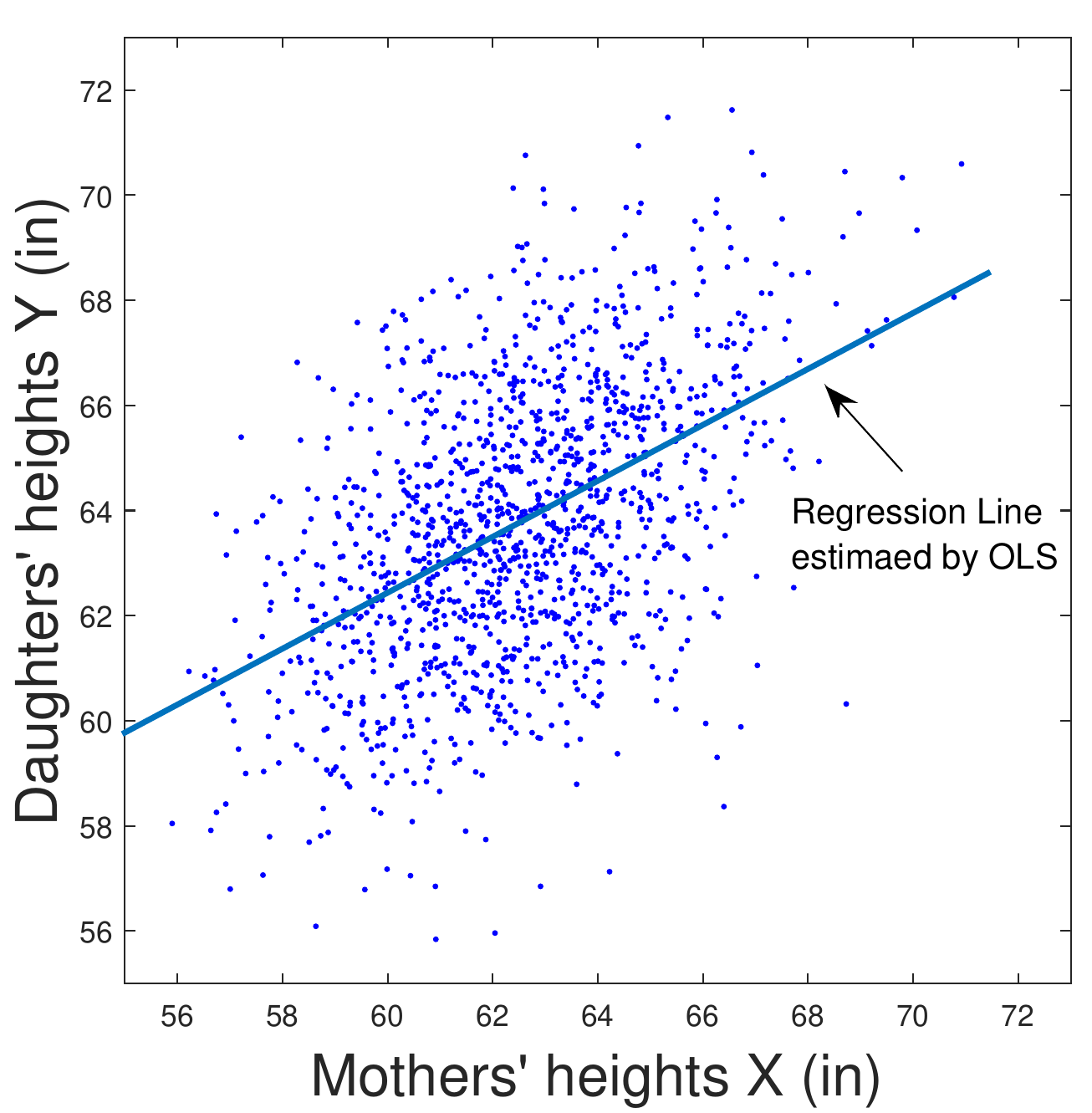}}\caption{The OLS regression line for the heights data, \href{http://www.its.caltech.edu/~zuev/teaching/2016Winter/heights.xlsx}{heights.xlsx}.}\label{fig:24OLS_heights}
\end{marginfigure}

Recall that in Lecture~\ref{ch:Regression}, we found the OLS  regression line for the Pearson's data, mothers' heights vs. daughters' heights, Fig.~\ref{fig:24OLS_heights}. The estimated valued of the slope is $\hat{\beta}_1=0.53$. The $p$-value is essentially zero,
\begin{equation}
p(X,Y)=9.7\times10^{-81},
\end{equation}
which means that Pearson's data provides extremely strong evidence against $H_0$. As expected.

\section{Confidence Intervals for $\beta_0$, $\beta_1$, and $\sigma^2$}

Now let us turn to constructing confidence intervals for the regression parameters, which can be used as a measure of the overall quality of the regression line. 

In the previous section, we have already established that, under normality assumption (\ref{eq:NormalSLR}), 
\begin{equation}
\left.\frac{\hat{\beta}_0-\beta_0}{\widehat{\mathrm{se}}(\hat{\beta}_0)}\right|\hspace{-1mm}\{X_i\}\hspace{1mm}\sim\hspace{1mm}t_{n-2}\hspace{3mm}\mbox{and}\hspace{3mm}\left.\frac{\hat{\beta}_1-\beta_1}{\widehat{\mathrm{se}}(\hat{\beta}_1)}\right|\hspace{-1mm}\{X_i\}\hspace{1mm}\sim\hspace{1mm}t_{n-2},
\end{equation}
where the estimated standard errors are 
\begin{equation}
\widehat{\mathrm{se}}(\hat{\beta}_0)=\hat{\sigma}\sqrt{\frac{1}{n}+\frac{\overline{X}^2}{S_{XX}}}\hspace{3mm}\mbox{and}\hspace{3mm}\widehat{\mathrm{se}}(\hat{\beta}_1)=\frac{\hat{\sigma}}{\sqrt{S_{XX}}},
\end{equation}
and $\hat{\sigma}$ is given by (\ref{eq:unbiased}). Therefore, a $100(1-\alpha)\%$ confidence interval for $\beta_i$ is given by
\begin{equation}\label{eq:CIbeta}
\hat{\beta}_i\pm t_{n-2,\frac{\alpha}{2}}\widehat{\mathrm{se}}(\hat{\beta}_i).
\end{equation}
The interpretation of this interval is the following: if we  
\begin{enumerate}
	\item fix $X_1,\ldots,X_n$,
	\item measure , for each $X_i$, the corresponding response $m$ times: 
	\begin{equation}
	X_i \longmapsto Y_i^{(1)},\ldots,Y_i^{(m)},
	\end{equation}
	\item construct $m$ intervals (\ref{eq:CIbeta}) from data $\{(X_i,Y_i^{(k)})\}$, $k=1,\ldots,m$,
\end{enumerate}
then approximately $(1-\alpha)m$ intervals will contain the true value of $\beta_i$ (assuming the the SLR model is correct).

\begin{marginfigure}
	\vspace{-7mm}
	\centerline{\includegraphics[width=\linewidth]{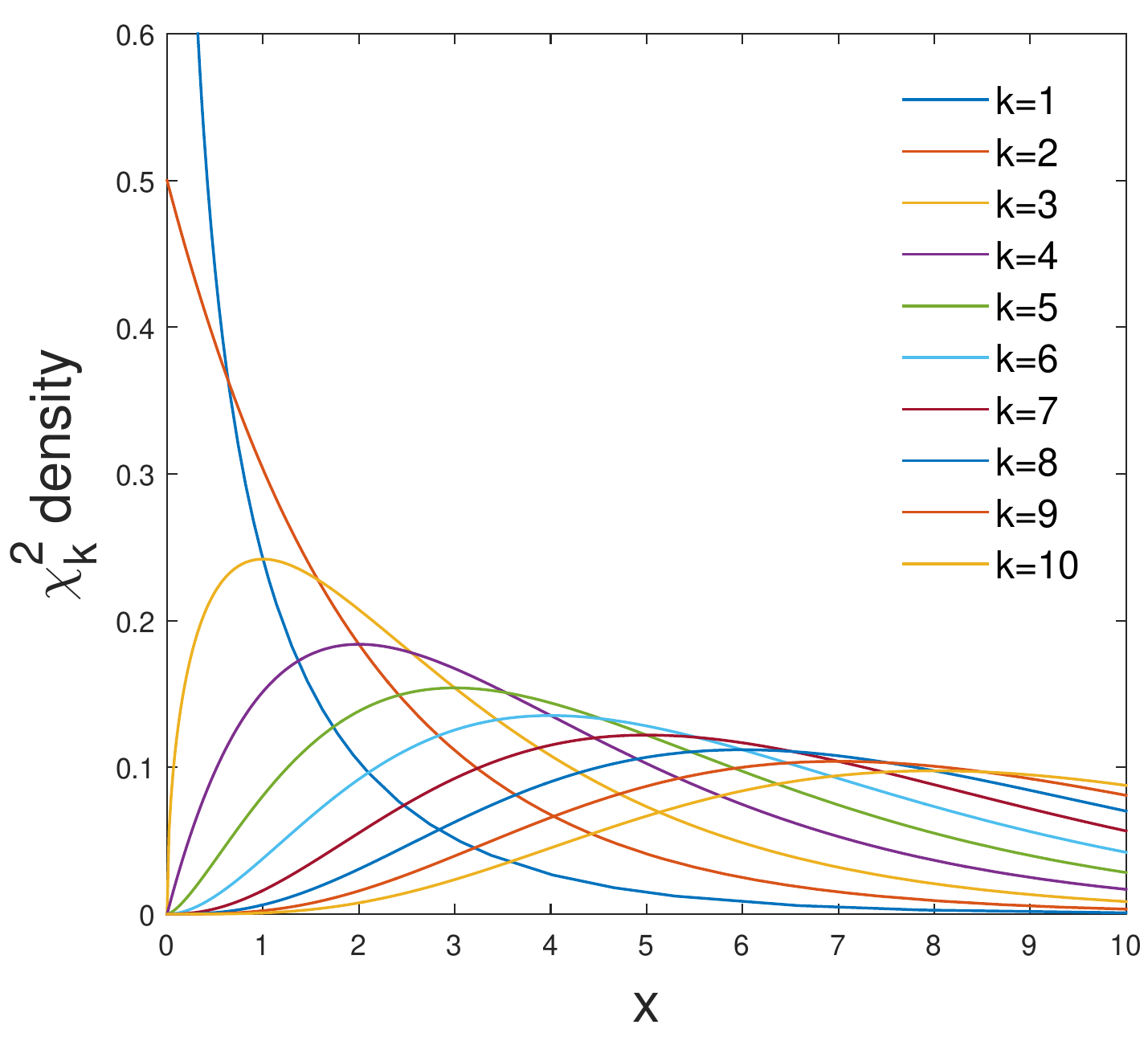}}\caption{The PDF of the $\chi^2$-distribution with $k$ degrees of freedom.}\label{fig:24chi}
\end{marginfigure}

To construct a confidence interval for the variance $\sigma^2$, we need to use the following technical result\footnote{For example, see Appendix C.3 in D.C.~Montgomery et al (2006) \textit{\href{https://books.google.com/books?id=0yR4KUL4VDkC}{Introduction to Linear Regression Analysis}}.}:
\begin{equation}\label{eq:24chi}
\frac{(n-2)\hat{\sigma}^2}{\sigma^2}\sim\chi^2_{n-2},
\end{equation}
where $\chi^2_q$ is the $\chi^2$-distribution with $q$ degrees of freedom\footnote{Recall the definition: if $Z_1,\ldots,Z_1$ are iid standard normal, then $$Q=\sum_{i=1}^qZ_i^2\sim\chi^2_q$$}. {Figure~\ref{fig:24chi}} shows the density of this distribution.

Introducing the standard notation, $\chi^2_{q,\alpha}$, for the point that defines the interval $(0,\chi^2_{q,\alpha})$ that contains probability mass $\alpha$, and using (\ref{eq:24chi}), we have:  
\begin{equation}
\mathbb{P}\left(\chi^2_{n-2,1-\frac{\alpha}{2}}<\frac{(n-2)\hat{\sigma}^2}{\sigma^2}<\chi^2_{n-2,\frac{\alpha}{2}}\right)=1-\alpha.
\end{equation}
This gives a $100(1-\alpha)\%$ confidence interval for $\sigma^2$:
\begin{equation}
\frac{(n-2)\hat{\sigma}^2}{\chi^2_{n-2,\frac{\alpha}{2}}}<\sigma^2<
\frac{(n-2)\hat{\sigma}^2}{\chi^2_{n-2,1-\frac{\alpha}{2}}}.
\end{equation}
\subsection{Example: Heights of Mothers and Daughters}
\begin{marginfigure}
%	\vspace{-20mm}
	\centerline{\includegraphics[width=\linewidth]{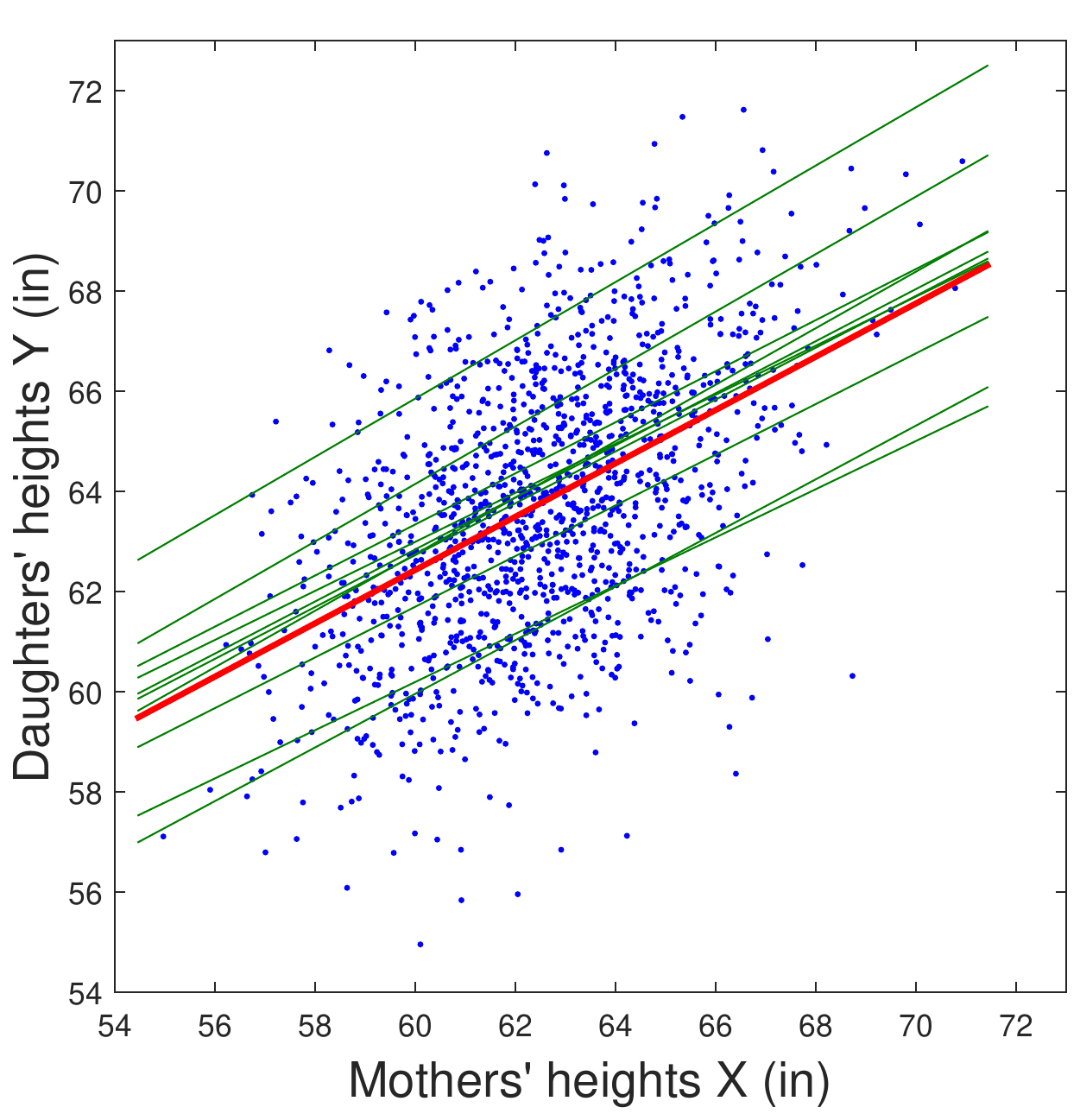}}\caption{The OLS regression line (red) and random regression lines (green) with slopes and intercepts chosen uniformly from the $95\%$ confidence intervals (\ref{eq:24CI}). }\label{fig:random_reg_lines}
\end{marginfigure}

The $95\%$ ($\alpha=0.05$) confidence intervals for $\beta_0$, $\beta_1$, and $\sigma^2$ are
\begin{equation}\label{eq:24CI}
27.3<\beta_0<33.7, \hspace{3mm} 0.48<\beta_1<0.58, \hspace{3mm} 4.9<\sigma^2<5.7.
\end{equation}
To visualize the uncertainty about the regression line, in Fig.~\ref{fig:random_reg_lines} we show the OLS regression line and 10 regression lines with slopes and intercepts chosen uniformly at random from the corresponding confidence intervals. Notice that the variability of the regression lines is quite large. It would be better to have green lines fluctuating more closely to the red line. Let us investigate what is going on. 

\begin{marginfigure}
%	\vspace{-40mm}
	\centerline{\includegraphics[width=\linewidth]{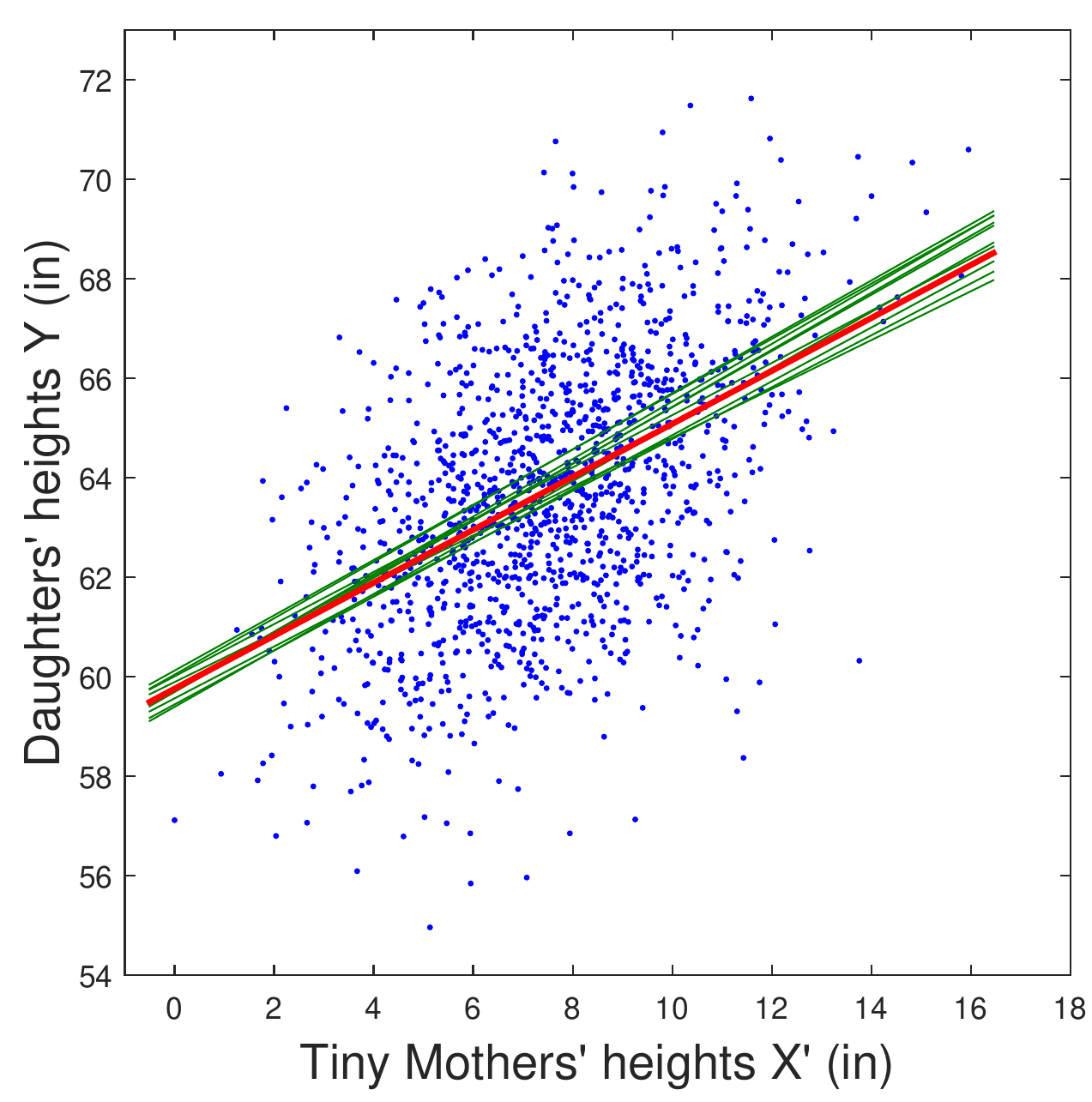}}\caption{The OLS regression line (red) and random regression lines (green) with slopes and intercepts chosen uniformly from the $95\%$ confidence intervals constructed for the data $\{(X_i',Y_i)\}$. }\label{fig:random_reg_lines_tiny}
\end{marginfigure}

High variability stems form the large size of the confidence intervals for $\beta_0$ and $\beta_1$ in (\ref{eq:24CI}), computed from (\ref{eq:CIbeta}). The size of the confidence intervals is $2t_{n-2,1-\frac{\alpha}{2}}\widehat{\mathrm{se}}(\hat{\beta}_i)$. The estimated standard errors are
\begin{equation}
\begin{split}
\widehat{\mathrm{se}}(\hat{\beta}_1)&=\frac{\hat{\sigma}}{\sqrt{S_{XX}}}=0.03, \\%\hspace{3mm}\mbox{and}\hspace{3mm}
\\ \widehat{\mathrm{se}}(\hat{\beta}_0)&=\hat{\sigma}\sqrt{\frac{1}{n}+\frac{\overline{X}^2}{S_{XX}}}\approx\frac{\hat{\sigma}\overline{X}}{\sqrt{S_{XX}}}=1.6.
\end{split}
\end{equation}
This shows that the high uncertainty is caused primarily by the high value of $\overline{X}=62.5$, which makes the confidence interval for $\beta_0$ large. If the average mothers' height were smaller, the uncertainty in the regression line were also smaller.  To confirm this observation, let us formally modify the data by reducing the mothers heights:
\begin{equation}
X_i\rightsquigarrow X_i'=X_i-\min\{X_i\}.
\end{equation}
The smallest mother in the new data has zero height (at least it is not negative!).
Fig.\ref{fig:random_reg_lines_tiny} is the same as Fig.\ref{fig:random_reg_lines}, but for data $\{(X'_i,Y_i)\}$. As expected, the uncertainty about the regression line is much smaller. 

\section{Further Reading}
\begin{enumerate}
	\item The analysis-of-variance (ANOVA) approach for testing significance of regression is an important alternative to the $t$-test considered in this lecture. For simple linear regression, the two methods are equivalent. ANOVA is especially useful in multiple regression. It is covered in depth in D.C.~Montgomery et al (2006) \textit{\href{https://books.google.com/books?id=0yR4KUL4VDkC}{Introduction to Linear Regression Analysis}}.
\end{enumerate}

\section{What is Next?} 
We will discuss the response prediction using the regression model. We will conclude these notes with a brief discussion of the residual plots, a useful tool or checking model assumptions. 

\chapter{Prediction \& Graphic Residual Analysis}\label{ch:Prediction}

\newthought{For the last time} in these notes, let us assume that we model data $(X_1,Y_1),\ldots,(X_n,Y_n)$ using the conditional normal SLR model:
\begin{equation}\label{eq:25NormalSLR}
\begin{split}
&Y_i=\beta_0+\beta_1X_i+e_i,  \\
&e_i|X_i\hspace{1mm}\sim\hspace{1mm}\mathcal{N}(0,\sigma^2), \hspace{5mm} i=1,\ldots,n,
\end{split}
\end{equation}
and let $\hat{\beta}_0$ and $\hat{\beta}_1$ denote the OLS estimates of the regression parameters. One of the main goals of  regression  is to predict the response to the future input. Let $X^*$ denote the future value of the predictor variable. The response, according to the model, is then
\begin{equation}
Y^*=\underbrace{\beta_0+\beta_1X^*}_{r(X^*)}+e^*, \hspace{5mm} e^*|X^*\sim\mathcal{N}(0,\sigma).
\end{equation}
We know\footnote{See Lecture~\ref{ch:Regression}.} that the optimal\footnote{In the MSE sense.} prediction for the response $Y^*$ is 
\begin{equation}
r(X^*)=\mathbb{E}[Y^*|X^*]=\beta_0+\beta_1X^*.
\end{equation}
If instead of normality assumption in (\ref{eq:25NormalSLR}) for the statistical errors, we make a weaker set of assumptions (\ref{eq:25SLR_Assumptions}), then pretty much all we can do is to report the point estimate of the optimal prediction, namely the fitted (or predicted value):
\begin{equation}
\widehat{Y}^*=\hat{r}(X^*)=\hat{\beta}_0+\hat{\beta}_1X^*.
\end{equation}

Under the normality assumption, we can do more: 
\begin{enumerate}
	\item[(a)] Construct a confidence interval for $r(X^*)$, which is a parameter of the model, and 
	\item[(b)] Construct a \textit{prediction interval} for $Y^*$, which is  an unobserved random variable.
\end{enumerate}
According to the model,
\begin{equation}
Y^*|X^*\hspace{1mm}\sim\hspace{1mm}\mathcal{N}(r(X^*),\sigma^2).
\end{equation}
So, (a) can be considered as the inference on the mean of the distribution, and (b) as the inference on the actual value. 

\section{Confidence Interval for $r(X^*)$}
Thanks to the normality assumption, the fitted value $\widehat{Y}^*=\hat{r}(X^*)$ is normally distributed, since it is a liner combination of $Y_i$, which are normal. Since the OLS estimates are unbiased, so is the fitted value:
\begin{equation}
\begin{split}
\mathbb{E}[\hat{r}(X^*)|\{X_i\},X^*]&=\mathbb{E}[\hat{\beta}_0+\hat{\beta}_1X^*|\{X_i\},X^*]\\ &=\beta_0+\beta_1X^*=r(X^*).
\end{split}
\end{equation}
Let us compute the variance: 
\begin{equation}
\begin{split}
&\mathbb{V}[\hat{r}(X^*)|\{X_i\},X^*]=\mathbb{V}[\hat{\beta}_0+\hat{\beta}_1X^*|\{X_i\},X^*]\\ &=\mathbb{V}[\hat{\beta}_0|\{X_i\}]+(X^*)^2\mathbb{V}[\hat{\beta}_1|\{X_i\}]+2X^*\mathrm{Cov}[\hat{\beta}_0,\hat{\beta}_1|\{X_i\}].
\end{split}
\end{equation}
Variance of both $\hat{\beta}_0$ and $\hat{\beta}_1$ we found in Lecture~\ref{ch:OLSproperties}\footnote{$\mathbb{V}[\hat{\beta}_1|\{X_i\}]=\frac{\sigma^2}{S_{XX}}$ and \\ $\mathbb{V}[\hat{\beta}_0|\{X_i\}]=\sigma^2\left(\frac{1}{n}+\frac{\overline{X}^2}{S_{XX}}\right)$.}. To find the covariance, we use $\hat{\beta}_0=\overline{Y}-\hat{\beta}_1\overline{X}$ and that $\mathrm{Cov}[\overline{Y},\hat{\beta}_1|\{X_i\}]=0$\footnote{Also, see Lecture~\ref{ch:OLSproperties}.}.
\begin{equation}
\begin{split}
\mathrm{Cov}[\hat{\beta}_0,\hat{\beta}_1|\{X_i\}]&=\mathrm{Cov}[\overline{Y}-\hat{\beta}_1\overline{X},\hat{\beta}_1|\{X_i\}]\\
&=-\overline{X}\mathbb{V}[\hat{\beta}_1|\{X_i\}]=-\frac{\sigma^2\overline{X}}{S_{XX}}.
\end{split}
\end{equation}
Thus,
\begin{equation}
\begin{split}
\mathbb{V}[\hat{r}(X^*)|\{X_i\},X^*]&=\sigma^2\left(\frac{1}{n}+\frac{\overline{X}^2}{S_{XX}}\right)+\frac{\sigma^2(X^*)^2}{S_{XX}}-\frac{2\sigma^2\overline{X}X^*}{S_{XX}}\\
&=\sigma^2\left(\frac{1}{n}+\frac{(X^*-\overline{X})^2}{S_{XX}}\right),
\end{split}
\end{equation}
and
\begin{equation}
\left.\frac{\hat{r}(X^*)-r(X^*)}{\sigma\sqrt{\frac{1}{n}+\frac{(X^*-\overline{X})^2}{S_{XX}}}}\right|\hspace{-1mm}\{X_i\},X^*\hspace{1mm}\sim\hspace{1mm}\mathcal{N}(0,1).
\end{equation}
If we replace $\sigma$ with its unbiased estimate $\hat{\sigma}$, then the distribution will be approximately normal. The exact distribution, as in the previous lecture, is $t$ with $(n-2)$ degrees of freedom. Consequently, a $100(1-\alpha)\%$ confidence interval for $r(X^*)$, mean response at $X^*$, is
\begin{equation}\label{eq:CI_r(X^*)}
r(X^*)\in\left(\hat{\beta}_0+\hat{\beta}_1X^*\pm t_{n-2,\frac{\alpha}{2}}\hat{\sigma}\sqrt{\frac{1}{n}+\frac{(X^*-\overline{X})^2}{S_{XX}}}\right).
\end{equation}
Note that the width of the confidence interval for $r(X^*)$ is, as expected, a function of $X^*$. The interval width is minimal if $X^*=\overline{X}$ and it increases as $X^*$ goes away from $\overline{X}$. Intuitively, this is reasonable: our prediction is most accurate near the center of the data, and as $|X^*-\overline{X}|$ increases, the prediction degenerates. 

\section{Prediction Interval for $Y^*$}
Now let us derive an interval estimate for the unobserved random response $Y^*$. Uncertainty in the value of the mean response $r(X^*)$ stems from the uncertainty in the regression coefficients $\beta_0$ and $\beta_1$. In the case of $Y^*$, we have an additional source of uncertainty coming from  the random statistical error $e^*$:
\begin{equation}
Y^*\neq r(X^*), \hspace{3mm}\mbox{but }\hspace{2mm}  Y^*=r(X^*)+e^*.
\end{equation}

Let $\hat{e}^*$ be the unobserved residual
\begin{equation}
\hat{e}^*=Y^*-\widehat{Y}^*=(\beta_0-\hat{\beta}_0)+(\beta_1-\hat{\beta}_1)X^*+e^*.
\end{equation}
Since both $Y^*$ and $\widehat{Y}^*$ are normal according to the model, so is $\hat{e}^*$. Its mean is zero:
\begin{equation}
\mathbb{E}[\hat{e}^*|\{X_i\},X^*]=\mathbb{E}[\hat{e}^*|X^*]=0,
\end{equation}
and the variance is 
\begin{equation}
\begin{split}
\mathbb{V}[\hat{e}^*|\{X_i\},X^*]&=\mathbb{V}[\hat{\beta}_0+\hat{\beta}_1X^*|\{X_i\},X^*]+\mathbb{V}[\hat{e}^*|X^*]\\
&=\sigma^2\left(1+\frac{1}{n}+\frac{(X^*-\overline{X})^2}{S_{XX}}\right).
\end{split}
\end{equation}
The rest is as above: 
\begin{equation}
\left.\frac{Y^*-\widehat{Y}^*}{\hat{\sigma}\sqrt{1+\frac{1}{n}+\frac{(X^*-\overline{X})^2}{S_{XX}}}}\right|\hspace{-1mm}\{X_i\},X^*\hspace{1mm}\sim\hspace{1mm}t_{n-2},
\end{equation}
and a $100(1-\alpha)\%$ prediction interval for the future response $Y^*$ to input $X^*$ is
\begin{equation}\label{eq:CI_Y^*}
Y^*\in\left(\hat{\beta}_0+\hat{\beta}_1X^*\pm t_{n-2,\frac{\alpha}{2}}\hat{\sigma}\sqrt{1+\frac{1}{n}+\frac{(X^*-\overline{X})^2}{S_{XX}}}\right).
\end{equation}
Notice that the intervals for the mean response $r(X^*)$ in (\ref{eq:CI_r(X^*)}) and the response $Y^*$ in (\ref{eq:CI_Y^*}) have the same center, the fitted value $\hat{\beta}_0+\hat{\beta}_1X^*$, but the width of the second interval is larger, since there is more uncertainty in the value of the response than in the value of its mean. 

\subsection{Example: Heights Data}
\begin{marginfigure}
	\vspace{-20mm}
	\centerline{\includegraphics[width=\linewidth]{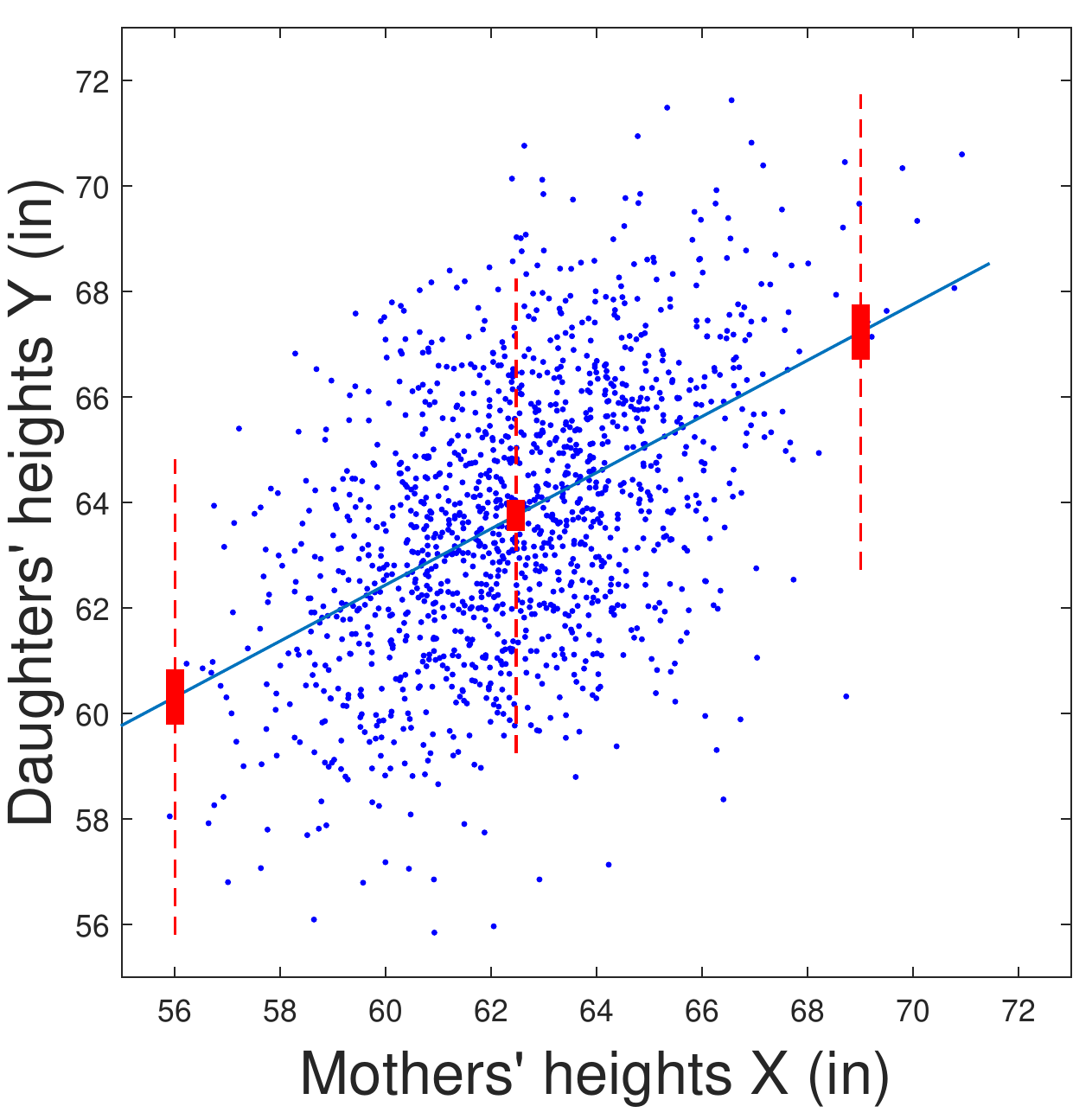}}\caption{Confidence intervals for $r(X^*)$ and prediction intervals for $Y^*$, for $X^*=56,62.5,$ and $69$.}\label{fig:CIs}
\end{marginfigure}

Figure~\ref{fig:CIs} shows six intervals: three for the future daughters' heights $Y^*$ (dashed) born by the mothers of heights $X^*=56, 62.5, 69$ and three for the mean daughters' height $r(X^*)$ born by the mothers of the same heights. In all cases, $\alpha=0.05$. In this example, it is especially clear that the uncertainty associated with future response $Y^*$ is much higher than that for the mean response.  

\section{Checking Model Assumptions using the Residual Plots}
There are two standard sets of assumptions in the SLR model: ``semi-parametric'' assumptions
\begin{equation}\label{eq:25SLR_Assumptions}
\begin{split}
1.\hspace{1mm}& e_i \mbox{ are independent,}\\
2.\hspace{1mm} & \mathbb{E}[e_i|X_i]=0,\\
3.\hspace{1mm} & \mathbb{V}[e_i|X_i]=\sigma^2,
\end{split}
\end{equation}
and a stronger parametric assumption
\begin{equation}\label{eq:25normality} e_i|X_i\hspace{1mm}\sim\hspace{1mm}\mathcal{N}(0,\sigma^2),
\end{equation}
which, in addition to point estimates, allows to test hypothesis and construct confidence intervals. 

The above assumptions are the statements about statistical errors,
\begin{equation}
e_i=Y_i-\beta_0-\beta_1X_i,
\end{equation}
which are unobserved\footnote{Since $\beta_0$ and $\beta_1$ are unknown.} random variables.  Strictly speaking, the residuals,
\begin{equation}
\hat{e}_i=Y_i-\hat{\beta}_0-\hat{\beta}_1X_i,
\end{equation}
are not realizations of errors, because the betas are hatted. Nevertheless, since $\hat{\beta}_0\approx\beta_0$ and $\hat{\beta}_1\approx\beta_1$, it is convenient to think of the residuals as the observed approximate realizations of random errors. Therefore, any departure from the assumptions on the errors should show up in the residuals. Plotting residuals is a very effective way to investigate how well the assumptions hold for the data in hand.  

Checking the independence assumption in (\ref{eq:25SLR_Assumptions}) using residuals is a bit tricky since the residuals are necessarily dependent\footnote{Recall than $\sum_{i=1}^n\hat{e}_i=0$.}. But if the statistical error are independent, then the dependence among the residuals is very low (especially if $n$ is large). One way to check the independence assumption is to use the \textit{lag plot} of the residuals, constructed by plotting residual $\hat{e}_i$ against residual $\hat{e}_{i-1}$, for $i=2,\ldots,n$. If the statistical errors are independent, there should be no pattern or structure in the lag plot and the point $\{(\hat{e}_{i-1},\hat{e}_{i})\}$ will appear to be randomly scattered. 

\begin{marginfigure}
	%\vspace{-25mm}
	\centerline{\includegraphics[width=\linewidth]{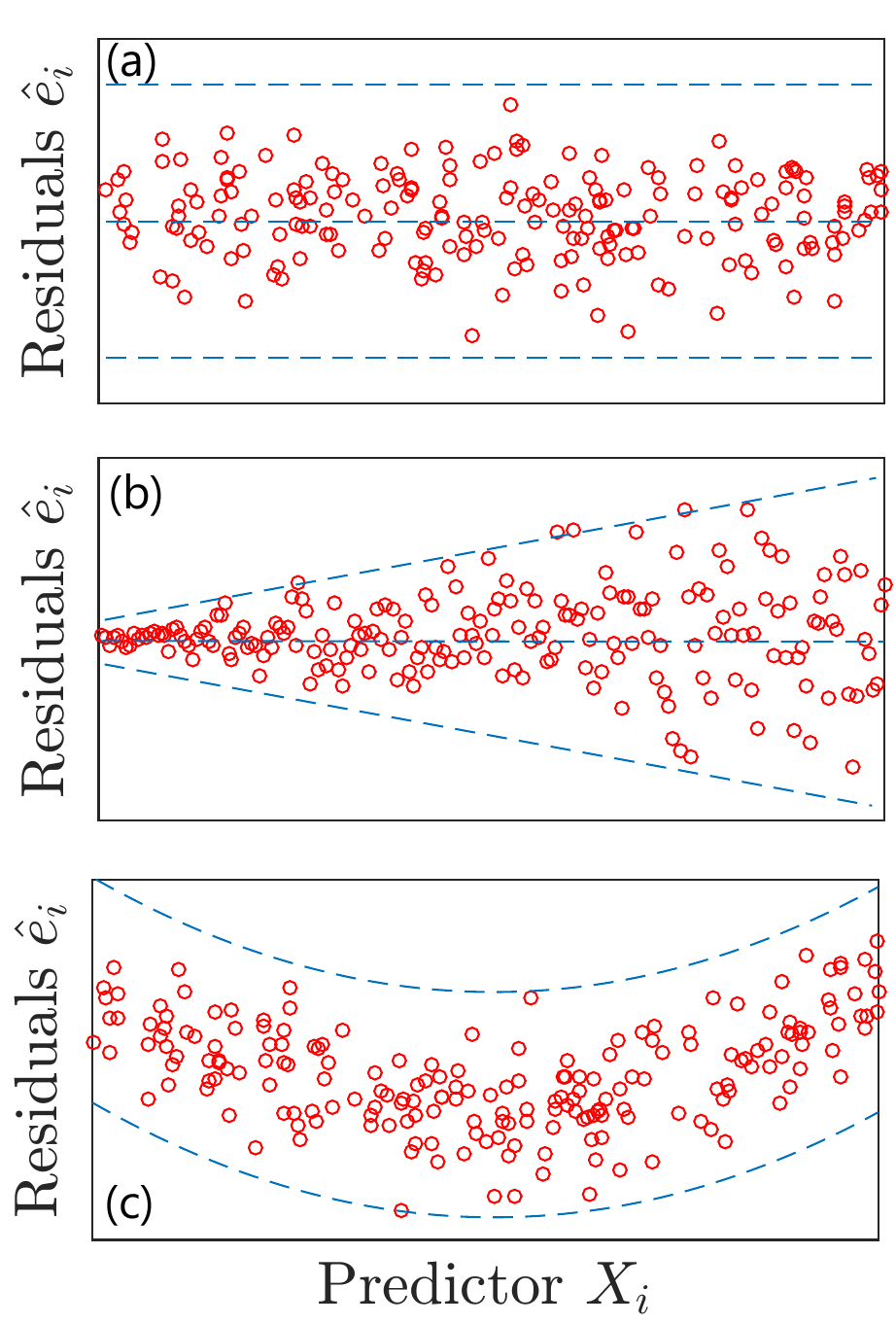}}\caption{Panel~(a): both zero mean and constant variance assumptions appear to be true; Panel~(b): homoscedasticity is violated; Panel~(c): linear model for the mean response is doubtful. }\label{fig:residuals}
\end{marginfigure}

The second and third assumptions in (\ref{eq:25SLR_Assumptions}) calls for plotting the residuals versus the predictor. A plot of $\hat{e}_i$ against $X_i$ is called a \textit{residual plot}. If all the assumptions are satisfied, then the residual plot should look like in Fig.~\ref{fig:residuals}(a), where all residuals are approximately contained in a horizontal band centered at $y$-axes. The pattern in Fig.~\ref{fig:residuals}(b) indicates that the assumption of constant variance (homoscedasticity) is violated. The presence of the curvature in Fig.~\ref{fig:residuals}(c) signals for nonlinearity: $\mathbb{E}[e_i|X_i]\neq0$, and, therefore, $\mathbb{E}[Y_i|X_i]\neq \beta_0+\beta_1X_i$. Nonlinearity can also be detected on the scatterplot of the original data, but residual plots often give a better ``resolution'' since a linear trend is removed.

Finally, to check the normality assumption (\ref{eq:25normality}), we can use a normal Q-Q plot, where we plot the ordered residual $\hat{e}_{(i)}$ against the corresponding normal quantiles\footnote{See Lecture~\ref{ch:Summarizing Data}.}. Under the normality assumption, the resulting points should lie approximately on a straight line. 

Please, keep in mind, however, that the discussed diagnostics are much better at indicating when the model assumption does not hold than when it does. For example, if you see Fig.~\ref{fig:residuals}(b), there is a  problem with the constant variance assumption, but if you see Fig.~\ref{fig:residuals}(a), it does not automatically mean that errors are homoscedastic. 

\section{Final Remarks on Simple Linear Regression}

Regression model is one of the most popular and widely used statistical models. As a result, often it is misused. Here is a list of most common mistakes.
\begin{enumerate}
	\item Regression model is often used for \textit{extrapolation}: predicting the response to the input which lies outside of the range of the values of the predictor variable used to fit the model. The danger associate with extrapolation is illustrated in Fig.~\ref{fig:extra}. The regression model is ``by construction'' an interpolation model, and should not be used for extrapolation, unless this is properly justified. 
	\begin{marginfigure}
		\vspace{-25mm}
		\centerline{\includegraphics[width=\linewidth]{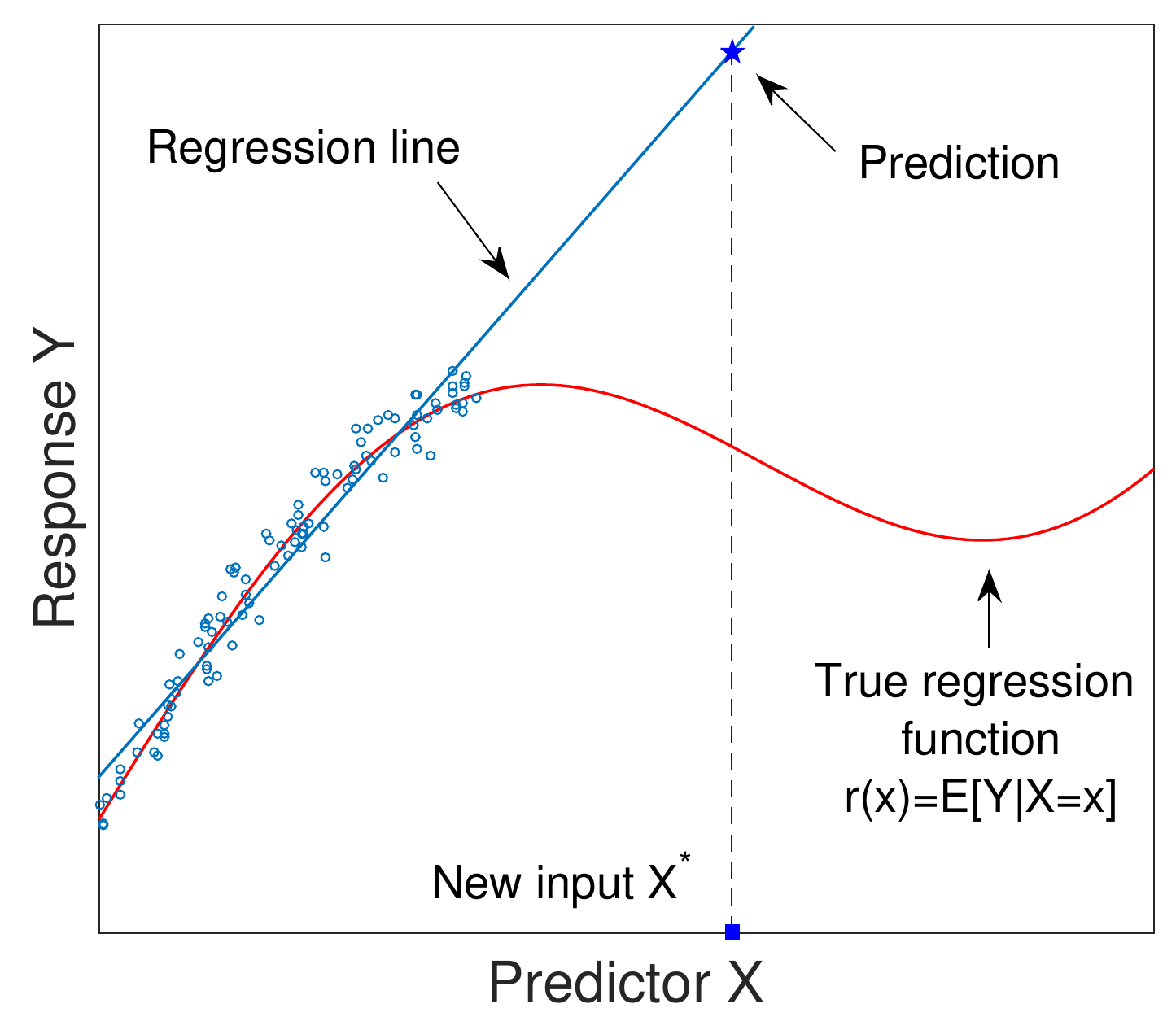}}\caption{Using the SLR model for extrapolation can lead to misleading predictions.}\label{fig:extra}
	\end{marginfigure}
	\item  \textit{Outliers} can strongly affect the OLS regression line\footnote{Recall the third examples of Anscombe's quartet in Lecture~\ref{ch:OLSproperties}.}, and yet they are often ignored and not taken care of. They can be detected either on the scatter plot of the original data or on the residual plot\footnote{The corresponding residuals are much larger in magnitude than all the others.}. If you see outliers, first, check that they are correct, \ie a true part of the system/phenomenon you study, and not a result of some measurement error. If this is indeed the case, set them for a separate study, which could be very interesting and rewarding. 
	\item Regression model studies dependence between input and response. Strong (linear or nonlinear) dependence suggests but does not imply that the variables are related in any \textit{causal sense}. For example, see Fig.~\ref{fig:pirates}: is the lack of pirates the real cause of global warming?  Correlation between the predictor and response is necessary for causation, but not sufficient. 
	\begin{marginfigure}
		\vspace{-20mm}
		\centerline{\includegraphics[width=\linewidth]{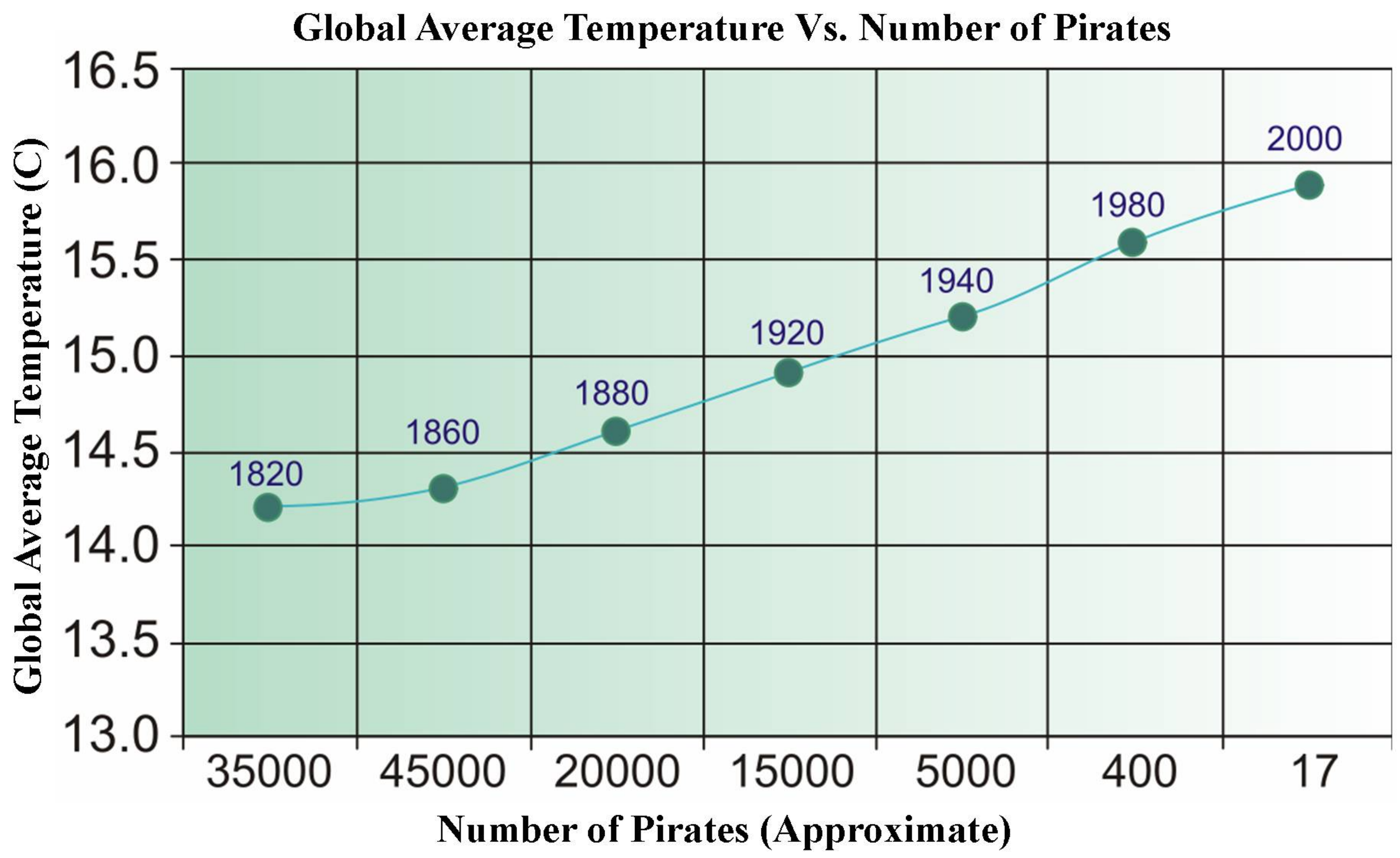}}\caption{The simple linear regression would fit almost perfectly here. Picture source: \href{https://commons.wikimedia.org/wiki/File:PiratesVsTemp(en).svg}{wikipedia.org}.}\label{fig:pirates}
	\end{marginfigure}
\end{enumerate}

\section{Further Reading}
\begin{enumerate}
	\item For other abuses of regression see G.E.P.~Box (1966) ``\href{http://www-stat.wharton.upenn.edu/~steele/Courses/Classics/BoxGeorge/BoxUseAbuse.pdf}{Use and abuse of regression},'' \textit{Technometrics}, 8(4): 625-629. 
	\item A comprehensive exposition of modern analysis of causation is given in a highly cited monograph by  J.~Pearl (2009) \textit{\href{https://books.google.com/books/about/Causality.html?id=LLkhAwAAQBAJ}{Causality: Models, Reasoning and Inference}}.
	\item Finally, if your statistical analysis does not bring your the desired results, I would recommend to use some techniques described in D.Huff (2007) \textit{\href{https://books.google.com/books/about/How_to_Lie_with_Statistics.html?id=5oSU5PepogEC&printsec=frontcover&source=kp_read_button}{How to Lie with Statistics}}. 
\end{enumerate}

\section{What is Next?} 
Many important areas of statistical inference --- Bayesian inference, causal inference, decision theory, simulation methods, to name but a few --- are not covered in these notes (but they are covered well in the texts listed in the Preface). I hope to find time in the future to extend the notes and make them more coherent. Any feedback\footnote{Emailed to kostia@caltech.edu.} on the current version would be greatly appreciated.

%%
% Start the main matter (normal chapters)
\mainmatter

%%
% The back matter contains appendices, bibliographies, indices, glossaries, etc.

\backmatter

%\bibliography{sample-handout}
%\bibliographystyle{plainnat}

\printindex

\end{document}